\documentclass[preprint,12pt]{aastex62}

\usepackage[normalem]{ulem}
\shorttitle{Orion Multiplicity}
\shortauthors{Tobin et al.}

\newcommand{\lsun}{\mbox{L$_{\sun}$}}
\newcommand{\msun}{\mbox{M$_{\sun}$}}
\newcommand{\mearth}{\mbox{M$_{\earth}$}}
\newcommand{\tbol}{\mbox{T$_{\rm bol}$}}
\newcommand{\lbol}{\mbox{L$_{\rm bol}$}}

\usepackage{ulem}

\begin{document}

\title{The VLA/ALMA Nascent Disk and Multiplicity (VANDAM) Survey of Orion Protostars V. A Characterization of Protostellar Multiplicity}
\author[0000-0002-6195-0152]{John J. Tobin}
\affiliation{National Radio Astronomy Observatory, 520 Edgemont Rd., Charlottesville, VA 22903, USA}
\author[0000-0003-1252-9916]{Stella S. R. Offner}
\affiliation{The University of Texas at Austin, 2500 Speedway, Austin, TX USA}
\author{Kaitlin M. Kratter}
\affiliation{University of Arizona, Steward Observatory, Tucson, AZ 85721}
\author{S. Thomas Megeath}
\affiliation{Department of Physics and Astronomy, University of Toledo, Toledo, OH 43560}
\author{Patrick D. Sheehan}
\affiliation{Northwestern University, Evanston, IL}
\author{Leslie W. Looney}
\affiliation{Department of Astronomy, University of Illinois, Urbana, IL 61801}
\author[0000-0001-9112-6474]{Ana Karla D{\'i}az-Rodr{\'i}guez }
\affiliation{Jodrell Bank Centre for Astrophysics \& UK ALMA Regional Centre Node, School of Physics and Astronomy, The University of Manchester, Oxford Road, Manchester, M13 9PL, UK}
\author[0000-0002-6737-5267]{Mayra Osorio}
\affiliation{Instituto de Astrof\'{\i}sica de Andaluc\'{\i}a, CSIC, Glorieta de la Astronom\'{\i}a s/n, E-18008 Granada, Spain}
\author[0000-0002-7506-5429]{Guillem Anglada}
\affiliation{Instituto de Astrof\'{\i}sica de Andaluc\'{\i}a, CSIC, Glorieta de la Astronom\'{\i}a s/n, E-18008 Granada, Spain}
\author{Sarah I. Sadavoy}
\affiliation{Department  for  Physics,  Engineering  Physics  and  Astrophysics, Queen’s University, Kingston, ON, K7L 3N6, Canada}
\author[0000-0001-9800-6248]{Elise Furlan}
\affiliation{IPAC, Mail Code 314-6, Caltech, 1200 E. California Blvd., Pasadena,CA 91125, USA}
\author{Dominique Segura-Cox}
\affiliation{Max-Planck-Institut f{\"u}r extraterrestrische Physik, Giessenbachstrasse 1, D-85748 Garching, Germany}
\author[0000-0003-3682-854X]{Nicole Karnath}
\affiliation{SOFIA Science Center, USRA, NASA Ames Research Center, Moffett Field, CA 94035, USA}
\author{Merel L. R. van 't Hoff}
\affiliation{University of Michigan, Department of Astronomy, 1085 S. University, Ann Arbor, MI 48109, USA}
\author{Ewine F. van Dishoeck}
\affiliation{Leiden Observatory, Leiden University, P.O. Box 9513, 2300-RA Leiden, The Netherlands}
\author{Zhi-Yun Li}
\affiliation{Department of Astronomy, University of Virginia, Charlottesville, VA 22903}
\author{Rajeeb Sharma}
\affiliation{Homer L. Dodge Department of Physics and Astronomy, University of Oklahoma, 440 W. Brooks Street, Norman, OK 73019, USA}
\author[0000-0003-2300-8200]{Amelia M.\ Stutz}
\affiliation{Departamento de Astronom\'{i}a, Universidad de Concepci\'{o}n,Casilla 160-C, Concepci\'{o}n, Chile}
\affiliation{Max-Planck-Institute for Astronomy, K\"{o}nigstuhl 17, 69117 Heidelberg, Germany}
\author{Łukasz Tychoniec}
\affiliation{Leiden Observatory, Leiden University, P.O. Box 9513, 2300-RA Leiden, The Netherlands}
\affiliation{European Southern Observatory, Karl-Schwarzschild-Strasse 2, 85748 Garching bei M\"unchen, Germany}

\begin{abstract}
We characterize protostellar multiplicity in the Orion molecular clouds
using ALMA 0.87~mm and VLA 9~mm continuum surveys toward 328 protostars. 
These observations are sensitive to projected spatial separations as small as $\sim$20~au, and we consider source separations 
up to 10$^4$~au as potential companions. The overall multiplicity fraction (MF) and
companion fraction (CF) for the Orion protostars are 0.30$\pm$0.03 and 
0.44$\pm$0.03, respectively, considering separations from 20 to 10$^4$~au.
The MFs and CFs are corrected for potential contamination 
by unassociated young stars using a 
probabilistic
scheme based on the surface density of young stars around each protostar.
The companion separation distribution 
as a whole is double peaked and inconsistent with the separation distribution of solar-type field stars, while the separation distribution of Flat Spectrum protostars is consistent solar-type field stars.
The multiplicity statistics and companion separation distributions
of the Perseus star-forming region are consistent with those of Orion.
Based on the observed peaks in the Class 0 separations at $\sim$100~au and $\sim$10$^3$~au,
we argue that multiples with separations
$<$500~au are likely produced by both disk fragmentation 
and turbulent fragmentation with migration,
and those at $\ga$10$^3$~au result primarily from turbulent fragmentation.
We also find that MFs/CFs may
rise from Class 0 to Flat Spectrum protostars 
between 100 and 10$^3$~au in regions of high YSO density. This finding
may be evidence for migration of companions from $>$10$^3$~au to $<$10$^3$~au, and 
that
some companions between 10$^3$ and 10$^4$~au must be (or become) unbound.

\end{abstract}

\section{Introduction}
Main sequence stars are frequently found in binary or higher order multiple 
systems with a strong correlation between multiplicity and stellar mass. The highest
mass stars are nearly always part of a multiple system, about 50\% of solar mass stars
are in a multiple system, and $\sim$25 - 30\% of lower-mass stars are part of a
multiple system \citep{Moe:2017,raghavan2010,sana2011,duchene2013,wardduong2015}. These multiple systems
are found at a range of separations from sub-au scales out to 10$^4$~au and beyond.
Solar-type companion stars have a mean separation of $\sim$50~au, while M class companions have a 
mean separation of $\sim$20~au \citep{raghavan2010,winters2019}.

The high frequency of multiplicity among all stellar spectral types and the typically close
separations of companion stars suggest that
the origin of
multiplicity 
is a direct result of
the physical conditions of star formation. Indeed, multiplicity
studies of pre-main-sequence (Class II and Class III sources) stars have frequently shown multiplicity fractions 
comparable to or in excess of main-sequence stars \citep{reipurth2007,kraus2011,Moe:2017}. 
Therefore, the multiplicity of main-sequence stars is primarily
established early in stellar evolution.
However, the populations of pre-main-sequence stars that have been examined 
are nearly fully formed stars
and may not reflect the actual multiplicity at the time of formation. 

Studies of Class I and more evolved Flat Spectrum protostars \citep{lada1987}
further extended the multiplicity characterization of 
young stars to earlier ages using infrared observations toward nearby star-forming regions
\citep{duchene2004, duchene2007,connelley2008,kounkel2016}. These studies found that 
protostars exhibit an equal or higher multiplicity fraction than pre-main-sequence
populations and solar-type field stars. Thus, these statistics provide
evidence that {\it most} stars form within multiple systems and that
the overall multiplicity (both in frequency and separation distribution)
evolves with protostellar evolution. 
While important in establishing that multiplicity properties
evolve as populations of young stars evolve, these studies still excluded
the youngest protostars, Class 0 systems \citep{andre1993}, 
where the dense infalling envelope of gas and dust limits the utility of
near infrared observations in most cases.

The youngest (Class 0) protostars are crucial to the study of multiplicity. This is because
the fragmentation mechanisms expected to produce multiple systems
are likely to be the most active during this phase when the
largest gas reservoir is available. The two most favored
mechanisms for the formation of multiple systems are disk fragmentation by gravitational
instability \citep{adams1989,kratter2010} and turbulent fragmentation within
protostellar cores \citep{padoan2002,Fisher:2004, offner2010}. Disk fragmentation 
preferentially operates on scales of $\sim$100~au and will initially produce
close multiples, while turbulent fragmentation initially produces
multiples separated by $\ga$500~au, which can migrate to closer separations
or become unbound \citep{offner2010,lee2019}. Thermal fragmentation and rotational fragmentation
of protostellar envelopes are also possible \citep{machida2008}, but are less likely based on cloud properties \citep{tohline2002,bate2012}.

Interferometry at submillimeter
to centimeter wavelengths is hence required to examine multiplicity during
the earliest stage of protostellar evolution where shorter wavelengths are
highly obscured and longer wavelength imaging has low angular resolution.
Several studies of multiplicity at submillimeter to centimeter wavelengths
have been conducted toward Class 0 and Class I protostars \citep[e.g.,][]{grossman1987,looney2000,
reipurth2002,maury2010, chen2013}.
While interferometry is capable of high angular resolution at 
long wavelengths, studies in the submillimeter and millimeter 
 observe dust emission surrounding the protostars, likely in 
the form of a disk, but at $\lambda$~$\sim$1~cm
the emission is a blend of free-free and dust emission. The
dust emission drops off rapidly with increasing wavelength, and the emission 
is dominated by free-free at wavelengths $>$2~cm. Thus, multiplicity
toward the youngest protostars is studied by detecting emission that is expected to be
associated with a protostar (e.g., dusty disks or 
compact free-free emission) rather than detecting direct stellar emission. 
A limiting factor of these pioneering studies was their
sensitivity, which resulted in small sample sizes. Moreover,
they all had spatial resolution limitations that prevented multiplicity searches
at separations 
less
than $\sim$100~au in most cases. Thus, these studies had
neither the statistics, resolution, nor sensitivity to examine multiplicity from 
$>$10$^3$~au down to $<$100~au, and they were unable to probe separations comparable
to the peak of the field separation distribution at $\sim$50~au \citep{raghavan2010}
and
the majority of the parameter space where disk fragmentation might operate.

The advent of the NSF's Karl G. Jansky Very Large Array (VLA) and the
Atacama Large Millimeter/submillimeter Array (ALMA) have changed this landscape dramatically.
The factor of $\sim$10 increases in sensitivity to continuum emission and routine
observations at high angular resolution enables all the practical limitations of
earlier multiplicity studies to be overcome. The first VLA/ALMA Nascent Disk And
Multiplicity (VANDAM) Survey was conducted toward the Perseus molecular cloud 
\citep{tobin2015,tobin2016a,seguracox2018,tobin2018}, 
characterizing the multiplicity toward all 80 known Class 0 and Class I protostars 
in the region at a wavelength of 9~mm and spatial resolution of $\sim$20~au (0\farcs07).
The VANDAM survey resolved multiples as close as 24~au, finding multiplicity and
companion fractions for Class 0 protostars of 0.57$\pm$0.09 and 1.2$\pm$0.2, 
respectively. 

While the VANDAM survey surpassed all previous studies of protostellar multiplicity, the
number of protostars and number of multiples was still low compared to the samples
achieved for main-sequence stars. Thus, the second VANDAM survey (VANDAM: Orion) was carried out
toward protostars in the Orion A and B molecular clouds 
\citep{tobin2019b,tobin2020a} using the
sample from the \textit{Herschel} Orion Protostar Survey \citep[HOPS,][]{fischer2010,stutz2013,
furlan2016}. 
VANDAM: Orion observed 328 protostars with ALMA at 0.87~mm and 0\farcs1 
($\sim$40~au) resolution, and 148 (mostly Class 0 protostars, a subset of the 328) 
were observed with the VLA at 9~mm and
0\farcs08 ($\sim$32~au) resolution (104 pointings). With these large (and nearly complete) 
samples of protostars, we are able to compare 
the multiplicity statistics between Perseus and Orion to determine if these
regions have similar multiplicity. Moreover, we
analyze the combined statistics to gain a refined perspective of multiplicity in the protostellar
phase with the largest sample currently available. We present the best possible statistics available to date and compare with predictions of multiple formation from theory and simulations.

This paper is organized as follows: the observations and data analysis are described in Section 2,
the multiplicity results are presented in Sections 3 and 4, we discuss our results in Section 5, and we
present our conclusions in Section 6.

\section{Observations and Data Analysis}

In this section, we describe the observations, sample selection, and data analysis methodologies
employed to characterize protostellar multiplicity. 
See also \citet{tobin2016a} and \citet{tobin2020a} (hereafter Paper I) for more detail. Readers chiefly interested in the multiplicity results may skip ahead to Section 3; however the details of how we arrive at these results depend upon some novel data analysis methods that are described starting at Section 2.4 and 
Appendix A.

\subsection{ALMA and VLA Observations}

The ALMA observations were conducted between 2016 and 2017 at 0.87~mm toward 328 protostars
in Orion and have a typical angular resolution of 0\farcs1.
The VLA observations were also conducted between 2016 and 2017 at 9.1~mm toward
148 systems with a typical angular resolution of 0\farcs08. The observations and data reduction were described in Paper I, and we
do not discuss these details further. The same data presented in Paper I
are used for the analyses in this paper. The reduced data are available
from the Harvard Dataverse\footnote{https://dataverse.harvard.edu/dataverse/VANDAMOrion}.

In addition to Orion, we also present
further analysis of the observations toward protostars in the Perseus 
molecular cloud. The details of these observations and data reduction were 
presented in \citet{tobin2016a}. The Perseus observations also had a typical angular
resolution of 0\farcs08, and the reduced data are available from the Harvard 
Dataverse\footnote{https://dataverse.harvard.edu/dataverse/VANDAM}.

\subsection{The Orion Sample}

The sample of Orion protostars is drawn from the 
HOPS Survey \citep[][]{fischer2010,stutz2013,furlan2016}.
The sample observed with ALMA is comprised of 94 Class 0 protostars, 
128 Class I protostars,
and 103 more evolved Flat Spectrum sources. These are a subset of the total sample of 
409 HOPS protostar candidates, since we required that they had reliable measurements
of bolometric temperature (\tbol) and bolometric luminosity (\lbol),
70~\micron\ detections,
and not be flagged as extragalactic contaminants. This sample also included a few
protostars that were not included in the HOPS sample but reside within the Orion
molecular clouds (HH270VLA1, HH270mms1, HH270mms2, HH212mms, and HH111mms). 
We also included
3 protostellar candidates from \citet{stutz2013} that are presumed to be 
Class 0 protostars (021010, 006006, 038002), but were also not detected 
at 0.87~mm. Within the sample, \lbol\
ranges from 0.1~\lsun\ to $\sim$1400~\lsun. There is also a known distance 
gradient across the Orion A and B molecular clouds \citep{kounkel2017,kounkel2018}.
The distance toward each system was estimated in Paper I using 
\textit{Gaia} data toward known young stellar objects in Orion. The distance
variation is within $\sim$10\% of the nominal 400~pc distance
toward the region. The input catalog for Orion is provided in Table \ref{input-table-orion}
listing the positions of all components, \lbol, \tbol, classes, distance, and surface
density of surrounding YSOs (see Appendix A).

Each HOPS protostar that was individually identified and classified by
\textit{Spitzer} and \textit{Herschel} was observed in an individual pointing
with ALMA, and each Class 0 protostar had an individual pointing with the
VLA. The Class I protostars that were detected by the VLA were those
that happened to fall within the VLA primary beam at 9.1~mm.

Not considering multiplicity, we detect continuum emission (ALMA and/or VLA) 
toward 86 Class 0 protostars, 111 Class I protostars, 
and 
92
Flat Spectrum protostars, constituting 
289
systems in total. 
Note that these numbers are slightly different from those 
in Paper I, reflecting a more refined
accounting of detections and non-detections associated with targeted systems.
An additional 18 continuum sources are detected but not associated with a \textit{Spitzer}
or \textit{Herschel} classified protostar. The lack of a counterpart in the near to far-infrared
may occur due to confusion with nebulosity, crowded sources, and/or saturation of the
\textit{Spitzer} detectors.
The total number of unique continuum sources detected by
the VLA and ALMA that are not associated with a known extra-galactic source is 432.
Of these, 395 are associated with a HOPS protostar and 142 are associated
with Class 0 protostars, 132 with Class I, and 121 with Flat Spectrum.

\subsection{The Perseus Sample}
The sample of protostars in the Perseus molecular cloud was selected from
\citet{enoch2009}, but we also included additional protostars that were
identified from the millimeter continuum \citep[e.g.,][]{looney2000} 
and \textit{Herschel} observations of the region \citep{sadavoy2014}. 
The protostars in the sample range in luminosity from $\sim$0.1 to 
$\sim$120~\lsun; the luminosity of the highest luminosity protostar, 
SVS13A (Per-emb-44) has some considerable uncertainty, however. Further details
of the sample selection are provided in \citet{tobin2016a}. Not considering multiplicity, we detect continuum emission 
toward 80
systems in total: 41 Class 0 protostars, 29 Class I and Flat Spectrum
protostars, and 10 Class II systems. We also detect continuum emission toward 2 additional unclassified
systems that may be YSOs.
The input catalog for Perseus is provided in Table \ref{input-table-perseus}
listing the positions of all components, \lbol, \tbol, classes, and surface
density of surrounding YSOs (see Appendix A).
The total number of unique continuum sources detected by
the VLA that are thought to be associated with YSOs is 106,
but only 104 are associated with
classified protostars. Of these 104 sources, 55 are associated with Class 0 protostars, 37 with Class I
or Flat Spectrum protostars, and 12 are associated with Class II YSOs.

\subsection{Data Analysis}
To characterize the protostellar multiplicity in Orion and Perseus, we first constructed 
a catalog of positions that were derived from Gaussian fits to the 
detected sources. We fit elliptical Gaussians
using the \textit{imfit} task of CASA 4.7.2, measuring position, flux density,
and source size. We use a merged catalog from the ALMA and VLA observations to ensure 
that we include sources from both sets of observations in the event that a source was 
detected with the VLA and not with ALMA (and vice versa). For the protostars in Perseus, 
we make use of the previously generated catalog from \citet{tobin2016a}, which used similar
methods to derive protostar positions with an earlier version of CASA.

Catalogs of protostars have been compiled for Orion and Perseus from
previous near-, mid-, and far-infrared imaging surveys using \textit{Spitzer} and \textit{Herschel}.
These surveys had best angular resolutions of $\sim$1\arcsec\ for \textit{Spitzer} 
and $\sim$5\arcsec\ for \textit{Herschel}. Thus, our much higher resolution 
observations from ALMA and the VLA can often resolve what appeared to be
single systems at lower resolution into multiple components that are assigned to the
same protostellar system and inherit the luminosity and classification of the
source from the infrared catalog. Thus, independent classifications for systems
is generally only possible for systems with separations $\ga$2000~au where they could
be resolved in the mid and far-infrared.
Moreover, for protostar systems it is difficult to denote a particular component as
the primary member. Generally, the protostar masses of each component are unknown, and the
main observable, dust emission, does not strictly relate to luminosity or protostar mass. Furthermore, 
\lbol\ is demonstrated to not directly relate to protostar mass. This is because most luminosity is likely 
produced from accretion processes and not from the protostar itself \citep{dunham2014,fischer2017}.
We also do not know whether a collection of continuum
sources constitutes a bound system or not. Thus, regardless of the inherent limitations, we
use spatial association alone to assign multiplicity. With only knowledge of projected
spatial association, there is the possibility of detected sources being counted as companions
when they are only line of sight projections. We will describe in a following section
how we select and count multiple systems, as well as account for possible contamination.

We also emphasize that we analyze the instantaneous projected separations of sources 
detected in our data, 
and we cannot infer the
true orbital semi-major axes because we do not know the orbital
plane or orbital phase for any companions observed. 
With knowledge of the underlying eccentricity distribution, corrections can be made
to an ensemble distribution of projected separations \citep[e.g.,][]{kuiper1935,brandeker2006}
to more closely reflect the distribution of semi-major axes. But given that
we are examining forming systems with an unknown eccentricity distribution, we
limit our analysis to projected separations as observed.

We limit our identification of multiple systems to separations as large as $10^4$~au, beyond
which the likelihood of associating a random YSO projected along the line of
sight becomes large (see Appendix). Also, the radius of the field of 
view for our observations is limited to $\sim$16000~au
for the VLA and $\sim$4000~au for ALMA toward Orion protostars, whereas the field
of view for Perseus is $\sim$12000~au for the VLA only. Thus, our
sample is going to be incomplete beyond separations of
$\sim10^4$~au. The 4000~au primary beam of the ALMA observations will not severely affect our
detection limits. This is because systems with $>$4000~au separation are resolved by the 
\textit{Spitzer Space Telescope} at 24~\micron\ and the \textit{Herschel Space Observatory} at 70~\micron.
They are then independently classified on the basis of their SEDs and therefore would be assigned a single ALMA pointing if they
are Class 0, I, or Flat Spectrum.
An additional motivation for our choice of $10^4$~au ($\sim$0.05~pc) as 
the maximum separation is because this is the typical radius of dense cores in which
protostars reside \citep{bm1989,bergin2007,lane2016,kirk2017}.

\subsubsection{Associating Multiple Systems}

To assemble the multiplicity statistics toward the Orion protostars, 
we utilize an iterative inside-out search approach. 
We search the ALMA+VLA catalog for the nearest neighbor to each continuum
source associated with a protostar targeted in the survey. 
Starting with 15~au as the smallest
 search radius, which is less than the most compact multiple in the sample such that we should not find companions, 
 we search for companions with increasing separation 
in 100 logarithmically spaced radial bins, ending with a maximum separation of $10^4$~au for association of two continuum sources.
When two sources
are associated, they are grouped as a multiple system, and we remove their
individual catalog entries and replace them with a single entry at their
average position, corresponding to the geometric midpoint between the components without
any weighting. These multiple systems can
then be further associated with other individual protostars or multiple systems, and the
removal of the individual sources from the catalog prevents the association of one component
to more than one multiple system. 
We do not limit the number of
possible associations (there is no upper limit on the order of multiple systems), other than the maximum separation of 10$^4$~au.

A simple example of our method is illustrated in Figure \ref{example}, starting
with a group of four protostars: A, B, C, and D. The algorithm
starts by searching for companions to A with separations $<$ d$_{lim}$ (where d$_{lim}$ is a number $>$ d$_{A,B}$ and $<$ d$_{A,C}$) and finding the nearest to be B with a separation
of d$_{A,B}$. The catalog entries for A and B are both removed, and an entry for AB is inserted
with the average position of A and B (AB$_{avg}$). Then the search for companions continues looping over all sources with separations $<$ d$_{lim}$.

The algorithm incrementally increases d$_{lim}$ after each loop in order
to search for progressively wider companions. Next, C will be identified as a companion to D with separation
d$_{C,D}$ once d$_{lim}$ $>$ d$_{C,D}$. Then, as before, C and D are removed from the search catalog and an entry
for CD is inserted with the average position of C and D (CD$_{avg}$). Then, as d$_{lim}$ is further increased, AB is found to be associated with CD with a separation d$_{AB,CD}$. 
The catalog positions for AB
and CD are then removed and the average position for ABCD is inserted (ABCD$_{avg}$). Thus, A, B, C, and D are 
together considered a single multiple system, despite D being greater than $10^4$~au 
from A or B. The separations that
are then included in our distribution of separations plots (histograms or cumulative distributions)
are d$_{A,B}$, d$_{C,D}$, and d$_{AB,CD}$. This hierarchical approach is similar to the one adopted
in the analysis of results from numerical simulations \citep{bate2012,lee2019}. While numerical simulations utilize known quantities like the center of mass and total binding energy, these are inaccessible in our observations. In particular, the lack of direct correlation between mm/cm flux and protostellar mass precludes the use of the ``center of light" as a center of mass proxy.

Our results would be different if we applied a simpler approach by assigning
A as the primary based on its luminosity or flux density of dust emission, as had been done in \citet{tobin2016a}.
We would then associate sources based on relative distance to A, and
we would have pairings of AB and AC, making this a triple system. The 
separation distributions would only have d$_{A,B}$, d$_{A,C}$.
D would not be included in this multiple system, because it is
greater than $10^4$~au from A, even though its distance from C is less than $d_{A,B}$ and $d_{B,C}$.

While neither of these methods distinguish bound pairs from chance 
alignments, our approach is
less prone to bias and individual judgment when assigning the multiplicity
status of particular systems. Moreover, it has greater reproducability, independent of millimeter flux density, and can easily be
checked against simulation data as well. 
We apply this method to both the Orion catalog and the Perseus catalog,
yielding a consistently derived distribution of separations for both datasets.

In light of the inherent limitations in only being able to characterize the multiplicity of protostars
using their projected separations, we make efforts to assess the probability of associated systems to be true
multiples using measurements of the surface density of surrounding YSOs that could yield false positives. 
We describe our analysis methods using probabilities for companion 
association in Appendix A.

\subsubsection{Multiplicity Statistics}
We calculate the multiplicity fraction and companion 
fractions for Orion, Perseus, and their combination
as metrics of multiple star formation in these regions.
The multiplicity fraction or frequency (MF) is the fraction of systems 
that are multiples (binary, triple, etc.) and is defined
as 
\begin{equation}
MF~=~\frac{B+T+Q+...}{S+B+T+Q+...}.
\end{equation}
The number of single systems is $S$, binaries is $B$, triples is $T$, quadruples is $Q$, etc. 
Then the companion fraction (CF), which provides the average number of
companions per system, is defined similarly as
\begin{equation}
CF~=~\frac{B+2T+3Q+...}{S+B+T+Q+...}.
\end{equation}
The uncertainties 
on the MF and CF are calculated using binomial 
statistics, specifically using the Wilson score interval \citep{wilson1927} 
\begin{equation}
\sigma_{MF}= \frac{1}{1+\frac{z^2}{N_{sys}}}\left (MF +\frac{z^2}{2N_{sys}}     \right )  \pm \frac{z}{1+\frac{z^2}{N_{sys}}}\sqrt{\frac{MF(1.0-MF)}{N_{sys}}+ \frac{z^2}{4N_{sys}^2}}
\end{equation}
where $N_{sys}$ is  the total number of systems ($S+B+T+Q+...$) (see Equations 1 and 2), 
and we adopt
$z$=1.0 such that the calculated uncertainties are 1$\sigma$. The uncertainty in the CF is calculated in the same manner, by substituting CF for the MF in Equation 3.
Note that if the CF $>$ 1.0 the number under the square root can be negative and even when the
CF approaches 1, the uncertainties can be inaccurate. Thus,
instead use Poisson statistics for calculating $\sigma_{CF}$ when CF $>$ 0.5.

We make use of companion probabilities (to correct for contamination
by unassociated YSOs), which are computed as described in Appendix A to determine
the order of a multiple system (binary, triple, etc.). For the example used in Figure \ref{example},
the 
system will have a probability associated with each member of the system 
\begin{equation}
[P_A,P_B,P_C,P_D] = [1.0,P_{A,B},1.0 \times P_{AB,CD},P_{C,D} \times P_{AB,CD}].
\end{equation}
Since A,B and C,D are initially associated as binaries, one component 
in both A,B and C,D are assigned
a probability of 1.0, this
is the primary (designated arbitrarily), and the other component has a probability of $P_{A,B}$ 
and $P_{C,D}$ (see Appendix A
for computation of individual probabilities). 
Then when the two binaries are associated with each other, one binary system
is assigned a probability of 1.0 (designated the primary) and the other binary has
a probability of $P_{AB,CD}$. Thus, when considering the system as a whole, the
component D will end up having the lowest probability because its overall probability
is $P_{C,D}$$\times$$P_{AB,CD}$. This process is continued whether singles are added to binaries,
binaries to triples, quadruples to binaries, etc. 

Then, in the final tabulation of multiplicity statistics, that is whether a system
is considered a binary, triple, etc., the multiplicity order is determined by the 
rounded sum of companion probabilities for all possible companions that 
comprise the system, 
i.e.,
\begin{equation}
N_{components} = \sum_{i}^{N} [P_A,P_B,P_C,P_D].
\end{equation}
If the sum of the companion probabilities is less than the total number
of companions included for a system, we check to see if the majority of the difference comes from
the addition of another multiple system to form a larger higher order system where
many components have a low probability.
If so, we split the two previously associated systems for the purposes of multiplicity
statistics and consider them as two (or more) lower-order systems, rather than a single
higher-order system. 
If only a single
component of a higher order system has low probability, then we split off that single component and
count it as single. The MF and CFs for the samples as a whole
are then constructed using Equations 1 and 2.

There are alternative methods to calculate the MF and CF
including the probabilities, and we describe one such method in Appendix A.4 that we used
as a sanity check. We prefer to use our method of rounding per system because it produces results that 
are more directly comparable to previous work, and the calculated MF and CFs are consistent between our
main method as described here and the alternative methods that are described in Appendix A.

\subsubsection{Comparing Multiplicity Properties}
The main quantities for comparing the multiplicity properties 
of different regions, classes, and samples
 are the calculated MFs, CFs, and separation distributions.
The comparison of separation distributions examines the relative shapes of the distributions.
The MFs and CFs on the other hand examine the total number of multiples
in a a given population. However, some spatial dependence of the MFs and CFs can be examined
by selecting on different ranges of separations. It is important to point this out
because the comparison of separation distributions is conducted via cumulative 
distribution functions (CDFs; Appendix A.2) and it is wholly independent of the 
MF and CF for a given
population. Thus, one can have a MF and CF that is consistent between samples, while the
separation distribution CDFs are inconsistent, and the converse can also be true as well. 
The independence of these quantities is important to keep in mind for Sections 4
and 5 where we make many comparisons of different samples and sub-samples.

\section{Observations of Multiple Protostars in Orion}
We provide an overview of the observations that detect the multiple 
protostar systems and a discussion of some specific protostars in the following subsections.
We also highlight regions 
where the VLA and ALMA yield different results and compare
the ALMA/VLA multiplicity detections to near-infrared detections.

\subsection{Overview of Multiplicity Detections}
The ALMA and VLA observations have enabled us to 
identify multiple protostar systems to separations as small
as $\sim$22~au toward protostars in Orion at a distance of 
$\sim$400~pc \citep{kounkel2017}, but only two systems are
detected at separations $<$ 40~au. For the Perseus observations from \cite{tobin2016a}, the revised distance of
$\sim$300~pc \citep{ortiz-leon2018} toward this region
places the smallest separations at $\sim$24~au.

We list the Orion multiple systems in Table \ref{orion-seps}, and we list the 
re-analyzed Perseus multiple systems in Table \ref{perseus-seps}. When multiple systems with compact separations are
paired with additional single sources or another multiple system, the separation
listed in Tables \ref{orion-seps} and \ref{perseus-seps} refers to the separation between the average position of 
one multiple system to either the position of the newly added source or other 
multiple system (see Section 2.4). Much of the new parameter space
we explore in Orion is at separations $<$2000~au, because scales larger
than this were studied by \textit{Spitzer} and \textit{Herschel}.
Nonetheless, some
new
wide multiples are found at larger scales toward 
regions with bright nebulosity that limited the sensitivity of IR
observations.

We have discovered a total of 
85
multiple systems (195 non-multiple) 
in the Orion Molecular clouds that have maximum separations less than
$10^4$~au, 58 multiples
with maximum separations less than
$10^3$~au, and 47 multiples
with maximum separations less than 500~au;
these numbers include the consideration of companion probabilities. Also,
the number of multiples specified for larger maximum separations also include those
at smaller separations.
These numbers reflect the multiplicity of systems that are classified as protostars and do not include
systems that are not classified as protostars by the HOPS project or other work.
Also, some systems that were regarded as separate HOPS sources
are now considered a single multiple system. Thus, the sum of the multiple
and single systems will not equal the number of detected systems.
These numbers also do not include widely separated sources that are 
spatially resolved in infrared observations but are not classified
as Class 0, I, or Flat Spectrum. Many of the Class I and Flat Spectrum systems
also had their multiplicity characterized on scales between 100 to 10$^3$~au by \citep{kounkel2016}.
The bulk of the new discoveries are toward Class 0 protostars at separations $\la$~4000~au and for Class I
and Flat Spectrum protostars at separations $\la$100~au.
We will discuss the ALMA/VLA results in the context of the HST observations in Section 3.6.

Images of each system are not included here, because they were presented 
in Paper I and \citet{tobin2016a}; 
instead we focus on interpreting the multiple system detections. 
We only show selected systems, some of which use different image
parameters as compared to Paper I to provide increased angular resolution
and better highlight the multiples.

\subsection{Multiple Systems with $<$ 500 au Separation}

The Orion A and B molecular clouds provide the largest available sample of protostellar multiples with 
separations less than 500~au. We show example images for some close multiple systems
with separations less than 500~au in Figure \ref{continuum_multiple}.
At scales less than 100~au, most systems represent new discoveries.

We detect 19 Class 0 multiple systems with separations less than 500~au (17 binary, 2 triple);
10 Class I binary systems, 18 Flat Spectrum binary systems, and 3 additional 
binary systems that are not classified. Table \ref{orion-seps} lists all of these multiple systems. One 
of the unclassified systems resides
in OMC1N \citep{teixeira2016} and is likely a protostar, but the other two are found toward
systems with near-infrared counterparts detected by the 2MASS survey \citep{skrutskie2006} and
are likely more evolved YSOs.

\subsection{Close Multiples Not Detected By Both ALMA and VLA}
The vast majority of multiple systems discovered in Orion were resolved and
detected independently by ALMA and the VLA. However, there are a few examples where
close multiplicity was detected only in the VLA observations or only in the ALMA observations.
We also discuss  examples where there is tentative evidence for a companion, 
but the evidence is not strong enough to merit inclusion in the sample of multiples.

\textbf{HOPS-361-C (also known as NGC 2071 IRS3): } 
Toward this 
protostar
we detect two sources separated by $\sim$46.5~au (0\farcs108) 
only in the VLA 9~mm observations (Figure \ref{continuum_multiple_vla}); 
the eastern source has a jet elongated 
nearly orthogonal to the position angle between the two components. The ALMA observations, 
on the other
hand, detect only a large disk surrounding the
two protostars, with some surface brightness variation. 
Extended emission from this disk is only visible at low S/N in the VLA data.
It is unclear why the two protostars do not stand out within this disk, but the circumbinary disk
could be optically thick at 0.87~mm, or the surface brightness of the circumbinary disk could be
comparable to the intensity from the circumstellar disks around each protostar. 

\textbf{HOPS-361-E: } This 
protostar
also falls within the HOPS-361 region 
($\sim$1140~au from HOPS-361-A) and
only has its multiplicity detected by the VLA (Figure \ref{continuum_multiple_vla}). This
source was recognized as a multiple in further analysis of the NGC 2071 IR 
region by Cheng et al. (in prep.). Its separation of 0\farcs055 ($\sim$22~au) 
is too small to be resolved in our 
ALMA observations, and only the higher resolution afforded by the VLA observations enabled
the system to be resolved. Thus, the compactness of this system resulted in it not 
being reported as a multiple in Paper I.

\textbf{HOPS-288: } We find another example
of compact multiplicity
toward this 
protostar,
shown in Figure
\ref{continuum_multiple_vla}, where two continuum sources are clearly 
detected by both ALMA and the VLA and are separated by $\sim$220~au (0\farcs542). The brighter, western
source is found to be extended in both ALMA and VLA images, but VLA 
imaging with higher resolution using a robust parameter of 0 (Briggs weighting) 
rather than Natural Weighting used in Paper I
reveals that there are two point sources within the brighter
source, separated by $\sim$54~au (0\farcs133). Imaging the ALMA data 
with lower values of the \textit{robust} parameter to
achieve higher resolution does not reveal the close companion source at 0.87~mm; this companion
may also be obscured by optically thick continuum emission. This close companion
is positioned along the expected axis of the disk of HOPS-288, orthogonal to the known molecular outflow
from this protostar \citep[][]{stanke2000,vankempen2016}.

\textbf{HOPS-384: }
This is also a close multiple system that was only detected by the VLA 
(Figure \ref{continuum_multiple_vla}). Similar to 
HOPS-288 and HOPS-361-C, HOPS-384 is also one of the highest luminosity HOPS protostars, 
as high as $\sim$1400~\lsun. The close companions are separated by 0\farcs11 in 
the VLA 9~mm image. The ALMA image at the highest resolution with superuniform weighting
has a protuberance in the direction of the companion and detection may simply require higher resolution.
There is a third companion to HOPS-384 located northwest by 3\farcs39 (not shown in Figure 3), but this source is not detected
by the VLA and appears as a near edge-on disk in the ALMA 0.87~mm image \citep{tobin2020a}.

The next two sources are each in their own category. The first has
companions detected by ALMA that were not subsequently detected by
the VLA, and the second was not recognized as a separate
protostellar system in Paper I, but is both a discrete protostellar
system and a binary system.

\textbf{HOPS-56} has companions detected by ALMA but not by the VLA. It is detected
as a close triple system in the ALMA image (Figure \ref{continuum_non_multiple}), both companions having 
separations of $\sim$0\farcs22. However, the VLA image toward HOPS-56 does not detect all three
components, failing to detect HOPS-56-A-B, likely due to poorer dust mass sensitivity
in the VLA data as compared to ALMA. Given the high-confidence detection
of all three sources with ALMA, we consider all three of these sources as companions.

\textbf{NGC 2024 FIR-3} was added to the sample in subsequent analysis of the data since 
the publication of Paper I. This source was detected in the VLA data toward HOPS-384
presented in Paper I, but it was not identified as a multiple system there because its nature was 
uncertain, and there were no ALMA data covering that region in our survey. An ALMA survey of the 
region by \citet{vanterwisga2020} associated the continuum at 1.3~mm emission with NGC 2024 FIR3, which was classified as a
Class 0 system by \citet{ren2016} with a bolometric
luminosity of 220~\lsun. Given the association with a bonafide protostar system, we include this detection in our
multiplicity statistics with a separation of 1\farcs46 ($\sim$586~au) and show the ALMA 1.3~mm image from \citet{vanterwisga2020} and the VLA images in 
Figure \ref{continuum_new_wide_multiple}. While we
zoom-in more closely on the source, the wider field image from \citet{vanterwisga2020} shows significant extended
structure associated with the envelope.

The following three protostars have possible companions where tentative evidence is found in the VLA data but not
the ALMA data. We do not include these companions in our multiplicity statistics given their tentative nature.

\textbf{HOPS-124: } This is a Class 0 protostar with a large disk that has asymmetric
dust emission with an apparent gap \citep{sheehan2020}. An obvious culprit that
can produce such features is a companion star that formed within the disk. Images produced
with Robust=-1 of only the highest frequency half of the VLA 9~mm dataset ($\lambda\sim$8.1~mm) 
detect a possible second source separated by $\sim$0\farcs08 (Figure \ref{continuum_possible_multiple}). 
However, this possible second
source is along the direction of the outflow and could be part of the jet since extended free-free emission
associated with the jet is seen in Ka-band toward some sources (see Figure \ref{continuum_multiple_vla}).
We are therefore hesitant to claim this as a companion since
it could also be an extension of the compact jet emission. This possible second source
is also coincident with the bright dust ring that appears in the ALMA images and the VLA image produced with 
robust=2 weighting \citep[Paper I;][]{sheehan2020}. Therefore, this second source could instead be a clump of dust in the
disk and not a true companion. Consequently, this possible source is not included in our multiplicity statistics.

\textbf{HOPS-403: } This is an extended source in both 0.87~mm and 9~mm dust continuum emission (Figure \ref{continuum_possible_multiple}). This protostar
is a member of a subset of the Class 0 protostars that are more deeply embedded than typical Class 0 protostars and may be one of the 
youngest protostars in Orion, which are known as PACS Bright Red Sources \citep[PBRS;][]{stutz2013,
tobin2015,karnath2020}. The ALMA 0.87~mm image only shows a fairly smooth surface 
brightness distribution with a $\sim$1\arcsec\ diameter,
while VLA 9~mm data reveal further sub-structure, including a possible second point source
with a separation of $\sim$0\farcs15, located west of the brighter source. 
However, its contrast with respect to the surrounding 
continuum emission is only $\sim$3$\sigma$. There is
no evidence for this second source in the ALMA 0.87~mm data due to the optical depth of
the continuum \citep{karnath2020}. Thus, we
are hesitant to definitively claim that this is a companion, and this source is not included in our multiplicity statistics.

\textbf{HH270VLA1: } Toward this Class 0 protostar, both VLA and ALMA detect two components 
separated by 0\farcs23 (Figure \ref{continuum_non_multiple}). However, there is a third component detected by the
VLA at Ka-band. This third source has a radio spectral index of $\sim$-0.4, indicating non-thermal
emission. Thus, this third source is either a shock in the outflow, similar to the extended
jets observed toward HOPS-370 \citep{tobin2019b,osorio2017} and HOPS-361-C (Figure \ref{continuum_multiple_vla}; \citep[][Cheng et al. in prep.]{carrasco2012}),  
or it is a background active galactic nucleus (AGN). The probability of a background AGN aligning so closely with a 
binary protostar is very small, $\sim$10$^{-5}$, see \citet{tobin2016a}.
For this reason, we do not consider the third source, denoted 
HH270VLA1-C in Paper I, a companion.

\subsection{Impact of Non-detections or Spurious Detections}

Given that the total sample size is $\sim$300 protostars for Orion, and there are
$\sim$100 in each protostellar class, the inclusion or exclusion of any particular source
will impact the resulting MFs and CFs by $\pm$0.01. With the typical
uncertainties of 0.03 to 0.05, we would have over or under count by $\sim$6 to 10 sources to end up with multiplicity statistics that are inconsistent by $>$1$\sigma$. 
Either way, if there
are undetected companions by either ALMA or the VLA (see Section 3.6) or the seven
possible transition disks detected by \citet{sheehan2020} are really circumbinary
disks, the statistics will be further underestimated.

\subsection{Dynamic Range of Detected Systems}

We examined the companion flux density ratios as a function of separation in Figure \ref{fluxratio}.
The flux density ratios are shown in separate panels for the ALMA 0.87~mm and VLA 9~mm. 
Tables \ref{orion-seps} and \ref{perseus-seps} list the ratios
corresponding to each system, but we only provide the ratios
for individual source pairs and not when a source is paired with an existing multiple 
or when two multiple systems are joined. We find that there
is no clear contrast limit for the detection of companions 
that is progressively lower as separation decreases.
The majority of companions tend to have flux density ratios that are within a factor of 10. 
Only a few have ratios greater than a factor of 100, and those are only seen in the ALMA data for companions
that have separations approaching 10\arcsec. Thus, it is not clear if dynamic range limits affect 
our ability to detect multiplicity in protostars with the current data. There are also additional factors to consider since
we are detecting emission from extended dusty disks rather than stellar point sources.
Source separation plays a role in defining the extent of their 
circumstellar disks \citep{lubow1994}. Then, the flux density that we observe is
determined from a combination of dust mass and disk radial 
size (or simply surface density), both of which are related to the dust continuum
opacity.

These flux density ratios have no relation to the underlying mass ratios of these systems because
there are clear examples where the brightest dust continuum source in a system is not the most 
massive \citep[e.g., L1448 IRS3B;][]{tobin2016b,reynolds2021}. The ratios
could be interpreted as a dust mass ratio of the circumstellar disks
around each component of a multiple system, with the caveat that the VLA data can have contributions
from free-free emission. However, such comparisons can be misleading if a substantial amount of the 
dust is optically thick. We avoid over-interpreting the flux ratios, but provide them as a
observational characteristic of the systems.

\subsection{Near-infrared vs. ALMA/VLA Detections}

The large sample of HOPS protostars observed with HST NICMOS and WFC3 \citep{kounkel2016,habel2021} 
enables us to compare
protostellar multiples that are detected from their direct stellar emission to those detected
via their circumstellar dust emission. This allows us to determine how much incompleteness there might be
in a particular type of observation. For the purposes of this analysis, we only consider
multiples that would have been within the range of detection by both projects, 
which limits the analysis to separations between $10^2$ and $10^3$~au ($\sim$0\farcs25 to $\sim$2\farcs5).
We emphasize that even if a companion is detected by HST and not ALMA/VLA, we do not add it back into
our current analysis, leaving our analysis based on ALMA/VLA data alone.

There were a total of 274 HOPS protostars observed by both ALMA/VLA and HST at 1.6~$\mu$m \citet{kounkel2016}. Neither ALMA/VLA nor HST detect companions toward 235
protostars, both ALMA/VLA and HST detect the same companions toward 19 protostars (2 Class 0; 6 Class I, and 11 Flat Spectrum), 
ALMA/VLA alone detects a companion toward 12 (9 Class 0 and 3 Class I) protostars, 
and HST alone detects a companion toward 8 (4 Class I, and 4 Flat Spectrum) protostars. 
We consider four of the HST-only companions as tentative given that they are very faint (HOPS-5,
HOPS-65, HOPS-86, and HOPS-281); HOPS-281 is also a tentative companion for the ALMA detection.

Between $10^2$ and $10^3$~au, we estimate a submillimeter/centimeter incompleteness of
$\sim$20\% (as given by 1-[Number of detected companions/Total companions]; Total companions
is the millimeter plus infrared companions). 
Considering only Class I and Flat Spectrum protostars, the incompleteness rises to 29\%.
The infrared-only incompleteness is 31\% if Class 0 protostars are included in the counting, but drops to
$\sim$11\% if only Class I and Flat spectrum protostars are included. Much of this incompleteness 
in the infrared is due to extinction; Class 0 protostars are rarely detected in the HST 1.6~$\mu$m data, while many
Class Is are only detected in scattered light \citep{habel2021}. The detection of companions in the HST data is
primarily toward the $\sim$30\% of the protostars that are visible as point sources \citep{kounkel2016}. Thus, both
techniques may have
comparable levels of incompleteness depending on the class of protostar observed. However, 
submillimeter/millimeter observations are clearly superior for characterizing Class 0 multiplicity,
while near-infrared observations appear superior for characterizing Class I and Flat Spectrum multiplicity
at separations $>$100~au. On the other hand, separations $<$100~au may be problematic in the near-infrared
for Class I protostars with significant envelopes due to significant scattered light confusion and 
dust opacity. Infrared observations at $<$100~au separations may be most effective for
the more evolved Flat Spectrum protostars.

A weakness of the near-infrared observations is that they also have a greater likelihood
of contamination as compared to submillimeter and millimeter observations. This is because contamination
can come from foreground or background stars, and contamination becomes extremely problematic at projected
separations $>10^3$~au \citep{kounkel2016}. Submillimeter and millimeter observations are also susceptible
to contamination, as described in Appendix A, but the requirement for them to have either dusty emission or
free-free emission significantly reduces the number of possible contaminating sources relative to near-infrared.

\section{Overall Multiplicity Characterization}

We describe the overall multiplicity results from our observations in the following
subsections. The Orion results are the main focus, but we also include details of 
Perseus where relevant and when the results are distinct from those of
\citet{tobin2016a}.

\subsection{Bolometric Luminosities and Temperatures}
We start by examining \tbol\ vs. \lbol\ for the 
single systems vs. the multiple systems as shown in Figure
\ref{lbol-tbol}. The figure shows all systems that are multiple from 20 to 10000~au. 
It is apparent that the luminosities of multiple systems, 
which implicitly includes the luminosities of all protostellar members, 
are systematically higher than the luminosities of systems that are single at our resolution limit. 
The median luminosities
for singles and multiples in Orion are 0.96 and 3.27~\lsun, respectively, and for Perseus 
they are 0.97 and 3.06~\lsun, respectively. The difference is quite obvious in Orion
and is also discernible by-eye in Perseus, despite a smaller sample size. 
We compared the \lbol\ distributions for single vs. multiples using the KS-test,
and for Orion the null hypothesis that singles and multiples are drawn from the same parent 
distribution is ruled out with a likelihood of $<$ 0.01; the null hypothesis cannot 
be ruled-out at the same likelihood level for Perseus, where we find a likelihood of 0.025. For both
Perseus and Orion, the \tbol\ distributions are consistent with having been drawn from the same
sample. We further note that 
the median luminosity differences and
inconsistency of the luminosity distributions persist for 
different separation ranges (20 to 500~au and 20 to 10$^3$~au, in addition
to the shown range of 20 to 10$^4$~au).

The skew toward higher luminosities for the multiples could be related to the fact
that multiple systems occur more frequently for higher-mass systems \citep[e.g.,][]{duchene2013}.
However, this is speculative given that the protostellar masses are not known for the majority of these
multiple systems. But, in any event, two (or more) protostars accreting at similar rates will
naturally have a higher luminosity than a single protostar accreting at the same rate.

\subsection{Separation Distributions}
A major goal of this study is to better determine the typical separations of 
companion stars, which can be connected to formation mechanism. We use the 
list of all separations measured (Tables \ref{orion-seps} and \ref{perseus-seps})
following the analysis methods described in Section 2.4 and 
Appendix A to
generate histograms and cumulative distributions for companion stars in different separation bins.
These histograms are generated with bin sizes of 0.25 for the log$_{10}$ of the projected separation in au.
The separation distribution histograms 
of the Orion results for the full sample, Class 0 protostars, 
Class I protostars, Flat Spectrum protostars,
and Class I and Flat Spectrum protostars considered together are shown in Figure \ref{separations_orion_all} .
We then show the Orion separation distributions
limited to only protostar separations of a particular class in Figure \ref{separations_orion_class}.
Finally, the separation distributions for Perseus are shown in Figure \ref{separations_perseus_all}.
In this and the following sections, and in Tables 
\ref{MF-CSFs-all}, \ref{MF-CSFs-comb}, and \ref{MF-CSFs-density}, we refer to the different multiple systems of
protostars from different classes in the following manner, with the first class listed representing
the majority classification of the system and the second item refers to the classes
of the other system components. We use the following classifications:
\begin{itemize}
\item Class 0--Class 0,
\item Class 0--(Class 0, Class I, Flat),
\item Class I--Class I,
\item Class I--(Class 0, Class I, Flat),
\item (Class I,  Flat)--(Class I, Flat),
\item (Class I, Flat)--(Class 0, Class I, Flat),
\item Flat--Flat
\item Flat--(Class 0, Class I, Flat).
\end{itemize}

For example, Class 0--Class 0 specifically refers to multiple systems that are only
composed of Class 0 protostars. These could be binaries, triples, or higher order, but 
all have the Class 0 classification. Then, Class 0--(Class 0, Class I, Flat) will include
multiple systems that include protostars of all Classes and not just Class 0s; however, the primary
classification of the system will be Class 0 due to a majority of its components 
being Class 0 (see Appendix A.3). This categorization does not result in double 
counting
of multiples between systems that are composed of different Classes,
but
Class 0--Class 0 and Class 0--(Class 0, Class I, Flat) do overlap in their samples. 
However,
one can see some differences visually in Figures \ref{separations_orion_all} and \ref{separations_orion_class}
between homogeneously classified multiples and
multiples with 
composite
classification.

We note that some composite categories, like (Class I, Flat)--(Class I, Flat), are needed
because the available Perseus classifications did not distinguish between Class I and 
the more evolved Flat Spectrum protostars. Thus, to compare with Orion, it is more appropriate
to consider the composite category.

The histograms generated for the Orion protostars fill the range of parameter 
space from 20~au to $10^4$~au. The histogram for the full sample (see Figure 
\ref{separations_orion_all}) shows some structure with a peak at 
$\sim$75~au
and another peak at $\sim$4000~au. Multiples exist between these two peaks, but
there is a local minimum 
that is visually apparent
at $\sim$300~au, and the histogram does not begin rising again 
until separations $\ge$10$^3$~au.
We note that the scheme for determining the probability of a detected continuum source to 
be a companion 
(Section 2.4.3)
lowers the significance of the peak at large separations but does not affect the histograms significantly
at less than $\sim$3000~au. The separation histograms for Perseus appear
similar to those of Orion (Figure \ref{separations_perseus_all}), with 
most of the histogram bins being within their 1$\sigma$ uncertainties.

The separation
distributions for Class 0 protostars (for both homogeneous and composite systems) 
appear the most distinct relative to the separation distributions of Class I and Flat Spectrum 
protostars. The principal difference is the bimodal appearance of the Class 0 
distributions, while the separation distributions of more evolved systems tend to be more flat and exhibit less structure.
Thus, it is the Class 0
protostars that are primarily responsible for this bimodal appearance in the full sample
(Figures \ref{separations_orion_all} and \ref{separations_orion_class}).
Because the trends in both regions are similar, we are motivated to combine the
datasets to improve the statistics of the separation distributions. 
We show the separation distribution histograms for the combined Orion and Perseus samples
in Figure \ref{separations_comb}. The combination of
the samples results in a smoother distribution
of separations due to the greater numbers, but is not
fundamentally different from Orion and Perseus on their own.

\subsection{Quantitative Separation Distribution Comparisons}

The distributions of separations, as shown in the histograms in Figures
\ref{separations_orion_all}, 
\ref{separations_orion_class}, \ref{separations_perseus_all}, 
and \ref{separations_comb}, are illuminating, 
but they do not
quantitatively demonstrate whether or not Perseus and Orion are statistically 
different
or
if analytic distributions are consistent with the observations. Statistical
tests are necessary to determine if the observed populations are likely drawn from the same parent distribution.
To evaluate this, we constructed (CDFs) from the separation 
distributions and used their respective companion probabilities to compare the distributions 
by randomly sampling the separation distribution CDFs 1000 times
and comparing each of these randomly sampled CDFs using 
the KS test as described in the Appendix A.2.
We only considered the full separation range (20 to 10$^4$~au), because the separation 
distributions from 20 to 10$^3$~au could not rule out the null hypothesis for any tests.
Figure \ref{cumulative_oriper} shows the CDFs for Orion and Perseus; the
upper left panel shows the effect of the companion probabilities on the CDFs for
the full sample. 

\subsubsection{Orion and Perseus Comparisons}
We first compared the Orion and Perseus separation distributions 
 or all possible combinations of the same classes.
The median likelihoods and their uncertainty (defined by quartiles) for the randomly sampled CDFs are used to evaluate the statistical tests.
We find that the null hypothesis cannot
be ruled out for any of the Orion and Perseus separation distributions, meaning that there is no statistical 
evidence for the Perseus and Orion separation distributions to be drawn from different samples. 
The sample with the lowest median likelihoods were Class I--Class I and (Class I, Flat)--(Class I, Flat).

\subsubsection{Comparison Between Classes}

 In Section 4.2, we highlighted some visual differences between the separation distributions
of different Orion protostar Classes. Here we  now evaluate whether they are 
statistically significant. Out of all the classes, only comparisons between
Class 0--(Class 0, Class I, Flat) vs. Flat--Flat and Class I--(Class 0, Class I, Flat) vs.
Flat--Flat had median likelihoods that were $<$0.01 along with more than half of their
realizations having likelihoods $<$ 0.01. One other comparison, Class 0--Class 0 vs.
Class I--(Class0, Class I, Flat) nearly made the cutoff for a statistically significant
difference with a median likelihood of 0.011.
Thus, the most significant difference
in separation distribution can be observed between the youngest multiple systems and the
most-evolved multiple systems in Orion. The principal difference in the histograms
shown in Figures \ref{separations_orion_all} and \ref{separations_orion_class} is
that the Class 0--(Class 0, Class I, Flat) distribution has significantly 
more companions at separations $>$1000~au as compared to the Flat--Flat distribution.

We also compared the separation distributions of the combined Orion
and Perseus samples. In this comparison we only find a statistically
significant difference for Class 0--(Class 0, Class I, Flat) vs. 
(Class I, Flat)--(Class I, Flat), with a median likelihood of 0.002.
Like the results from Orion only, we find
a statistically significant difference between the youngest systems and 
those that are the most evolved.

\subsubsection{Comparison with a Log-Flat Distribution}

We also compared our separation distribution to an analytic distribution that is flat
in $log$ separation space, a log-flat distribution,
which is commonly referred to as \"{O}pik's law \citep{opik1924} and has
been found to describe populations of companion separations
\citep[e.g.,][]{kouwenhoven2007}.
We show an example log-flat separation distribution
in Figure \ref{cumulative_oriper_logflat_field}.
Comparing with a log-flat distribution also
indirectly checks whether the bimodal appearance is statistically robust. We performed a
one-sided KS-test with an analytic log-flat distribution. The results from 1000 KS-tests, sampling the source separation distribution according
to the probabilities are comparable to using a 2-sample KS-test with a log-flat distribution with a sample size 100$\times$ the size of the input sample.
Only the Orion Class 0--(Class 0, Class I, Flat) sample
and the Class I--(Class 0, Class I, Flat) sample 
were able to reject the null hypothesis with a median likelihood
$<$ 0.01.

The Perseus separation distributions were also tested against
a log-flat separation distribution, but none of the Perseus
sub-samples nor overall sample are able to reject the log-flat separation
distribution with a likelihood $<$ 0.01. This is contrary to the
result from \citet{tobin2016a} which found evidence that a log-flat distribution
could be ruled-out for the full sample, with a likelihood $<$ 0.1.
However, in this study none of the Perseus samples are able to
rule-out the null hypothesis even for likelihoods of 0.1. The sample that has
the lowest likelihood is the Perseus Class 0--(Class 0, Class I, Flat)
sample with a likelihood of 0.11. 
To investigate if our companion probability scheme is the cause of the difference, we also
performed the comparison with a distribution of separations without probabilities (assuming
all are companions), and the lowest likelihood was 0.135, also for Class 0--(Class 0, Class I, Flat) (see Appendix B). 
The reason why we do not find the same
result as the previous study is because
 the separation distributions are constructed differently and  
 use automated methods rather than manual associations. In this
 case, the main source of difference arises from the higher order systems whereas the previous study
 chose the primary and separations were all calculated with respect to the primary.

We also compared the combined sample of Orion and Perseus with the
log-flat distribution. Similar to the case for Orion alone,
both Class 0--(Class 0, Class I, Flat) and Class I--(Class 0, Class I, Flat) samples
are able to reject a log-flat distribution, 
having median likelihoods $<$~0.01.
Ruling-out this commonly observed distribution of separations provides evidence
that the structure in those separation distributions is real and is
not the result of statistical uncertainty or histogram binning. In particular, 
for the Class 0--(Class 0, Class I, Flat) distribution, ruling-out
the log-flat distribution suggests that the bimodal appearance may be real.

\subsubsection{Comparison Field Solar-type Multiples}

We next compared the observed separation distributions to the distribution of separations 
for field solar-type stars \citep{raghavan2010} using the KS test. We make use of the
observed separation distribution from 24 to 10$^4$~au from \citet{raghavan2010}, sampling
the same range of separations for which we have detections. We show the distribution
from \citet{raghavan2010} along with the observations of the cumulative distribution for
the full samples for Orion and Perseus in Figure \ref{cumulative_oriper_logflat_field}.
While the field solar-type stars are a well-characterized sample to compare with, it is important
to highlight that most of the protostars in the HOPS sample are likely to become
M-stars rather than F, G, or K-type stars that make up the \citet{raghavan2010} sample.

For Perseus alone, we can only reject
the null hypothesis for the Class 0--(Class 0, Class I, Flat) sample.
Then, for Orion, the null hypothesis is ruled out for all
Classes 
except for the Flat--Flat, Flat--(Class 0, Class I, Flat), and Class I--Class I samples.
Only the separation distributions for the more evolved protostars in Orion cannot be distinguished from
the separation distribution of field solar-type stars. The comparison
of the \citet{raghavan2010} distribution to the combined Orion and Perseus samples yields 
similar results to those for Orion alone.

\subsection{Overall Multiplicity Statistics}

We compute the MFs and CFs for the protostars in Orion,
Perseus, and their combined sample. The statistics are examined for a variety of 
separation ranges to both illustrate the size scale on which most multiples are found and evaluate whether there are differences
between classes for the different separation ranges. 
The MFs and CFs for Orion, Perseus, and their combined
samples are provided in Tables \ref{MF-CSFs-all} and \ref{MF-CSFs-comb} and will
contain information
related to the formation mechanism(s) and the evolutionary paths of multiplicity.

The Orion and Perseus samples on scales from 20~au to 10$^4$~au have MFs 
of 0.30$\pm$0.03 and 0.38$\pm$0.07, respectively. The respective
CFs are then 0.44$\pm$0.03 and 0.57$\pm$0.07. 
For the separation range from 20~au to 10$^3$~au,
the MFs drop to 0.17$\pm$0.02 and 0.26$^{+0.06}_{-0.05}$, and the CFs
drop to 0.19$\pm$0.02 and 0.28$\pm$0.06, respectively. This shows numerically
that about 1/3 of the multiples in Orion and Perseus are found at 10$^3$ to 10$^4$~au separations.
The CFs are nearly identical to the MFs on 20~au to 10$^3$~au scales because there are few
triples and high-order multiples detected on these smaller scales. 
The MFs and CFs for the full samples of protostars in Orion and Perseus are consistent 
within their uncertainties. The MFs and CFs computed here for
Perseus are lower than previously presented in \citet{tobin2016a}. These differences
are largest for the MFs and CFs computed using the companion probabilities but
are still present even if the companion probabilities are not taken into account (see Appendix A.5). Thus,
the differences result from both the companion probabilities and our new method for associating multiples.

The typical architectures of the multiple systems are also found in Tables \ref{MF-CSFs-all} and
\ref{MF-CSFs-comb}, where the number of systems that correspond to binaries, triples,
quadruples, etc. is provided. For separations from 20 to 500~au and 20 to 1000~au, the most common architecture
is binaries, outnumbering triples by a factor of $\sim$20 for separations between 20 and 500~au,
and a factor of $\sim$15 for separations between 20 and 1000~au. Binaries are still the most
common type of multiple system for separations between 20 and 10$^4$~au, but the ratio of
binaries to higher order systems is now only a factor of $\sim$4. In fact, some systems that
were binaries at smaller separation ranges are part of higher-order systems when larger
separations are considered. Systems higher order than triple are relatively uncommon and 
the number of triples found within the separation range between 20 and 10$^4$~au,
is comparable to the total number of all systems with 4 or more components. However, in Orion, the
total number of individual components within quadruples or higher-order systems (44) is larger than the 
number of individual components in triples (27).

We plot the MFs and CFs as a function of protostellar class\footnote{For the purposes of these figures, we refer to the majority classification of the
system (Appendix B), but they will include companions of any class. Thus, `Class 0s' 
refers to Class 0--(Class 0, Class I, Flat), see Section 4.2.} in Figure \ref{mf_csf} for Orion and Perseus in three separation ranges: 20 to 10$^4$~au, 20 to 10$^3$~au,
and 20 to 500~au. 
On the 20 to 10$^4$~au range, 
the MFs and CFs for Class 0s are systematically
higher than those for Class Is and Flat Spectrum protostars, similar to 
the results from previous studies \citep[e.g.,][]{chen2013,tobin2016a}. 
However, 
only the differences in the
CF 
approach statistical significance
in the separation range of 20 to 10$^4$~au:
2.7$\sigma$ for Class 0s relative to
Class I and 2.8$\sigma$ for Class 0s relative
Class I and Flat Spectrum protostars. Then the difference is 
2$\sigma$ for Class 0 to Flat Spectrum protostars, while the differences
in the MFs and CFs between other classes are not statistically significant.
This tells us that 
there are more higher-order companions to Class 0 protostars than more evolved protostars
when separations out to 10$^4$~au are considered. Our observed CFs for Class 0s
are similar to those reported in \citet{chen2013}, but our MFs for Class 0 protostars are $\sim$30\% lower 
than \citet{chen2013}. We suspect that this is due to sample bias in \citet{chen2013}, given that some of the
archival data studied were previously known to be multiple systems
and our larger samples are balanced both by detection of new multiples and also 
non-multiples.

The MFs and CFs computed in the 20 to 10$^3$~au and 20 to 500~au ranges tell a somewhat different story.
The overall MFs and CFs for both Perseus and Orion decrease, in part because 
continuum sources that are part of a multiple system at separations
out to 10$^4$~au are considered as independent, single sources for the 500 and 10$^3$~au statistics.
 This increases
the number of singles counted overall and for each class, resulting in a reduced MF and CF.
Even if we did not count the singles in this way, the MFs and CFs would still decrease due to fewer multiples and higher-order multiples on these scales.

The overall trend in the MFs and CFs may show that multiplicity decreases from
Class 0 to Class I and then rises from Class I to Flat Spectrum. However, 
this apparent trend is not statistically robust with the data in hand given
that 
neither the MFs nor CFs
have differences $>$3$\sigma$.

The differences in the CFs for Class 0 protostars with respect to
Class I and Flat Spectrum protostars
result from there being more companions at $>$1000~au 
toward Class 0 protostars
than the Class I and Flat Spectrum protostars.
The separation distributions of Class 0 protostars relative
to Class I and Flat Spectrum protostars also show large differences
at $>$1000~au, and
some of the Class 0 separation distributions are statistically 
inconsistent with the distributions of more-evolved classes, see Section 4.3.2.
These results imply that the principal change in multiplicity with protostellar 
evolution primarily affects companions with separations between 
10$^3$ and 10$^{4}$~au because the MFs and CFs are comparable
for Class 0 and Flat Spectrum protostars in the 20 to 10$^3$~au and 20 to 500~au ranges.

The sizes of the Class I and Flat Spectrum samples in Orion are similar; thus, there is no clear
sample bias that would produce a difference in the multiplicity statistics between
Class 0 and more-evolved protostars. Flat Spectrum protostars 
are drawn from a parameter space that overlaps Class I protostars in terms of T$_{\rm bol}$. 
The Orion Flat spectrum protostars have T$_{\rm bol}$ values that are skewed toward
larger values of T$_{\rm bol}$ than typical Class I protostars (Paper I). Such a signature could not be
searched for in Perseus due to a lack of necessary mid-infrared spectroscopy data
from \textit{Spitzer} toward the Perseus protostars as compared 
to the Orion protostars. 

We also examine the MFs and CFs for Orion and Perseus as a combined sample as a function of protostellar
class and different ranges of separations in Figure \ref{mf_csf_combined}. 
However, due to the lack of distinction between Class I and Flat spectrum in Perseus,
the combined sample is only relevant for the full sample, Class 0, and Class I + Flat Spectrum samples.
For Class I and Flat Spectrum, the combined statistics simply use those from Orion.
Using the data from \citet{raghavan2010}, we calculate the
field star MFs and CFs for system 
separations of 20 to 10$^4$~au 
(0.28$\pm$0.02 and 0.32$\pm$0.02, respectively), 20 to 
10$^3$~au (0.22$\pm$0.02 and 0.23$\pm$0.02), and 20 to 500~au (0.19$\pm$0.02 and 0.20$\pm$0.02).

The MFs and CFs for the Class 0 and Flat Spectrum protostars for separations between 20 and 10$^3$~au and 
20 to 500~au are consistent with the field solar type stars. The Class I protostars 
have MF and CF values below that of
 the field solar-type stars within these two ranges of separations, but the 
differences are $<$~2$\sigma$ for the 20 to 500~au separation range
 and are $\sim$2.3$\sigma$ for the 20 to 1000~au separation range.
 The Class I + Flat sample 
is also lower than that of the field stars but
is consistent within the 2$\sigma$ uncertainties. At separations from 20 to 10$^4$~au, only the
Class 0 MF and CF disagree with those of solar-type field stars; the MF,
however, has differences $<$2$\sigma$
but the CF difference is greater than 4.6$\sigma$.
The higher multiplicity statistics of Class 0 protostars indicate both that 
stars form with higher multiplicity and that this early multiplicity also 
tends to comprise higher order systems.
Based on the comparison of the MFs, CFs, and separation distributions in Section 4.3.3,
the multiplicity properties of more evolved protostars tend to be the most similar to the
field solar-type stars, and the Class 0 multiples tend to be the most dissimilar from those of field stars.
We do note, however, that it is expected that most of the protostars studied in this work will not become
solar-type stars based on the expected initial mass function.

\subsection{Relationship Between Multiplicity and YSO Density}

Previous multiplicity studies have examined the relationship between MF/CF
and the local stellar density. In the Orion Nebula Cluster (ONC),
\citet{reipurth2007} found a deficit
of companions within a separation range of 150 to 675~au relative to T Tauri
associations \citep{reipurth1993}. However, in a study of Class I and Flat Spectrum
protostellar multiplicity in the Orion A and B molecular clouds, \citet{kounkel2016} found the opposite,
in that protostars and pre-MS stars had {\it larger} CFs from 100 to 10$^3$~au in regions
with higher YSO density, where they adopted 45~pc$^{-2}$ as the boundary between low
and high YSO density. It is important to point out that the YSO densities 
examined here and in \citet{kounkel2016}
are lower than those of the dense center of the ONC, where the \citet{reipurth2007} study was conducted.
It is thought that dynamical stripping of wide companions
in the dense center of the ONC is responsible for the deficit.

We examined our Orion sample for differences in the multiplicity
statistics between regions of high and low YSO density.
YSO density is determined using 
YSO catalogs that are constructed from infrared samples \citep{megeath2012,pokhrel2020}
and completeness corrected using X-ray catalogs where available \citep{megeath2016}.
We set our boundary between
low and high density regions at a YSO density of 30~pc$^{-2}$ such that there were comparable
numbers of protostars in the regions assigned as high and low YSO densities.
These MFs and CFs
are listed in Table \ref{MF-CSFs-density}, and we show them graphically for
different ranges of separations in Figure \ref{mf_csf_yso_density}. 
There are apparent differences in the MFs and CFs in the 20 to 10$^4$~au separation range. 
The differences in the CFs between high and low YSO density regions are
1.8$\sigma$ for Class 0, 3$\sigma$ for Class I, 4$\sigma$ for
Class I with Flat Spectrum, and 1.9$\sigma$ for Flat Spectrum on its own, while the 
differences in the MFs are not statistically significant. The difference in the CFs is also $>$3$\sigma$ when considering all classes together.

We now shift our focus to 
the smaller separation range between 100 to 10$^3$~au in order to
compare with \citet{kounkel2016},
although we also examine 20 to 10$^3$~au because
our data allow for smaller minimum separations.
We find a tentative difference between the
CFs for high and low YSO density regions
for both the full sample and the combined Class I and Flat Spectrum sample. 
The Class I + Flat Spectrum sample has a 1.9$\sigma$ difference in the CFs between
high and low YSO densities, while the difference in the CF for the Flat Spectrum 
sample alone is 1.5$\sigma$.
The MF differences are less significant (at most 1.6$\sigma$) because the MFs are smaller than the CFs
in this range of separations. If separations from 20 to 10$^3$~au are instead considered, 
the differences 
are 1.4$\sigma$ and 1.8$\sigma$ for the Class I + Flat Spectrum and Flat Spectrum
samples, respectively.

Finally, we note that only the Class 0 protostars do not exhibit 
significant differences in their MFs and CFs in separation ranges of 
100 to 10$^3$~au, 20 to 500~au, or 20 to 10$^3$~au. 
Thus, the Class I and Flat Spectrum 
protostars are the only ones whose multiplicity properties may be
to be sensitive to YSO density for the 100 to 10$^3$~au separation range.

\section{Discussion}

The combined results for Orion and Perseus demonstrate the relative 
consistency for the same protostar classes.
There are some differences in the separation distributions and MFs/CFs between
protostar classes and with respect to high and low YSO densities.
Furthermore, while not analyzed 
as part of this work, the MFs and CFs observed in Ophiuchus are found to be
comparable to Orion and Perseus \citep{encalada2021}. Characterization of the
evolutionary classes in the Ophiuchus sample remains challenging, however
\citep{mcclure2010}, and there are substantially fewer Class 0 protostars
compared to Orion and Perseus.
Nevertheless, the consistency of multiplicity properties measured in
different regions hints at common physical processes
at work 
to give rise to the observed multiplicity, and
the statistics afforded by these surveys enable further constraints on
the evolution of multiplicity within the protostellar phase and the role
multiplicity might play in protostellar evolution.

\subsection{Formation Mechanisms of Multiple Systems}

We detect multiple protostar systems 
with separations
from 10s to 1000s of au
(Figures \ref{separations_orion_all} - \ref{separations_comb}),
encompassing the range of physical scales where the disk ($\sim$10s to 100s of au) and 
infalling envelope (1000s of au) are the dominant physical structures.
The most favored mechanisms for multiple star formation are 
disk fragmentation from gravitational instability (GI), 
operating on $\la$500 au scales, and turbulent fragmentation operating on 100s of au scales
and larger. Although turbulent fragmentation occurs on relatively larger scales, companions formed from this process can migrate from $\ga$10$^3$~au to less than 100~au
in a few 100~kyr \citep{offner2010,lee2019}, and there is also evidence for smaller scale 
turbulent fragmentation in some numerical simulations \citep{bate2012}.
Thus, it is possible that both mechanisms produce
companions with separations $\la$500~au, while a single mechanism 
populates $\ga$500~au. Dynamical interactions within higher order 
multiples can also contribute to substantial evolution of system separations \citep{kroupa1995,bate2002,sadavoy2017,cournoyer2021}, 
either via the ejection of one component, or longer timescale secular 
mechanisms such as Kozai-Lidov oscillations \citep{fabrycky2007,reipurth2012}, 
though this is likely a smaller contribution \citep{moe2018}.

For the turbulent fragmentation and disk fragmentation mechanisms to operate,
certain physical conditions are required. The main ingredients for turbulent 
fragmentation is dense gas and a turbulent velocity spectrum. The turbulence
will create local regions of higher density, which can collapse to form
stars if a region becomes becomes gravitationally bound and collapses. Simulations
can produce this type of fragmentation in a variety of conditions with driven or
decaying turbulence and in the presence or absence of magnetic fields \citep{offner2010,bate2012,li2018,lee2019}.

The fragmentation of disks to form companion stars broadly requires the presence
of massive disks during the star formation process. The typical criterion used to
describe disk stability is the Toomre Q parameter
\begin{equation}
Q = \frac{c_s\Omega}{\pi G\Sigma}.
\end{equation}
Where $c_s$ is the sound speed of the gas, $\Omega$ is the angular velocity set by Keplerian
rotation, $G$ is the gravitation constant, and $\Sigma$ is the gas surface density of the disk.
Values of $Q$ near or below 1 have strong enough self-gravity relative to the sound speed and rotational
shear that can lead to the formation of fragments in the disk as well as spiral arm formation
\citep[e.g.,][]{kratter2016}.
Also, Q is a function of radius, so portions of the disk can be unstable while
other portions remain stable. Fragmentation will thus generally be most feasible in the outer disk
where rotation is slower and the gas is cooler, so long as the surface density is not too low.

While turbulent fragmentation can produce multiples that are ultimately found on
both large and small spatial scales, it is unclear
whether this mechanism can create enough multiples 
that migrate to
$<$500~au separations to reproduce the observations.
Recent simulations by \citet{lee2019} examined the formation
of multiple systems from turbulent fragmentation and their subsequent evolution. The 
simulations did not have the resolution to resolve fragmentation within disks,
but could track sink particle migration down to 10s of au scales. Thus, these simulations 
provide an estimate for the fraction of companions at $\la$500~au separations
formed via turbulent fragmentation with migration. Our observations for the combined 
Orion and Perseus samples find that the observed fraction of companions 
are $\sim$2-3 times larger on $\la$500~au scales (depending on whether the full sample or only
Class 0s are compared) than the fractions found
in the simulations by \citet{lee2019}, see their Figure 20 vs. our Figure \ref{separations_comb}.
Furthermore, studies by \citet{cournoyer2021} with different models
indicate that companions formed through dynamical capture in a forming cluster
typically have separations $>$10$^3$~au. 
This provides some evidence that at least 
half of the companions found on scales $\la$500~au are likely to have formed via disk fragmentation. 
This agrees with $\sim$66\% of close multiples in Perseus
that were consistent with having formed via disk fragmentation based on their observed
disk/envelope rotation on $<$500~au scales \citep{tobin2018}. 

While this comparison seems to suggest that
disk fragmentation is important to forming multiples at $<$500~au, a simple characterization of
gravitational instability for the disks in Orion in Paper I did not indicate
a large number of Class 0 or Class I disks near the instability limit. 
Also, the average disk radius for Class 0 and I protostars in Orion is $\sim$40~au,
rather than $\sim$100~au. This could 
pose a challenge to populate the $<$500~au multiples from disk fragmentation, 
unless gas masses were underestimated (which is certainly possible). Also, fragmentation can
happen very quickly, on an orbital timescale ($\sim$10$^3$~yr for $M_*$=1.0~\msun\ at 100~au).
Therefore, the close multiples we 
observe
could be the result of the disks that 
already fragmented, and the
disks may not remain in a near-unstable state for long enough time to enable 
many of them to be discovered. For example, we did not detect obvious analogs 
to the fragmenting triple system L1448 IRS3B in Perseus \citep{tobin2016b,reynolds2021}
within Orion, and spatial resolution was too low to detect analogues to [BHB2007] 11
\citep{alves2019}. 
We do note however that the instability calculations presented in Paper I made many simplifying
assumptions to arrive at a disk-averaged limit, so those values should not be considered absolute.

As we discussed previously, contamination by chance alignments at large separations can 
artificially inflate the number of wide companions.
This effect increases
the apparent importance of turbulent fragmentation. Despite our attempts to 
mitigate the contribution of likely false associations (see light gray bars vs. dark
gray bars in Figure \ref{separations_comb}), some contamination may still be present
at large separations.
The simulations from \citet{lee2019} also characterize the number of true multiples versus
the number of line of sight associations. They found that on scales $\ge$10$^3$~au at least
50\% of companions could be line of sight associations. \citet{lee2019} note, however, that this result
is partially due to a filamentary cloud that is oriented along one of the Cartesian axes, resulting
in a large number of false associations, so 50\% could be a 
worst case scenario. Furthermore, the numerical 
simulations did not have accompanying synthetic observations, but rather analyzed sink particle
positions. Our
observations here use the dusty circumstellar disks as signposts of multiplicity, which will
not detect companions with 100\% efficiency, as demonstrated in Section 3.6. Therefore, we expect
that the simulations (when observational bias is not applied) will include more companions
than observations detect. 
Consistent with this expectation, the histogram of separations from the
Perseus protostars
appears to agree well qualitatively with the Lee et al.
distribution of true companions
(see their Figure 20 on scales from $\sim$300~au to $\sim$3000~au.)
Our estimates of contamination do not exceed $\sim$30\%
until separations $\ga$3000~au (see Figures \ref{separations_orion_all}-\ref{separations_comb}),
and that contamination is not significant
on scales $<$3000 au.
This result is
largely consistent 
with turbulent fragmentation simulations at these separations. Also,
the use of dust continuum
and free-free continuum as a probe of multiplicity implicitly creates a selection bias 
that may reduce our susceptibility to contamination by effectively 
requiring that protostars have a certain amount of circumstellar dust (or jet emission) to be detectable. 

In summary, existing simulations of multiplicity originating from turbulent fragmentation alone
do not produce sufficient numbers of companions at separations $<$~500~au to explain the observations.
Thus, we suggest that disk fragmentation 
is
responsible for $\sim$50\% of multiple systems with separations
$<$~500~au. Beyond 500~au separations, simulations of turbulent fragmentation are able to reproduce 
the observations. Therefore, we expect that the observed separation distributions result from both
disk fragmentation and turbulent fragmentation, and disk fragmentation is expected to be most likely
in the Class 0 phase when the disks have the most mass \citep{tobin2020a}.

\subsection{Future Evolution of the Companions}

The view of multiplicity obtained toward the protostars in our sample is a single snapshot 
in the evolution of these systems, and the separations of the observed systems will 
continue to evolve during protostellar evolution and beyond. Here we discuss the evolution
that we expect to occur in the context of our observational knowledge of multiplicity
in later stages of the star formation process and field stars.

\subsubsection{Evolution within the Protostellar Phase}
While 
our observed multiplicity properties
may not have significant contamination from unrelated protostars
at scales $<$3000~au, our observed 
distribution of separations for all protostar classes does not imply that all of these
associated protostars are going to remain bound systems and/or at their current
separations. The functional form
of the separation distribution of field solar-type stars from \citet{raghavan2010} is 
overlaid on the separation distributions in Figures \ref{separations_orion_all}~-~\ref{separations_comb} (also see Figure \ref{cumulative_oriper_logflat_field}). The histograms of separations
are quite inconsistent with the field solar-type stars on scales $\ga$10$^3$~au, and
we can rule out that they are drawn from the same parent distribution for 
most protostar classes, except the most evolved sub-samples, see Section 4.3.4.

Thus, either a large number of the current protostellar companions 
between 10$^3$ to 10$^4$~au
must migrate to smaller separations
or they must become unbound. It is also important to point out that most protostars
observed in this study will become M-stars rather than solar-type stars,
and main-sequence M-stars (M$\sim$0.3~\msun) have a separation distribution that is even more 
skewed to smaller separations. Solar-type stars have a mean separation of 50~au \citep{raghavan2010}, 
while M-stars have a mean separation of $\sim$20~au \citep{winters2019}. Therefore, it is likely that
most of the systems that we study here--and
remain multiples into the main sequence--will ultimately have a mean separation
of $\sim$20~au.

It is noteworthy that the study by \citet{lee2019} found a similar fraction of companions
from $\sim$500 to 3000~au as our observations in Perseus, in addition to Orion.
However, the companions studied in \citet{lee2019} were indeed bound
when only considering the sink particle mass and did not need to consider the
surrounding gas mass potential. Thus, companions must either continue to evolve to
smaller separations, become unbound through $\ge$3-body interactions, or have formed with
unbound relative velocities. 
\citet{lee2019} noted that only when the gas
mass is comparable to or exceeding the sink mass was the boundedness affected by the
surrounding gas potential. 
Class 0 protostars tend to have a larger number of companions separated by 
greater than 10$^3$~au (Figures \ref{separations_comb} and \ref{mf_csf_combined}),
and these systems are the most likely to have an envelope mass 
greater than the stellar mass. We therefore argue that in order for the separation distributions of
Class 0 protostars to evolve toward the field solar-type distribution 
(Class I and Flat Spectrum as well),
under the assumption that most companions from 500 to 3000~au are bound,
two processes must be at work. First, protostellar  
multiples with current separations greater than 10$^3$~au may have companions 
that are lost due to either being unbound or 
becoming dynamically unbound due to loss of gas mass by outflow expulsion 
or interactions \citep[e.g.,][]{offner2014,sadavoy2017}. 
Second, many of those that do not become unbound must migrate to smaller separations. 
Some may migrate from $>$~10$^3$~au to $\sim$100s of au, while those that may have formed
at $\sim$100~au can migrate to $<$~30~au. The migration to $<$~30~au is necessary
to dilute the peak in separations at $\sim$100~au observed for Class 0 protostars and
produce the larger number of multiples found toward main sequence stars at separations $<$~50~au.
The statistical significance in the
difference of the separation distributions for Class 0 protostars with 
respect to 
the more evolved
Flat Spectrum is further evidence
that separations evolve with protostellar evolution (Figures \ref{separations_orion_all}
and \ref{separations_orion_class} (Section 4.3.2). This will be discussed further in the following sections.

At present, it is unclear how quickly protostellar separations can evolve to
$<$~30~au  given that our spatial resolution 
is limited to $\sim$20~au in Perseus and Orion.
Companions with sub-au separation likely
exist by the Class I phase 
from radial velocity observations of embedded sources \citep{vianaalmeida2012}
and of some exotic 
phenomena like pulsed accretion \citep{muzerolle2013},
but statistics are clearly lacking. 
ALMA can probe scales 
$<$~10~au at the distances to nearby star-forming regions, 
but such observations are only available for a handful of protostars.

Even with better spatial resolution, there are a number of complicating issues
that 
may cause incompleteness in multiplicity characterization
at $<$~30~au scales when using observations of dust continuum.
As companions with smaller separations
are examined, the amount of dust mass that can reside in circumstellar disks
decreases, rendering them less detectable due to disk truncation \citep{lubow1994}. 
Also, even if
a small circumstellar disk has a high surface density, dust opacity will limit
the emergent flux density, depending on wavelength. This could cause some circumstellar disks to blend with their circumbinary disks
which could be the case of HOPS-361-C and 
HOPS-288 (Figure \ref{continuum_multiple_vla}). 
Thus, centimeter wavelengths remain important
given than they can pierce through highly obscuring dust, and free-free emission
at these wavelengths can provide another signpost of multiplicity that will be useful toward smaller
separations.

\subsubsection{Evolution Beyond the Protostellar Phase}

The results from this study as a whole, and the Class 0 protostars
in particular, establish the starting point for multiplicity
evolution as a bimodal distribution of separations.
This distribution of separations for Class 0 protostars 
is inconsistent with solar-type field stars and
more evolved Flat Spectrum protostars.
However, it is not possible to rule out that Flat spectrum protostars are drawn from the same parent distribution as field stars,
affirming that 
the log-normal fit to the \citet{raghavan2010} sample appears similar to the most-evolved protostar samples.
Furthermore, the MFs and CFs
for the most evolved protostars on all scales are consistent with those of the field stars,
while the CF of Class 0s is inconsistent by $>$ 3$\sigma$
(Figure \ref{mf_csf_combined}).

Observational evidence from our survey alone
therefore indicates that an initially bimodal distribution
of separations can evolve toward the observed field distribution. 
Moreover, models that include prescriptions for turbulent fragmentation, 
disk fragmentation, gas-driven migration, and 
dynamical upheaval can reproduce the observed separation distribution
of field multiples, when assuming an initially bimodal distribution of 
protostellar companions \citep{moe2018,tokovinin2020}. Given that the most evolved
protostars cannot be statistically distinguished from the field
separation distribution, the question arises as to whether most of the multiplicity
evolution occurs during the protostellar phase. If true, this implies
that the presence of circum-multiple material is the primary driver of 
multiplicity evolution and once the gas is accreted or expelled, rapid reorganization
is no longer possible and must occur via n-body interactions.

Thus, more evolved Class II and III YSOs represent the obvious population 
to study in order to answer the aforementioned question. 
For separations $<$10~au, Class II and Class III
YSOs are consistent with the field star multiplicity \citep{kounkel2019}. Such
close systems cannot typically be detected toward protostars, aside from a 
few special cases \citep[e.g.,][]{ortiz-leon2017,maureira2020}. For separations between
$\sim$100~au to 1000~au, the MFs/CFs and separation distributions of most YSO populations agree well with the field (aside from Taurus-Auriga) 
\citep{reipurth2007, kraus2011,king2012,duchene2018},
and the MFs/CFs of protostars in our sample are also in good agreement for this range.

On the other hand, the MF and CF in the separation range of
$\sim$10-60~au for YSOs in Orion can exceed those of the field by a factor of $\sim$2 (CF = 0.218$^{+0.08}_{-0.051}$), with 
1.7 to 2.7$\sigma$ significance relative to field solar-type stars and M-type stars,
respectively \citep{duchene2018}. We note, however, that \citet{defurio2019}
did not confirm this result for M-stars in Orion.
The separation distributions of YSOs in different
star-forming regions sampled from 19 to 100~au may also not be statistically
consistent with each other or the field \citep{king2012}. This range of separations 
is unique because systems with such separations will not be dynamically altered once the disk(s) has 
dissipated in all but the highest density cluster environments.
Tables \ref{orion-seps} and \ref{perseus-seps} show that
there are only 13 multiples (10 Orion, 3 Perseus) with separations between 10 and 60~au (MF/CF=0.027~$\pm$~0.01)
and 34 (25 Orion, 9 Perseus) companions from 10 to 100~au (MF/CF=0.07~$\pm$~0.01).
Thus, from 10 to 60~au our observations are statistically inconsistent with those
of \citet{duchene2018} at $>$3$\sigma$ significance, then compared to
\citet{king2012} we are at most 2$\sigma$ different from the regions with the highest
CFs. 
We did not compare separation distributions given the small number of separations
for the protostar samples.

Taken at face value, the discrepancy between our survey and that of \citet{duchene2018}
is quite large and could imply that more companions must migrate inward
to 10 to 60~au separations following the Flat Spectrum phase. However, there are no extremely
large excesses of companions in the separation distributions (Figures \ref{separations_orion_all}-\ref{separations_perseus_all} for Class I and Flat spectrum
protostars except beyond 3000~au and the migration timescales will likely not
align with the evolution timescale from Flat Spectrum to Class II.

We instead argue that incompleteness in our survey between 10 and 60~au 
enhances the disagreement.  
Our raw angular resolution limit is $\sim$20-30~au and to reach
the highest resolutions the noise is increased and the emission from companions
must be bright enough to be detected. Thus, millimeter/centimeter measurements of
multiplicity also likely suffer from a bias analogous to the Branch bias
\citep{branch1976} known in visible-light studies where stellar  twins are easier to detect and more complete than systems with fainter companions. But, unlike visible light studies, we cannot
easily correct our data for this bias, because there is no expected
distribution of companion flux densities.
Also, at our observed resolution, many sources are marginally-resolved (Paper I)
due to the dust emission probing their circumstellar disks), which will
further reduce sensitivity to close companions. This is in addition to 
other mitigating factors discussed in Sections 3 and 5.1 which
could limit our ability to detect companions at millimeter/centimeter 
wavelengths due to dust opacity and contrast of circumstellar
emission to optically thick circumbinary emission. 
Further investigation is therefore required to determine if there
is truly an inconsistency between our protostellar sample and 
more-evolved YSOs in the separation range 10 to 100~au.

\subsection{Multiplicity Variation with YSO Density: A Sign of Migration?}

The observed differences in the separation distributions, MFs, and CFs as a function
of protostellar class, suggest that the 
number of (detectable) multiples and shape of the separation distributions
evolve over time
(Figures \ref{separations_orion_all}~-~\ref{separations_comb}, \ref{mf_csf}, and \ref{mf_csf_combined}).
This implies that there may be more direct evidence for
how and when companion migration occurs in our sample statistics.
We specifically focus on the 100 to 10$^3$~au separation range. This is the range 
of separations that must be traversed if companions formed at $>$10$^3$~au separations
migrate
inward to further populate separations $<$100~au.

We then further examine the MFs and CFs as function of protostellar class separately
for regions of high and low YSO density (Figure \ref{mf_csf_yso_density}). 
Within the 100 to 10$^3$~au separation range, there are higher MFs and CFs in Orion for 
the 
combined Class I + Flat Spectrum sample and the Flat Spectrum sample
alone
in regions of high YSO density versus low
YSO density; however, we note that the significance the CF differences are only
1.9$\sigma$ and 1.5$\sigma$, respectively, and the MF differences are at most 1.6$\sigma$.
Then Class 0s exhibit comparable MFs and CFs 
at both high and low YSO densities for the same separation range. 
This finding for Class I and Flat Spectrum protostars
is consistent with a previous near-infrared study in Orion by \citet{kounkel2016}. 
The difference between Class I and Flat Spectrum 
MFs and CFs in regions of high vs.~low YSO density, 
while the Class 0s statistics remain similar, may be a
sign of migration.

The fact that the MFs or CFs for the Class 0 samples at high and low YSO densities 
are consistent in the separation ranges
from 20 to 500~au, 20 to 10$^3$~au, and 100 to 10$^3$~au implies that
the observed companions in these ranges are
mostly primordial (Figure \ref{mf_csf_yso_density}), 
meaning that they most likely formed near where we currently observe them. 
However, on 20 to 10$^4$~au scales, the CF of Class 0 
protostars is $>$2$\times$ higher at high YSO densities than at low YSO densities,
with a 1.8$\sigma$ difference. This leads us to suggest that 
Class 0 protostars have a reservoir of young, higher-order companions that 
could migrate to smaller separations through the 100 to 10$^3$~au
separation range at later times, particularly in the regions of high YSO density.

We suggest that the overall larger number of wide
companions in high YSO density regions for Class 0 protostars results in larger MFs/CFs
in the 100 to 10$^3$~au separation range for Class I and Flat Spectrum protostars
at high vs. low YSO density.
This is only true for more evolved protostars because of the time lag between formation
and migration to smaller separations. We provide a sketch of our proposed scenario in
Figure \ref{multiplicity_evolution}.
Thus, if our interpretation is correct, 
we can constrain the typical migration timescale to be longer than the Class 0 lifetime, but 
shorter than the lifetime of Class I and Flat Spectrum protostars.
The Class 0 phase is only expected to
last $\sim$160~kyr \citep{dunham2014}, but it could be shorter \citep{kristensen2018}. 
Class I and Flat Spectrum protostars, on the other hand, are expected to be at 
least $\sim$100 to 300~kyr older than Class 0 protostars \citep{kristensen2018,dunham2014},
allowing a longer timescale for migration to take place. Therefore, a typical timescale
for migration from $>$10$^3$~au to $<$10$^3$~au is between $\sim$0.2 and 0.5~Myr.
This proposed timescale based on the evolutionary classes is consistent 
with the migration timescale from 
$>$10$^3$~au separation to $\la$100~au predicted in numerical simulations by \citet{offner2010},
when magnetic fields were not included, and \citet{lee2019} with magnetic fields. Both
indicate that there would be too little time for significant migration 
from $>$10$^3$~au during the Class 0 phase.

Finally, prior observations suggest that companions are also migrating 
to separations smaller than 20~au (separations below our resolution limit) during this timeframe 
\citep[see Figure \ref{multiplicity_evolution}, and][]{muzerolle2013,moe2018,kounkel2019}. However, our
results indicate that companions 
with separations between 100 to 10$^3$~au are replenished
via migration faster than they are depleted by migration to separations $<$~20~au.
We do not expect significant
formation of companions \textit{in situ} for Class I and Flat Spectrum protostars
because their envelopes are significantly reduced in mass and density \citep{furlan2016},
and their disks are systematically lower in mass, making disk fragmentation at the Class I and
Flat Spectrum stage unlikely (Paper I). Thus, inward migration from initial separations
$>$10$^3$~au is a likely explanation for the increased MFs/CFs from 100 to 
10$^3$~au for Class I and Flat Spectrum protostars. This inward migration from $>$10$^3$~au
is primarily caused by dynamical friction with the gas surrounding the protostars \citep{lee2019}.
It is possible that close dynamical interactions may produce
ejections from small separations to larger separations, but the small number of
triples with separations $<$~500~au makes this mechanism unlikely.
Capture during dispersal of small clusters is also possible \citep{moeckel2011,cournoyer2021}, 
but this mechanism is relatively inefficient, not likely to produce the increases 
in MF and CF observed in the Class I and Flat Spectrum protostars, and does
not produce the separation distributions we observe.
Finally, our previous analysis suggests
the likelihood of contamination by non-physically associated sources in
the 100 to 10$^3$~au separation range is low, providing further evidence 
in favor of this interpretation.

Consequently, the 
increase in the MF and CF between 100 and 10$^3$~au for the Class I+Flat
Spectrum sample (and the full sample) for protostars found in regions of high 
YSO density versus low YSO density points to the fragmentation properties of the molecular cloud
indirectly playing a role in setting the MF and CF on this spatial scale in particular. 
This is because regions of higher YSO density have a strong correlation
with higher gas surface density \citep{gutermuth2011,pokhrel2020}. 
Then, turbulent fragmentation of this dense gas likely plays a role in 
creating YSOs that have initially wide separations ($\ga$10$^3$~au).
This means that in regions of high YSO density, there will be more companions
with wide separations with respect to regions with low YSO density. 
The regions with lower YSO density can be assumed to have had less 
dense gas, making those regions less susceptible to
the formation of companions with initial separations $\ga$10$^3$~au
\citep[e.g.,][]{padoan2002,bate2012,offner2016,krumholz2016}. 
Thus, there will then be a smaller population of $>$10$^3$~au companions initially that would be able to
migrate to smaller orbits in regions of low YSO density.

To summarize, Class 0 protostars in regions of high and low YSO density have similar
multiplicity fractions, indicating that companions on 
scales less than 10$^3$~au tend to form with the same frequency. There is little
time for migration during the Class 0 phase, so the MFs and CFs are comparable
in regions of high and low YSO density for separations $<$10$^3$~au. However,
the Class 0 CF from 20 to 10$^4$~au at high YSO densities is $>$2$\times$
(and 1.8$\sigma$) larger than at low YSO densities 
(Figure \ref{mf_csf_yso_density}). Then, the larger
MFs and CFs for Class I and Flat Spectrum protostars 
for regions of high versus low YSO density (and the lack of variation for Class 0 
MFs/CFs) suggest that the differences may arise
during evolution (Figure \ref{multiplicity_evolution}).
Furthermore, regions of high YSO density are more likely to form more higher-order systems with
$>$10$^3$~au initial separations than low YSO density regions, and the 
time it takes for these protostars to migrate to the 100 to 10$^3$~au separation 
range is comparable to the expected age of protostars in the Class I and Flat Spectrum sample.
Thus, the increase in MFs and CFs of 
Class I + Flat Spectrum
protostars 
and Flat Spectrum Protostars alone
at high vs. low YSO densities could be
evidence that migration is occurring and populating the range of intermediate separations.
We caution, however, that the CFs of these samples for high and low YSO density regions only 
differ by 1.9$\sigma$ and 1.5$\sigma$, while the MFs differ by $\sim$1$\sigma$ to 1.6$\sigma$. Thus, while the proposed
scenario is plausible, better statistics are required to reach higher statistical confidence.
Better statistics would likely require combining multiple star-forming regions, since the sample of Orion protostars is highly complete.

\subsection{Multiplicity of Class I and Flat Spectrum Protostars and Impact on Class Lifetime}

Throughout this paper we have discussed the multiplicity statistics of the different protostellar 
evolutionary classes, but we have not discussed whether multiplicity affects 
protostellar evolution between classes. We highlighted in Section 4.4 that Flat 
Spectrum protostars are more often found in binary or multiple systems than Class I protostars.
The MFs and CFs of Orion Flat Spectrum protostars are all larger than those of Class Is at all ranges of
separations. However, we do note that the differences are generally only 1-1.5$\sigma$, so the 
trend is not statistically robust. The more evolved Flat Spectrum protostars 
are not significantly different from the Class I protostars in terms of luminosity or 
0.87~mm flux density (though they are slightly fainter systematically; Paper I). The classification of
Perseus protostars does not distinguish between Class I and Flat Spectrum protostars \citep{enoch2009}, 
so it is unclear whether the difference exists there as well, but the numbers are smaller in Perseus,
so differences will be 
more uncertain.

Given the estimated total lifetime of the protostellar phase ($\sim$1~Myr; all class lifetimes combined), we do not expect 
that star formation conditions change significantly over 
this short timespan in a way that could affect the number of multiples formed
for Class Is (younger) relative to Flat Spectrum (older).
Therefore, we instead suggest that multiple star formation within a core itself
and/or the likely migration of companions (Figure \ref{multiplicity_evolution}) could speed up the transition between 
classes, resulting in a later evolutionary class having more multiples than the earlier
classes. This could happen in a few ways, some of which would happen separately and
some could be happening simultaneously, compounding the effect. 
First, outflows from two protostars within a dense 
core will be able to entrain and expel the dense gas faster. Second,
consider a protostar formed via turbulent
fragmentation that migrates closer to a companion. If the two protostars have misaligned outflows  \citep[e.g.,][]{yildiz2012,offner2016,lee2016,tobin2018}, 
the outflows can more efficiently clear envelope gas as their separation contracts. Furthermore,
the interaction of companions migrating in this way may cause the outflows to change
direction over time leading to even more rapid clearing of envelope material than 
two misaligned outflows whose directions are not changing. Third, protostars separated by a few~$\times$~100~au
(whether formed there or migrated to that location) will accrete from a larger volume of the
envelope, reducing the central envelope density more quickly than a single protostar. 
These protostars can also have misaligned outflows,
even if they both formed near those locations originally. Indeed, the systematically higher
luminosities for multiples (Figure \ref{lbol-tbol}) is evidence that 
accretion 
may be more rapid in multiple systems.

All these scenarios will impact the evolution of the gas distribution 
in the inner envelope, thereby reducing the material responsible for
emission between $\sim$10 to $\sim$160~\micron\ (from reprocessed shorter wavelength radiation from
the protostar and accretion). SED modeling of all the HOPS protostars found
that Flat Spectrum protostars have lower envelope densities than Class Is.
Thus, the SED slope from the near-infrared to the mid-infrared of multiple 
protostar systems could be intrinsically more shallow. 
This may be true whether the components of the multiple systems
are coeval or not \citep{murillo2016}.
We conclude by suggesting that multiplicity and the reorganization of multiple
systems that likely follow the Class 0 phase causes the protostar system to more rapidly evolve
through the Class I phase to become a Flat Spectrum protostar. 

In addition to the impact the multiples have on the envelope,
they will impact the disks as well. \citet{Harris2012} found that
disks toward Taurus multiples are detected less frequently than toward
single stars. Then, \citet{akeson2019} find that binary Class II systems
follow a similar dust disk mass to stellar mass relationship as single
Class II systems, but this relationship is shifted to lower disk masses.
Furthermore, the disk radii in close binaries are also found to be tidally
truncated \citep{manara2019}, consistent with theoretical predictions \citep[e.g.,][]{lubow1994}.
These findings indicate that the presence of a companion will likely
increase the rate of disk dispersal. Such a mechanism may explain the lack
of detection by ALMA toward a small number of Class I and Flat Spectrum
multiples that are detected by HST (see Section 3.6).

\subsection{Coevality of Multiples}

In this section, we examine the
co-evality of multiple systems in Orion by considering systems composed of different
protostellar classes. It is important to stress, however, that while class is used as a 
proxy for age, the absolute ages of protostar systems are unknown.
\citet{murillo2016} examined the coevality of multiples in Perseus, finding
that 33\% of multiple systems are likely non-coeval, as measured by the fraction of systems  with mixed
evolutionary classes. A limitation to this analysis, however, 
is that independent class identifications for individual members can only be made for
multiples with separations $>$10$^3$~au and companions $<$10$^3$~au are assumed to be in
the same class.

To estimate the number of systems with a non-coeval member, we count the relative number of multiple
systems that contain members with different evolutionary classes and those that contain members of the same class (see Table \ref{MF-CSFs-all}). 
Of the multiple systems identified, 42\% of Class 0 multiple systems have a non-coeval member compared
to 22\% of Class I systems, and just 4\% of Flat Spectrum systems have a non-coeval member. In total, the number of systems with a non-coveal
member is 25\%, which is comparable to results from Perseus.

One interesting point is that the majority of non-coeval systems are Class 0 systems. This implies that non-coeval systems are frequently associated with
the most recent epoch of protostar formation, and hence, the presence of dense gas. However, the 
estimated lifetimes of the Class 0, I, and Flat Spectrum phases are 
$\sim$0.15, 0.35, 0.4~Myr \citep{dunham2014}, so the estimated age difference
between Class 0 and Flat Spectrum systems is not large.
In Sections 5.1 and 5.2, we discussed in
how turbulent fragmentation most likely forms widely separated systems, which
migrate to different separations over time. Furthermore, the free-fall time of
gas with a typical dense core density of $\sim$10$^{5-6}$~cm$^{-3}$ \citep{bergin2007} is $\sim$0.05-0.15~Myr. Thus, it is plausible for
star formation to occur with asynchronicity over the size scale
of a star-forming core due to turbulent fragmentation, which can lead to
the appearance of non-coeval systems.

The apparent lack of non-coeval members within systems dominated by Flat Spectrum protostars
can be understood in a couple ways. First, star formation in regions dominated by Flat Spectrum protostars
may have largely finished. In addition, the Flat Spectrum lifetime is expected to be 
longer than the Class 0 lifetime, 
so the evolution of one member could catch up with the evolutionary state of other members.
Another possibility is that Flat Spectrum protostars 
have more time to migrate away from their birth sites and are more likely to leave a particular
system (e.g., Figure \ref{multiplicity_evolution}), potentially leaving systems with less Flat Spectrum members.
Thus, the various outcomes of turbulent fragmentation can naturally explain both the formation of
and the prevalence of apparent non-coeval systems in those dominated by Class 0 protostars, as well as the lack
of non-coeval Flat Spectrum systems.

Overall, taking the class assignments of protostars at face-value, our data also seem 
to support the idea of non-coeval formation of companions. However, a combination
of effects like gas configuration, envelope asymmetry, efficiency of gas removal
by outflows, and relative inclinations/orientations can lead to systems appearing
non-coeval, even when they are indeed coeval. Furthermore, accepting that some systems
are indeed non-coeval, the absolute
age differences that correspond of these systems are small
compared to the expected lifetimes of the stars. So whether a multiple system
was coeval or not (having the same evolutionary class at the observed time, by our definition) 
is not likely affect stellar evolution. 
Moreover, such small age
differences, like the difference in expected ages between Class 0 and Flat Spectrum
protostars ($\sim$0.35~Myr), will not impact later diagnostics of pre-main sequence
age like Lithium abundances and differences in location on pre-main sequence evolutionary
tracks will not be significant.

\section{Conclusions}

We have conducted a multiplicity analysis of protostars in the Orion A and B
molecular clouds using data taken with ALMA (0.87~mm) and the VLA (9~mm) toward 328 protostars
that have been classified with both \textit{Spitzer} and \textit{Herschel}. We have
also reanalyzed previously obtained data from the Perseus molecular cloud \citep{tobin2016a}
to enable a consistent comparison of these data and Orion.
Our main results are as follows:
\begin{itemize}
\item We have characterized the multiplicity fractions (MFs) and companion fractions (CFs) for Orion, Perseus, and the two samples combined, we also sub-divide the MFs and CFs by separation and protostellar class.
Using the full range of separations (20 to 10$^4$ au) the MFs of Orion and Perseus are 
0.30$\pm$0.03 and 0.38$\pm$0.07, respectively, and the CFs
are 0.44$\pm$0.03 and 0.57$\pm$0.07, respectively. The combined MF and CF within the
same separation range are 0.32$\pm$0.02 and 0.46$\pm$0.03, respectively.

\item We find that MFs and CFs for 20 to 10$^4$~au separations
decrease from Class 0 to Class I, and Flat Spectrum protostars
are consistent with Class I.
However, at separations less than 10$^3$~au Flat Spectrum protostars can have higher MFs and CFs
relative to Class Is in Orion as a whole and in each sub-region of Orion,
but the differences are only at the 1 to 1.5$\sigma$ level. If the difference is real this
could be evidence that
multiplicity speeds up evolution through Class I to Flat Spectrum.

\item We characterize the separation distributions for Orion and Perseus, both as a whole and 
divided by protostellar class. The overall and Class 0 samples appear bimodal with
peaks near $\sim$100~au and $\sim$3000~au. This bimodal
appearance is driven by the Class 0 sample, since the Class I and Flat Spectrum samples do
not appear bimodal on their own.
Thus, Class 0 protostars have the most companions between 20 to 500~au separations and
at $>$10$^3$~au. A statistical comparison between the Orion and Perseus separation
distributions does not reveal any statistically significant differences between the two regions.

\item Statistical evidence shows that the separation distributions for Orion Class 0s 
and Orion Class Is [Class 0--(Class 0, Class I, Flat) and Class I--(Class 0, Class I, Flat)], 
are inconsistent
with being drawn from a log-flat distribution of separations. 
Class 0 protostars [Class 0--Class 0 and Class 0--(Class 0, Class I, Flat)] 
are also statistically inconsistent with the separation distribution
for field solar-type stars. This indicates that the bimodal separation distribution 
apparent in histograms
may  be robust and that significant evolution of these separation
distributions is required to produce the field separation distributions for solar-type and M-type stars
from the Class 0 separation distributions. The separation distributions for Flat Spectrum protostars, 
however, are statistically consistent with the separation distribution of field
solar-type stars.

\item
We find that protostars with higher surrounding YSO densities have correspondingly higher MFs/CFs,
in agreement with
previous results from \citet{kounkel2016}.
We find that the MF and CF for Class 0 protostars on 20 to 10$^4$~au separations is higher
relative to Class I and Flat Spectrum protostars at both high and low YSO densities.
This is due the larger frequency of higher order multiples present in Class 0 protostars.
However, the MF differences are only at the 1$\sigma$ level, while the CF differences
are at the 1.5$\sigma$ to 1.9$\sigma$ level.
Some of 
these companions at $>$10$^3$~au are expected to migrate through
the 100 to 10$^3$~au range and contribute to the elevated MFs ($\sim$1$\sigma$ differences) and CFs (1.5$\sigma$ - 1.9$\sigma$ differences)
observed toward Class I + Flat spectrum protostars and Flat Spectrum protostars alone
on these scales in regions with high YSO densities.
Also, the MFs and CFs for Class 0 protostars showed no differences between high and low YSO densities
for the separation ranges of 20 to 500~au, 20 to 10$^3$~au, and 100 to 10$^3$~au. We therefore suggest
that the Class 0 MFs and CFs within these ranges of separation are primordial and the MF/CF increase 
of Class I + Flat Spectrum and Flat Spectrum protostars for 100 to 10$^3$~au separations 
could be due to migration.

\item We find more companions with separations from 20 to 500~au 
than predicted by numerical simulations that form multiples only via turbulent fragmentation
\citep[e.g.,][]{lee2019}. This 
suggests that both disk 
fragmentation and turbulent fragmentation with migration 
are needed to produce the observed population of close multiple systems.

\item The distribution of bolometric luminosities for single and multiple protostars
are found to have statistically significant differences. The median
luminosity of single protostars is found to be 0.96~\lsun, while the median
luminosity of multiple protostars is 3.27~\lsun. The higher luminosities could
result from both higher accretion rates in multiples and/or multiple protostars accreting at the
same rate. 

\item We compare near-infrared and millimeter detection statistics for those multiple systems
in Orion with overlapping observations with \textit{Hubble Space Telescope} observations at 1.6~\micron\
and ALMA and/or the VLA. Comparing the sample of protostars observed at both
near-infrared and radio/sub-millimeter wavelengths 
and in the same 100 to 10$^3$~au separation range. We find that the incompleteness of 
the millimeter-only studies is 20\% for samples of Class 0, Class I, and Flat Spectrum protostars,
and this rises to 29\% if the sample is limited to Class I and Flat Spectrum protostars.
The near-infrared is 31\% incomplete for samples of Class 0, Class I, and Flat Spectrum protostars,
but if only Class I and Flat Spectrum protostars are sampled, the incompleteness drops to 11\%. This
finding is in part due to the inability of HST to directly detect Class 0 protostars and many Class I
protostars at 1.6~\micron
due to their embedded nature.

\item Dust opacity at millimeter/submillimeter wavelengths can play a role in the ability
to detect companions with separations $<$200~au.
We find that three systems only have VLA (9~mm) detections of their companions that were undetected
by ALMA at 0.87~mm. The non-detection by ALMA is most likely due to the high 
optical depth of the surrounding dust emission,
obscuring the companion at shorter wavelengths.
There are further two tentative companions that are detected by the VLA but not by ALMA, likely
for the same reason.
Future studies must remain cognizant of this potential observational bias when observing at
submillimeter and millimeter wavelengths.
\end{itemize}

The statistics on protostellar multiplicity from this survey have
resulted in a robust characterization of the starting point for multiplicity evolution.
The data
indicate that similar processes produce the observed populations of protostellar
multiples in Orion and Perseus; however, it remains to be seen if all nearby star-forming regions and isolated cores have consistent multiplicity properties.
The most populous regions without comprehensive surveys
are Serpens/Aquila and the California Molecular Cloud. In addition to
obtaining greater statistics in the nearby regions at similar spatial scales, 
it is also important to probe closer separations in the protostellar phase in 
order to understand how close multiples form and
evolve. ALMA, in addition to future instruments like the ngVLA, SKA, and 30~m class telescopes 
will be essential for characterizing multiplicity at the closer separations.

\acknowledgements
The authors thank the anonymous referee for a constructive report that helped improve
the quality of the manuscript.
J.J.T. acknowledges support from  NSF AST-1814762 and HST-GO-15141.018-A.
SSRO acknowledges support from NSF Career grant 1748571.
LWL acknowledges support from NSF AST-1910364. G.A. and M.O. acknowledge support from MINECO (Spain)
AYA2017-84390-C2-1-R grant (co-funded by FEDER) and financial support from the State Agency for Research of the
Spanish MCIU through the ``Center of Excellence Severo Ochoa" award for the Instituto de Astrof{\'i}sica de
Andaluc{\'i}a (SEV-2017-0709). 
ZYL is supported in part by NASA NSSC18K1095 and NSF 1815784.
AS gratefully acknowledges funding support through Fondecyt Regular (project code
1180350) and from the Chilean Centro de Excelencia en Astrofísica y Tecnologías Afines (CATA) BASAL grant AFB-170002. 
This paper makes use of the following ALMA data: ADS/JAO.ALMA\#2015.1.00041.S.
ALMA is a partnership of ESO (representing its member states), NSF (USA) and 
NINS (Japan), together with NRC (Canada), NSC and ASIAA (Taiwan), and 
KASI (Republic of Korea), in cooperation with the Republic of Chile. 
The Joint ALMA Observatory is operated by ESO, AUI/NRAO and NAOJ.
The PI acknowledges assistance from Allegro, the European ALMA Regional
Center node in the Netherlands.
The National Radio Astronomy 
Observatory is a facility of the National Science Foundation 
operated under cooperative agreement by Associated Universities, Inc.
This research made use of APLpy, an open-source plotting package for Python 
hosted at http://aplpy.github.com. This research made use of Astropy, 
a community-developed core Python package for 
Astronomy (Astropy Collaboration, 2013) http://www.astropy.org.

 \facility{ALMA, VLA}
\software{Astropy \citep[http://www.astropy.org; ][]{astropy2013,astropy2018}, 
APLpy \citep[http://aplpy.github.com; ][]{aplpy}, CASA \citep[http://casa.nrao.edu; ][]{mcmullin2007}}

\appendix
\section{Use of Companion Probabilities in Multiplicity Characterization}

\subsection{Probability of a Detected Source being a Companion}
The multiple systems in our sample may be contaminated by chance alignments with other YSOs in the Orion cloud, with the probability of contamination increasing as separation increases and higher local YSO surface densities. 
We attempt to account for this effect in our algorithm as follows.
We begin by estimating the typical YSO surface densities. We used the  
surface density of YSOs measured toward each protostar position, using the 11th nearest
neighbor, and then dividing 10 by the area of a circle at the radius of the
11th nearest neighbor to determine the local YSO 
surface density \citep[$\Sigma_{YSO}$; e.g.,][]{gutermuth2005,gutermuth2011,megeath2016}.
The protostar, whose surrounding $\Sigma_{YSO}$ is being measured, is not included in the count
of the 11 nearest neighbors.
This results in an uncertainty of 33\% in the $\Sigma_{YSO}$ measurement \citep{casertano1985}; equations describing the
surface density estimator are provided in \citet{megeath2016}. The YSO surface densities for Orion 
and Perseus are provided with the input catalogs in Tables \ref{input-table-orion} and \ref{input-table-perseus}.

For the protostars in Orion, we used the catalog of probable YSOs from \citet{megeath2012},
and then for Perseus we used the YSO catalog of Perseus from the \textit{Spitzer} Extended 
Solar Neighborhood Archive (SESNA) catalog \citep[][Gutermuth et al. in prep.]{pokhrel2020}. If there was no source in
the catalog within 2\farcs5 of an ALMA or VLA detection, the ALMA or VLA source was added to the catalog
for computation of the YSO surface density.
The YSO surface densities of Orion were corrected for completeness using 
X-ray catalogs when available \citep{megeath2016}; the Perseus catalog did not undergo such a correction. 
For cases where a YSO was more isolated or did not have enough data of its surrounding 
region, we adopted the nearest measured YSO surface density value in our sample.

The YSO surface density toward each protostar provides an expectation value for the number of
YSOs present within a given area around the protostar. Since nearly all the protostars in Perseus
and Orion are contained within our samples, contamination is expected to come from disk-bearing
YSOs. Based on our survey dust mass sensitivity of $\sim$1~\mearth, compared to deeper surveys
of YSOs in Lupus \citep{ansdell2016}, we would detect
75\% of disk-bearing YSOs.
The detection probability follows a Poisson distribution
\begin{equation}
P(k) = \frac{\lambda^k e^{-\lambda}}{k!}
\end{equation}
where $\lambda$ = 0.75$\Sigma_{YSO}$ $\pi d^{2}$ and $k$ is the number of YSOs
expected. $\Sigma_{YSO}$ is the number of YSOs per square parsec, and $d$ refers
to the projected sky radius considered for calculating the probability of a random YSO. Thus,
the probability of detecting $\ge$~1 unassociated YSOs is 1 - P(0), where
\begin{equation}
P(\ge 1~YSO) = 1.0 - e^{-0.75 \Sigma \pi d^2}.
\end{equation}

We ultimately need to determine the probability of whether 
a detected source is a companion or not. For this we make use of 
Bayes Theorem to determine 
\begin{equation}
P(companion|detection) = \frac{P(detection | companion) P(companion)}{P(detection)}. \label{bayes}
\end{equation}
Simply stated, given a detection, Equation \ref{bayes} yields the
probability of that detection being a companion. $P(detection|companion)$ refers to the 
probability that we would indeed detect a companion when there is one present.
We assume that $P(detection|companion)$ = 0.75, because, as mentioned above, 
our sensitivity is expected to detect 75\% of YSOs, which also applies to protostars, under
the assumption that they are drawn from the same general population of dusty disks.
We find that if we used values anywhere between 0.25 to 0.9, the probability of a 
given continuum detection being a companion is not significantly changed; in any 
event 0.75 is likely a reasonable estimate.

$P(detection)$ is the likelihood that we detect a source 
regardless of it being a companion
or an unassociated source.
Thus, we have
\begin{equation}
P(detection) = 0.75~P(companion) + (1-e^{-0.75 \Sigma \pi d^2})(1-0.75~P(companion)).
\end{equation}
The first part of the equation, $0.75~P(companion)$, is the probability of detecting a companion, 
while the latter part of the equation is the probability of finding a 
detection if there is no companion (i.e., an unassociated source). 
The factor of $(1-0.75~P(companion))$ is important, because this will balance
the probability of detection in the event that $\Sigma$ is very large. If we did not
include the 0.75 factor, $P(detection)$ would always be less than 1 when $\Sigma$~$\rightarrow$~$\inf$
and $P(companion)$~$>$~0.
We thus require an estimate of $P(companion)$, and we must use the companion 
fraction (CF) computed from our data to do this (see Section 2.4.3). Our
data without accounting for contamination by unassociated YSOs will provide 
an overestimate of $P(companion)$, so we must then recompute $P(companion)$
using the CF derived with contamination taken into account. This is an iterative process since the CF will
slightly decrease when contamination is considered, and we recompute 
our multiplicity statistics for both Orion and Perseus
until the multiplicity fraction (MF) and CF (Section 2.4.3) are changing by less than 0.005 relative to the previous calculation. We find that running the multiplicity statistics three times
is sufficient to reach convergence, such that the calculated MFs and CFs are no longer changing. 
$P(companion)$ depends on separation, so for separations less than 500, $10^3$, and $10^4$~au,
$P(companion)$ = 0.14, 0.19 and 0.46, respectively; these values reflect the combined sample of Orion
and Perseus. Only the value of $P(companion)$ for separations out to $10^4$~au changes
with iteration because contamination does not significantly affect separations less than $10^3$~au.

We use the calculation of $P(companion|detection)$ as the probability of the companion being a real companion vs. a
line of sight association. If a system has a probability below 0.001, then it is not
included as a potential multiple because the probability is too low. The probabilities between 0.001 and 1.0 
are also utilized in the histograms of companions versus separation and in the cumulative distributions. 
We also use the rounded sum of the probabilities for the entire multiple system to 
determine the degree of multiplicity (binary, triple, etc.), this is further described in Section 2.4.3.

\subsection{Comparing Separation Distributions}

The distribution of projected companion separations is one of the key observables from multiplicity
studies of protostars. Thus, we compare the distributions of Orion and Perseus, sub-regions within Orion,
and between evolutionary Classes to determine if they are 
inconsistent with each other.
While histograms are used to create demonstrative plots of the separations where most companions
are found, statistical tests are typically performed using cumulative distribution functions (CDFs).
The CDFs we create in this paper use the companion probabilities described in the
preceding section.

To create the CDFs with companion probabilities, we begin with a list of companion probabilities 
associated with the separation of each component in a multiple system defined as 
\begin{equation}
[P_A,P_B,P_C,...,P_Z].
\end{equation}
For the multiple system shown in Figure \ref{example}, we would have the following probabilities associated with each
component
\begin{equation}
[P_A,P_B,P_C,P_D] = [1.0,P_{A,B},(1.0\times P_{AB,CD}), (P_{C,D}) \times P_{AB,CD})].
\end{equation}
Thus, for the three separations that are defined as part of this multiple system, they would have probabilities of
\begin{equation}
[P_{A,B},P_{C,D},P_{C,D}\times P_{AB,CD}]
\end{equation}
for the associated separations
\begin{equation}
[d_{A,B},d_{C,D},d_{AB,CD}].
\end{equation}

The CDF at separation $d$ is defined as
\begin{equation}
CDF(d)=\frac{\sum_{n=0}^{d} P_n}{\sum_{n=0}^{N} P_n}.
\end{equation}
The numerator is the cumulative sum of the companion probabilities for each companion 
within distance $d$ and the denominator is 
the sum of the probabilities for all separations. This method reduces to a 
standard CDF if all the probabilities are unity. While this creates a CDF that is appropriate for visualization,
it is not appropriate for statistical comparison between different subsamples, or other CDFs (measured or analytic).

We statistically compare
the distributions using a Kolmogorov-Smirnoff (KS) test. 
The KS test enables us test the null hypothesis that
the two distributions being compared are drawn from the same parent distribution.
However, due to each separation in the CDFs having a probability associated 
with it, we could not simply run a single KS test
of our companion separations CDF against another sample or an analytic CDF. We instead
construct a randomly sampled CDF from the full list of possible separations. For each
separation in the full list, we draw a random number between 0 and 1 from a uniform distribution,
and if a separation probability is greater than the random
number, it is included in the CDF, otherwise it is excluded. The KS test is then run using
the randomly sampled CDF and the reference CDF, then the likelihood is recorded. We then repeat this
random CDF sampling and KS test 1000 times, recording the median likelihood, the quartiles
of the distribution, and the fraction of realizations that have likelihoods in excess of our cutoff
value for significant rejection of the null hypothesis that the two distributions are drawn from the
same parent distribution. We consider low median likelihoods (p~$\le$~0.01) resulting from the KS test to be 
evidence that the two populations of multiples are not drawn from the same parent distribution,
rejecting the null hypothesis.

In the case of comparing an analytic CDF to an observed CDF, we made use of the
one-sided KS test and we used the two-sided KS-test when directly comparing
observed samples. We make use of the \textit{scipy} implementation of the
KS test. We also examined the results using a two-sided KS test for comparisons with an analytic function
where we created our own CDF of the analytic function. The one-sided KS test agrees with the two-sided
KS test when the number of samples in the analytic CDF are much larger.

\subsection{Multiplicity Statistics Reporting Per Protostar Class}

\begin{figure}
\begin{center}
\includegraphics[scale=0.4]{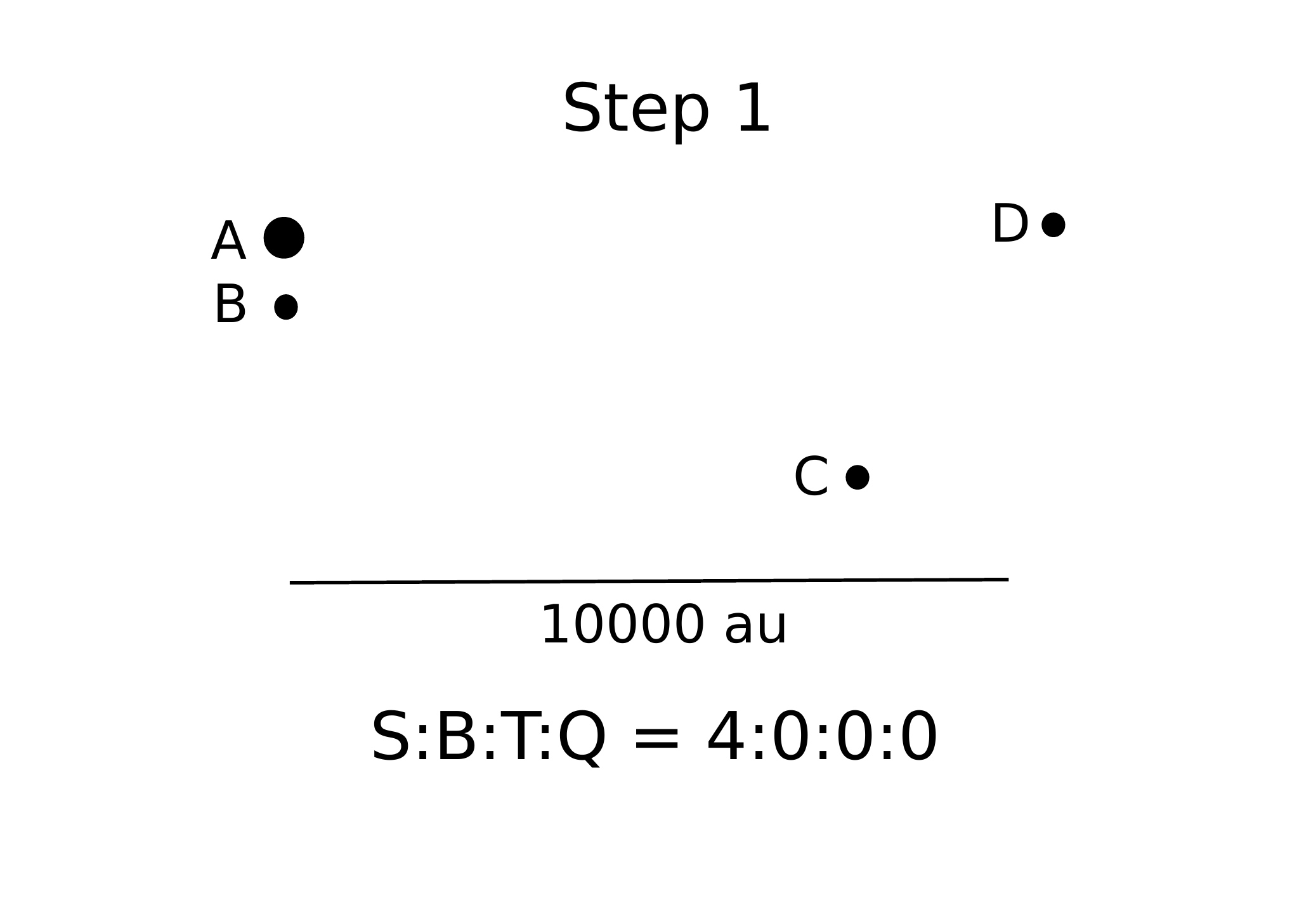}
\includegraphics[scale=0.4]{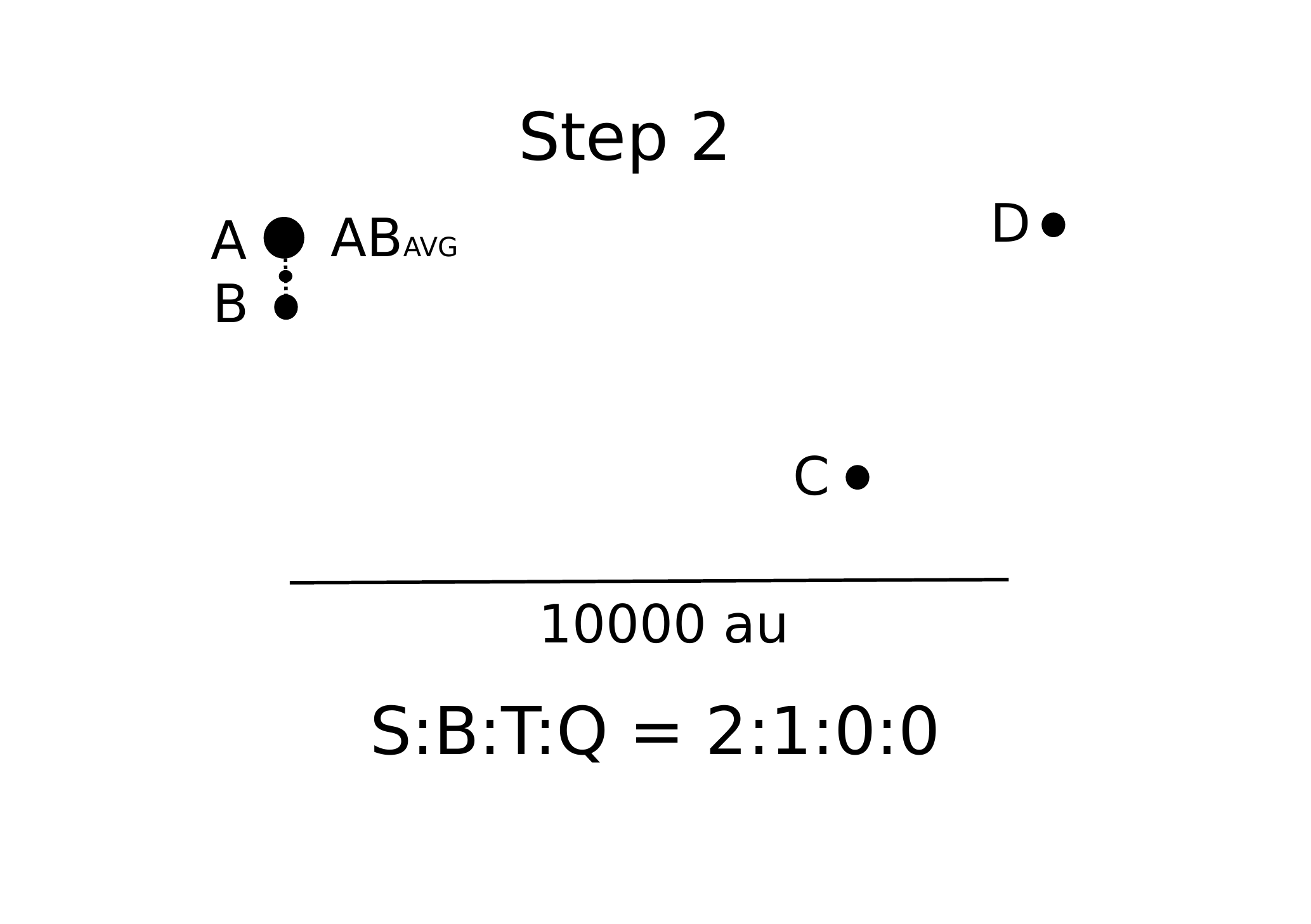}
\includegraphics[scale=0.4]{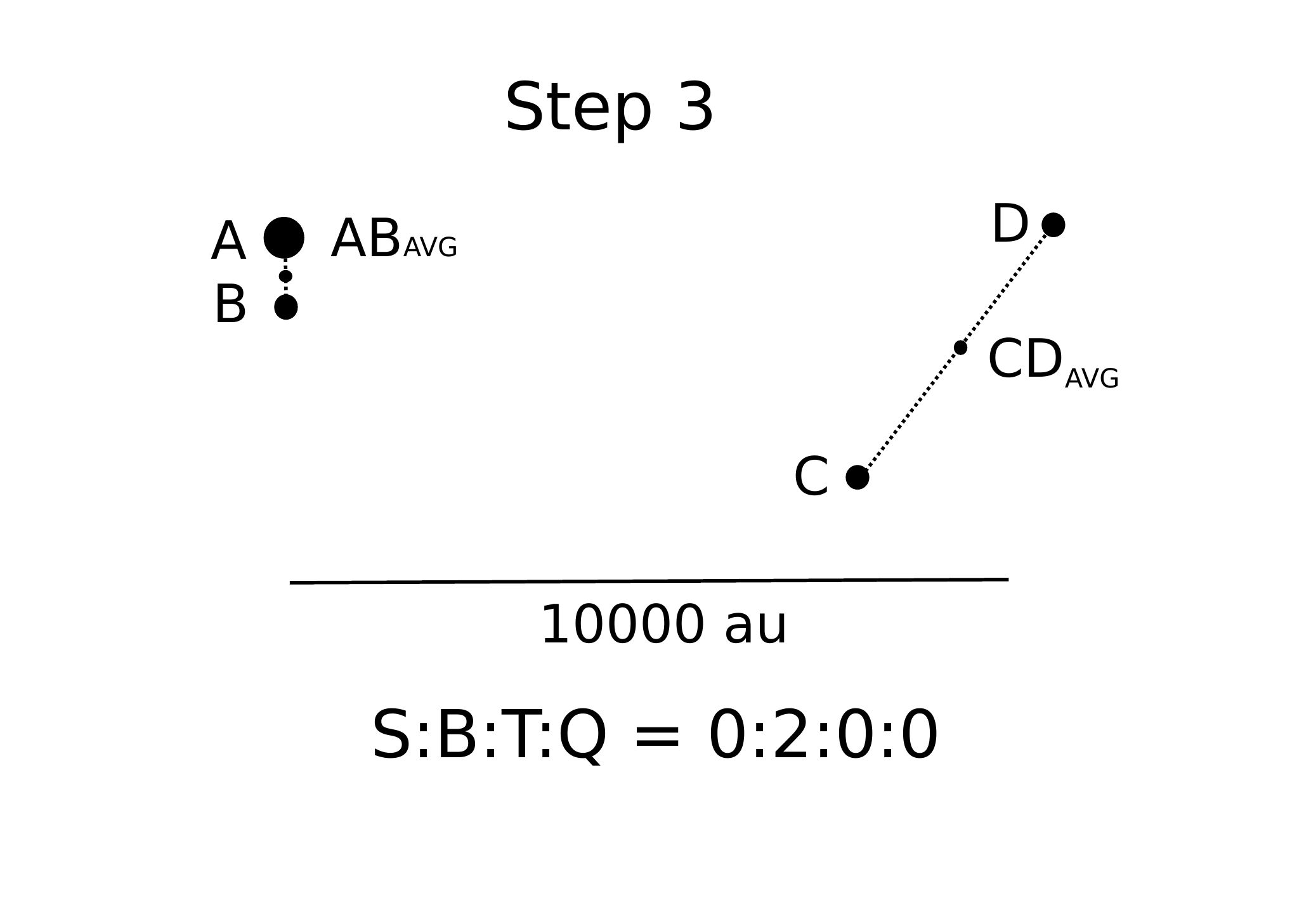}
\includegraphics[scale=0.4]{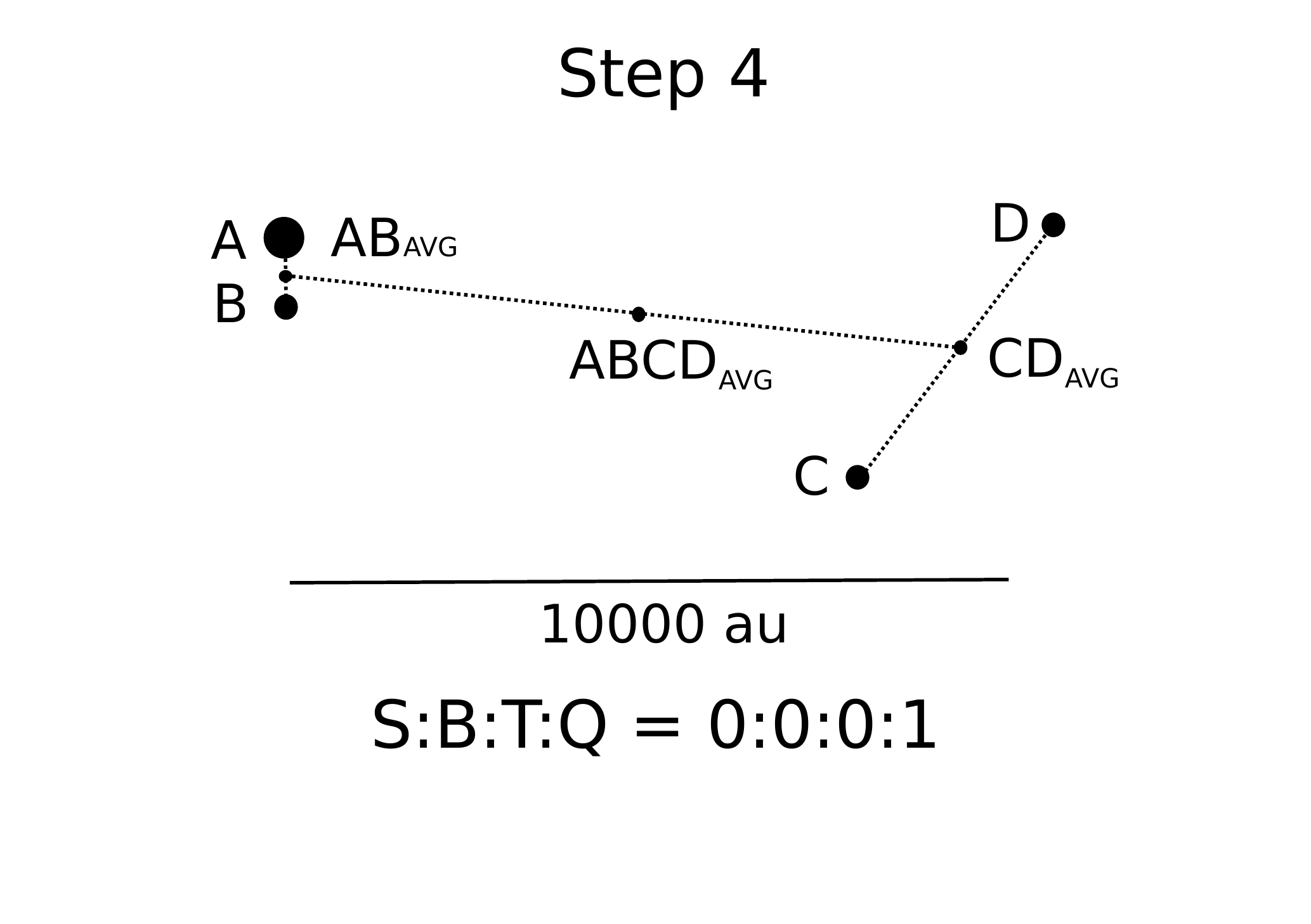}
\end{center}
\caption{Graphical demonstration of the construction of multiple systems from our method.
The size of the dot is 
proportional to the luminosity of a component, but this quantity is not used.
Our modified nearest neighbor method iteratively searches for companions at progressively larger radii.
Once a pair is created, the position is updated to be the average position and additional
companions are searched for relative to the average position of the pair. 
The pairing AB is found first since 
they have the closest separation, then CD is found later, and finally, the average positions 
for AB and CD are found to be associated, forming a quadruple system ABCD. Thus, 
component D is part of the system, despite having a distance 
greater than 10$^4$ au from A, because the average
position of CD is within 10$^4$ AU of AB.
A more conventional approach would assign
A to be the primary, creating pairings (A,B), and A,C); (A,D) would be ignored because the separation
is greater than 10$^4$~au. The S:B:T:Q at the bottom of each panel refers to how we could
count the individual continuum sources in our multiplicity statistics (single, binary, triple, quadruple) that comprise this 'system' depending on what separation range the statistics are being considered.
}
\label{example}
\end{figure}

\begin{figure}
\begin{center}
\includegraphics[scale=0.4]{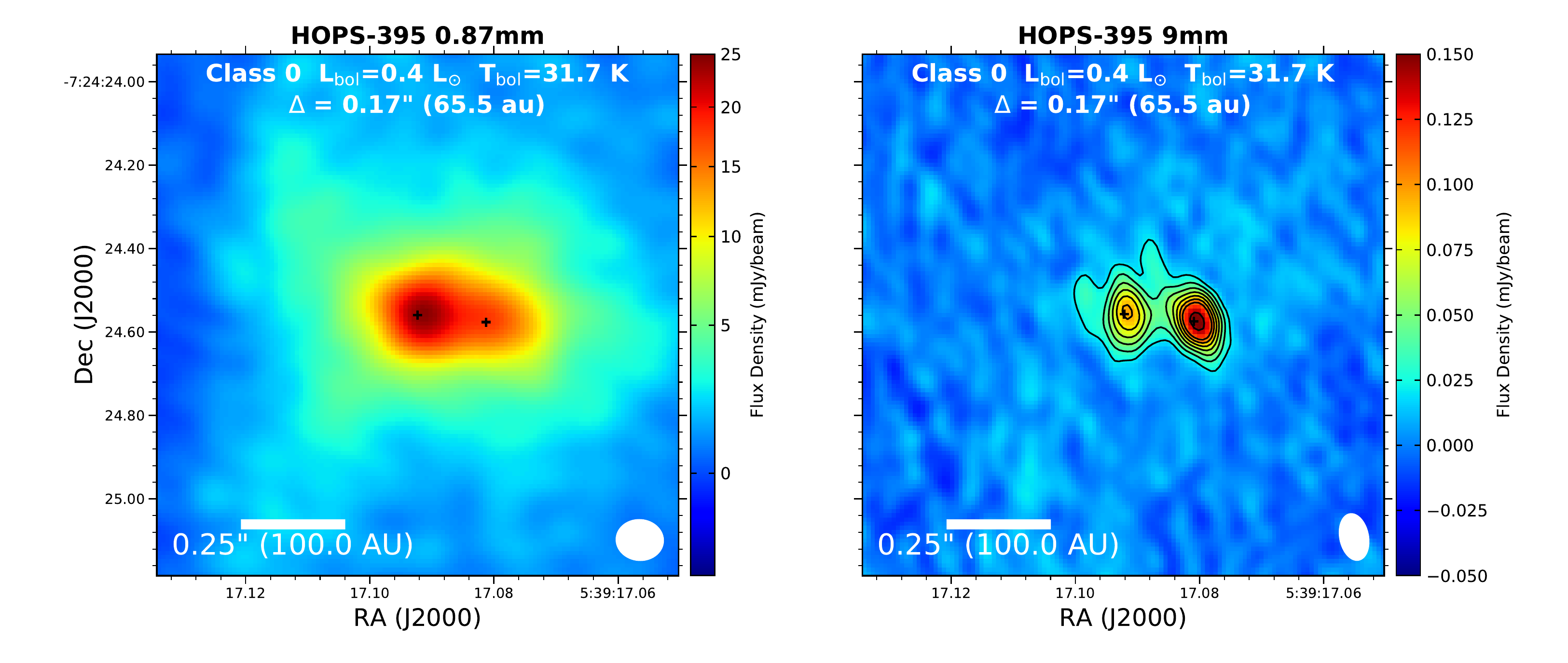}
\includegraphics[scale=0.4]{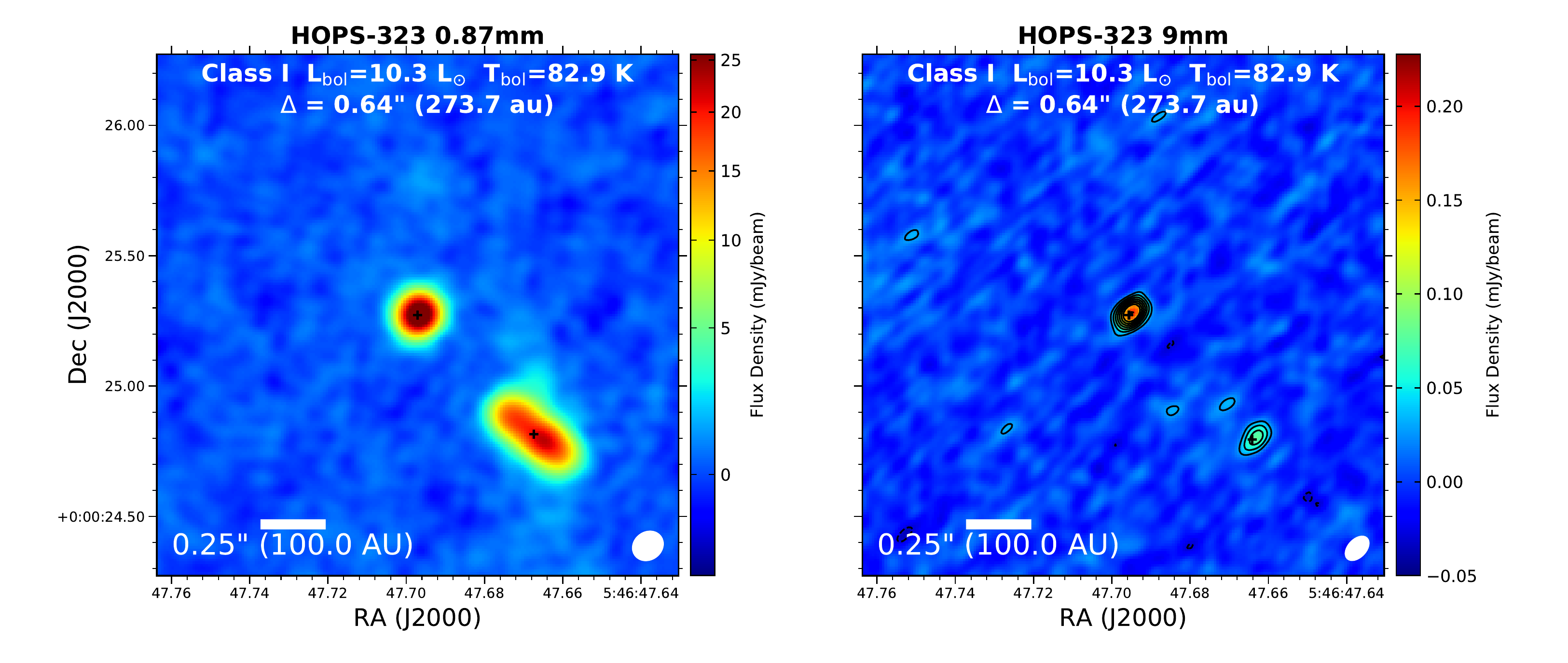}
\includegraphics[scale=0.4]{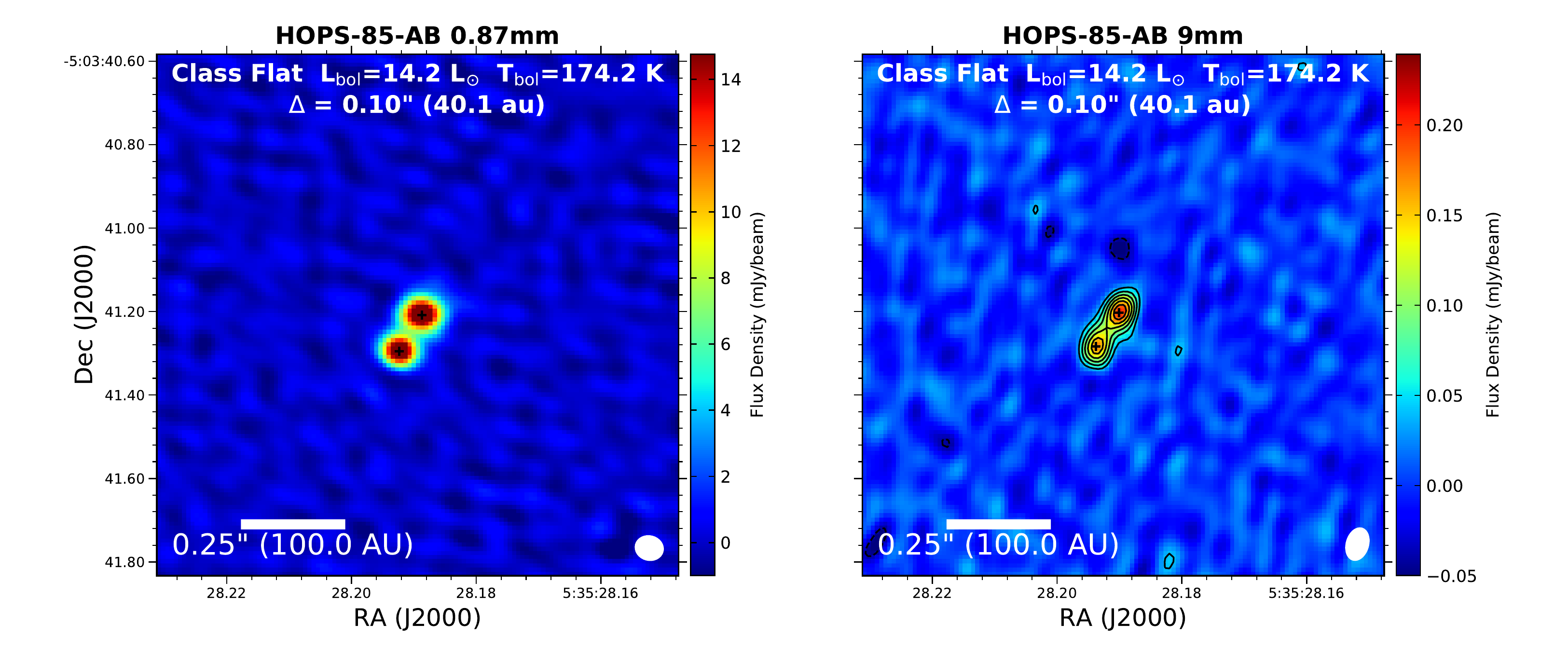}

\end{center}
\caption{
Examples of Class 0, Class I, and Flat Spectrum multiples from Orion; the ALMA images are shown on the
left, and the VLA images are shown on the right. The Class 0 and Class I
images are reproduced from Paper I, but the HOPS-85 images are produced
with superuniform weighting (ALMA image, left) and robust=0 weighting 
(VLA image, right) showing the binary more clearly
than the images from Paper I. HOPS-395 may have a shared inner disk or envelope surrounding the binary, 
the southern component of HOPS-323 appears to be an edge-on disk, and we only  detect 
the individual circumstellar disks
toward HOPS-85. The contours on the HOPS-85 VLA image start at $\pm$3$\sigma$ and increase
by $\pm$2$\sigma$, where $\sigma$=11.3~$\mu$Jy/beam. Black crosses mark the source positions from
the Gaussian fitting to the VLA data. The ALMA and VLA beam sizes are $\sim$0\farcs11 and $\sim$0\farcs08, 
respectively, while for HOPS-85 the respective beams are $\sim$0\farcs08 and 0\farcs06.
}
\label{continuum_multiple}
\end{figure}

\begin{figure}
\begin{center}
\includegraphics[scale=0.35]{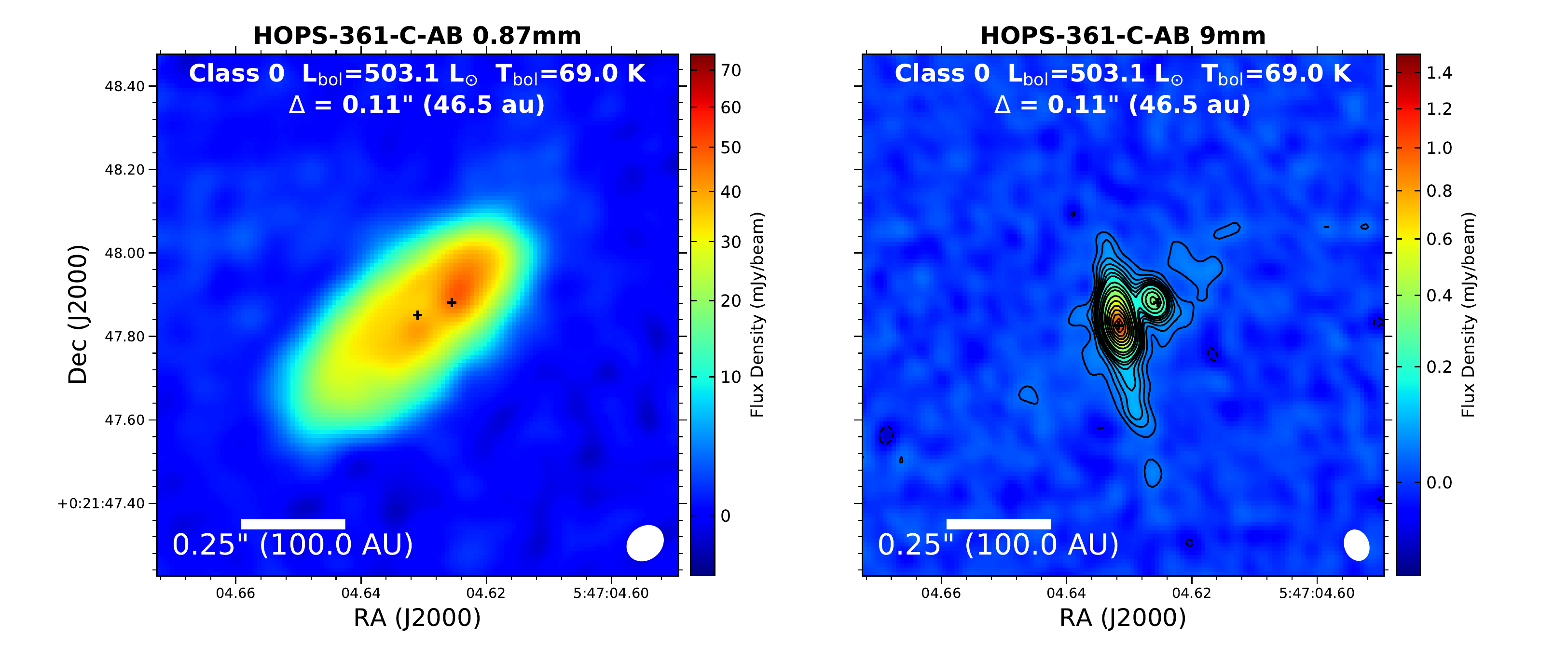}
\includegraphics[scale=0.35]{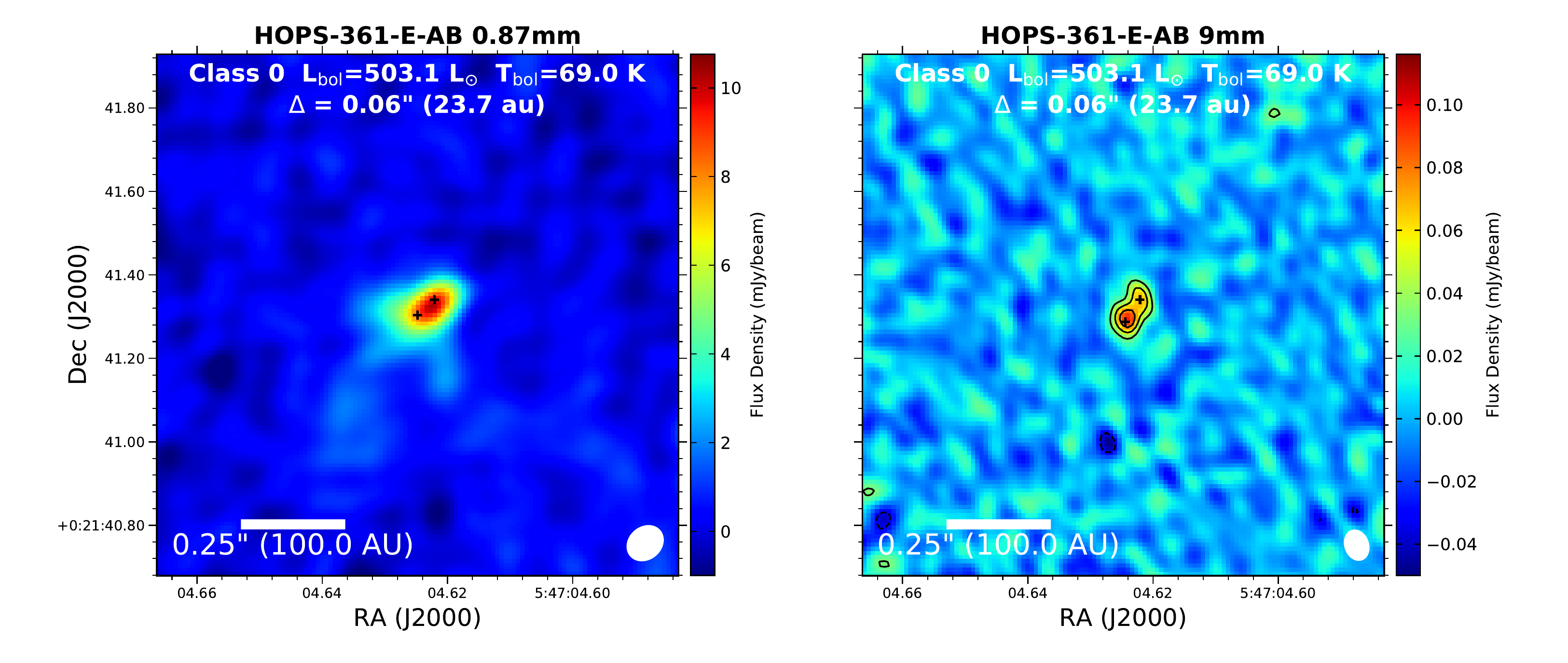}
\includegraphics[scale=0.35]{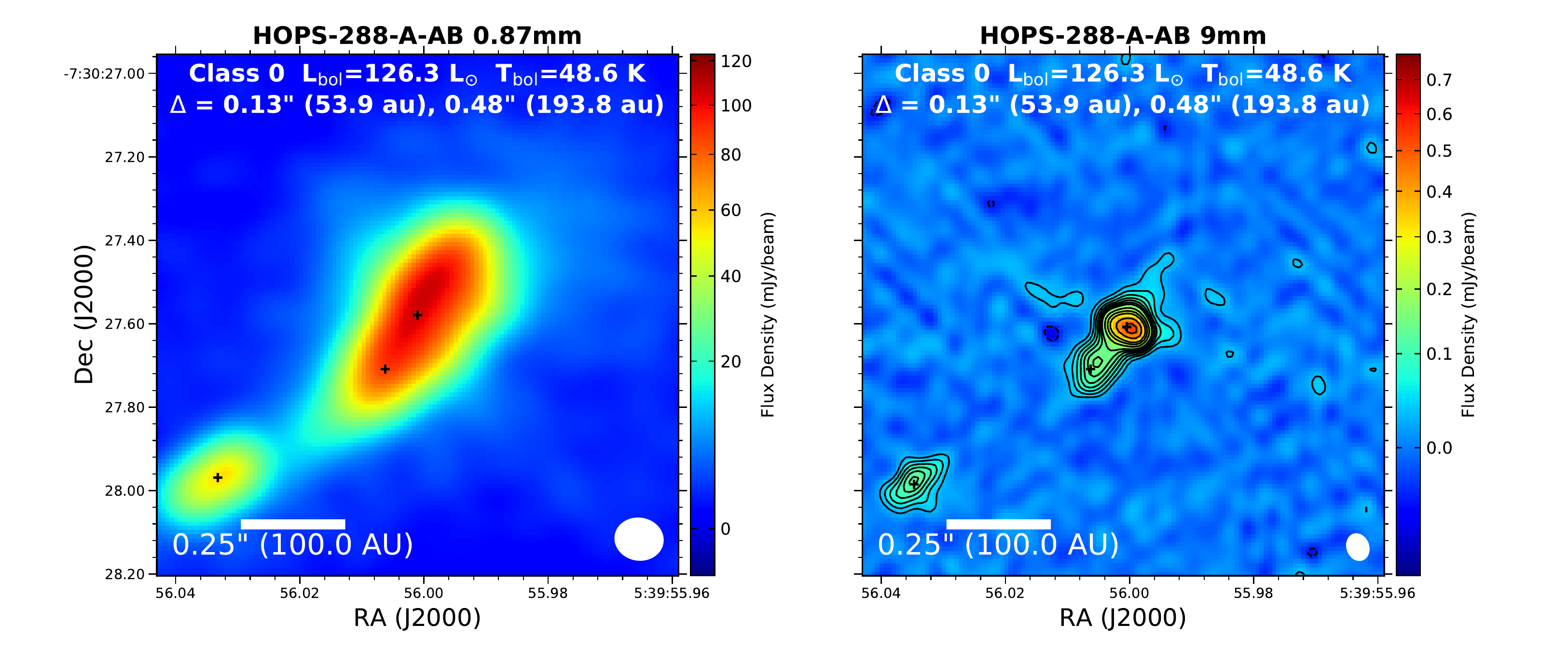}
\includegraphics[scale=0.35]{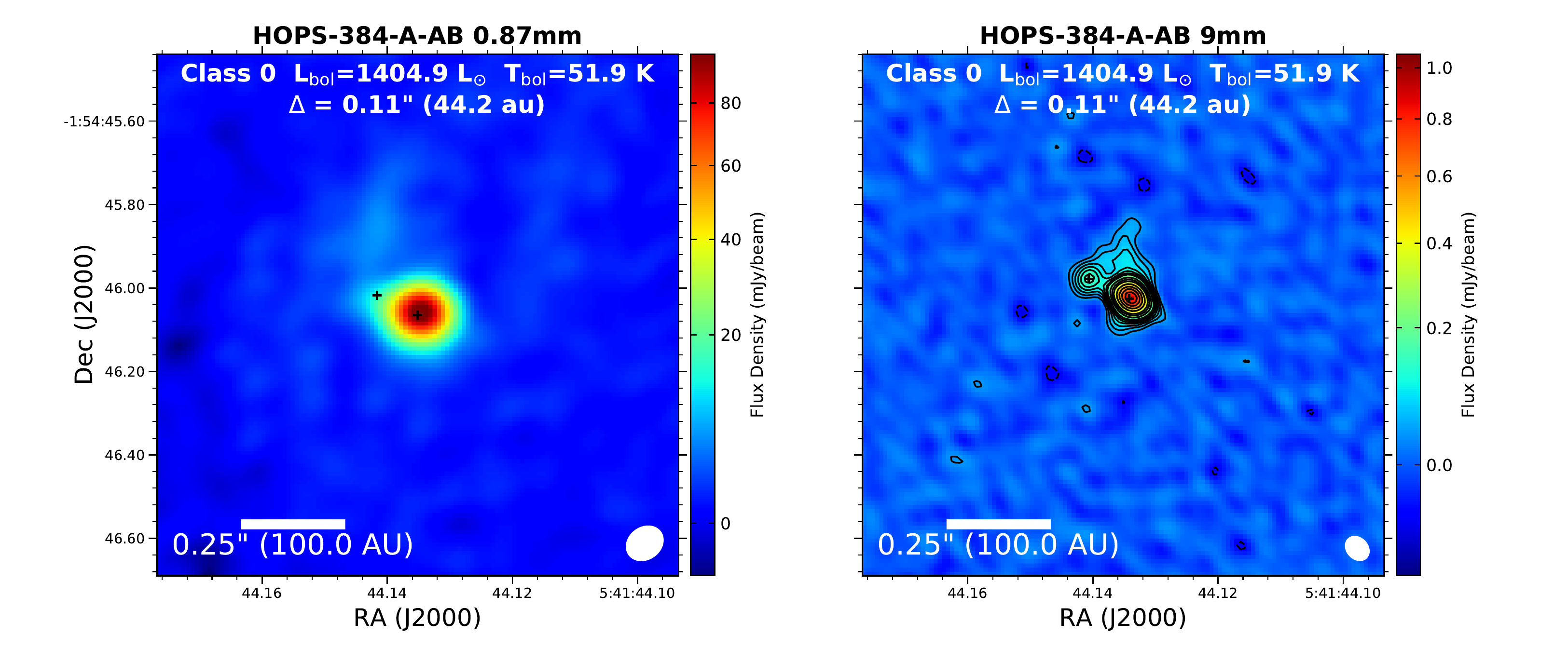}
\end{center}
\caption{
Multiple systems that are not detected by ALMA but are resolved by the VLA imaging. The
lack of ALMA detections may be a combination of resolution and dust opacity.
The VLA images are produced with robust=0 weighting. The contours start at 3$\sigma$, increase
by 2$\sigma$ until 15$\sigma$, then increase by 5$\sigma$ until 30$\sigma$, and then
increase by 10$\sigma$. The values for $\sigma$ are 10.8, 10.0, and 10.9~$\mu$Jy/beam for 
HOPS-361, HOPS-288, and HOPS-384, respectively. Black crosses mark the source positions from
the Gaussian fitting to the VLA data. The ALMA  and VLA beam sizes are $\sim$0\farcs11 and $\sim$0\farcs08, 
respectively.
}
\label{continuum_multiple_vla}
\end{figure}

\begin{figure}
\begin{center}
\includegraphics[scale=0.4]{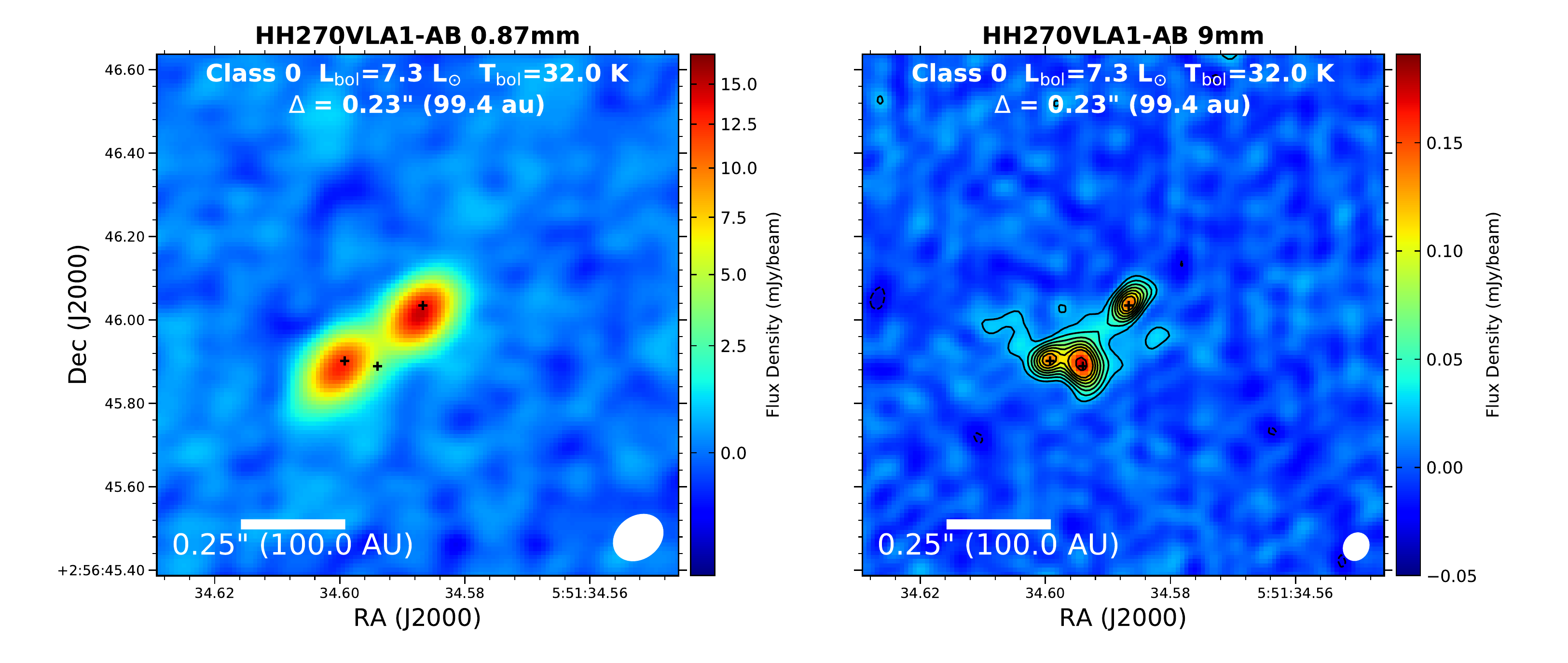}
\includegraphics[scale=0.4]{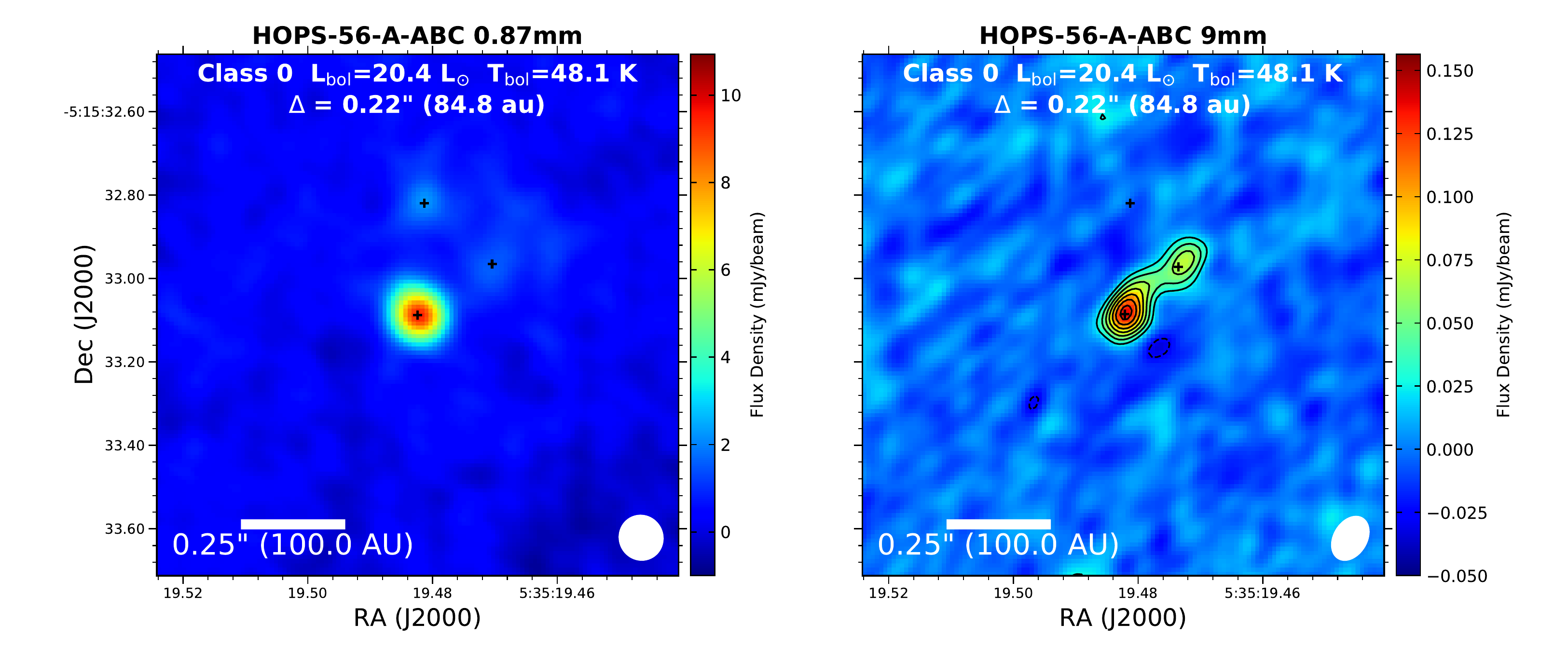}
\end{center}
\caption{
ALMA and VLA continuum images of multiples detected by VLA and ALMA, but not all sources
are present at both wavelengths. Three sources of continuum emission are detected with ALMA for HOPS-56, but only two
are detected by the VLA. All three are considered to be companions since the VLA is less sensitive to dust
emission and the non-detection does not rule out one of the sources. For HH270VLA1 on the other hand, three
sources are detected by the VLA and only 2 sources are detected by ALMA. The source not detected by ALMA
is likely an outflow shock feature or a background AGN because of its strongly negative spectral index. 
The contours start at 3$\sigma$ and increase
by 2$\sigma$ until 15$\sigma$, then increase by 5$\sigma$ until 30$\sigma$, and then
increase by 10$\sigma$. The values for $\sigma$ are 8.1 and 8.5~$\mu$Jy/beam for HH270VLA1 and HOPS-56, respectively.
The black crosses mark the source positions from
Gaussian fitting to the VLA data for HH270VLA1 and the ALMA data for HOPS-56.
The ALMA and VLA beam sizes for HH270VLA1 are $\sim$0\farcs11 and $\sim$0\farcs06, 
respectively, and for HOPS-56 the respective beam sizes are $\sim$0\farcs11 and $\sim$0\farcs08.
}
\label{continuum_non_multiple}
\end{figure}

\begin{figure}
\begin{center}
\includegraphics[scale=0.4]{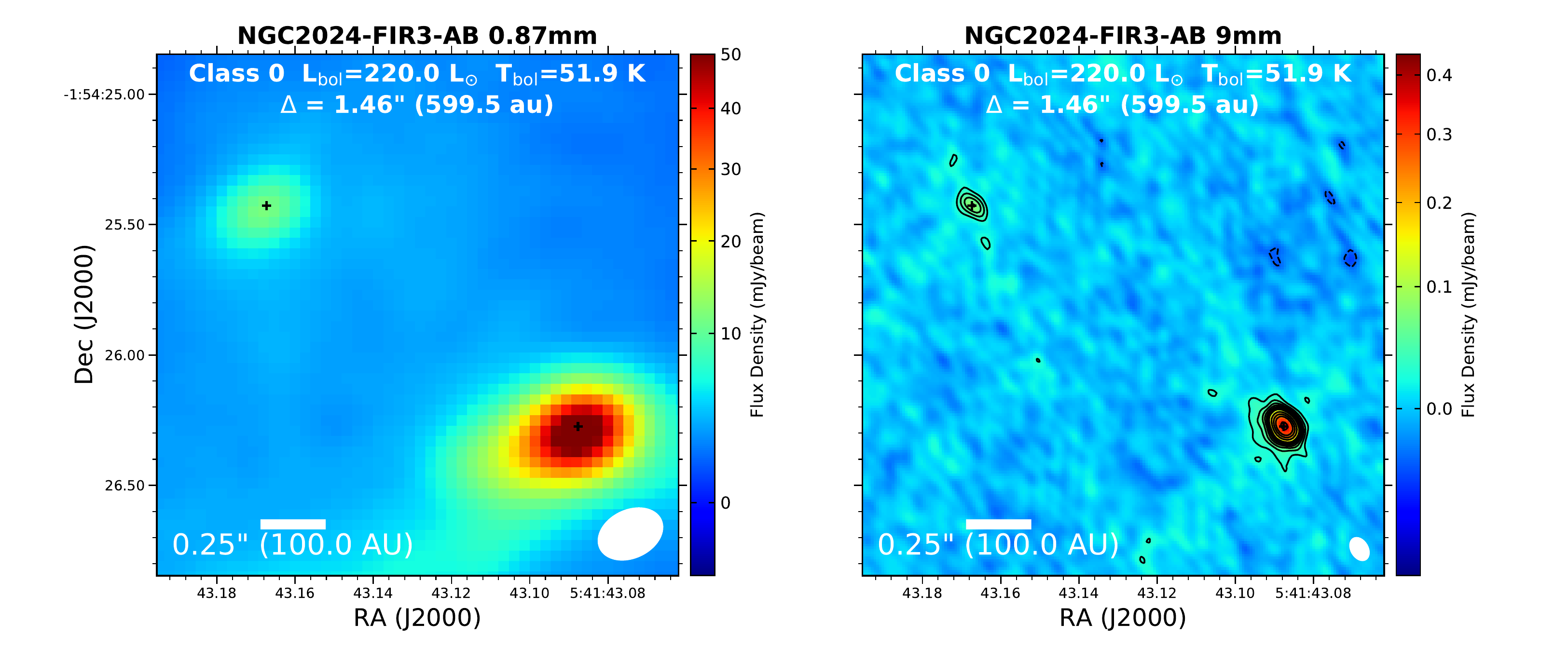}

\end{center}
\caption{
ALMA (left) and VLA (right) continuum images of the Class 0 system NGC 2024 FIR 3. The ALMA image
was not observed by our program, but is an ALMA 1.3~mm image 
from
\citet{vanterwisga2020}.
The contours start at 3$\sigma$ and increase
by 2$\sigma$ until 15$\sigma$, then increasing by 5$\sigma$ until 30$\sigma$, and then
increasing by 10$\sigma$. The value for $\sigma$ is 8.5~$\mu$Jy/beam.
The black crosses mark the source positions from
Gaussian fitting to the VLA data.
The ALMA beam is $\sim$0\farcs25, and the VLA beam is $\sim$0\farcs08.
}
\label{continuum_new_wide_multiple}
\end{figure}

\begin{figure}
\begin{center}
\includegraphics[scale=0.4]{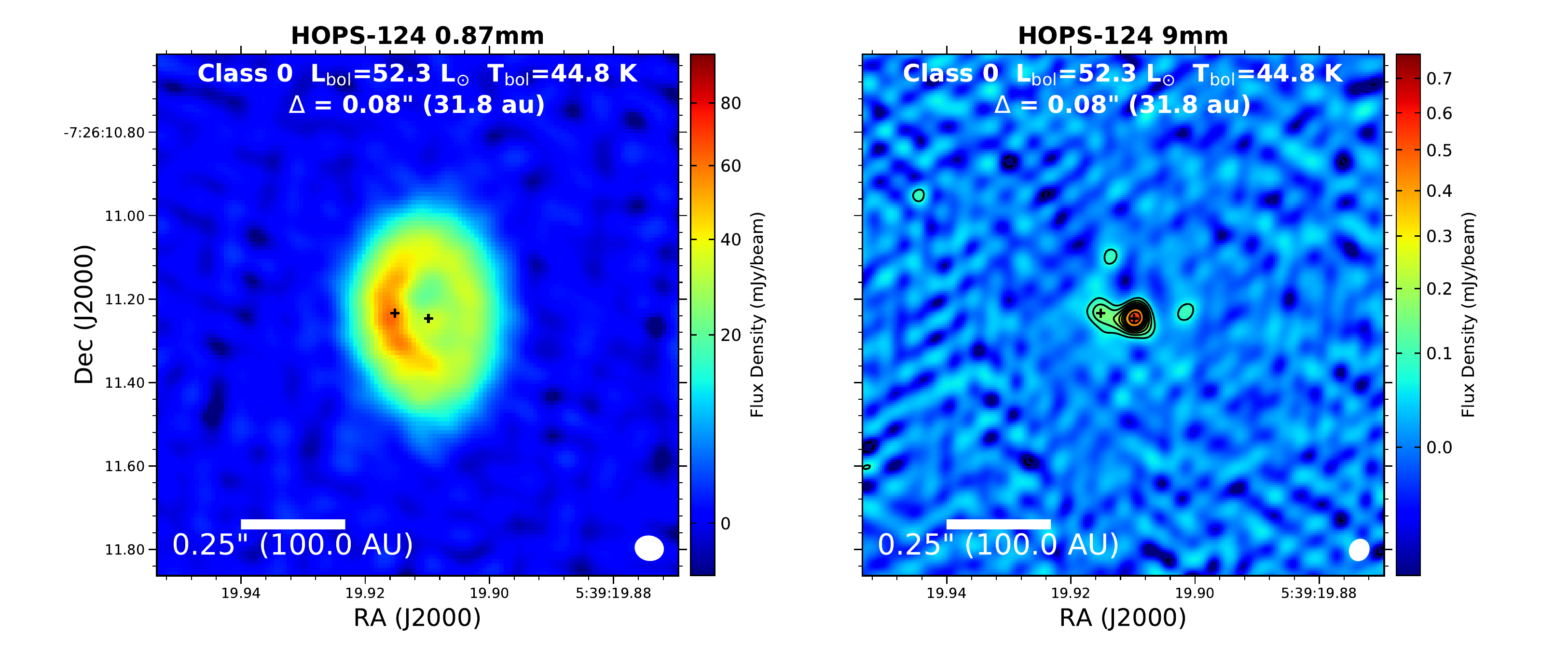}
\includegraphics[scale=0.4]{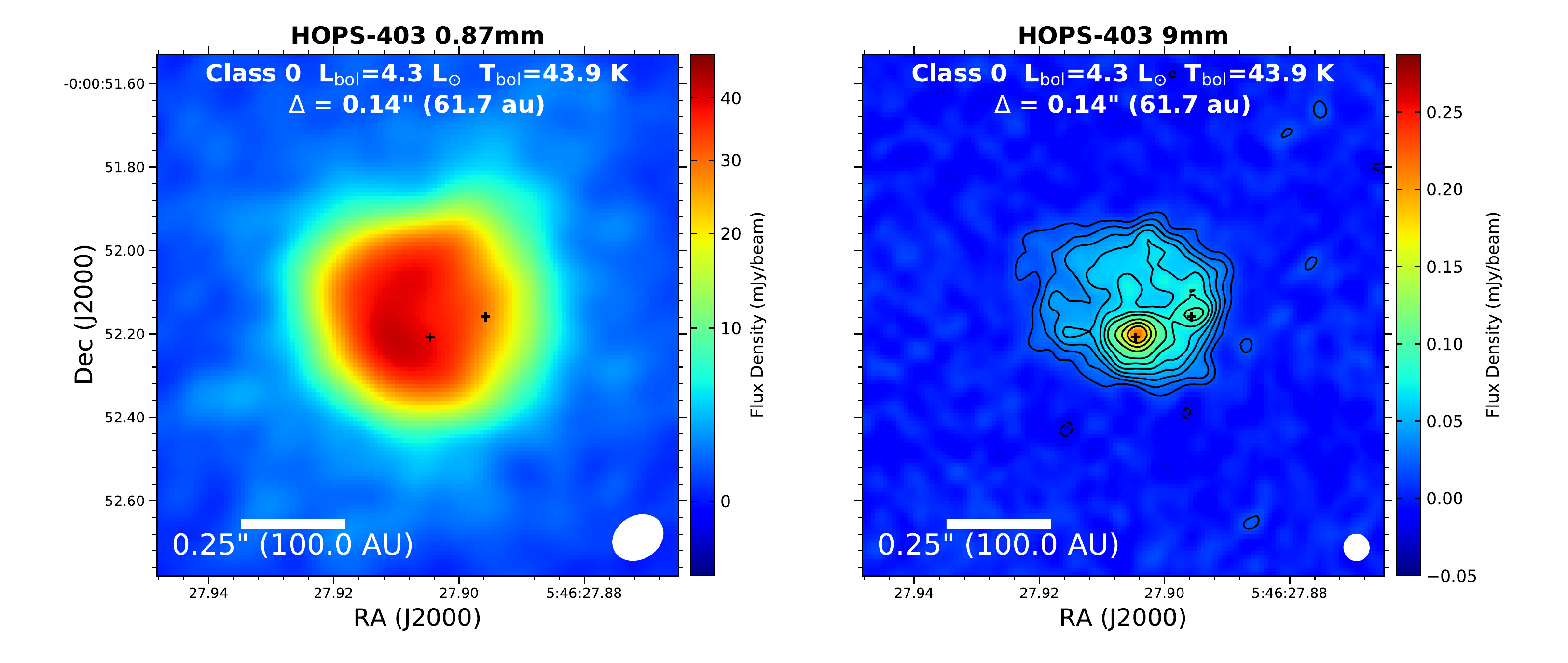}

\end{center}
\caption{
ALMA and VLA continuum images of possible multiples detected in VLA imaging but not
in the ALMA imaging. These are classified as possible multiples due to the low amplitude of their
peak intensity above the noise, so they are not included in the 
multiplicity statistics. In the case of HOPS-124 (top row), the possible companion
is also detected in the outflow direction, corresponding to the near center of the crescent-shaped
dust emission from the ALMA image. The ALMA HOPS-124 image is produced with superuniform weighting,
while the VLA image is produced with robust=-1 and only for the 8.1~mm portion (the highest frequency half of the VLA data) of the data.
The 0.87~mm continuum emission for HOPS-403 is likely optically thick \citep{karnath2020}, which is
probably the reason for the lack of detection toward the two peaks by ALMA.
In the VLA images, the contours start at 3$\sigma$, increase
by 2$\sigma$ until 15$\sigma$, then increase by 5$\sigma$ until 30$\sigma$, and then
increase by 10$\sigma$. The values for $\sigma$ are 6 and 24~$\mu$Jy/beam for HOPS-403 and HOPS-124, respectively.
The black crosses mark the source positions from
Gaussian fitting to the VLA data.
The ALMA and VLA beam sizes for HOPS-124 are $\sim$0\farcs09 and $\sim$0\farcs06, 
respectively, and for HOPS-403 the respective beam sizes are $\sim$0\farcs11 and $\sim$0\farcs07.
}
\label{continuum_possible_multiple}
\end{figure}

\begin{figure}
\begin{center}
\includegraphics[scale=0.3]{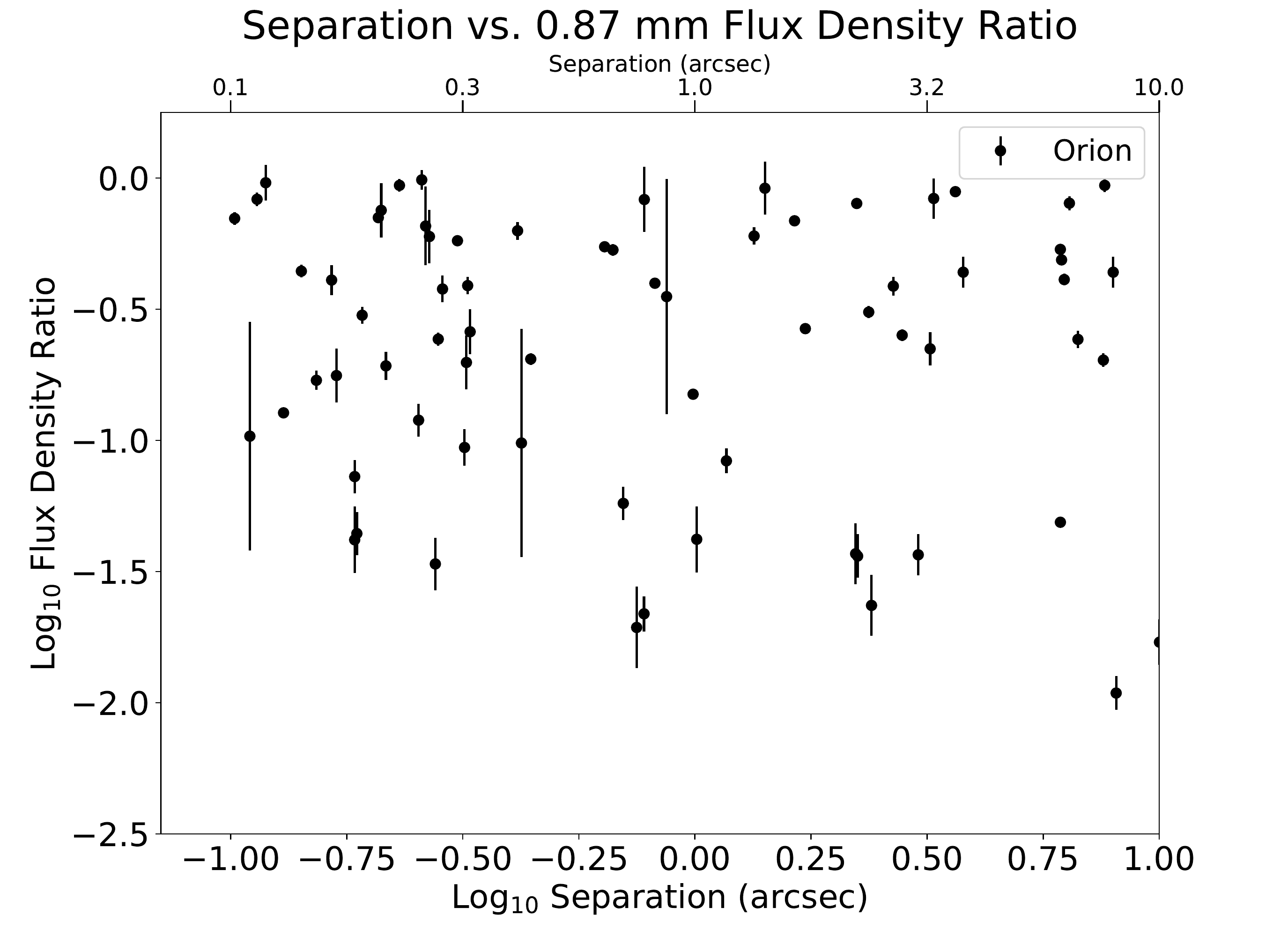}
\includegraphics[scale=0.3]{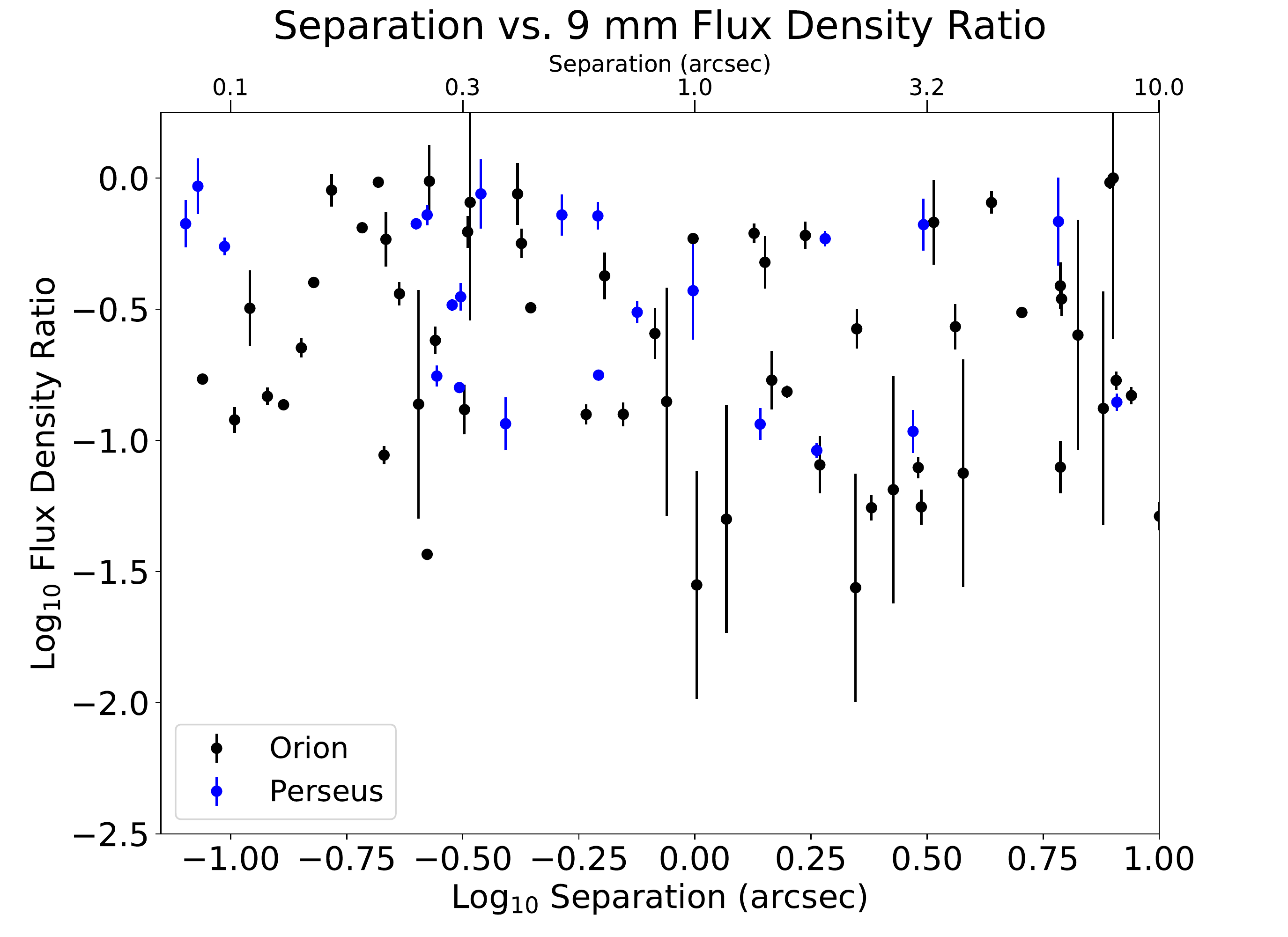}
\end{center}
\caption{
Flux density ratio versus separation for the companions
observed at 0.87~mm (left) and 9~mm (right). It is not clear if there is an intrinsic
contrast limit at smaller separations, and the contrast limits may be limited by the physical disk
structures within an individual system, in addition to instrumental
dynamic range limits. The uncertainties here are statistical only
and do not include absolute flux calibration uncertainty.
However, the vast majority of points plotted are within the same
field of view and would not have absolute calibration that is
systematically different.
}
\label{fluxratio}
\end{figure}

\begin{figure}
\begin{center}
\includegraphics[scale=0.25]{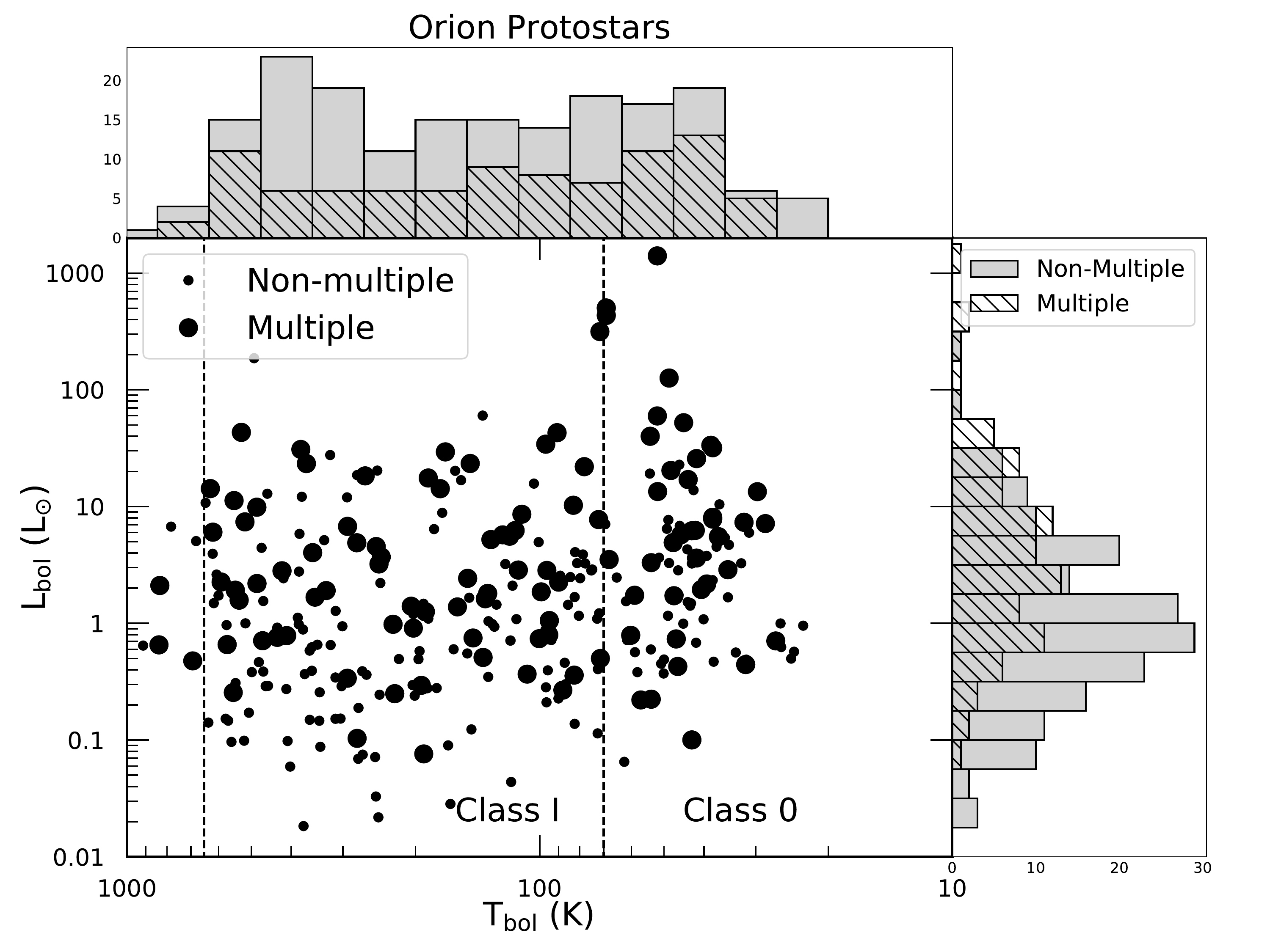}
\includegraphics[scale=0.25]{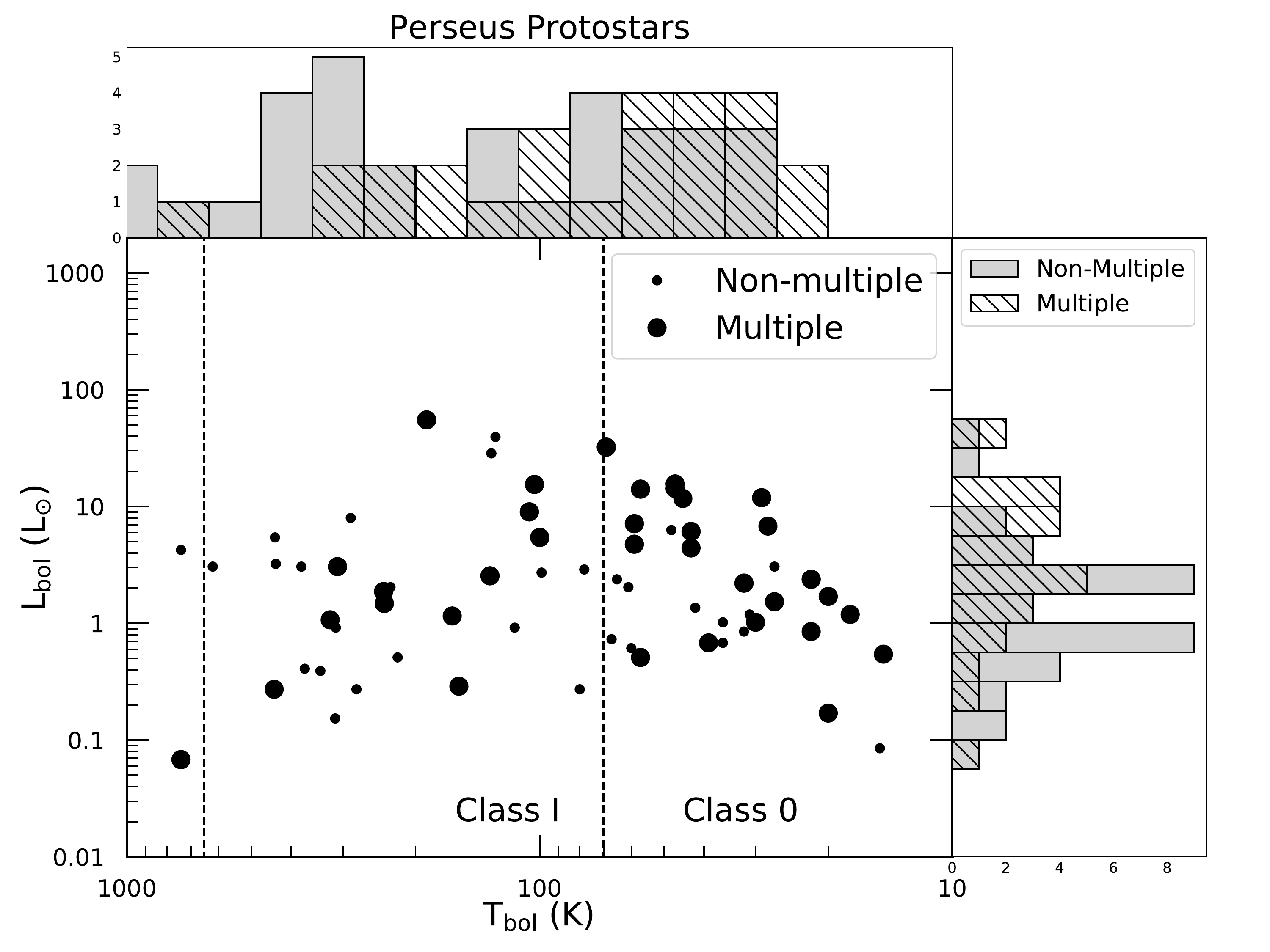}
\end{center}
\caption{
Bolometric temperature vs. bolometric luminosity plots for both Orion and Perseus.
Single systems are represented by small points, while the
large points represent the composite luminosity of each multiple system, with separations up to 10$^4$~au. Histograms 
are shown along the top and right axes for the scatter plots along those directions. It
is apparent that the \tbol\ distributions are quite similar for singles vs. multiple systems, while
the luminosity distributions appear different.
For Orion the median luminosities of single and multiple systems are 0.96 and 3.27~\lsun, respectively, and for Perseus the median luminosities for single and multiples systems are 0.97 and 3.06~\lsun, respectively. 
The difference between the two distributions (singles
vs. multiples) is statistically significant
for Orion,
while for Perseus the difference is not quite
statistically significant (see Section 4.1).
}
\label{lbol-tbol}
\end{figure}

\begin{figure}
\begin{center}
\includegraphics[scale=0.3]{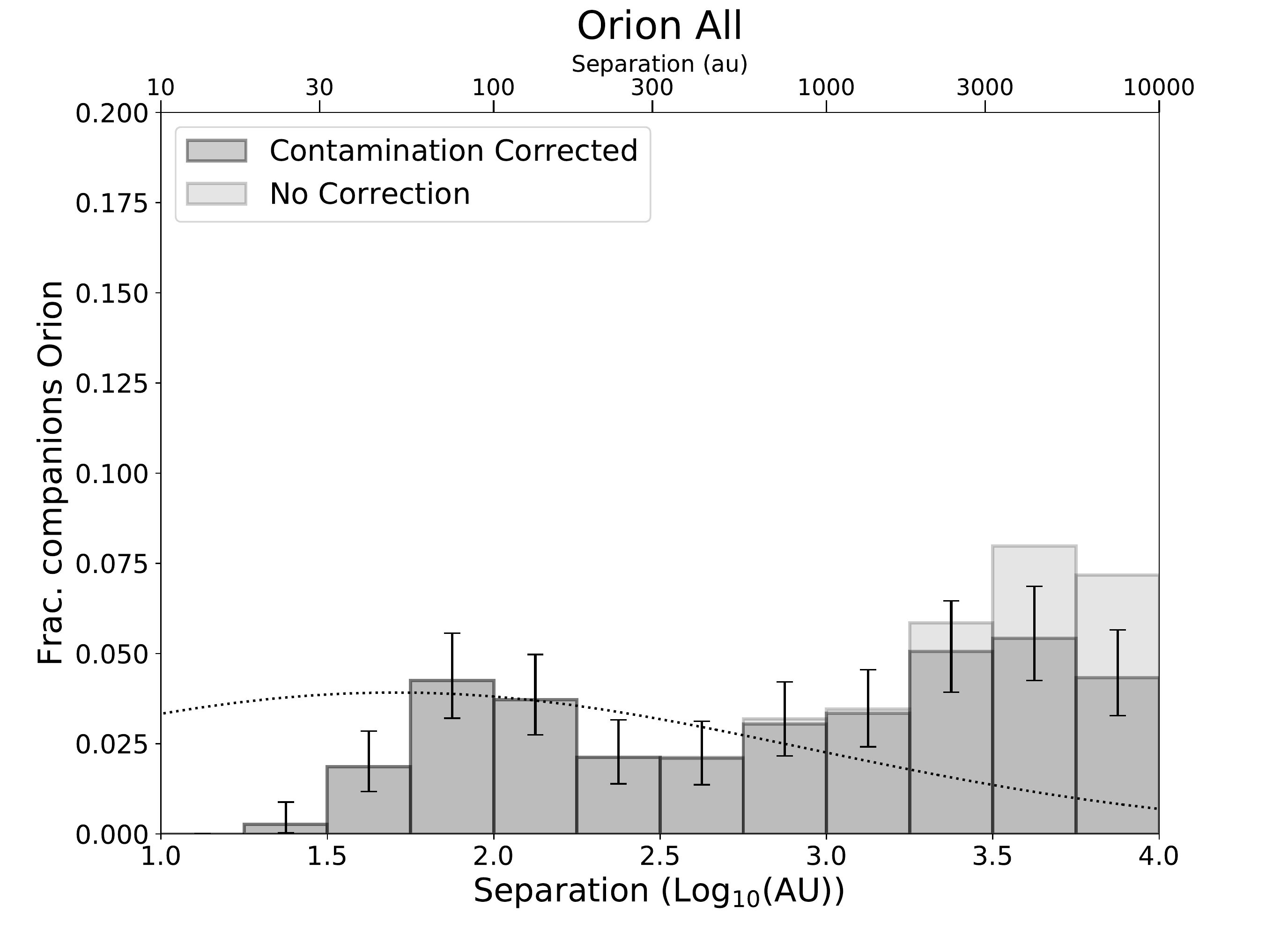}
\includegraphics[scale=0.3]{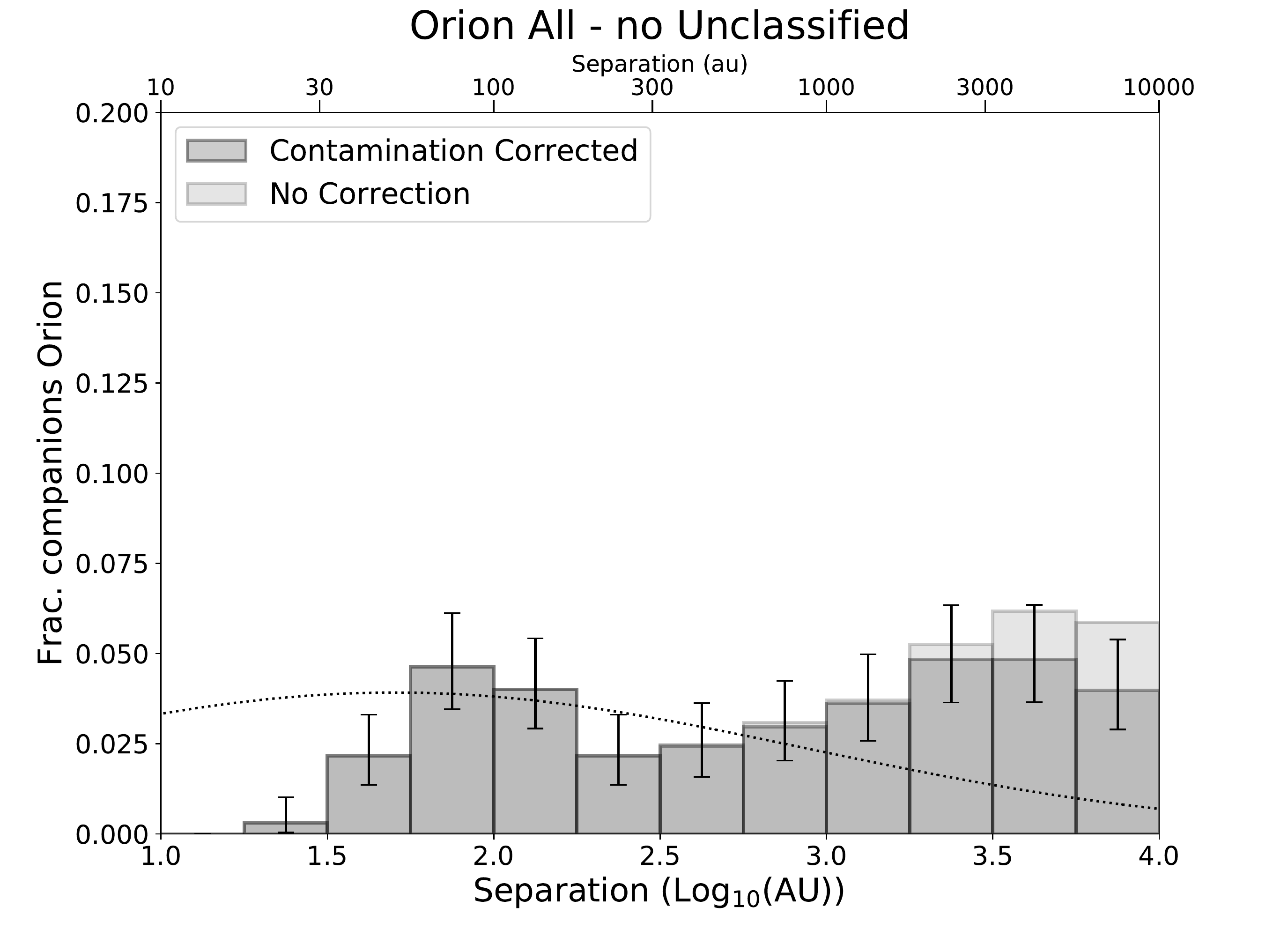}
\includegraphics[scale=0.3]{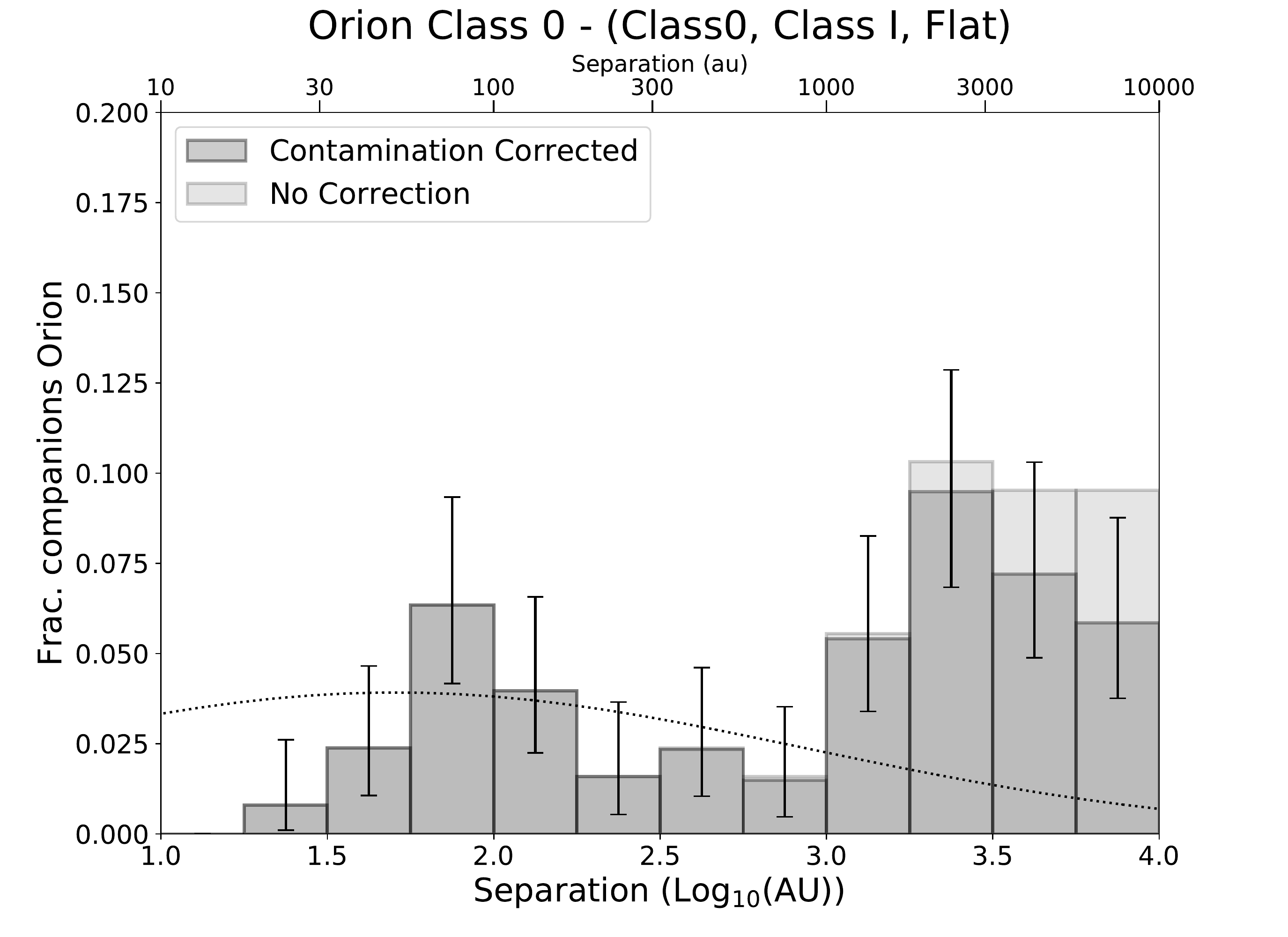}
\includegraphics[scale=0.3]{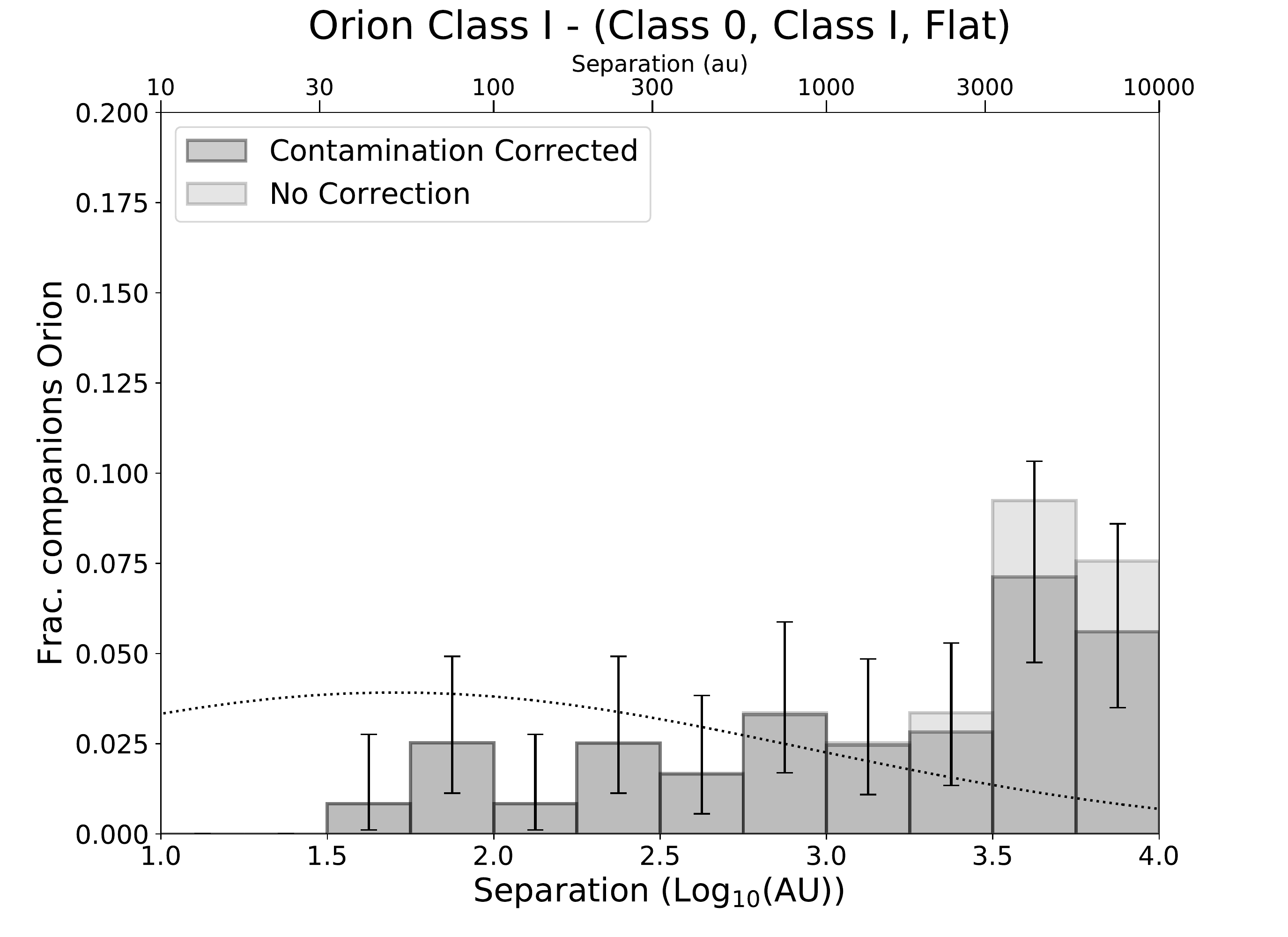}
\includegraphics[scale=0.3]{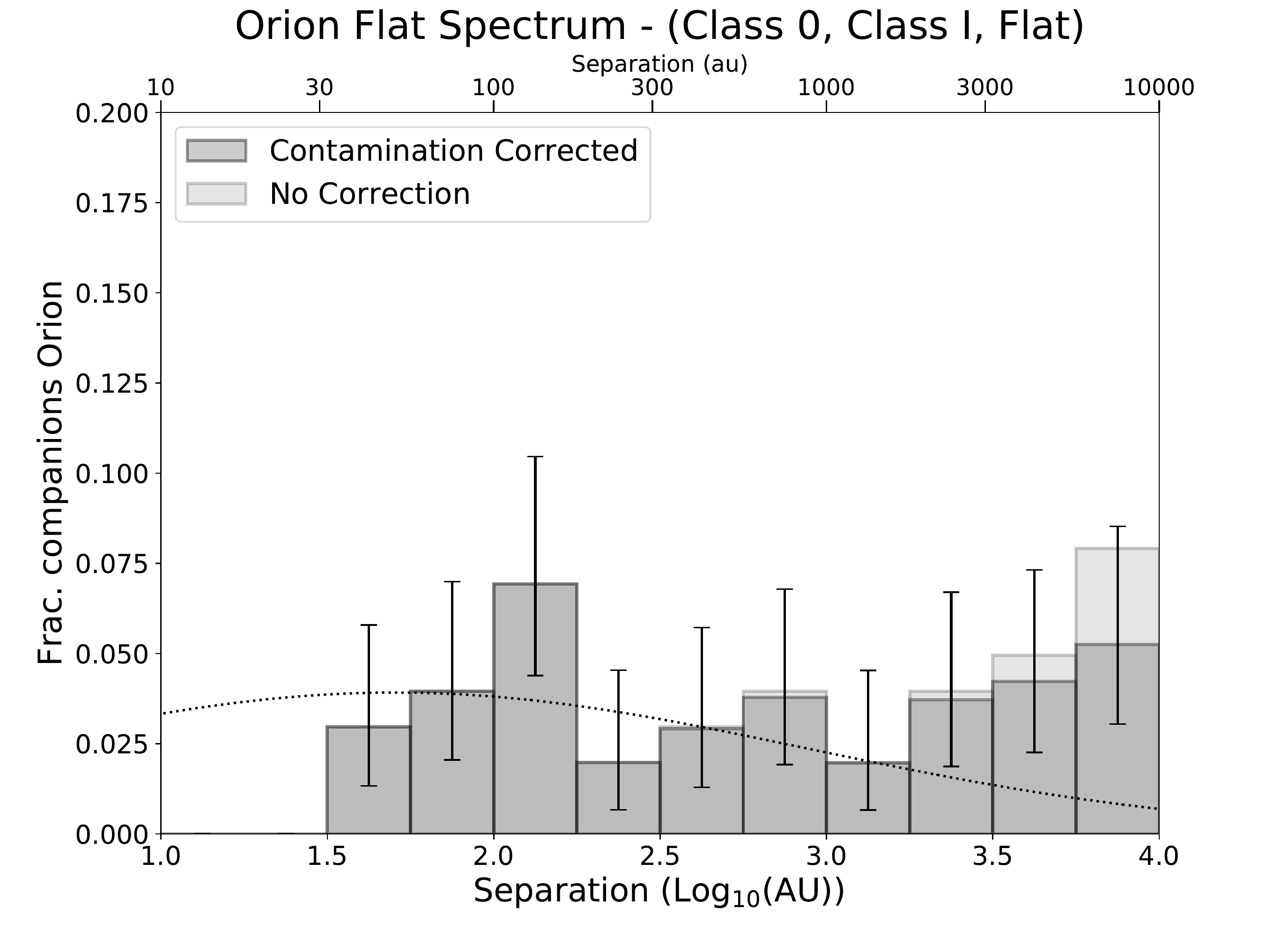}
\includegraphics[scale=0.3]{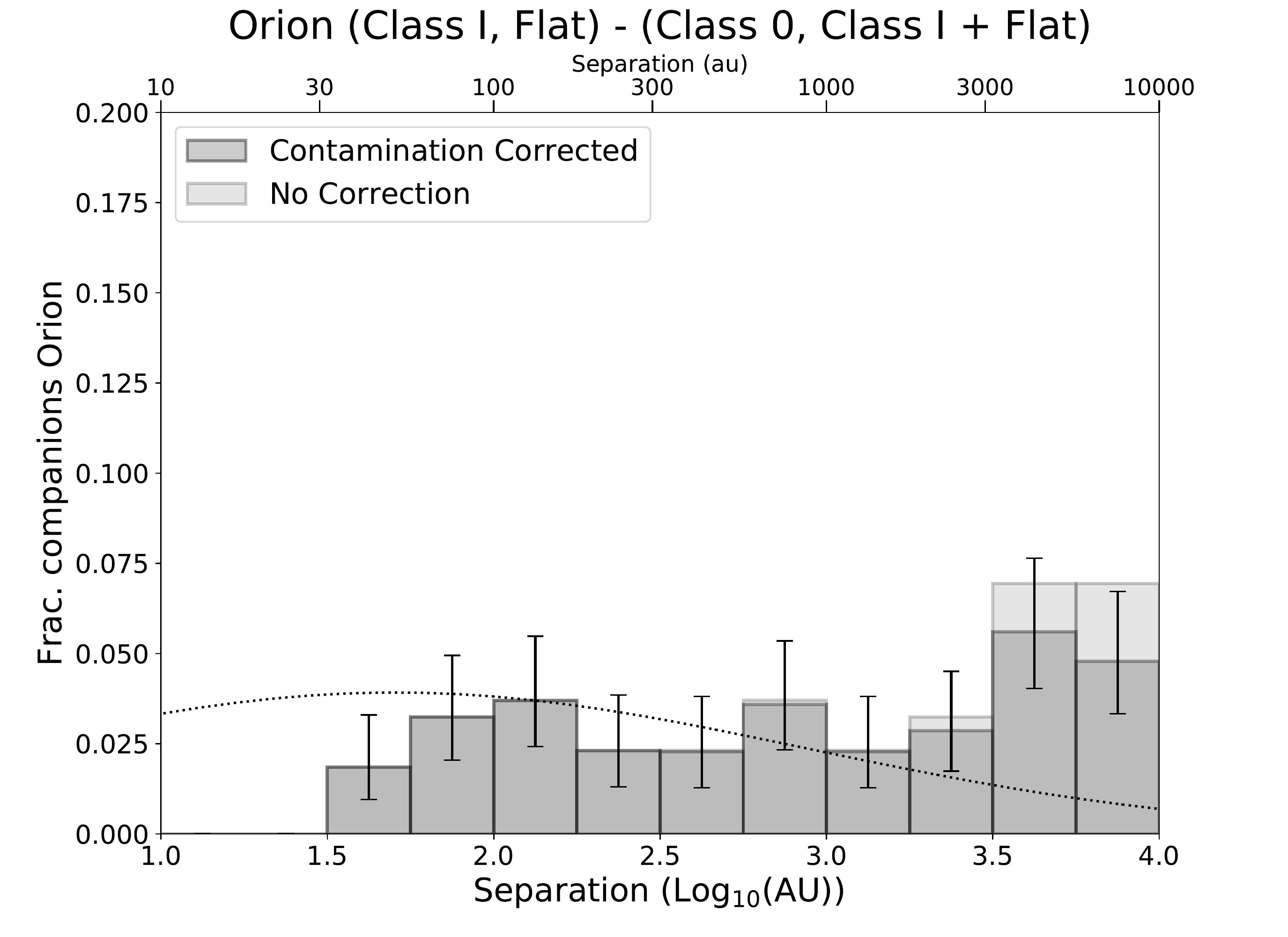}
\end{center}
\caption{Histograms of companion separations for different
selections from the Orion sample. 
The histograms specifying a particular
protostar class only include separations from each protostar to the specified class.
Separations that include protostars from different classes or more than one unclassified source
are only shown in the panel labeled `All'. The lighter gray
histograms do not include the
companion probabilities as determined from the YSO surface density. This shows
that contamination is likely only significant for separations greater than 3000~au.
The thin dotted curve in each panel
is the Gaussian fit to the separation distribution of solar-type field stars \citep{raghavan2010}.
The resolution limit for Orion and Perseus is $\sim$20~au (1.3 in log units).
The error bars for each bin are computed using binomial statistics as discussed in 
Section 2.4.3. The number of separations contributing to each bin are then divided by
the total number of singles and the total number of separations displayed in the plot.
}
\label{separations_orion_all}
\end{figure}

\begin{figure}
\begin{center}
\includegraphics[scale=0.3]{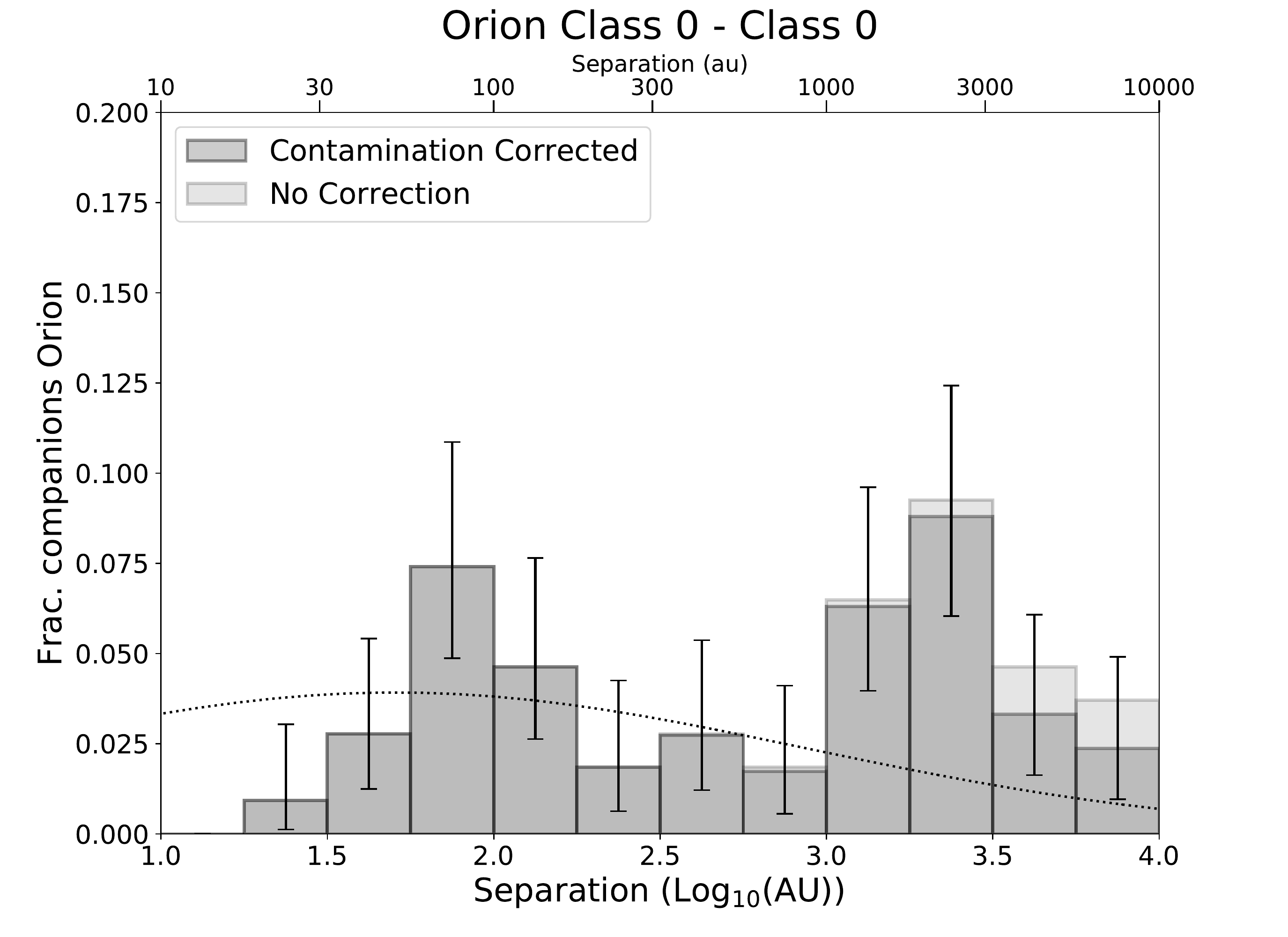}
\includegraphics[scale=0.3]{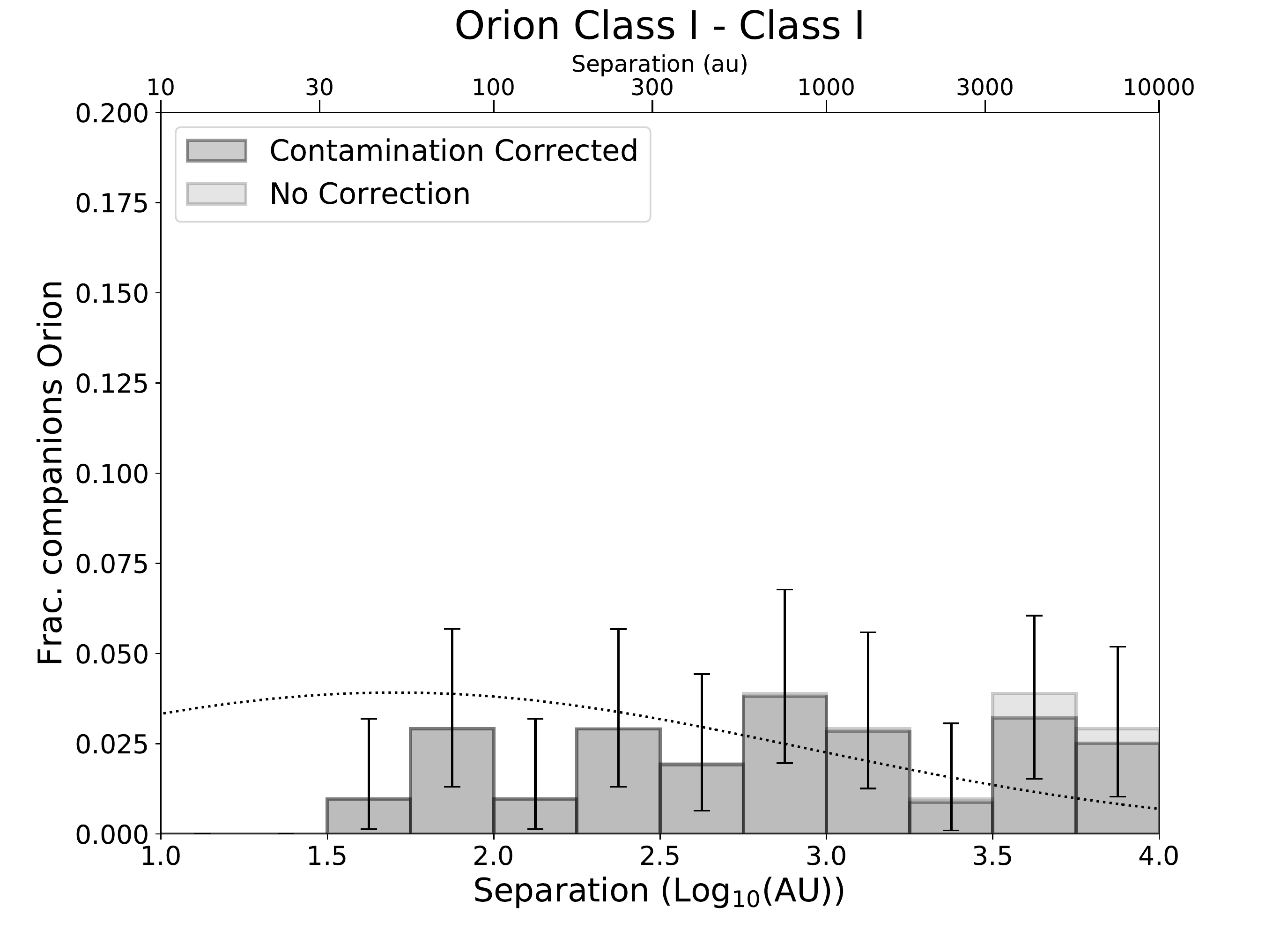}
\includegraphics[scale=0.3]{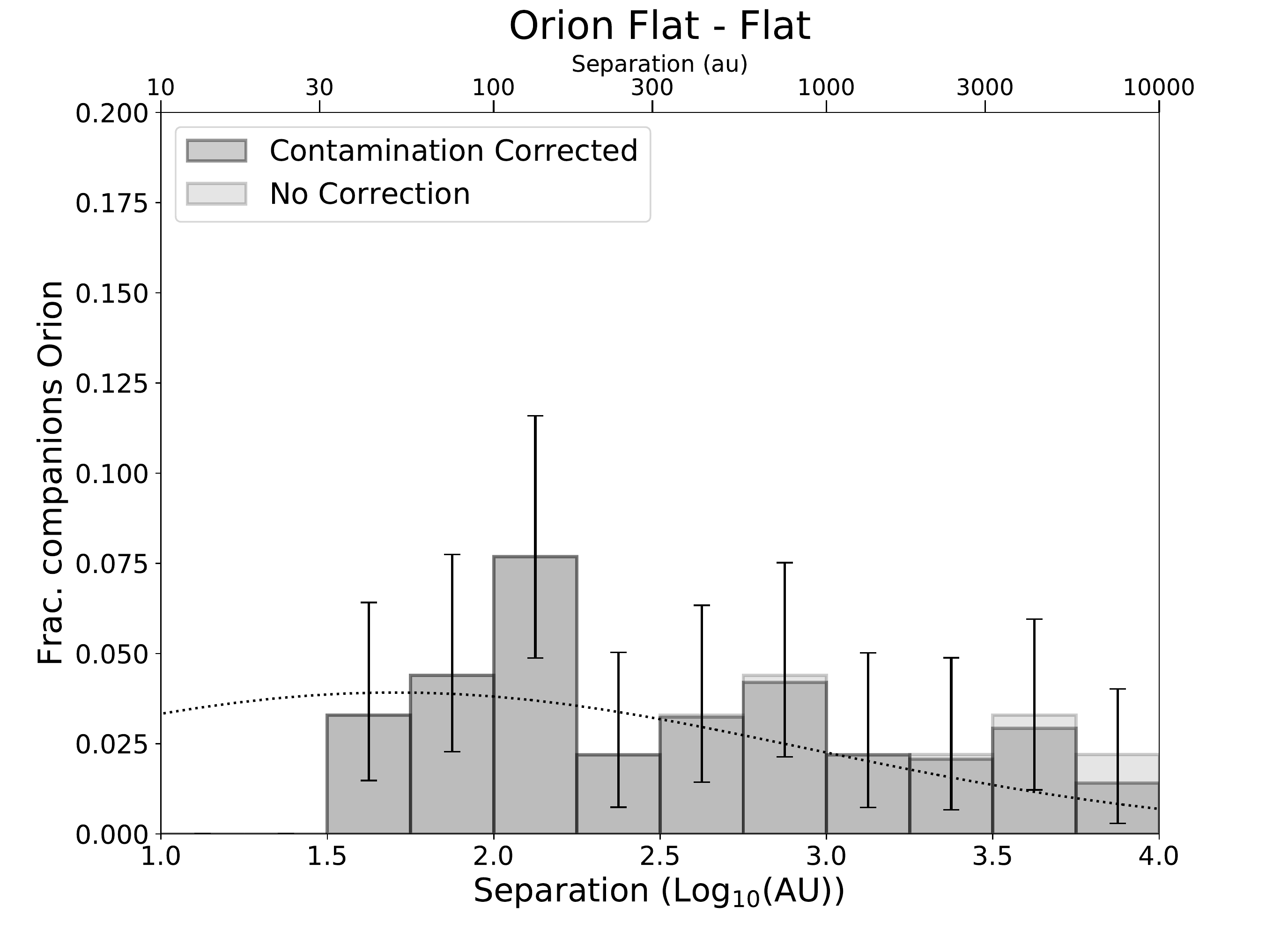}
\includegraphics[scale=0.3]{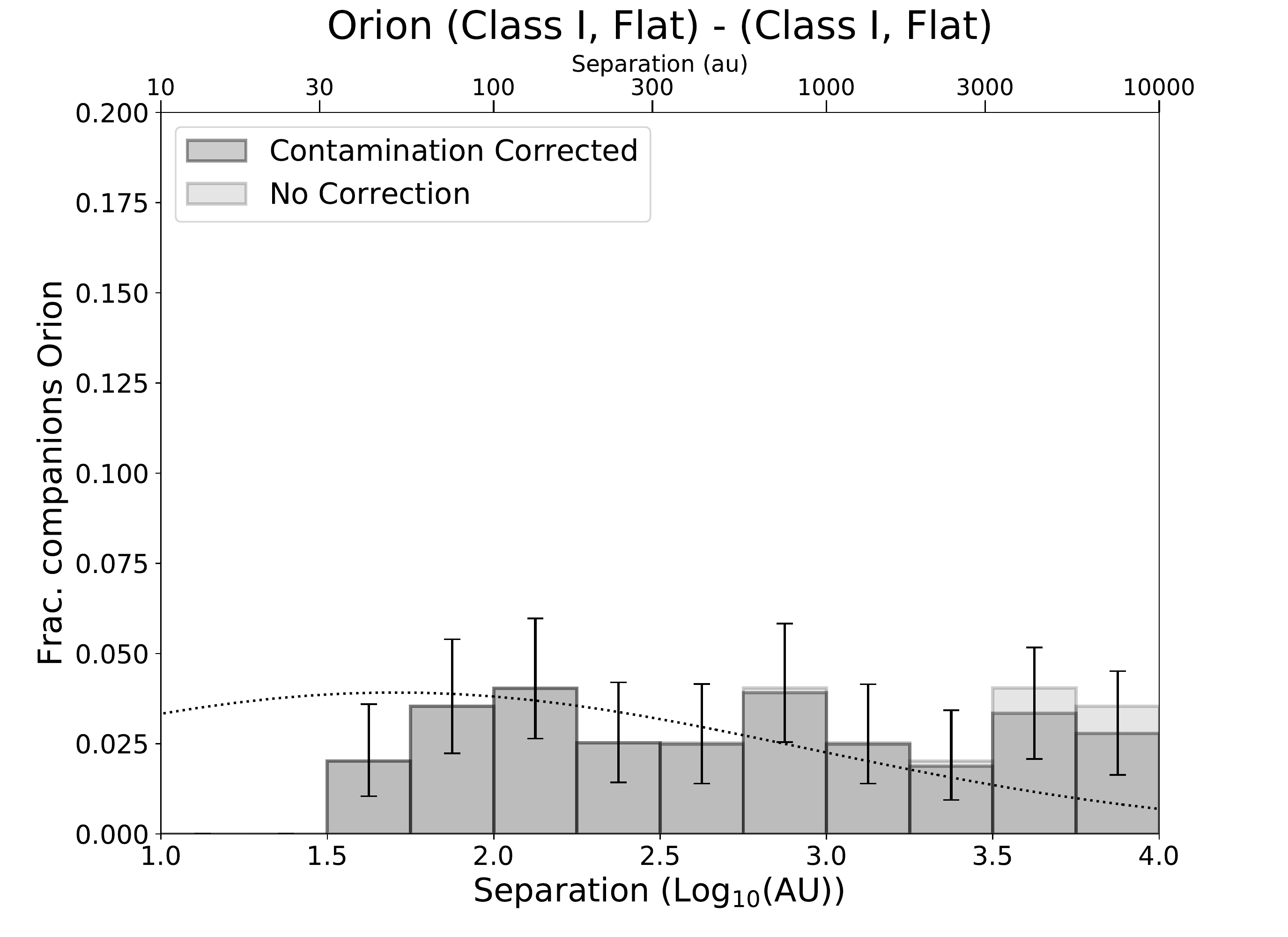}
\end{center}
\caption{Same as Figure \ref{separations_orion_all}, but for histograms constructed
only for protostars of the same class within Orion.
}
\label{separations_orion_class}
\end{figure}

\begin{figure}
\begin{center}
\includegraphics[scale=0.3]{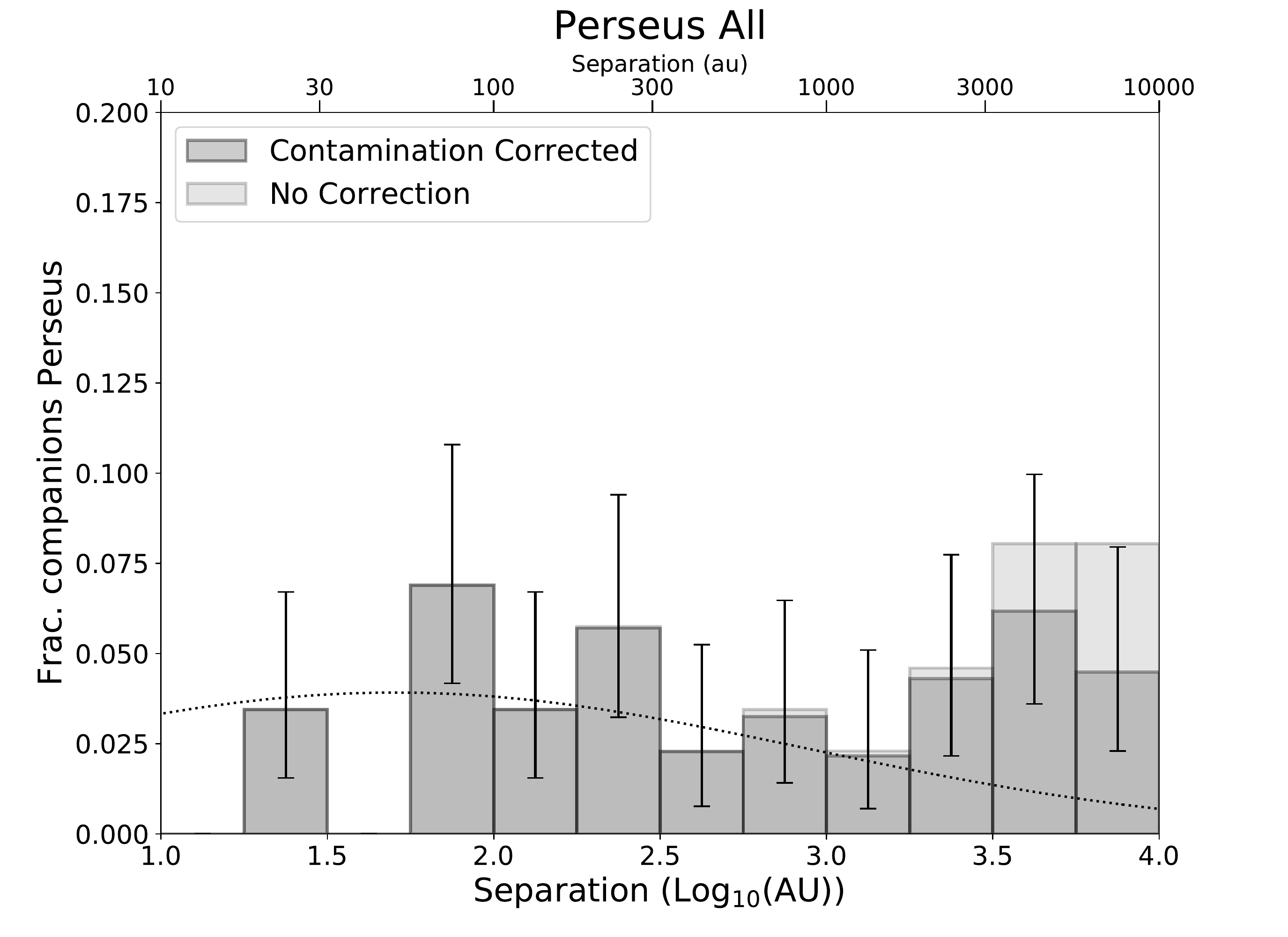}
\includegraphics[scale=0.3]{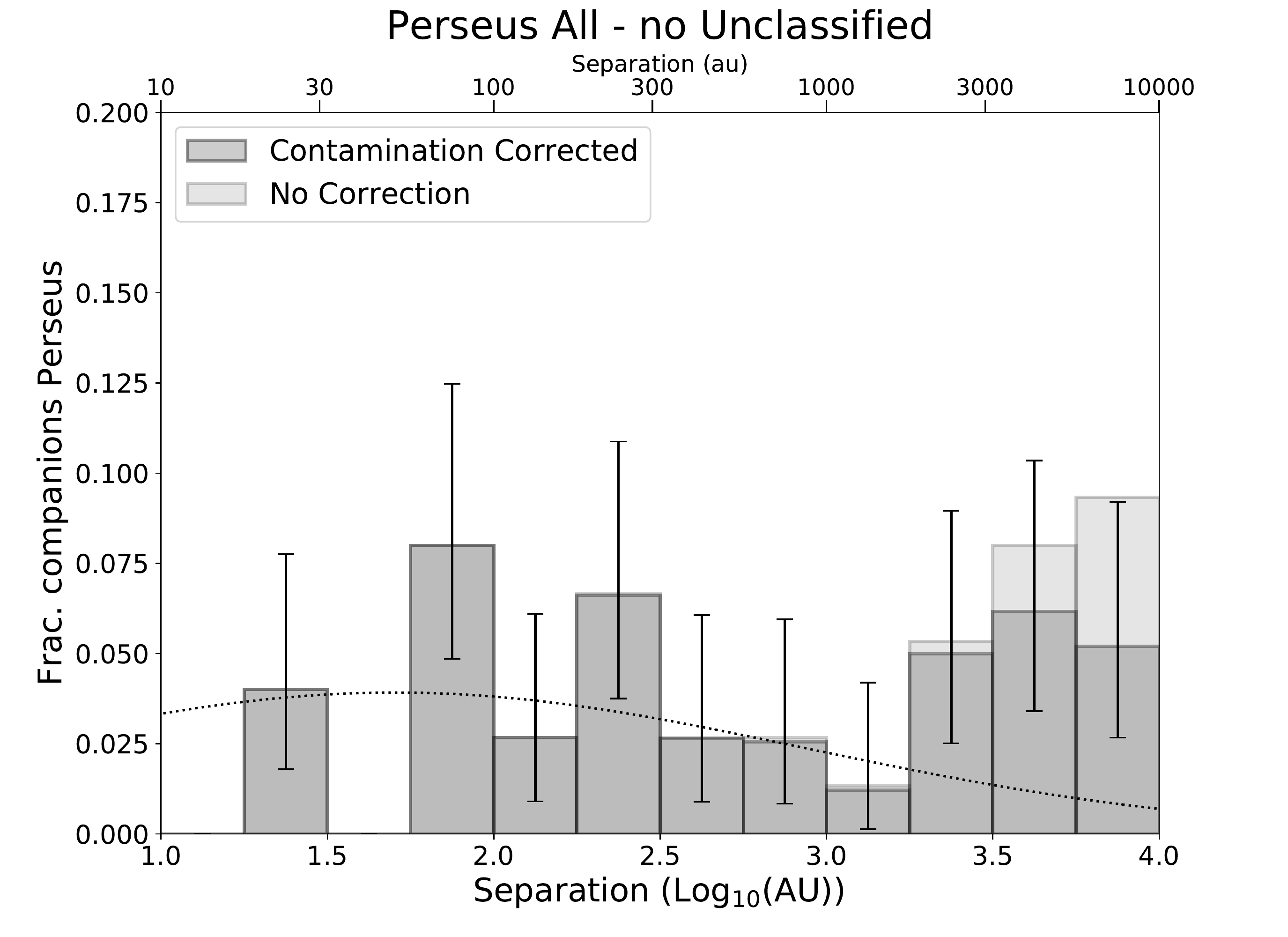}
\includegraphics[scale=0.3]{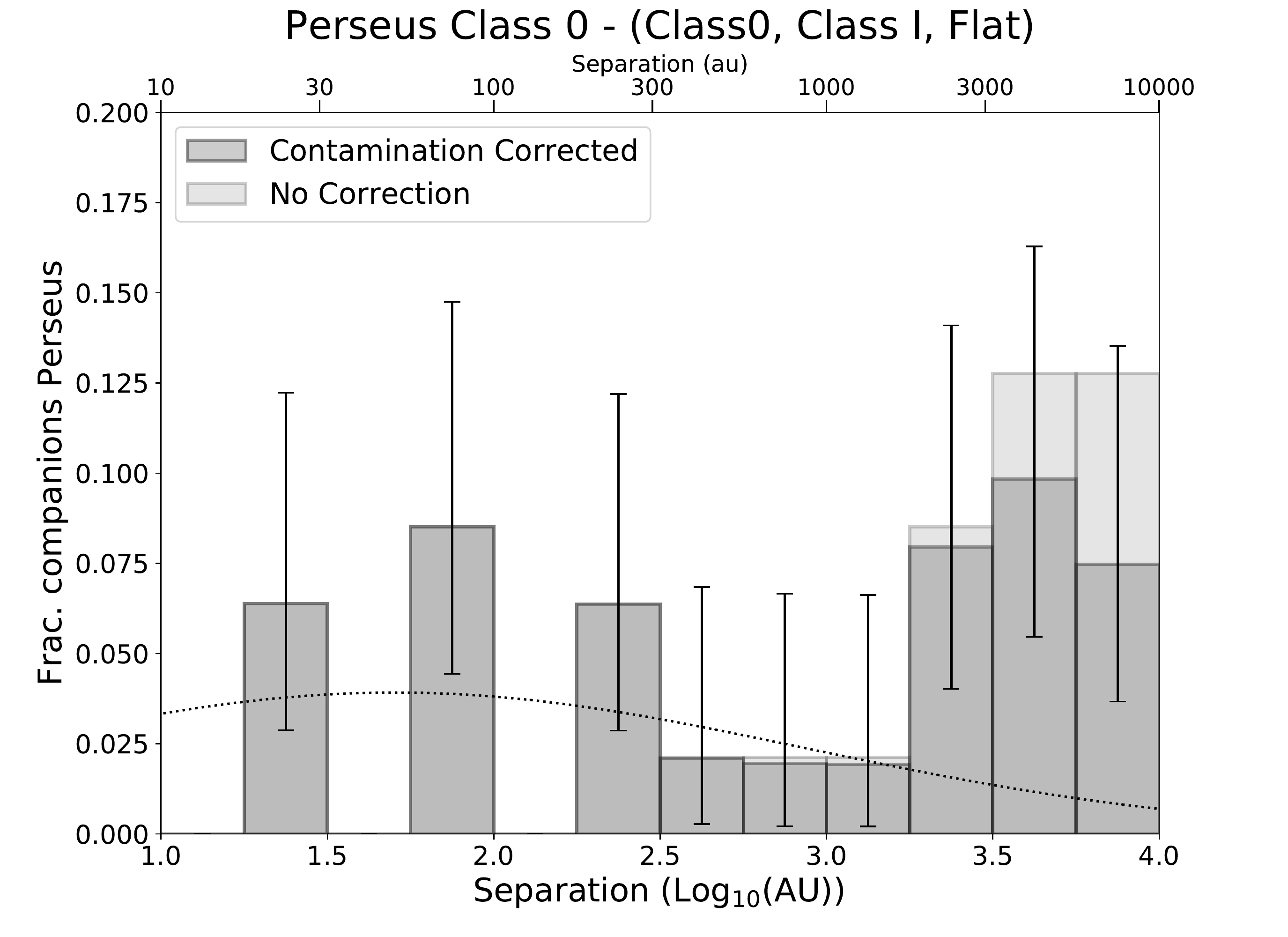}
\includegraphics[scale=0.3]{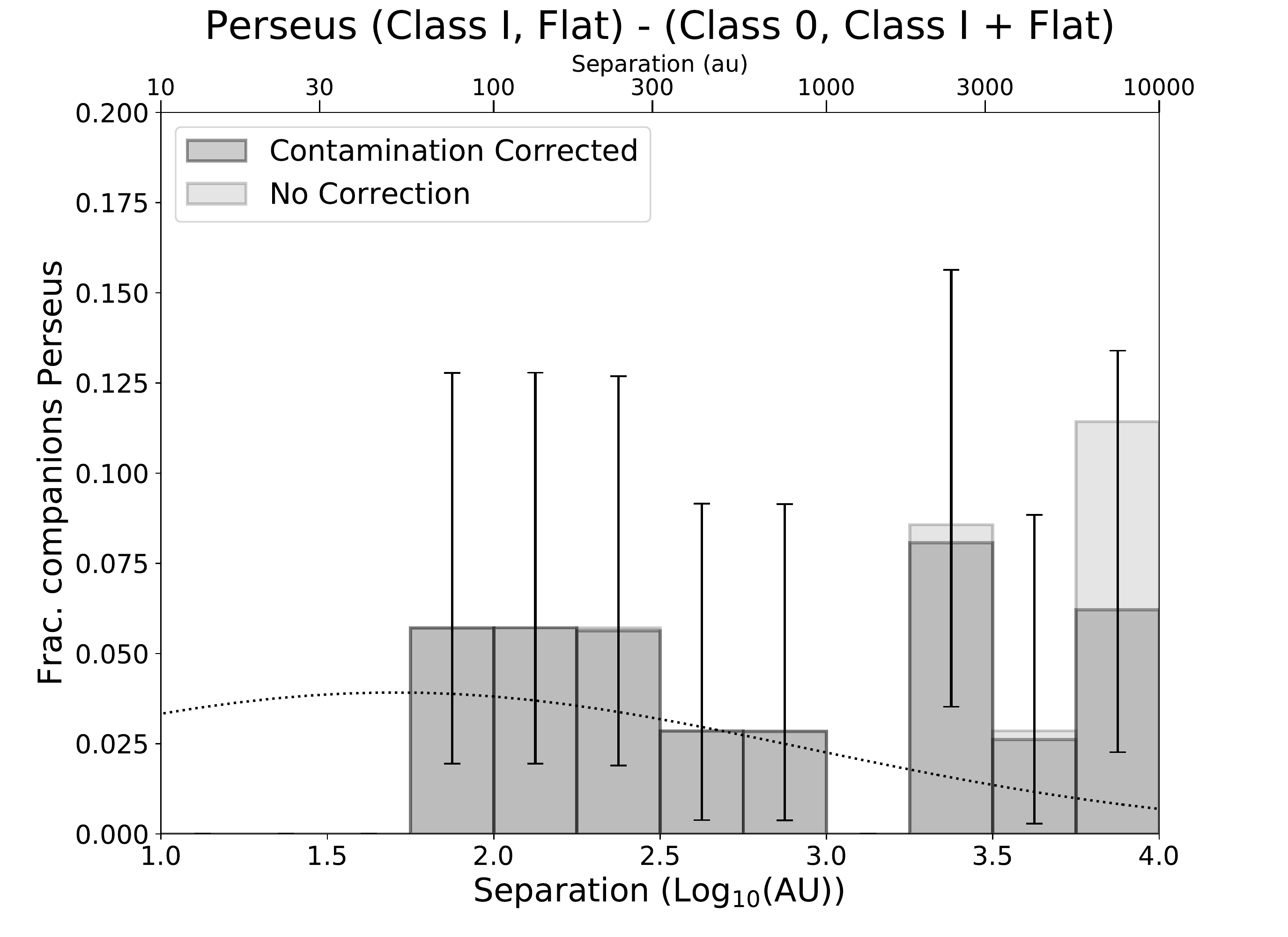}
\includegraphics[scale=0.3]{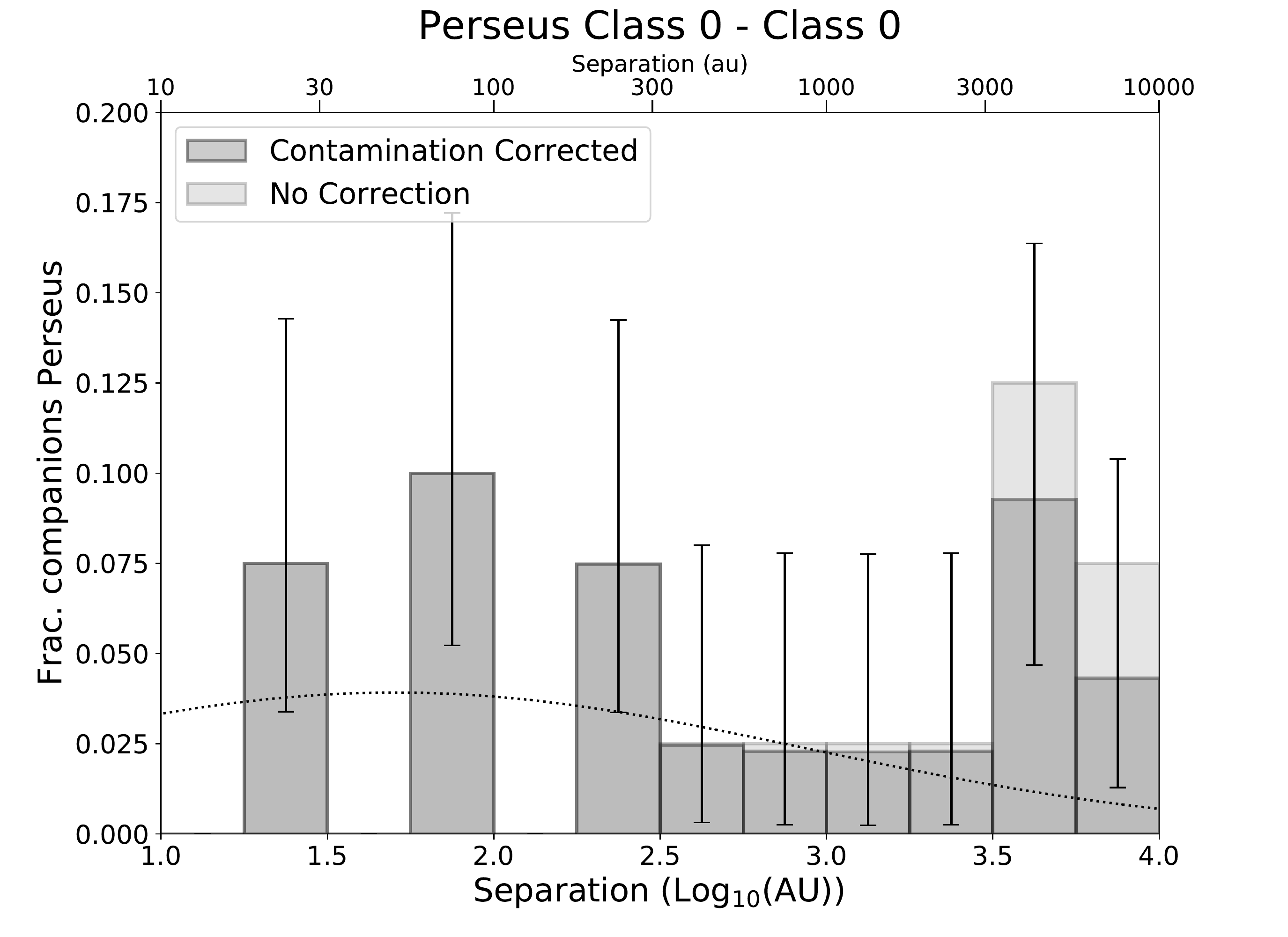}
\includegraphics[scale=0.3]{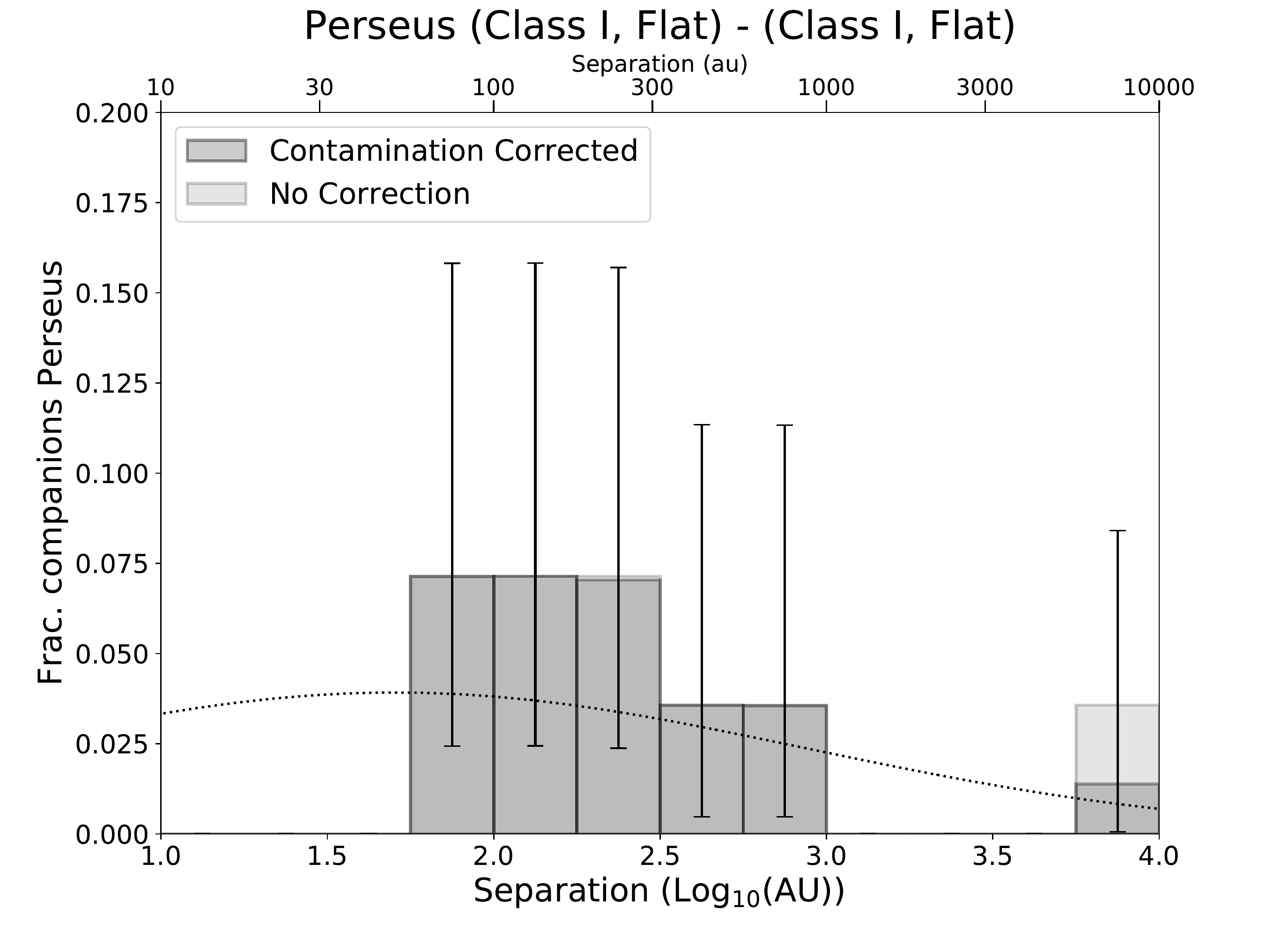}

\end{center}
\caption{Same as Figure \ref{separations_orion_all} but for Perseus.
}
\label{separations_perseus_all}
\end{figure}

\begin{figure}
\begin{center}
\includegraphics[scale=0.3]{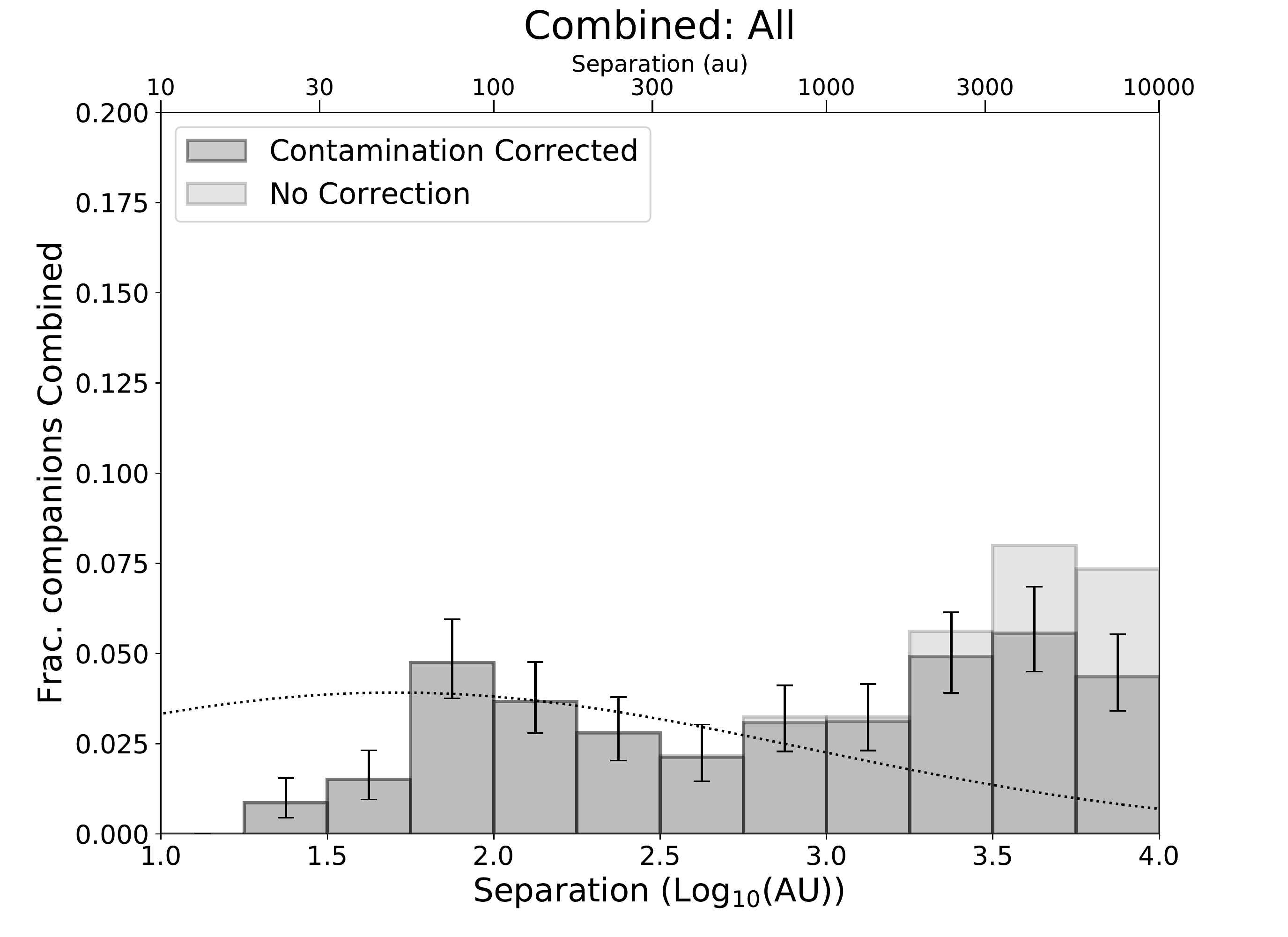}
\includegraphics[scale=0.3]{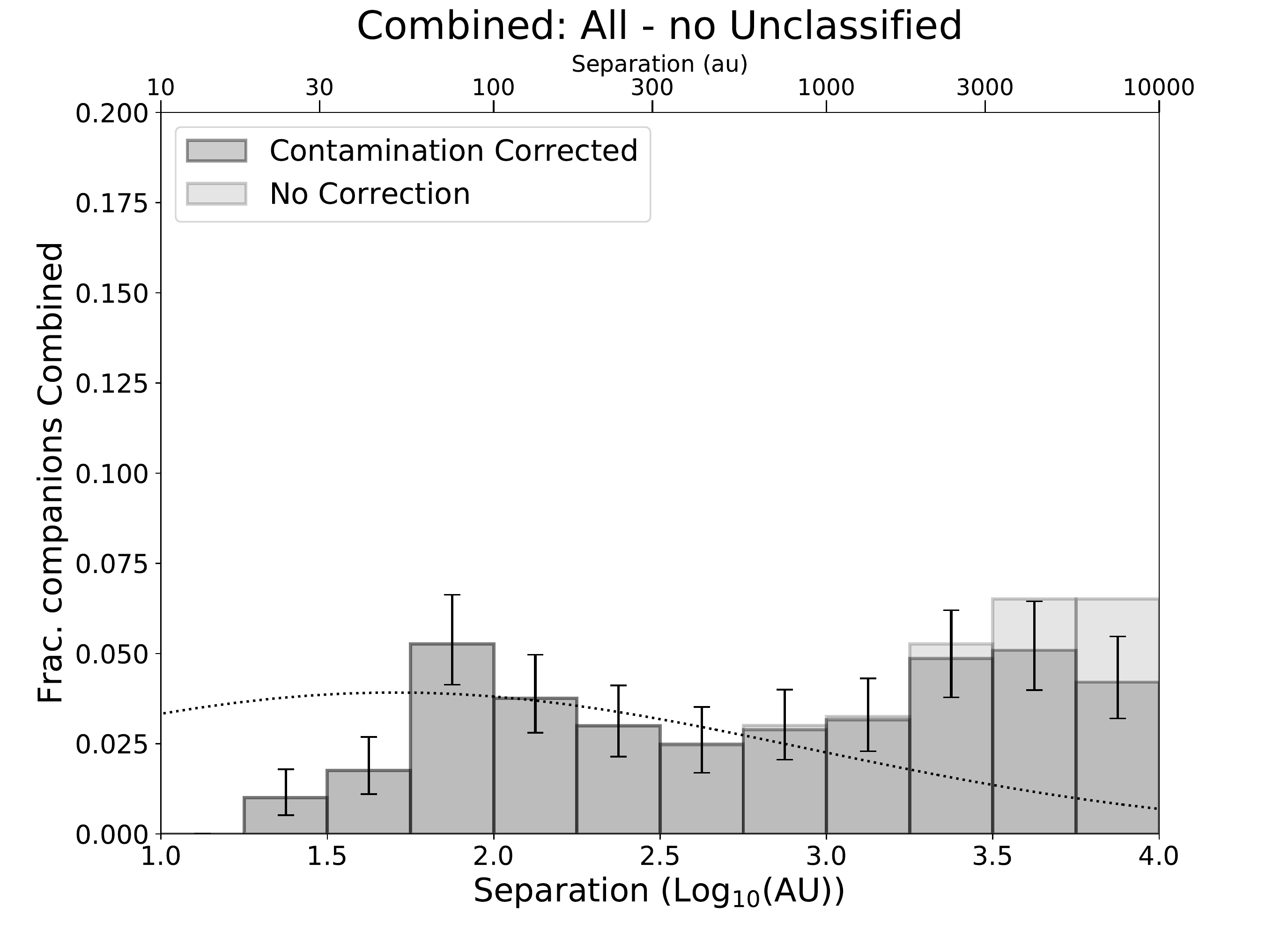}
\includegraphics[scale=0.3]{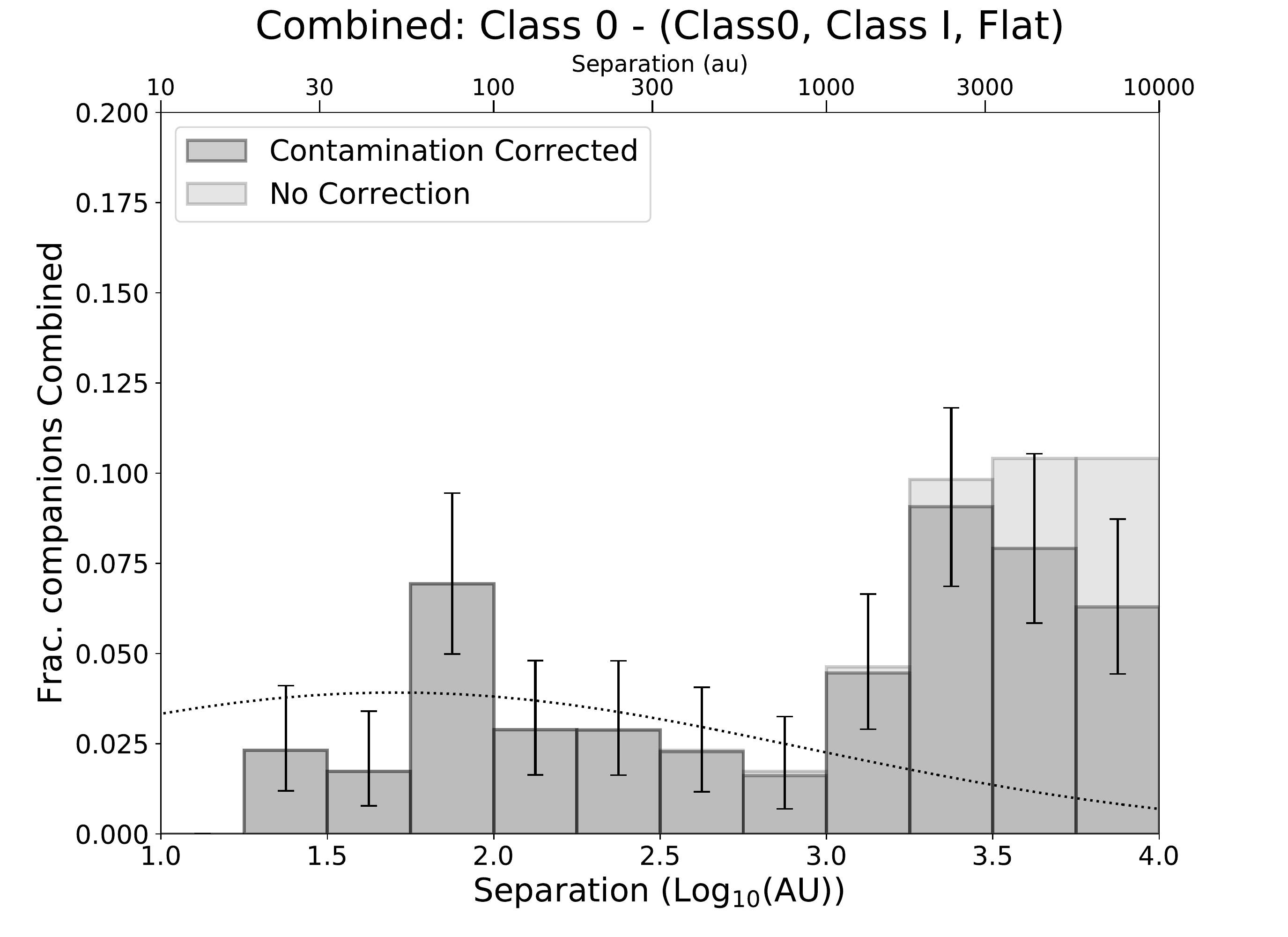}
\includegraphics[scale=0.3]{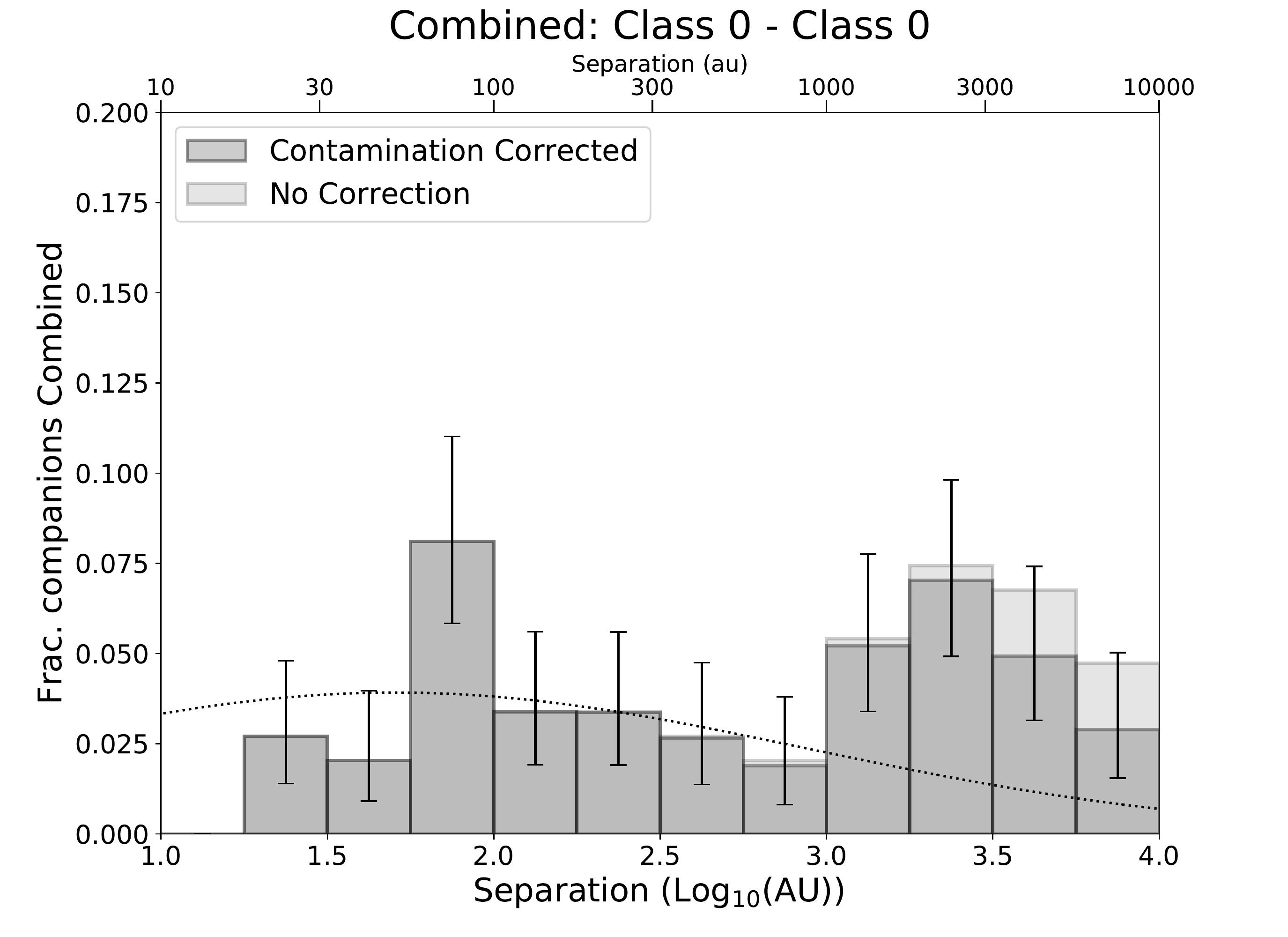}
\includegraphics[scale=0.3]{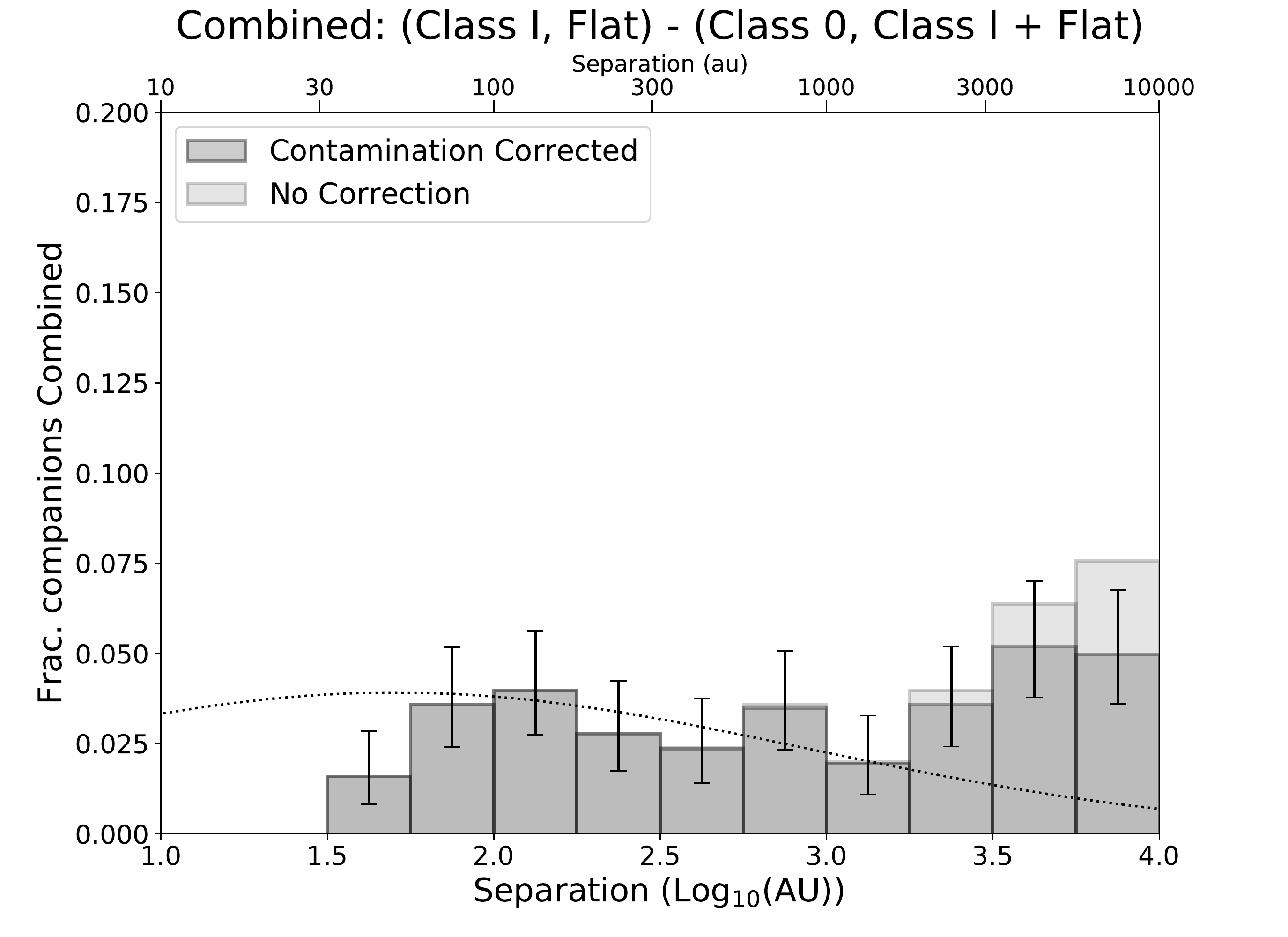}
\includegraphics[scale=0.3]{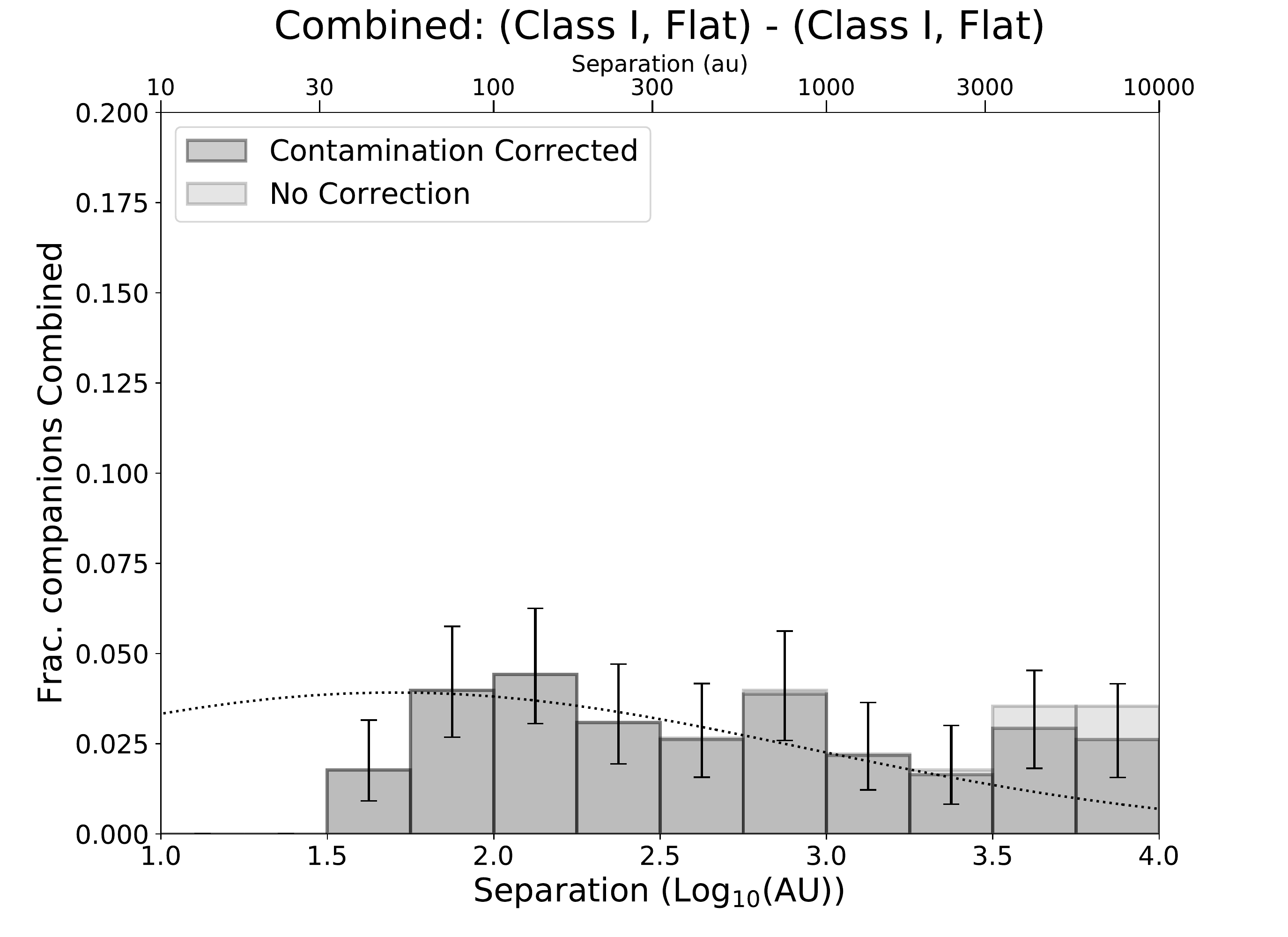}
\end{center}
\caption{Same as Figure \ref{separations_orion_all} and \ref{separations_perseus_all} but for histograms constructed from the combined sample of
Orion and Perseus. Class I and Flat Spectrum are not differentiated here because the Perseus sample does not distinguish between Flat Spectrum and Class I sources.
}
\label{separations_comb}
\end{figure}

\begin{figure}
\begin{center}
\includegraphics[scale=0.3]{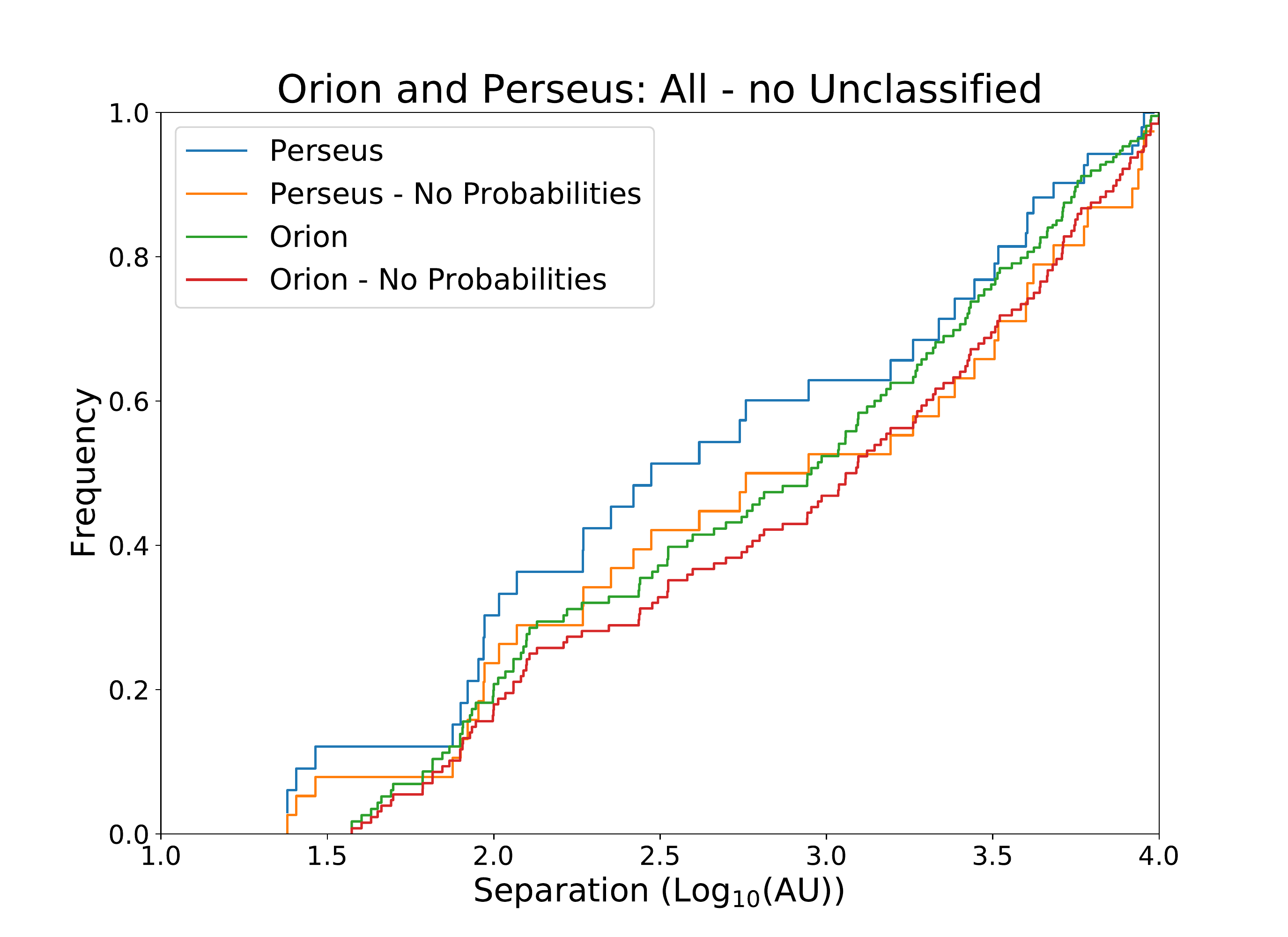}
\includegraphics[scale=0.3]{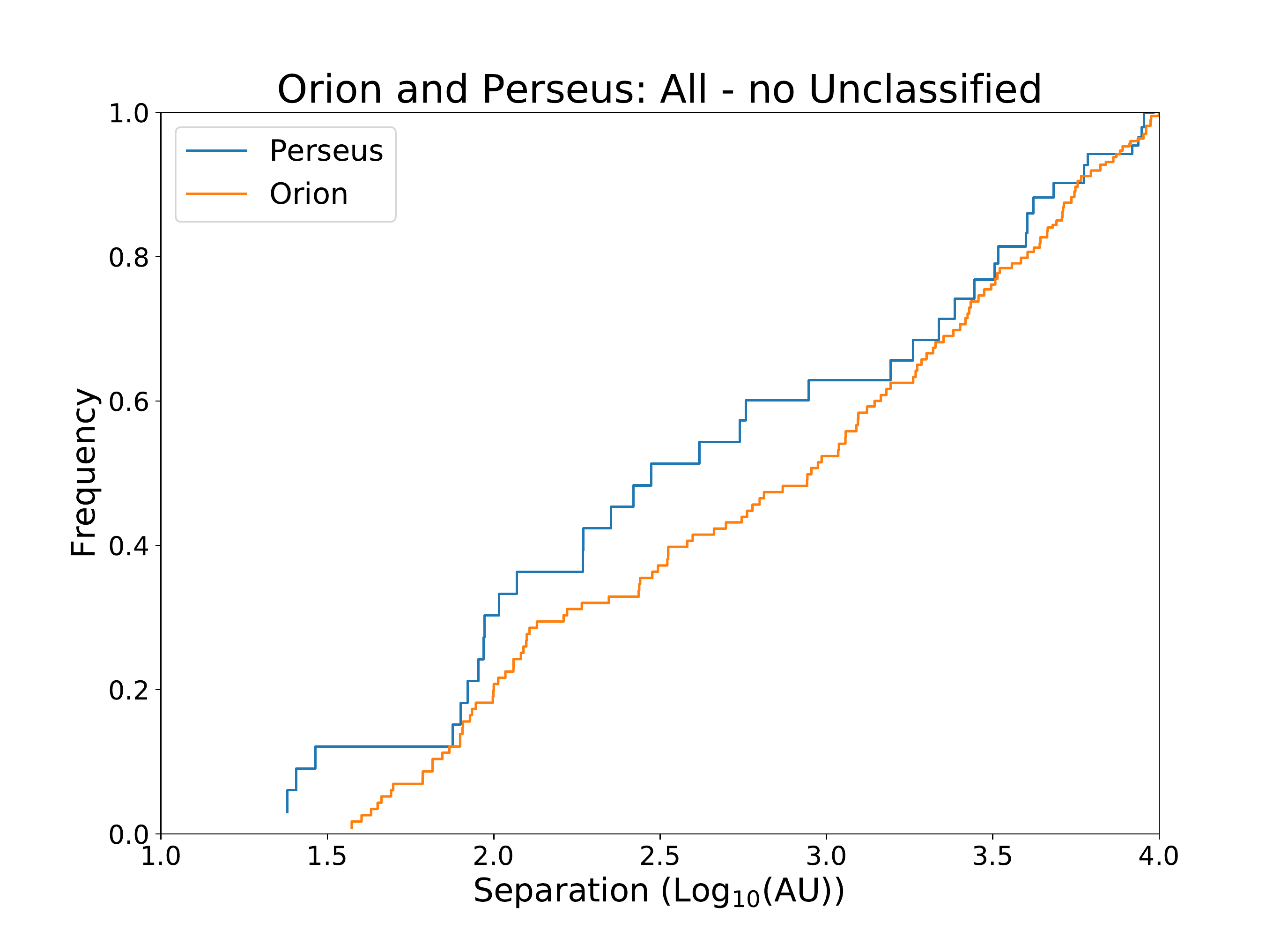}
\includegraphics[scale=0.3]{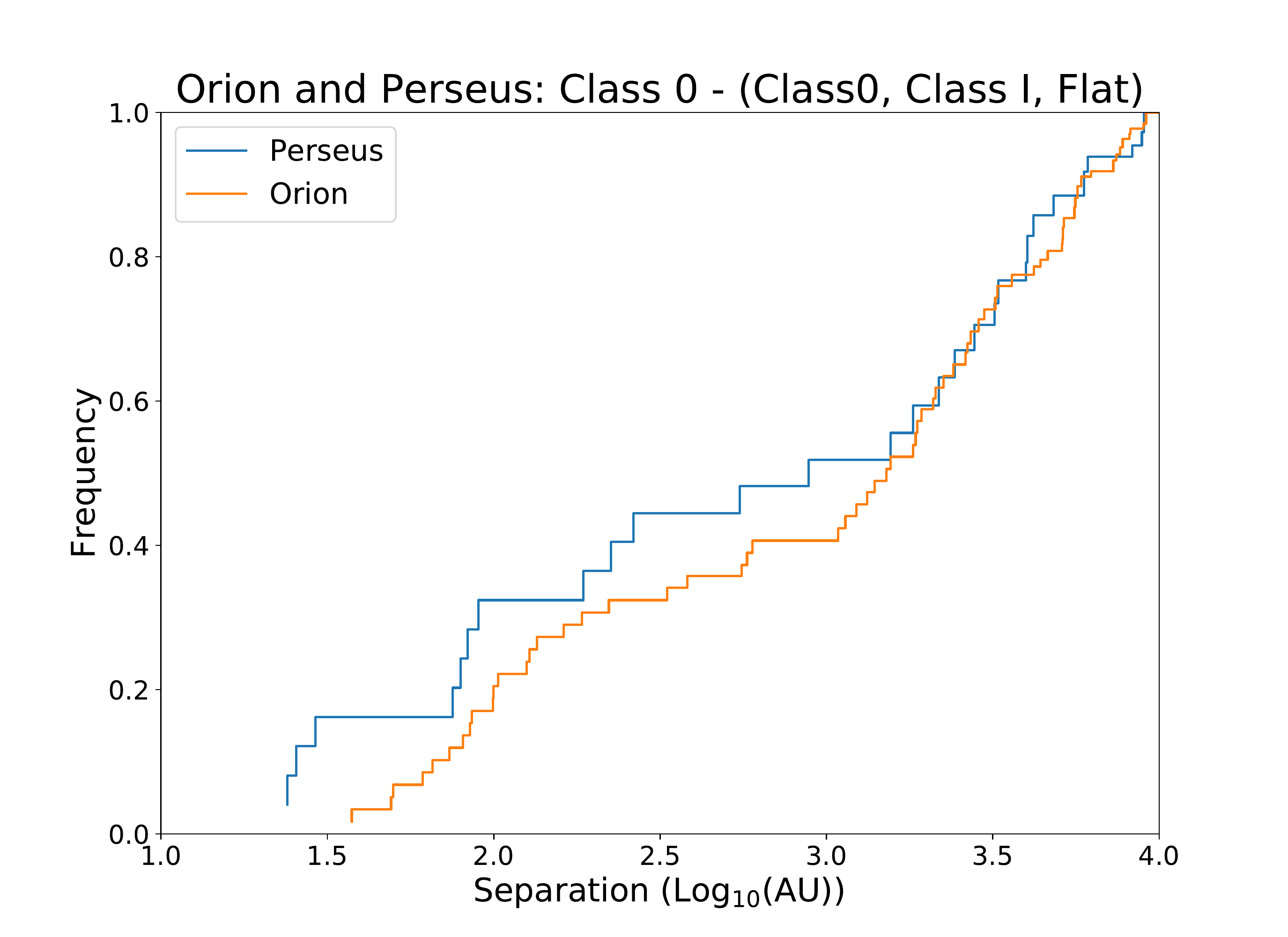}
\includegraphics[scale=0.3]{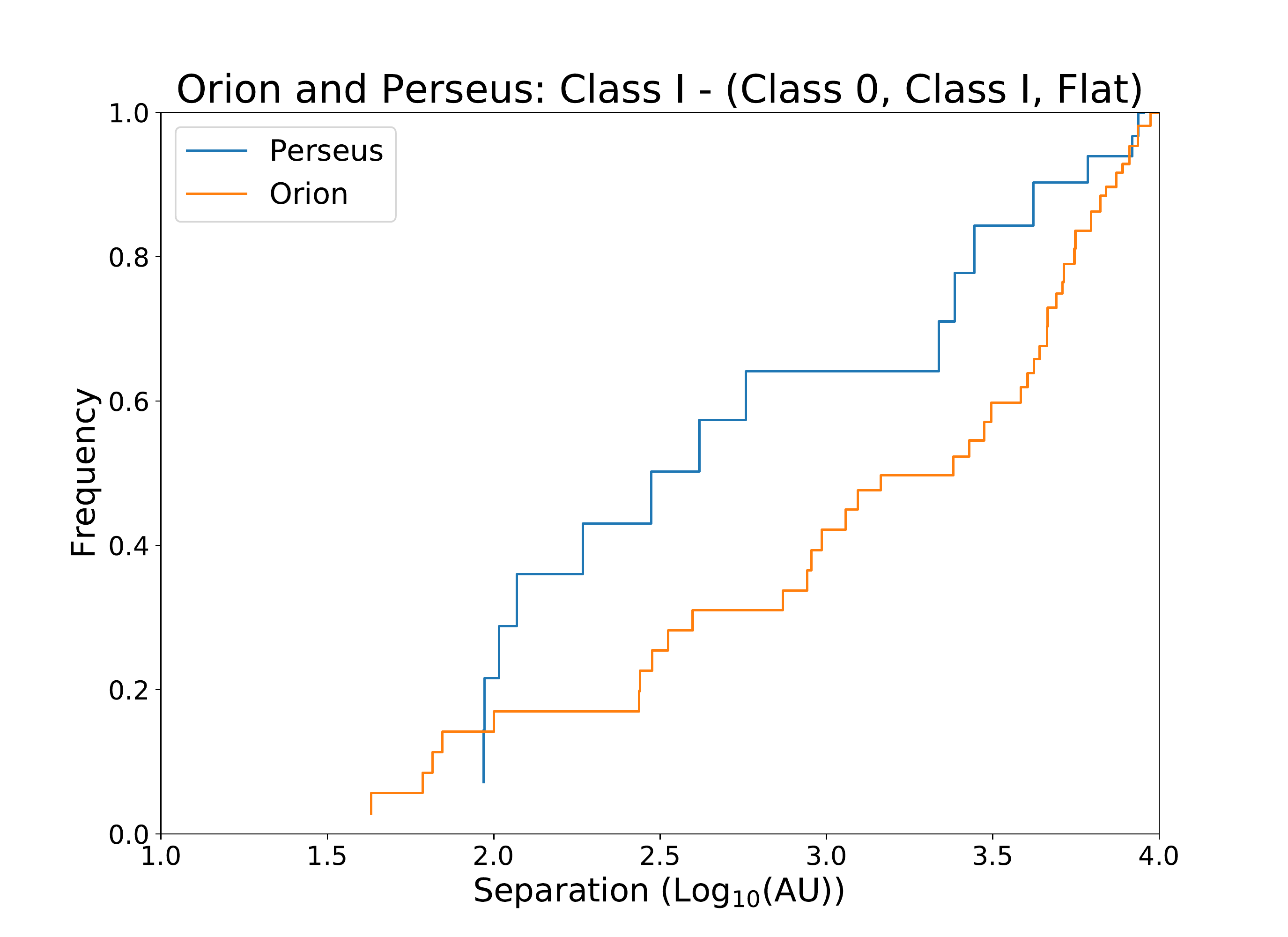}
\includegraphics[scale=0.3]{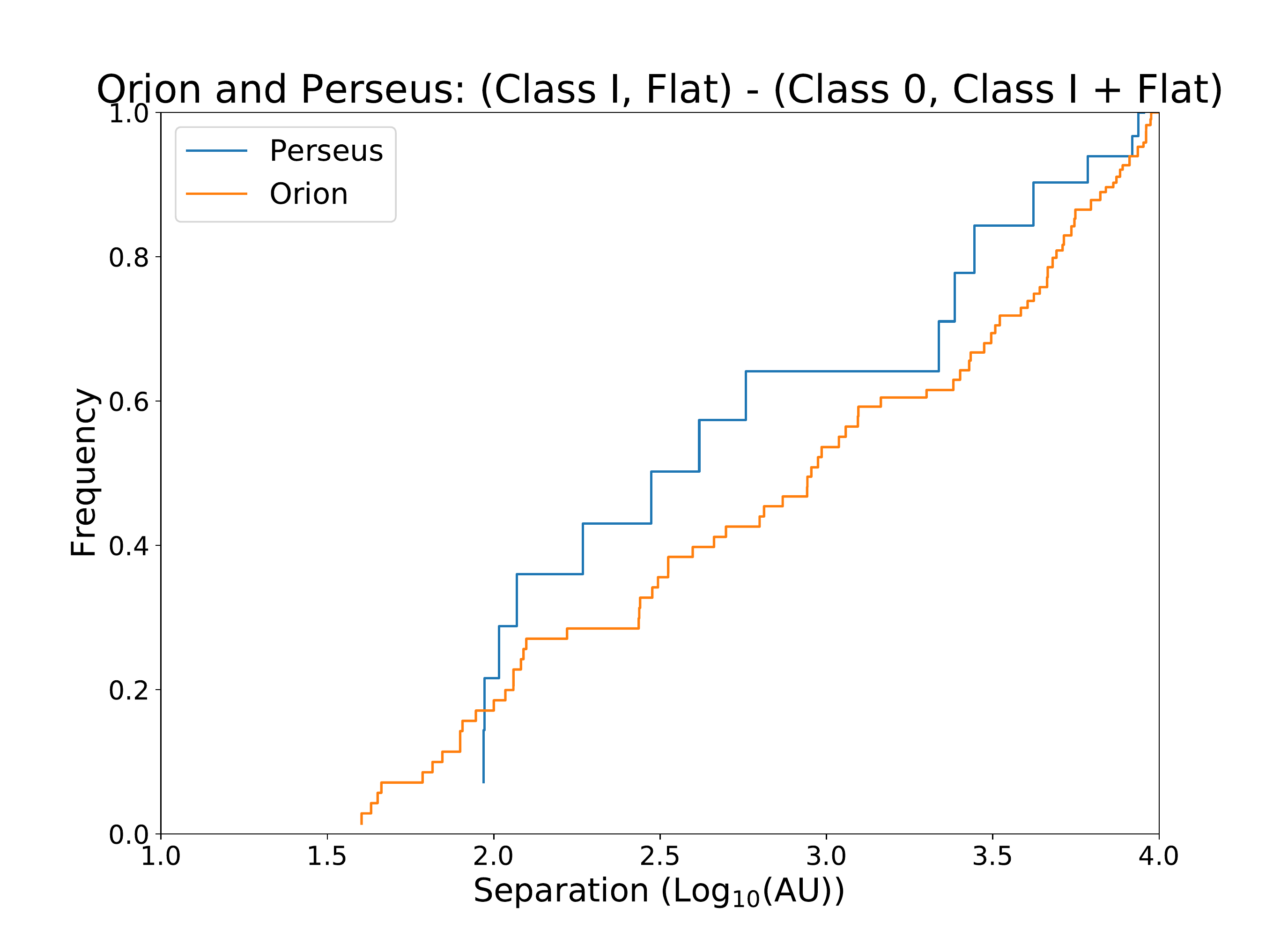}
\includegraphics[scale=0.3]{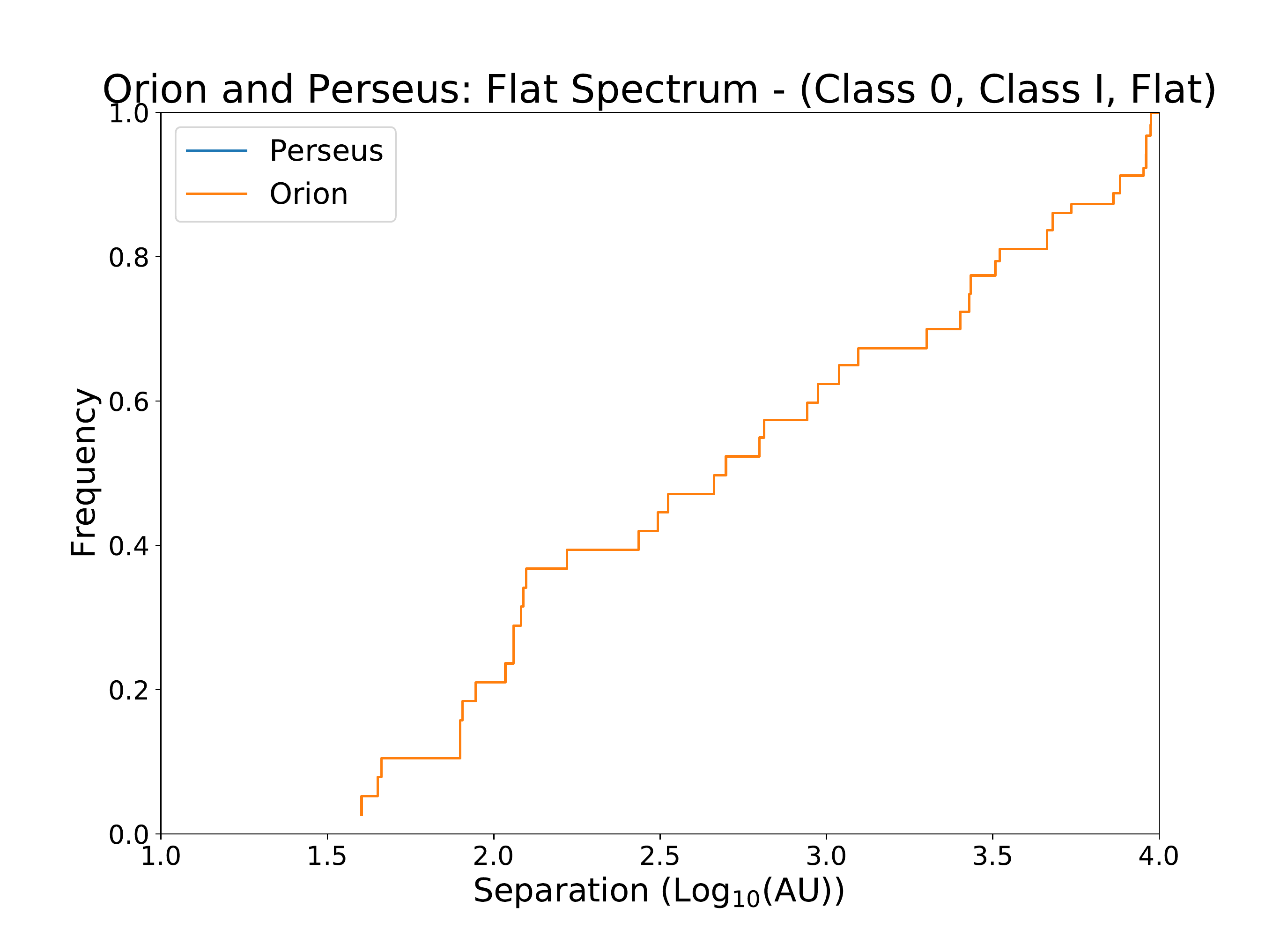}
\end{center}
\caption{Cumulative separation distributions for Orion and Perseus protostars
for different protostar classes. The samples for different protostar
classes are all consistent with being drawn from the same parent distribution.
The upper left panel shows the impact of our companion
probabilities on the CDFs for the full sample.
}
\label{cumulative_oriper}
\end{figure}

\begin{figure}
\begin{center}
\includegraphics[scale=0.3]{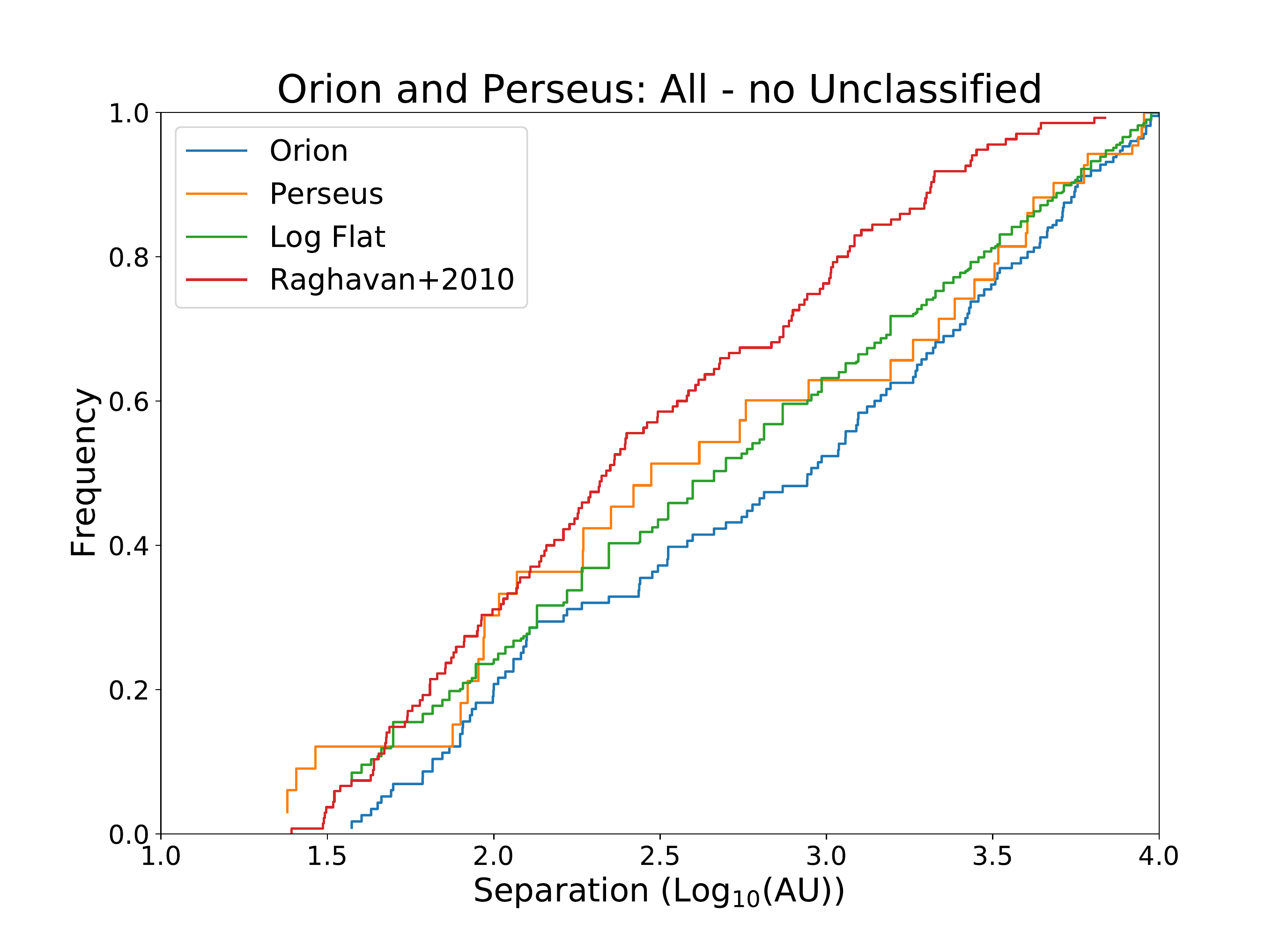}

\end{center}
\caption{Cumulative separation distributions for Orion and Perseus protostars 
shown relative to a log-flat separation distribution and the separation distribution
for solar-type field stars from \citep{raghavan2010}. The log-flat distribution shown 
is calculated with respect to the Orion separations.
}
\label{cumulative_oriper_logflat_field}
\end{figure}

\begin{figure}
\begin{center}
\includegraphics[scale=0.3]{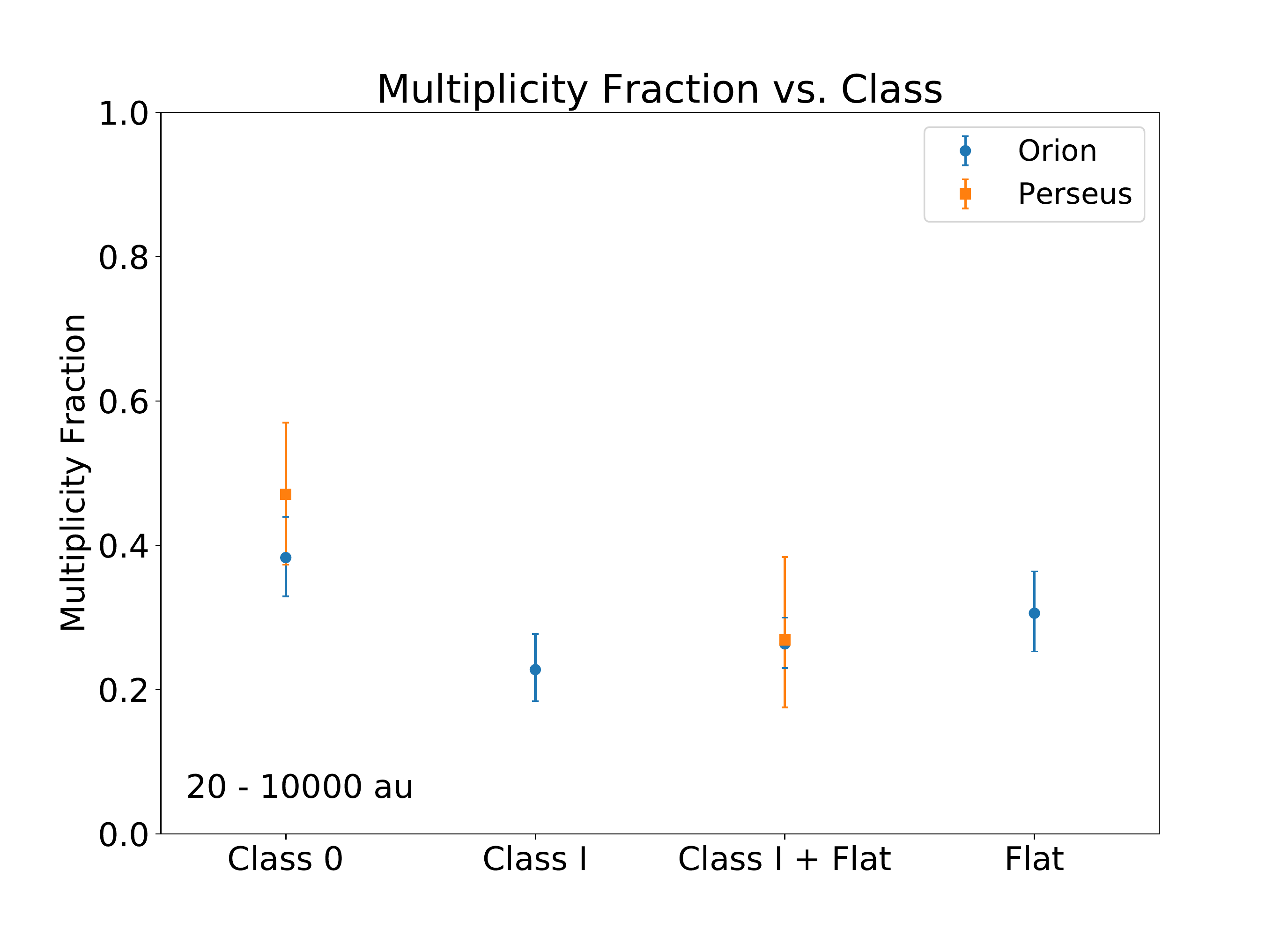}
\includegraphics[scale=0.3]{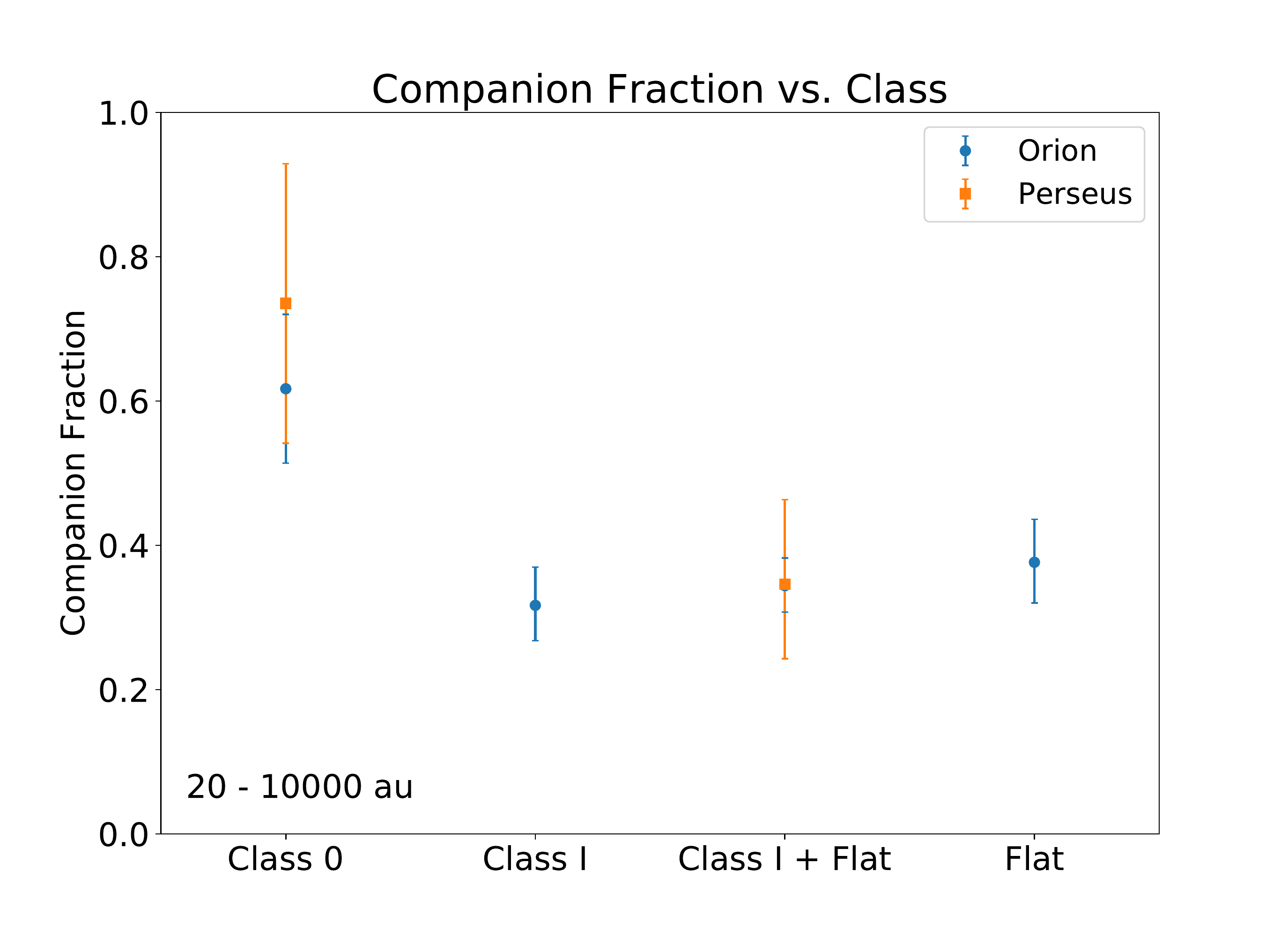}
\includegraphics[scale=0.3]{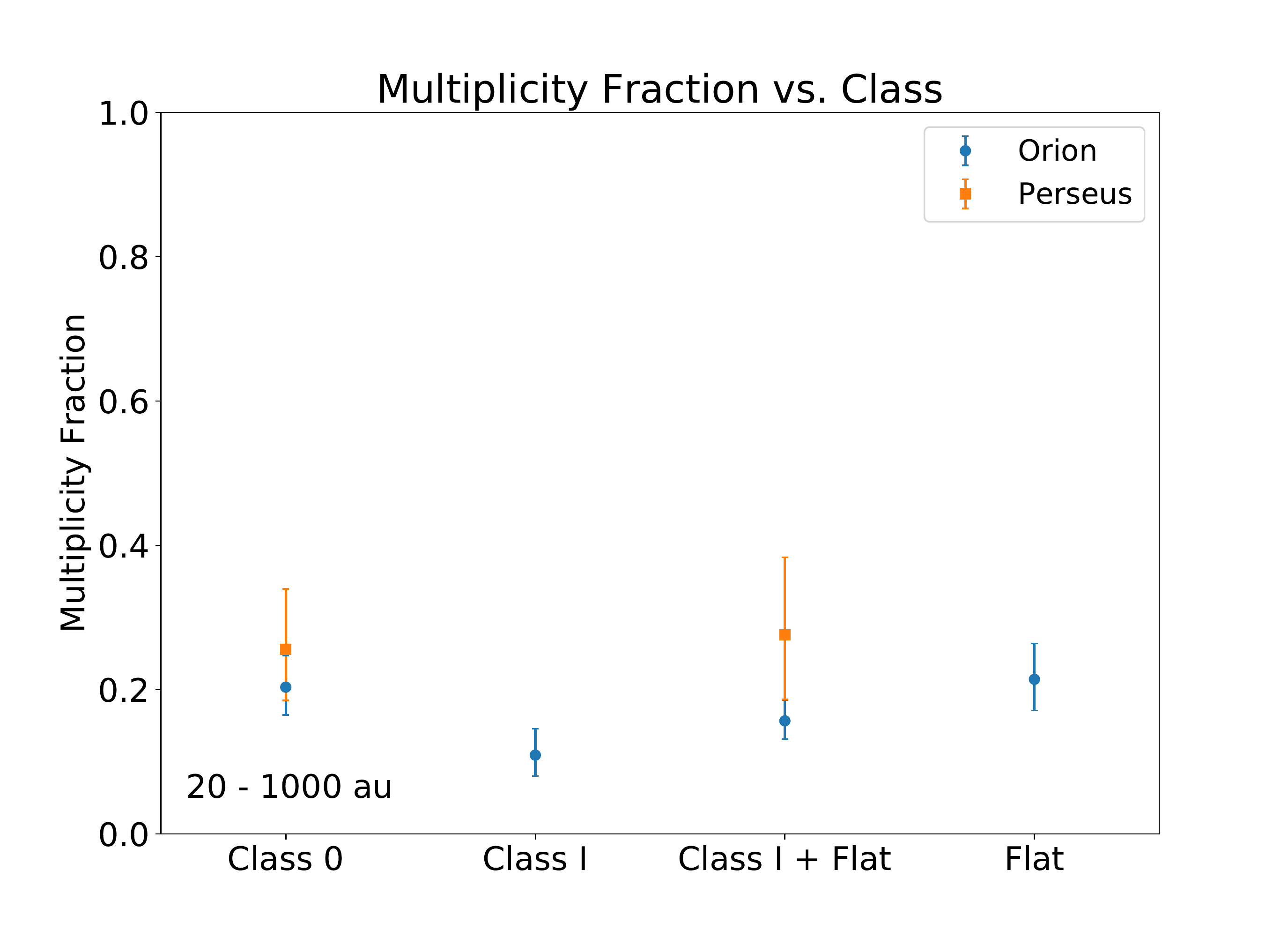}
\includegraphics[scale=0.3]{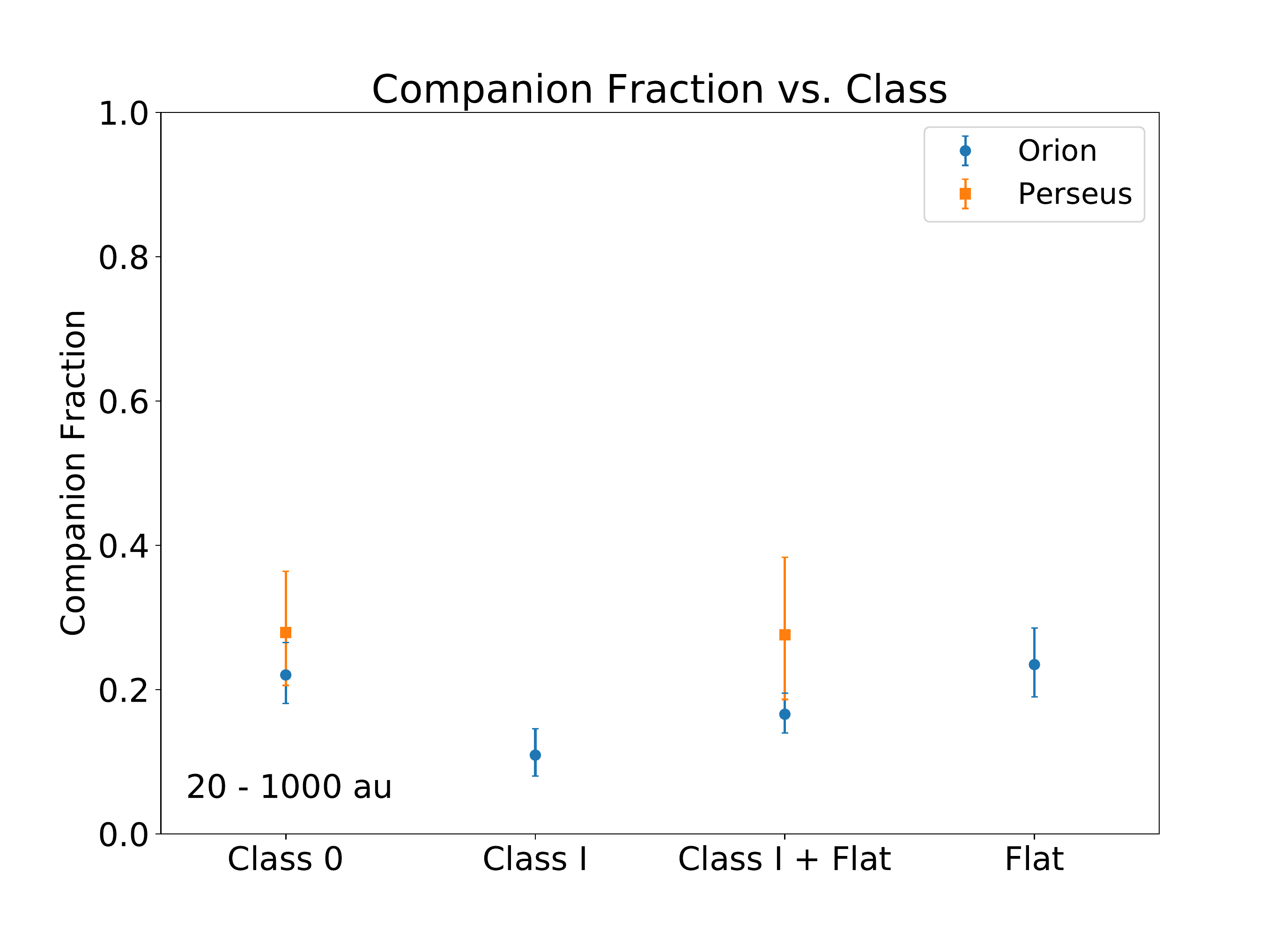}
\includegraphics[scale=0.3]{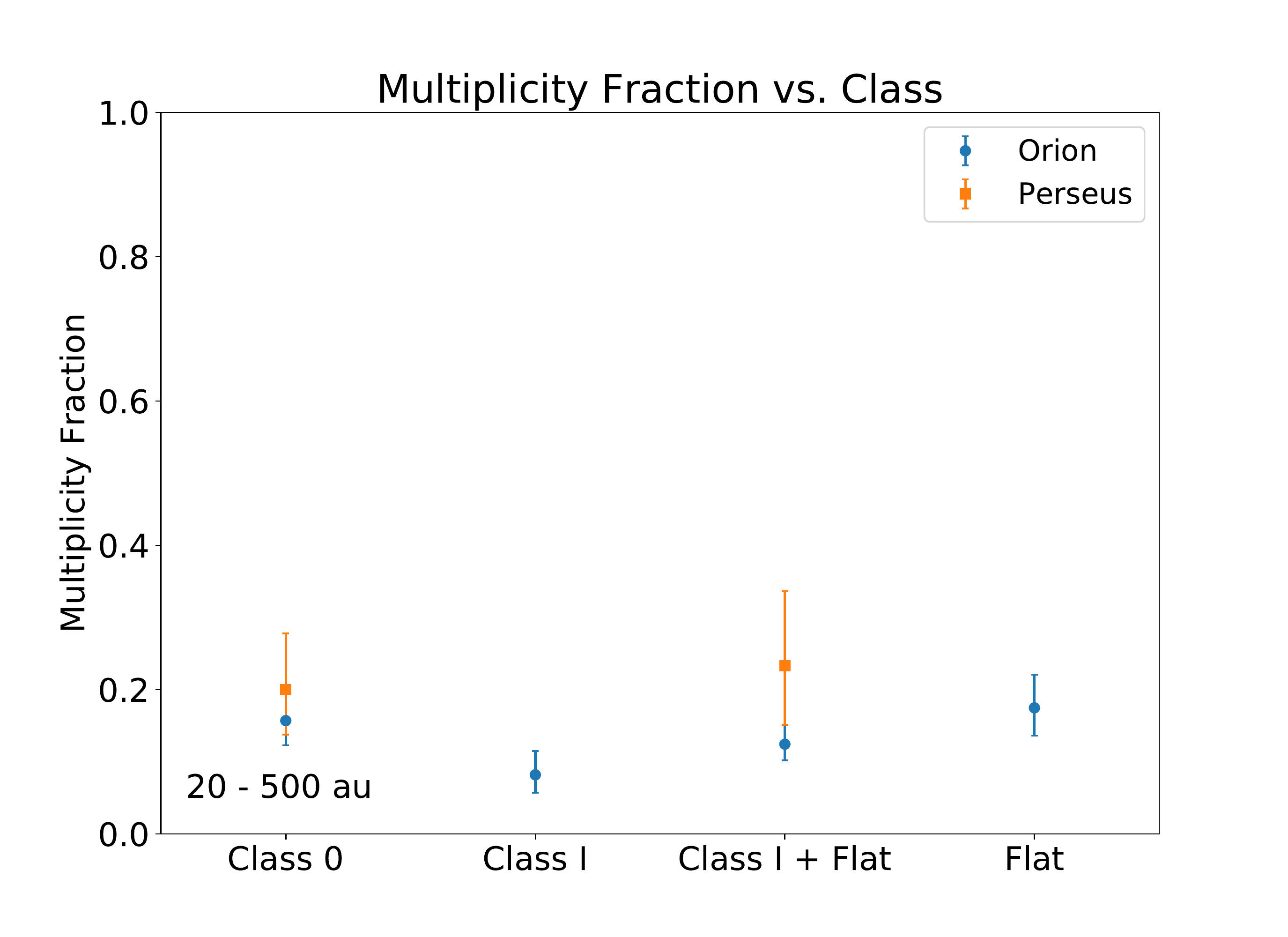}
\includegraphics[scale=0.3]{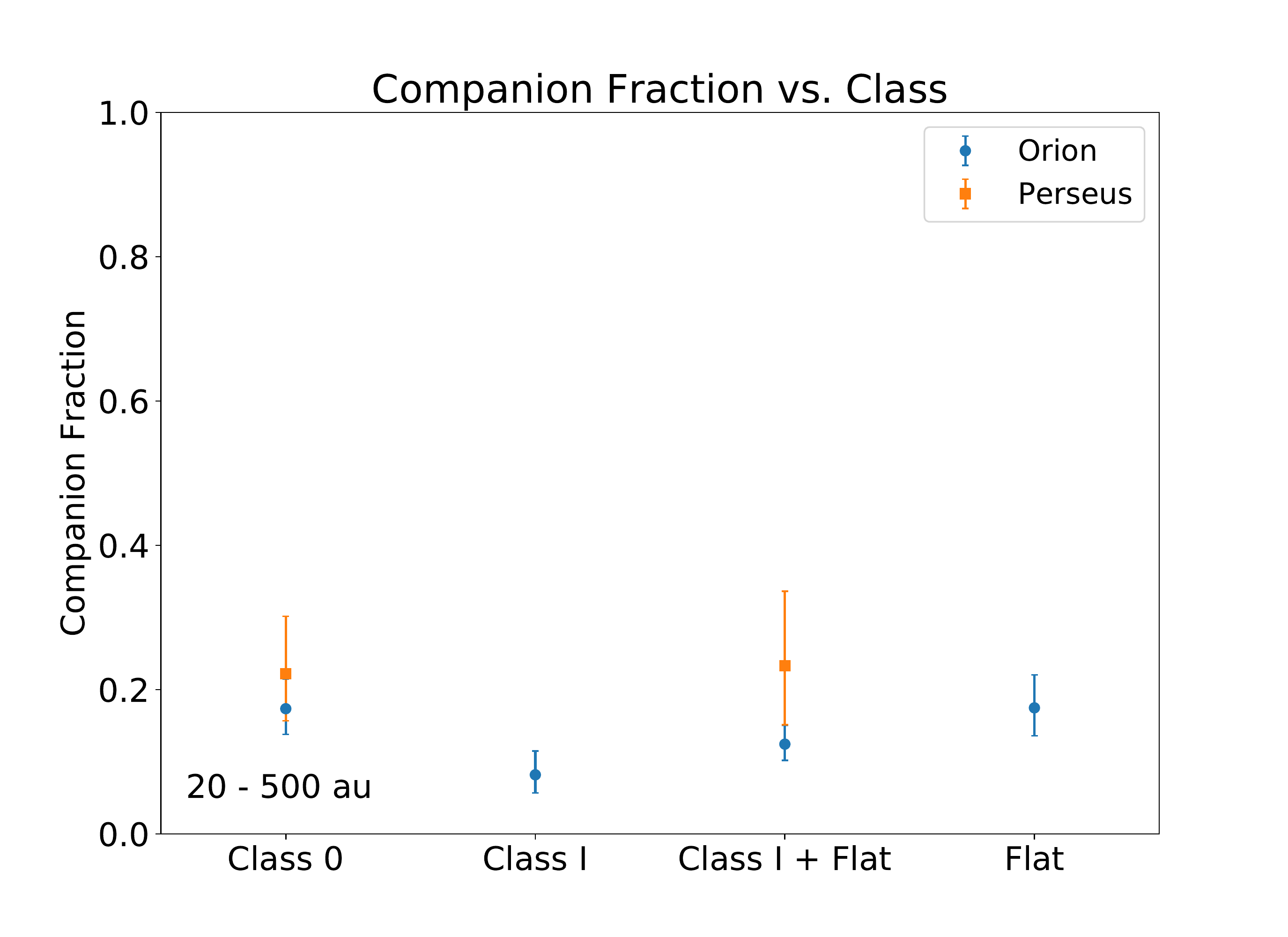}
\end{center}
\caption{
Plots of Multiplicity Fraction (left) and Companion 
Fraction (right)
for different ranges of separations and plotted as a function of protostellar class for
Orion and Perseus.
The CFs on 20 to 10$^4$~au scales for Class 0 protostars differ by $\ge$2$\sigma$, 
but less than 3$\sigma$,
with respect the other Classes.
}
\label{mf_csf}
\end{figure}

\begin{figure}
\begin{center}
\includegraphics[scale=0.3]{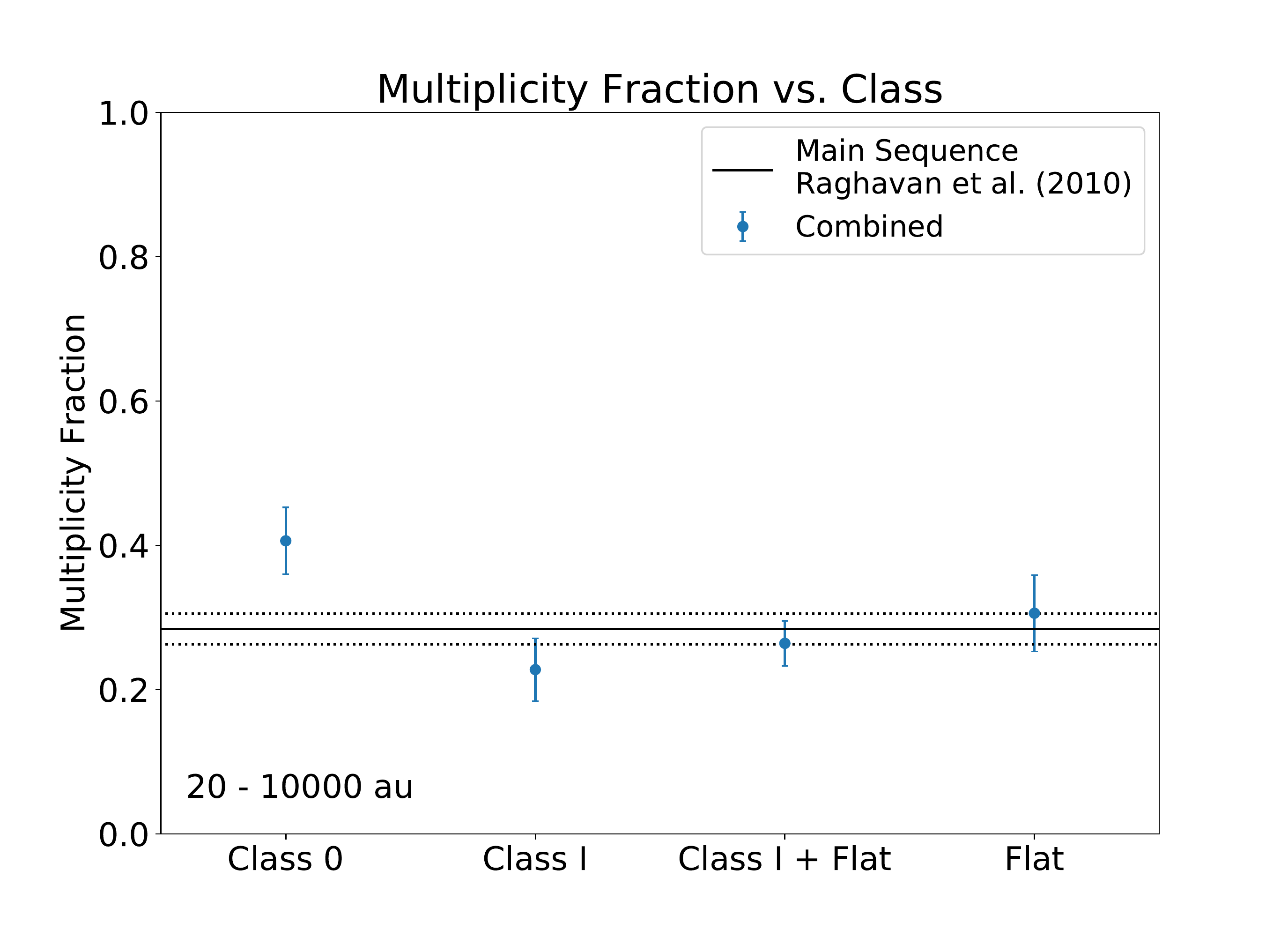}
\includegraphics[scale=0.3]{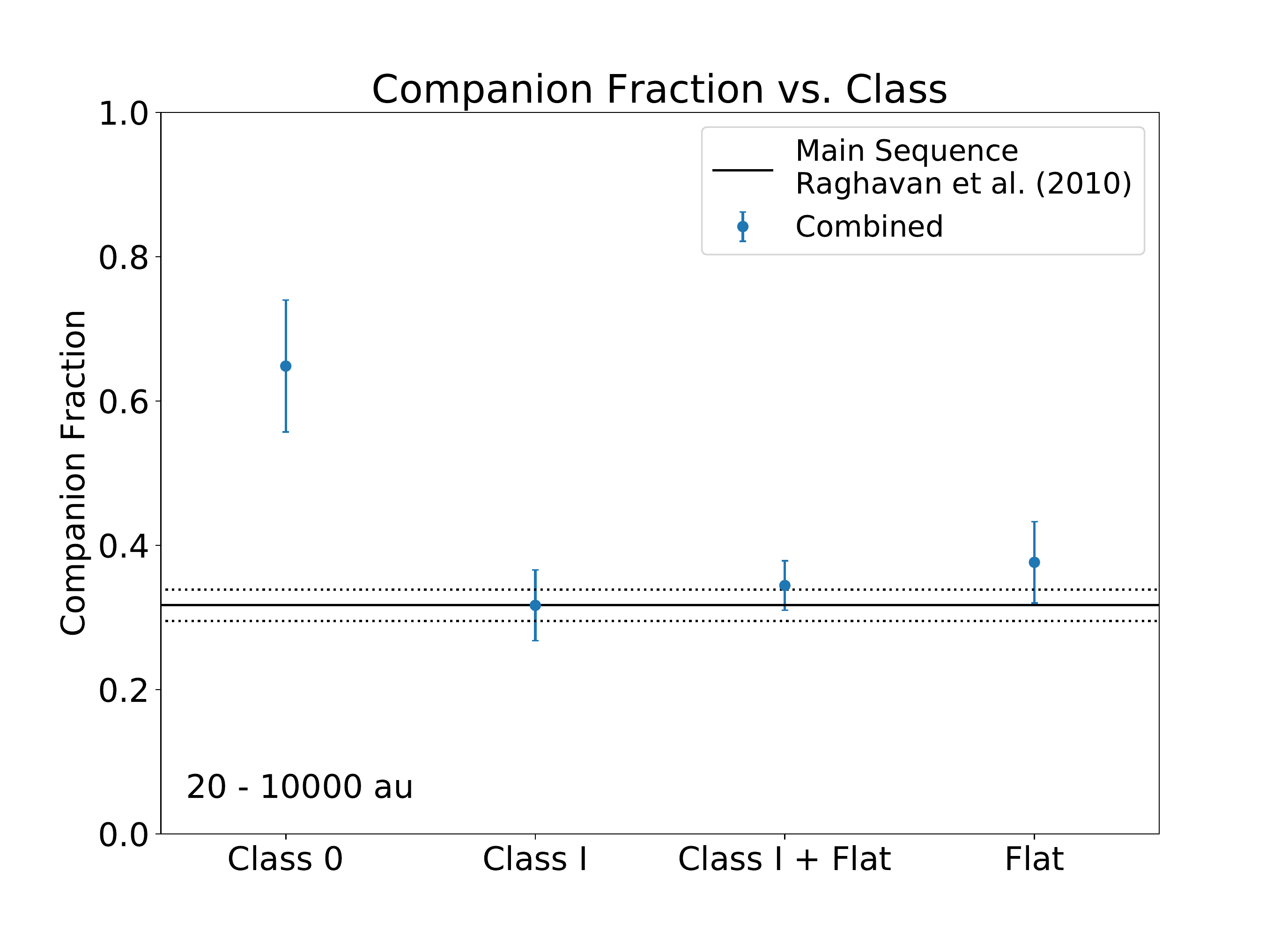}
\includegraphics[scale=0.3]{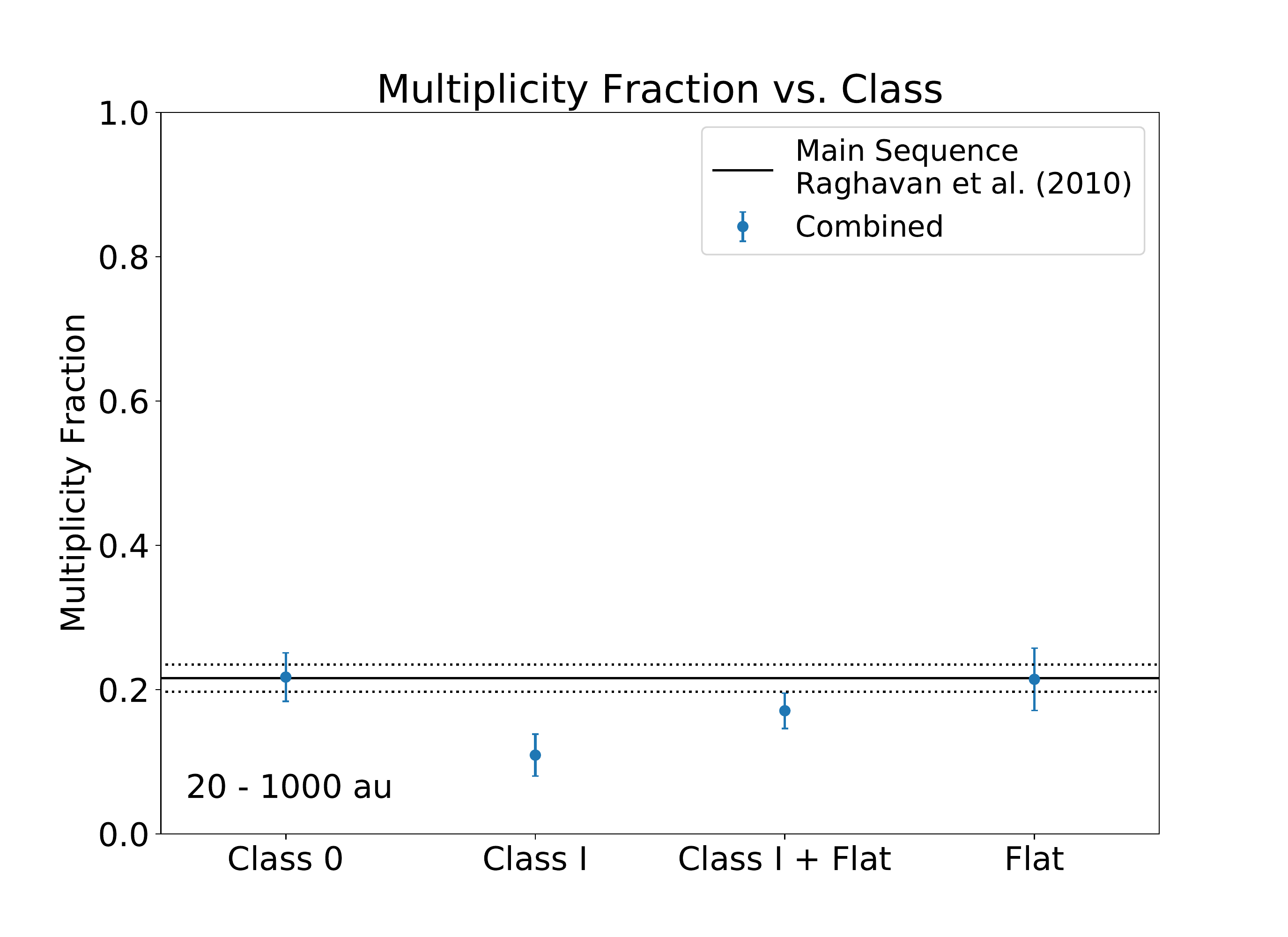}
\includegraphics[scale=0.3]{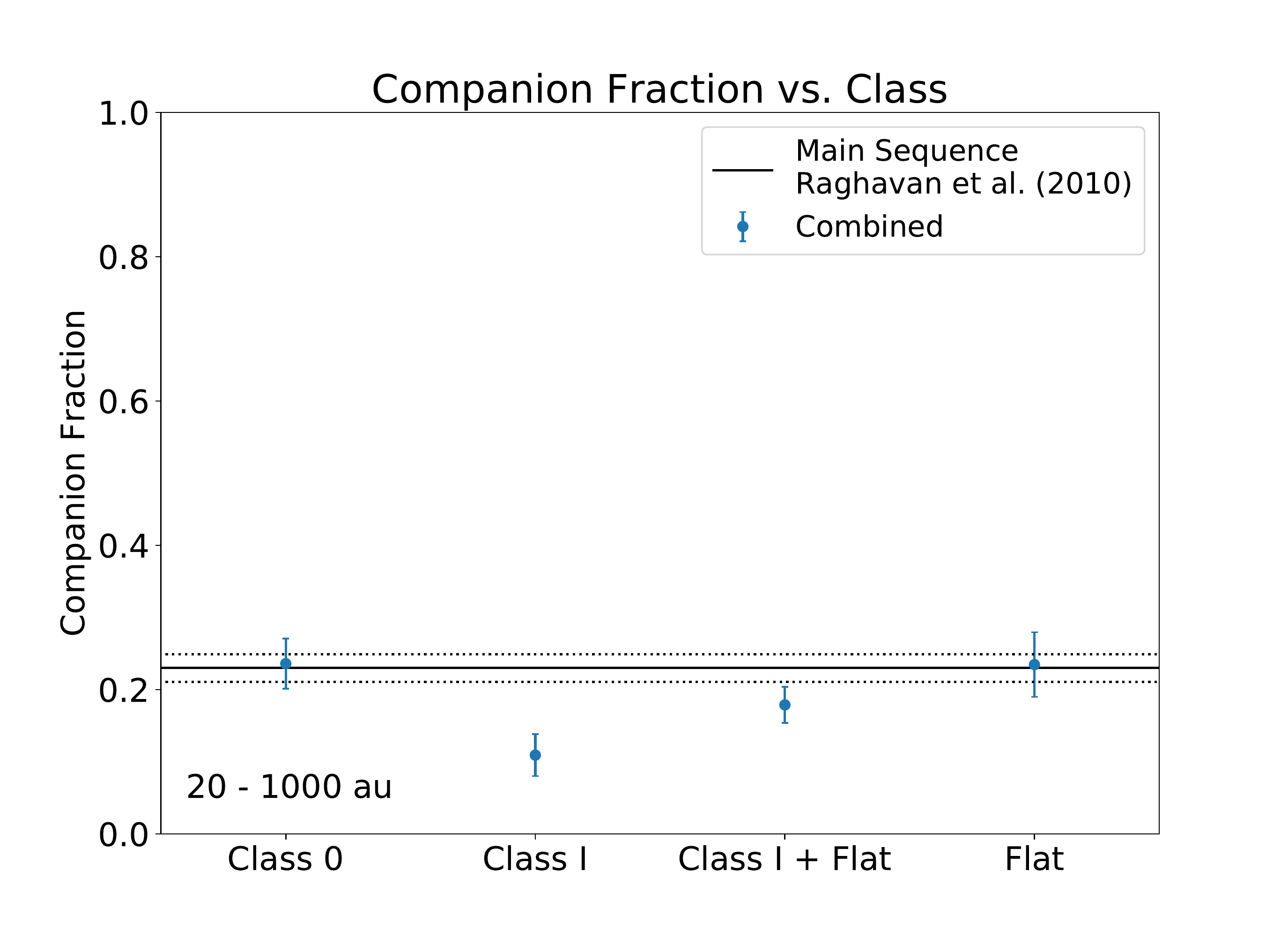}
\includegraphics[scale=0.3]{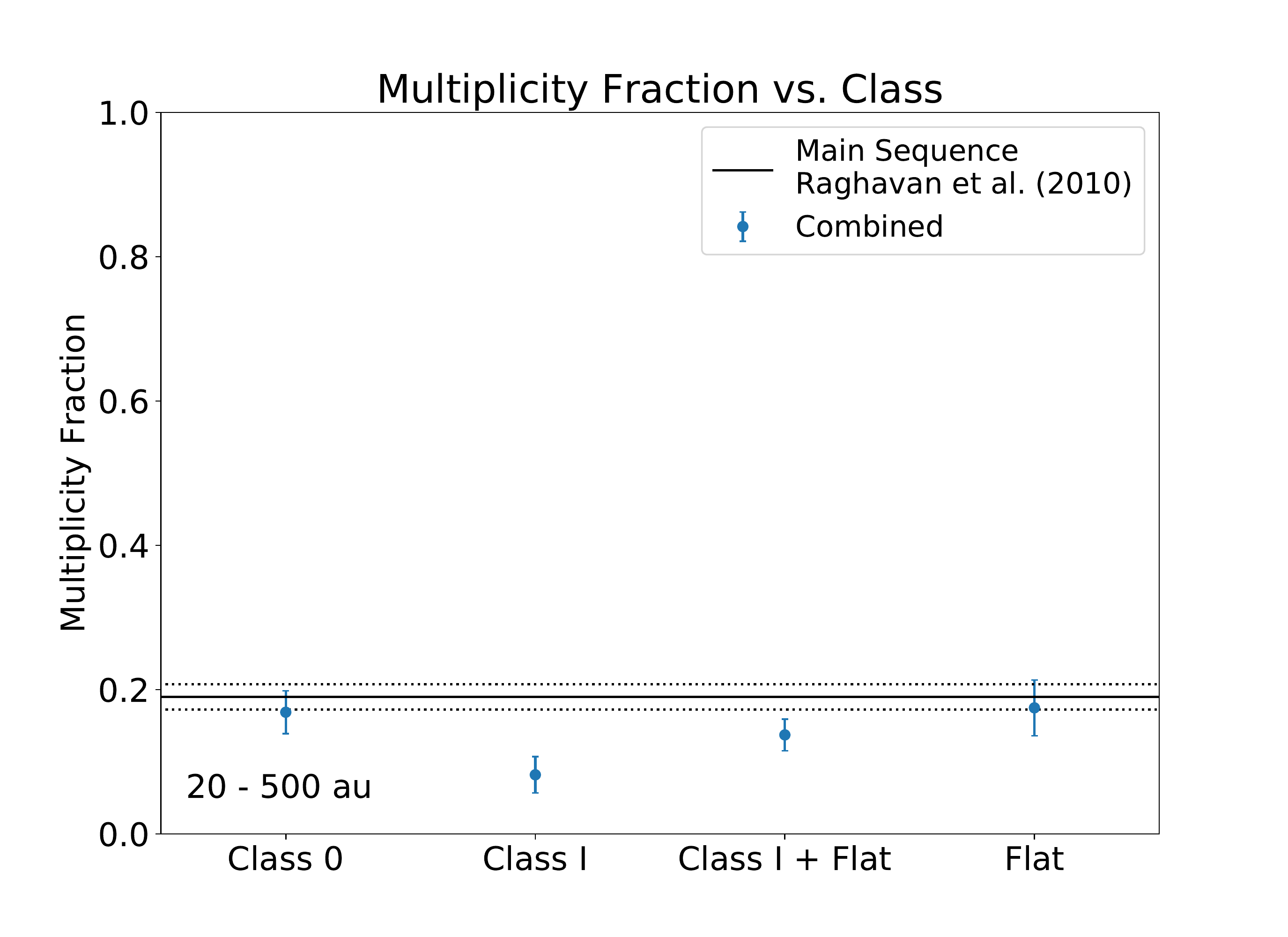}
\includegraphics[scale=0.3]{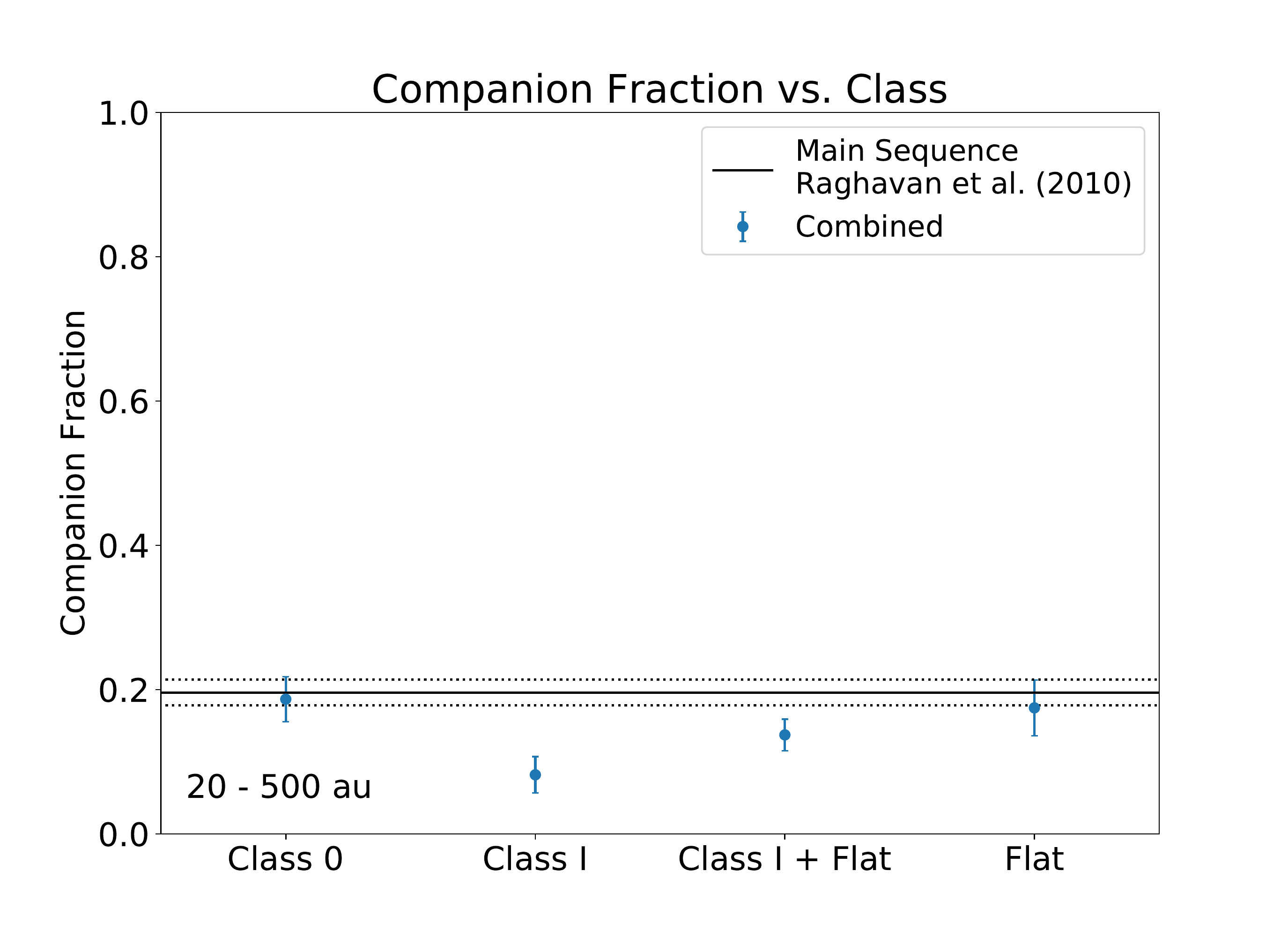}
\end{center}
\caption{
Same as Figure \ref{mf_csf} but for the combined sample of Orion and Perseus. The
MF and CF for solar-type field stars from \citet{raghavan2010},
calculated for the same ranges of separations as the protostar
data, are 
also included for comparison. The Class 0 protostars are
inconsistent with field stars for separations of
20 to 10$^4$~au but are consistent for separations of 20 to
10$^3$~au and 20 to 500~au. The most evolved
protostars (Flat Spectrum) are consistent with the field at 
all ranges of separations.
}
\label{mf_csf_combined}
\end{figure}

\begin{figure}
\begin{center}
\includegraphics[scale=0.3]{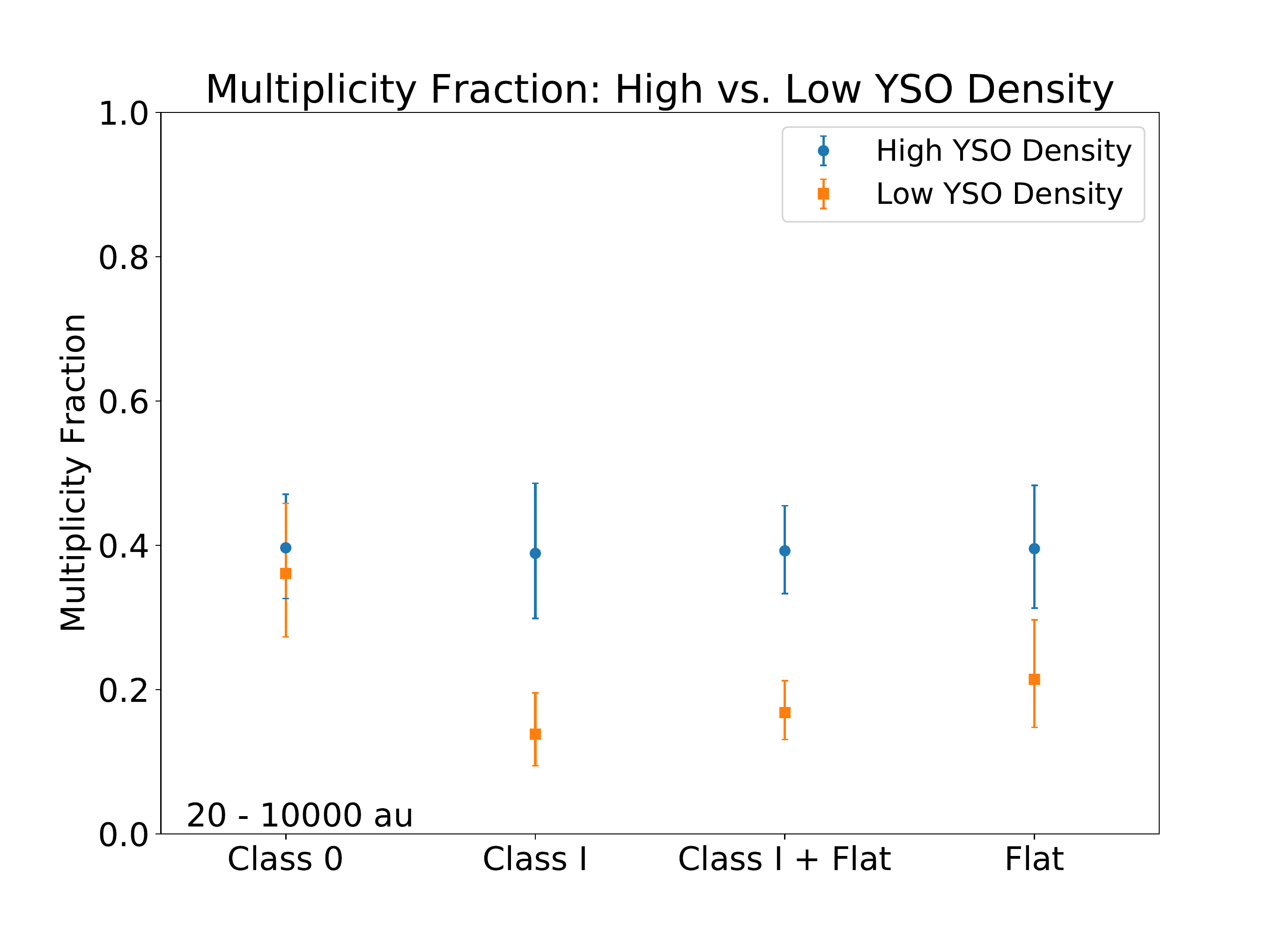}
\includegraphics[scale=0.3]{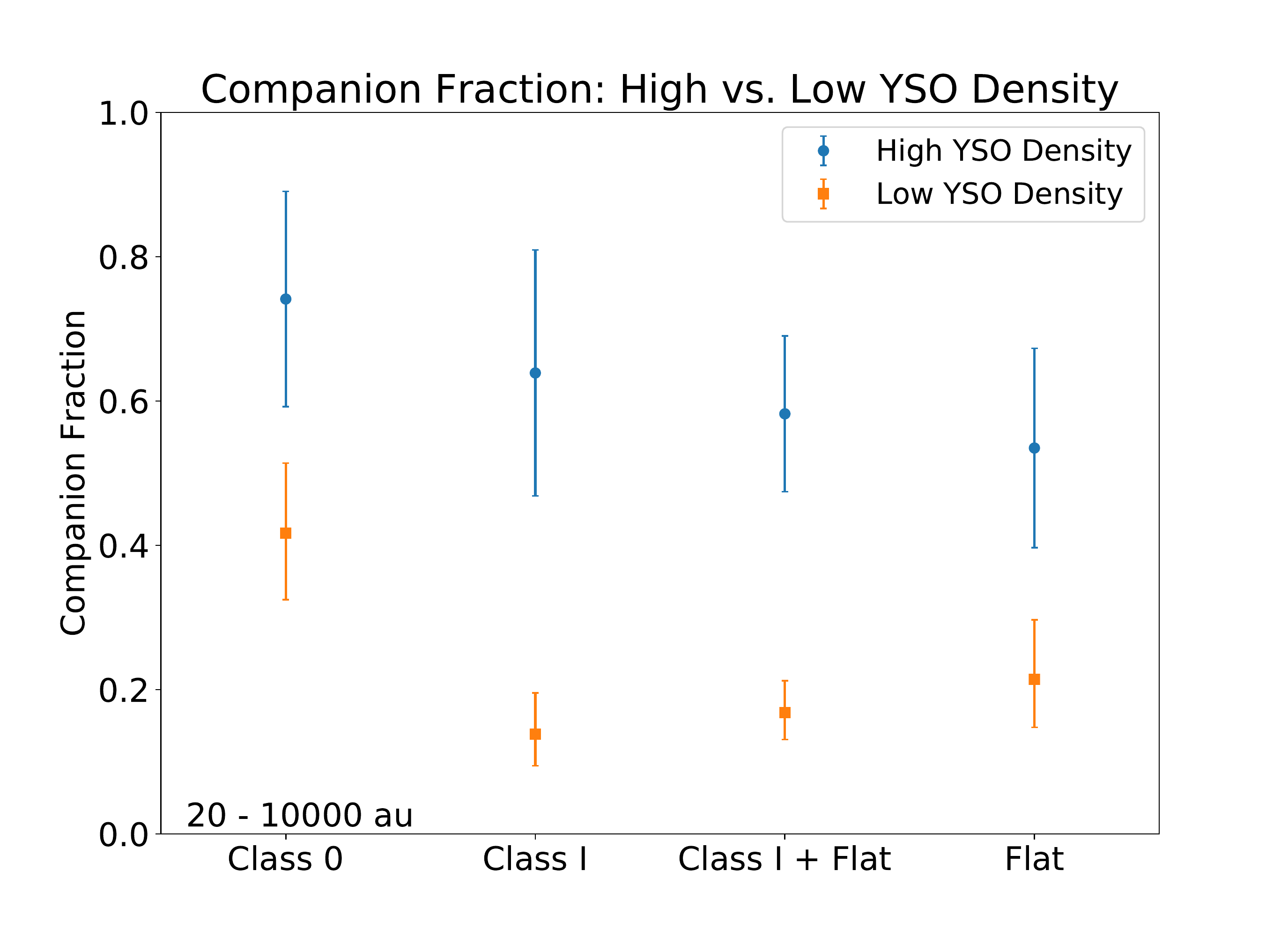}
\includegraphics[scale=0.3]{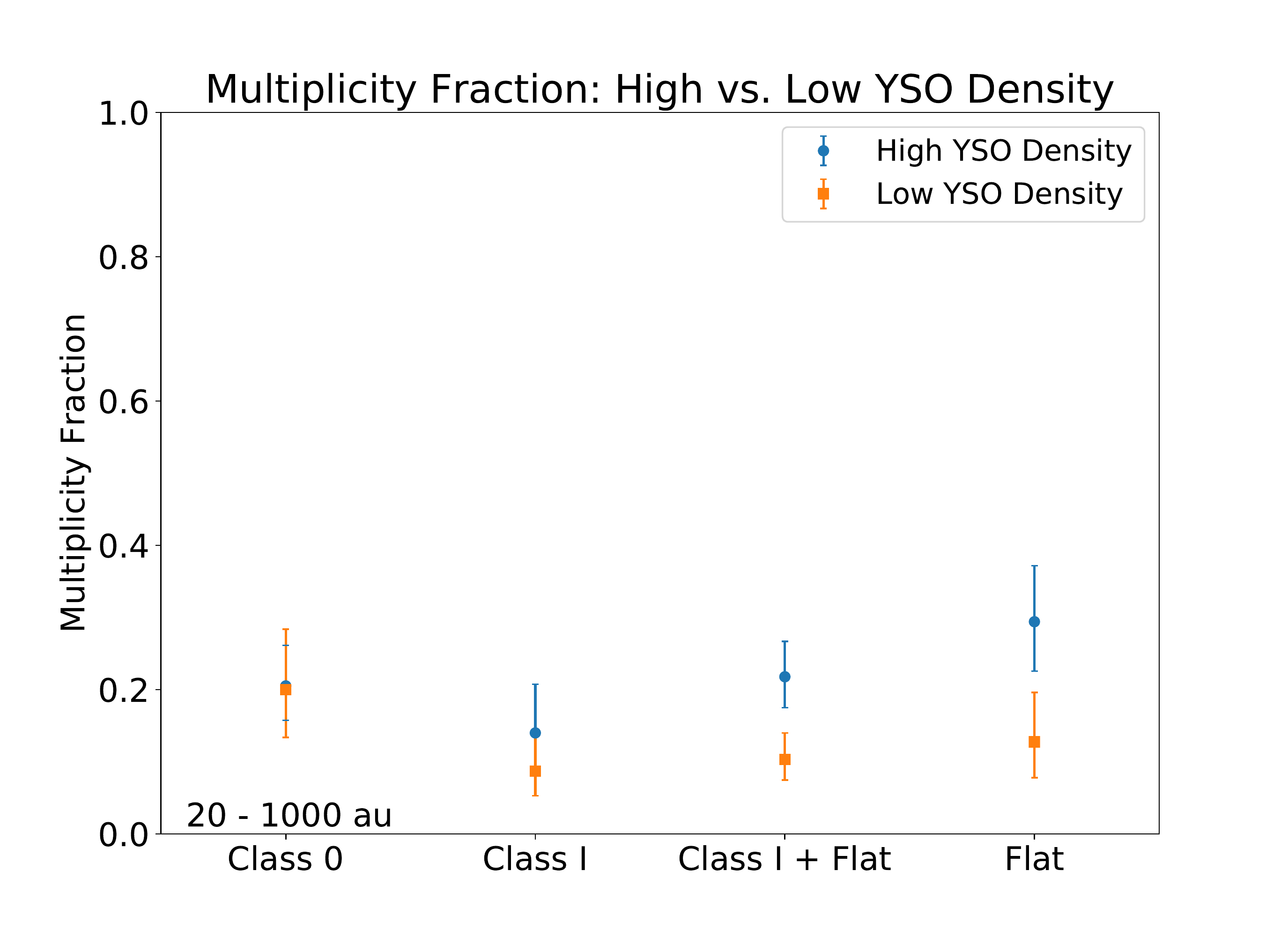}
\includegraphics[scale=0.3]{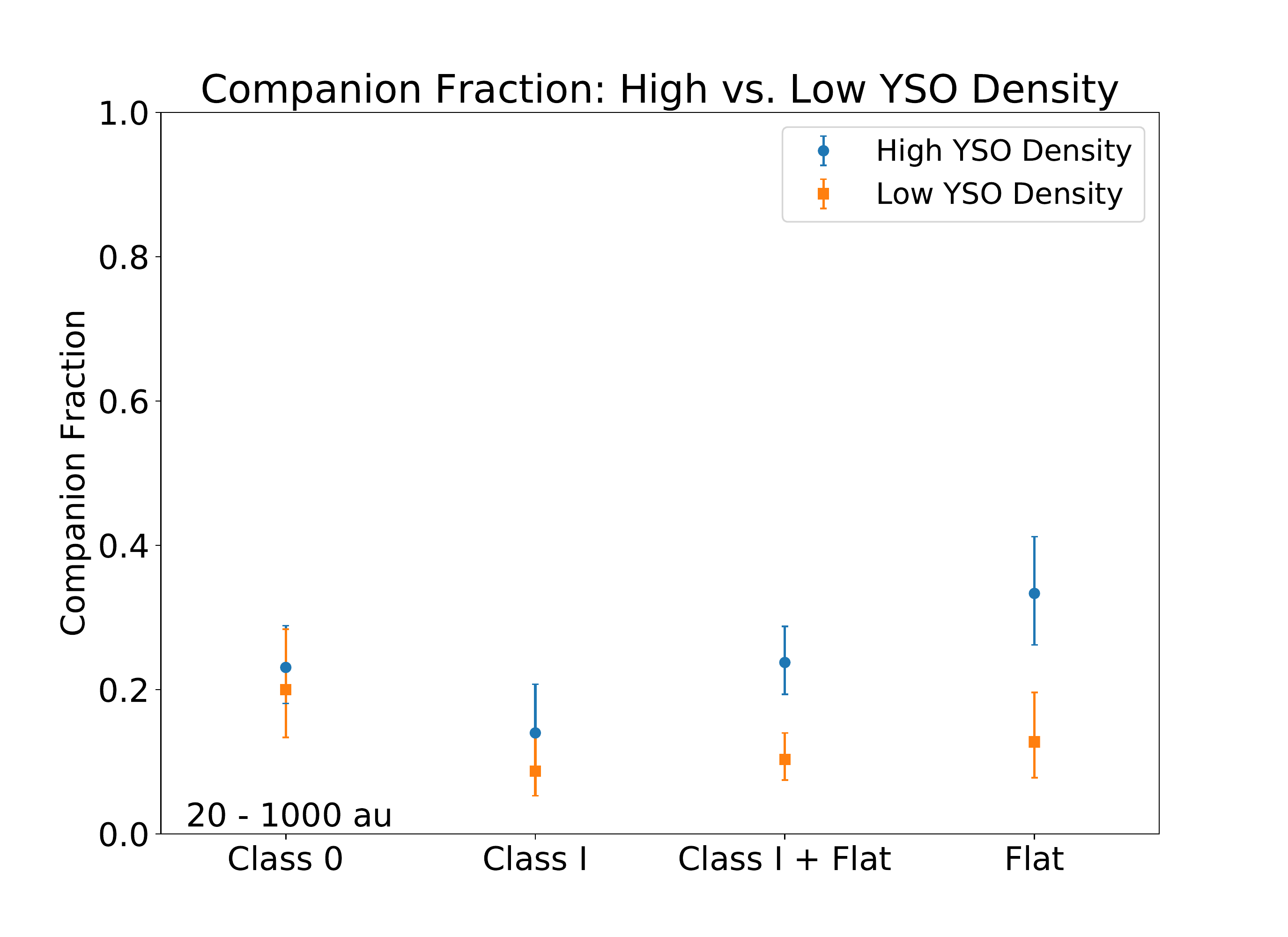}
\includegraphics[scale=0.3]{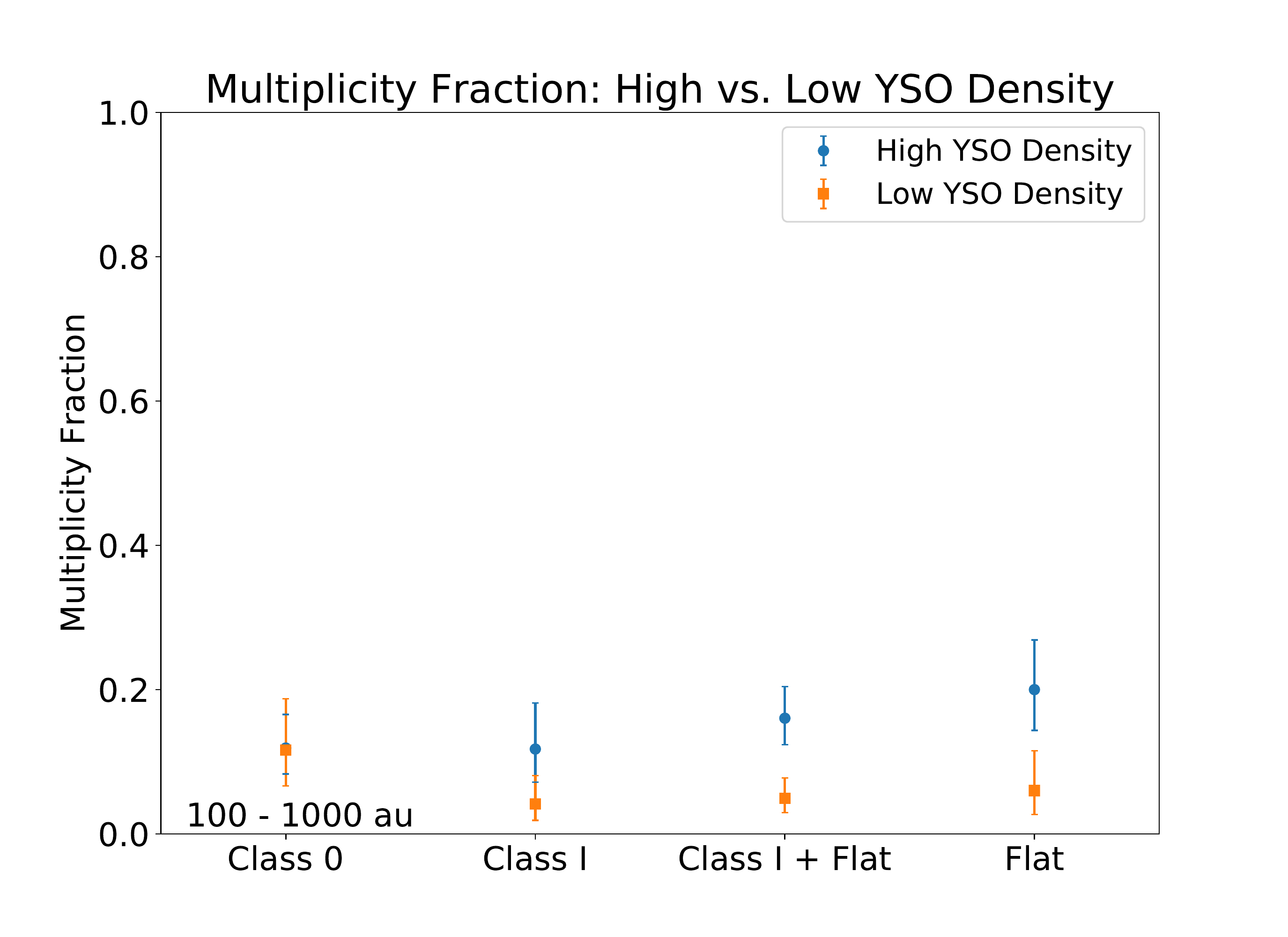}
\includegraphics[scale=0.3]{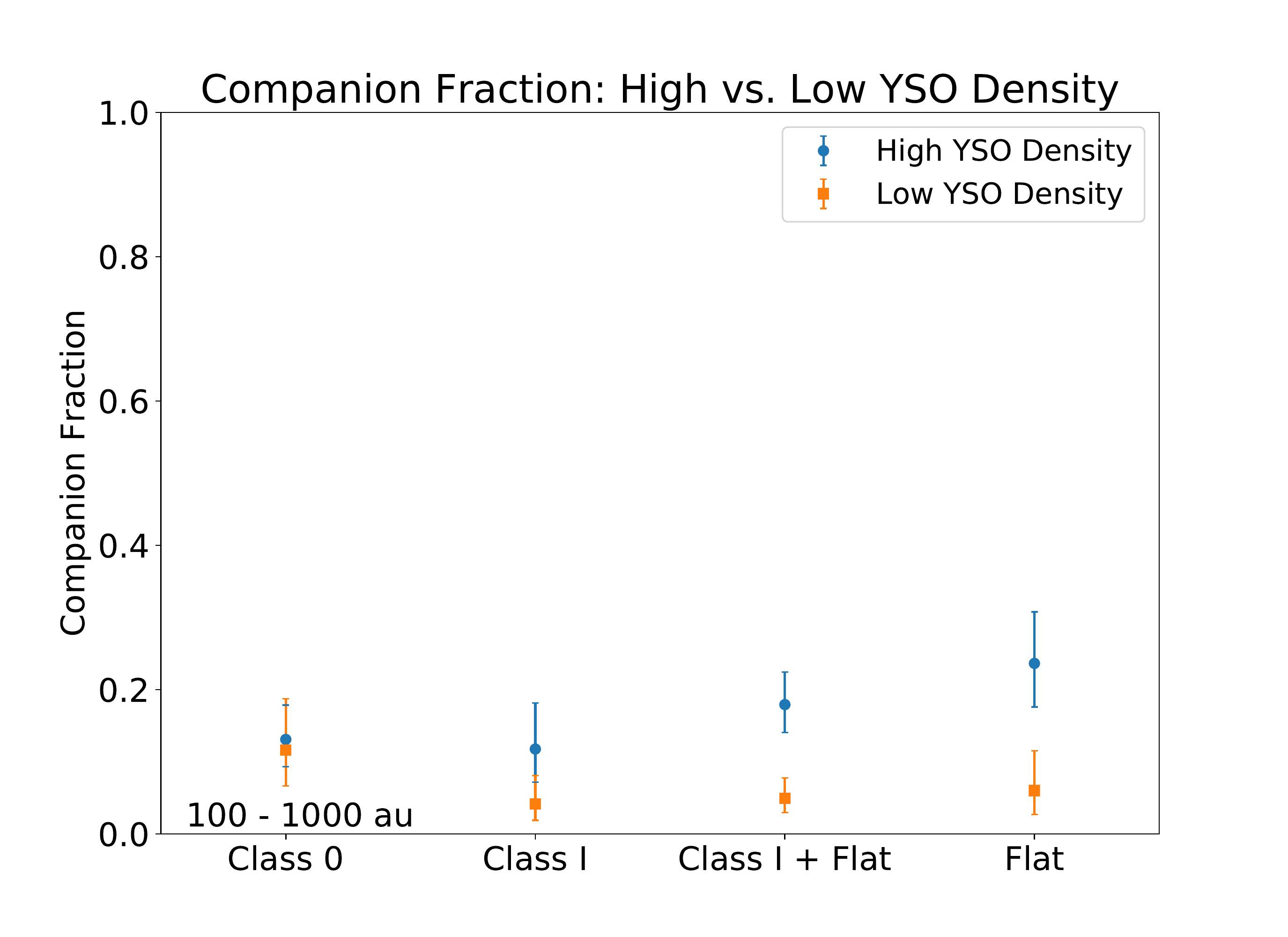}

\end{center}
\caption{
MFs and CFs on different scales as a function of protostar class for protostars
residing at high ($\ge$30~pc$^{-2}$) and low ($<$30~pc$^{-2}$) YSO surface density.
Regions with high YSO density have systematically larger CFs, and the MFs are also larger for
Class I and Flat Spectrum protostars. 
For 20 to 10$^4$~au separations, the MF differences between high and low YSO
density are all $<$2$\sigma$, while the CF differences are $>$3$\sigma$ for Class I and 
Class I + Flat Spectrum and $\sim$1.8$\sigma$ for Class 0 and Flat Spectrum protostars.
Class 0 protostars have comparable MFs and CFs 
for 20 to 10$^3$~au and 100 to 10$^3$~au, indicating primordial similarity of the multiplicity
statistics in the Class 0 phase. The MFs and CFs are rising in the 20 to 10$^3$~au and 100 to 10$^3$~au 
ranges for the more-evolved protostars, which is interpreted as a sign of 
companion migration, see Section 4.5 and 5.3. However, the significance of the MF and CF
differences are all $<$2$\sigma$ (see Sections 4.5 and 5.3).
}
\label{mf_csf_yso_density}
\end{figure}

\begin{figure}
\begin{center}
\includegraphics[scale=0.6]{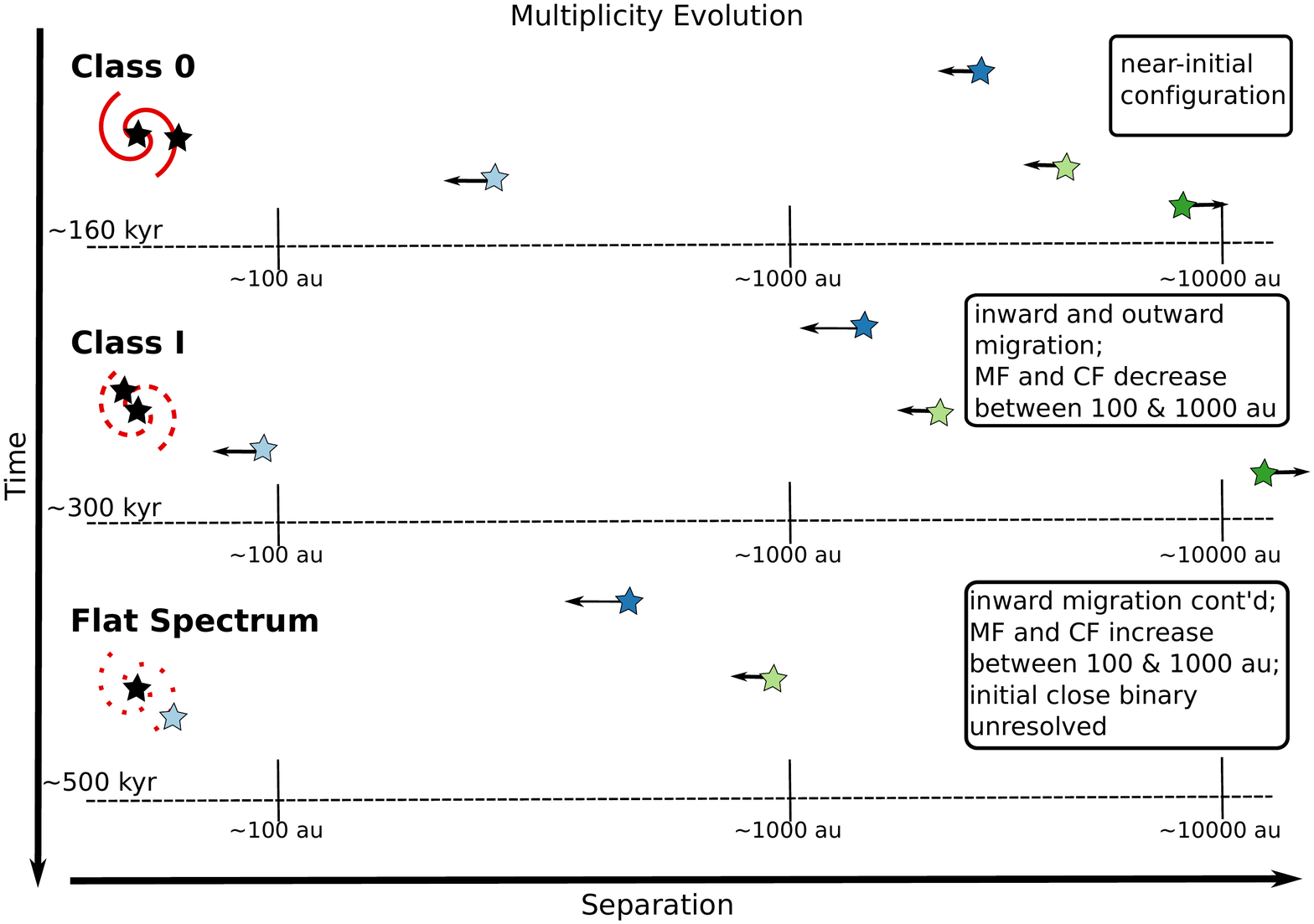}
\end{center}
\caption{
Illustration of our proposed scenario for migration to explain the MF and CF changes from 
Class 0 to Flat Spectrum on 100 to 1000~au separations as shown in Figure
\ref{mf_csf_yso_density} for YSOs in regions with high YSO density.
We begin with a Class 0 system at the top that forms an initial binary within
its disk (black stars and red spiral), but in its vicinity another protostar formed
between 100 and 1000~au and three more between 1000 and 10000~au. As an example,
the three nearest protostars at $>$100~au have net inward velocities, while the
outermost star (red) has an outward net velocity. By the Class I phase, after $\sim$160~kyr,
the inner binary (black stars) has become closer and the protostar that
was originally between 100 and 1000~au is now at $<$100~au. Then, two of the
widely separated protostars have begun to migrate inward, but are still at $>$10$^3$~au,
and one protostar (red) has moved to $>$10$^4$~au and is no longer considered a companion. 
This scenario could explain the slight drop in MF and CF between 100 and
10$^3$~au for Class Is. Finally, by the Flat Spectrum phase, after $\sim$300~kyr, the initial 
inner binary shrunk to $<$30~au and is no longer resolved, but a companion that was
previously between 100 and 10$^3$~au is now orbiting the close binary. Then the
two protostars that were initially $>$10$^3$~au (but were migrating inward) are now
located between 100 and 10$^3$~au. Thus, this could explain the rise in MF and CF for 
Flat Spectrum protostars that are expected to be among the most evolved in our sample.}
\label{multiplicity_evolution}
\end{figure}

It is desirable to report our multiplicity statistics for several different subsets of sources, 
selecting on separation range, protostellar class, and region. As described
in Section 2.4.1, we iteratively search for multiples from our resolution limit out to 
a maximum separation of 10$^4$~au in evenly spaced logarithmic bins. At each 
step we compute the multiplicity statistics, the MF, and the CF, as well as breaking down the results by evolutionary class.
We treat each continuum source as a discrete source if it has not been paired with another source
on the separation range currently being tabulated. This is illustrated in Figure \ref{example},
where we show the number of singles, binaries, triples, and quadruples in each panel; refer to Section 2.4.3
for more detail on the multiplicity statistics. If this hypothetical
system was a single HOPS source in our catalog, it would be counted as 4 singles if the maximum
scale being examined was smaller than the separations of any pair of protostars.  After A and B are paired
the configuration is counted as 2 singles and 1 binary in the statistics, and for the
largest separation range, it is considered a single quadruple system. The individual 
continuum sources that comprise a single HOPS or Perseus
catalog entry will inherit the protostar class,
$L_{bol}$, and $T_{bol}$ that they are associated with from the input catalog.

When considering higher order systems, especially when separations become wide, more than one
source from the input catalog could be considered part of a multiple system. Thus, multiple systems may contain a mix
of protostar classes and sometimes unclassified continuum sources. 
The class that has the largest number of components will determine which class group
the statistics are added to. 
If there are equal numbers of a 
particular class, for example two Class 0s and two Class Is,
then the earlier evolutionary classes will take precedence 
and the system would be considered Class 0. 
Because
we report the statistics using the most common Class in the multiple system,
adding up 
the numbers of components (e.g., N singles + 2$\times$Binaries, + 3$\times$ Triples) 
of each Class will not equal the combined number of 
components for all ranges of separations considered. Put simply, 
the total number of components listed in the multiplicity 
tabulation of Class 0 protostars from 20 to 10$^3$~au will not be equivalent 
to the total number of Class 0 components from 20 to 10$^4$~au 
because some of these may be considered Class I or Flat Spectrum
protostars as a whole, depending on the classification and number of wide companions. 

If a system has unclassified sources, we still include them in the per-class
statistics if there is only a single unclassified source. Unclassified systems can
come from regions that did not have valid infrared source detections due to confusion or high
extinction. Systems that include 
more than one unclassified continuum source are left out of the per-class 
statistics, but these systems are still counted in the full sample statistics that 
include unclassified sources. The association of unclassified sources is fairly 
rare and the most substantial impact is in the field around HOPS-384 where 
numerous additional sources were detected by the VLA.
The systems including HOPS-394, HOPS-370, HOPS-108, and HOPS 92 each have a single 
unclassified source associated with their higher-order systems. Then HOPS-56 
and NGC2024 FIR3 both have two unclassified sources associated with them. 

It is important to point out that these `unclassified' sources are likely 
Class II or Class III YSOs and not background galaxies. Two
background radio galaxies were in the VLA observations toward 
HOPS-173 and HOPS-168, but they clearly had 
negative spectral indices and had no associated emission from 1 to 24~\micron.
Class III YSOs can also be detected by the VLA at 9.1~mm from their free-free emission, which 
tends to have a flat spectral index rather than a negative spectral index, 
even if their dust emission is too weak to be detected by ALMA.
Finally, the sources within OMC1N are listed as unclassified but are deeply embedded
and are presumed to be Class 0 protostars. In any event, the OMC1N sources are only 
in close enough proximity to each other to yield pairs with sources also within OMC1N.

The separation distribution histograms (Figures \ref{separations_orion_all}-\ref{separations_comb})
show all of the separations for each pairing
of individual continuum sources and paired/grouped continuum sources (see Section 4.1). For the example shown in 
Figure \ref{example}, the histogram of separations and CDF would contain three 
source pairs with separations, d$_{AB}$, d$_{CD}$, and d$_{AB,CD}$. The same
criteria mentioned in the previous paragraphs are also used here to determine the Class a particular
separation belongs to. However, there is a bit more ambiguity in the distribution of separations, because when 
continuum sources are paired, their catalog entries are removed and replaced with a new catalog entry for the paired
sources. Using the same example from Figure \ref{example}, if A is Class 0 and  B is Class I, the
combined entry AB is considered Class 0. Then if C and D are both Class I, their combined entry
will be Class I. However, when AB and CD are paired, their separation would also be considered Class 0, 
even though this quadruple system is comprised of three Class Is and one Class 0. 
This situation of mixed evolutionary classes is common for
separations $\ga$~3000~au and is an
artifact of the preference we have for assigning class based on the youngest
component. We tested creating our separation distributions where the more evolved
Class was selected, rather then the less-evolved Class
and the separation distributions do not change significantly.

\subsection{Alternative calculation of MF and CF with Probabilities}

In Section 2.4.3, we described our adopted method of calculating the multiplicity degree of a
system by the rounded sum of probabilities for all components. We then use the degree of multiplicity for
each system to calculate the MF and CF for the samples as a whole. However, we can also directly
calculate the MF and CSF using the probabilities of each system without rounding using the 
methods described below.

We first need to calculate the number of multiples, which can be computed via
\begin{equation}
N_{multiples} = N_{P_i > 2.0} + \sum_{i}^{N} (P_i(2>P_i>1)- 1.0 ).
\end{equation}
$P_i$ refers to the sum of the probabilities for each multiple system, with the probabilities
of the individual components determined as described in Appendix A.1, and
$N_{P_i > 2.0}$ is the total number of definite multiple systems, where the sum of the probabilities are 
greater than 2. Then systems 
whose companion probability sums
are between 2 and 1, which could either be either a single or a multiple 
(see section 2.4.3), 
are accounted for by adding their excess probability above 1.0.

Then the total number of companions is expressed by
\begin{equation}
N_{companions} = \sum_{i}^{N} (P_i - 1.0),
\end{equation}
because the CF counts the total number of companions in each system, which is equivalent to
the total number of components, minus 1.

Finally, to arrive at the MF and CF the number of systems is equal to the number of singles plus the sum of
the number of possible members in each system minus the sum of 
the probabilities for each system, thereby accounting for the low probability pairs in the
total number of systems. This can be expressed as 
\begin{equation}
N_{systems} = N_{singles} + \sum_{i}^{N} (N_{comps,i} - P_i ),
\end{equation}
where $N_{comps,i}$ is the total number of possible components of a multiple system, and $P_i$ refers to the 
sum of the companion probabilities for each multiple system $\sum_{j}^{N_{comps,i}}$~$P_j$, where $N$ is the total number
of components. Then we can compute the MF and CF for the chosen sample where MF = $N_{multiples}$/$N_{systems}$ 
and CF = $N_{companions}$/$N_{systems}$. When we compute the MF and CF from this method, 
the values are consistent with the method
outlined Section 2.4.3 where we round to the nearest
integer to count the number of components that make up a multiple system.
Thus, we make use of that method given that it provides the degree of multiplicity
for each system, comparable to previous work. However, the method described in this section provides 
a way to verify those results without the possible loss of information from rounding.

\subsection{Multiplicity Statistics and Statistical Tests Without Companion Probabilities}

In the main text, we described statistical tests performed on the cumulative
distribution functions in Section 4.2.
There were not significant differences between the statistical 
tests that used and those that did not use companion probabilities.
Then, we also computed the MFs and CFs without considering the companion
probabilities and list them in Tables \ref{MF-CSFs-all-noweight}, and
\ref{MF-CSFs-comb-noweight} for reference. We only include the 20 to 10$^4$~au
separation ranges because the companion probabilities did not affect the 20 to 500~au
and 20 to 10$^3$~au ranges.

\section{Comparison of Regions Within the Orion Complex}

The large sample of Orion protostars also enables analysis of multiplicity properties
from distinct regions within the Orion Molecular cloud complex. We have divided the 
Orion sample into three regions based on Declination: the northern Integral-Shaped 
Filament (ISF) including OMC-1, OMC-2, and OMC-3 (-4.5\degr~$>$~$\delta$~$>$~-5.5\degr), southern
ISF and L1641 (hereafter L1641; $\delta$~$\le$~-5.5\degr) both part of the filamentary Orion A cloud, and Orion B ($\delta$~$\ge$~-4.5\degr). 
Our sample contains 45 detected protostar systems in the ISF (17 Class 0, 12 Class I, and 17 Flat Spectrum), 
168 detected protostar systems in L1641 (37 Class 0, 68 Class I, and 63 Flat Spectrum), and 
75 detected protostar systems in Orion B (32 Class 0,
31 Class I, and 12 Flat Spectrum). This selection
divides the Orion protostars a bit unevenly, since L1641 contains more protostars
than Orion B and the ISF combined. However, the northern ISF stands out, since it has
significantly higher protostellar density than anywhere else in Orion and hosts the Orion
Nebula Cluster \citep[e.g.,][]{carpenter2000,megeath2016}. Orion B is notable for is 
low star formation efficiency, low fraction of high density filaments, and high fraction 
of Class 0 protostars \citep{stutz2015,megeath2016,orkisz2019,karnath2020}. In comparison, the L1641 
clouds contains primarily smaller groups and clusters of young stars and has a lower fraction 
of Class 0 protostars than the ISF, and has lower gas column densities 
\citep{allen2008,hsu2013,stutz2015,megeath2016}. In total, these three regions sample very different environments.

\begin{figure}
\begin{center}
\includegraphics[scale=0.3]{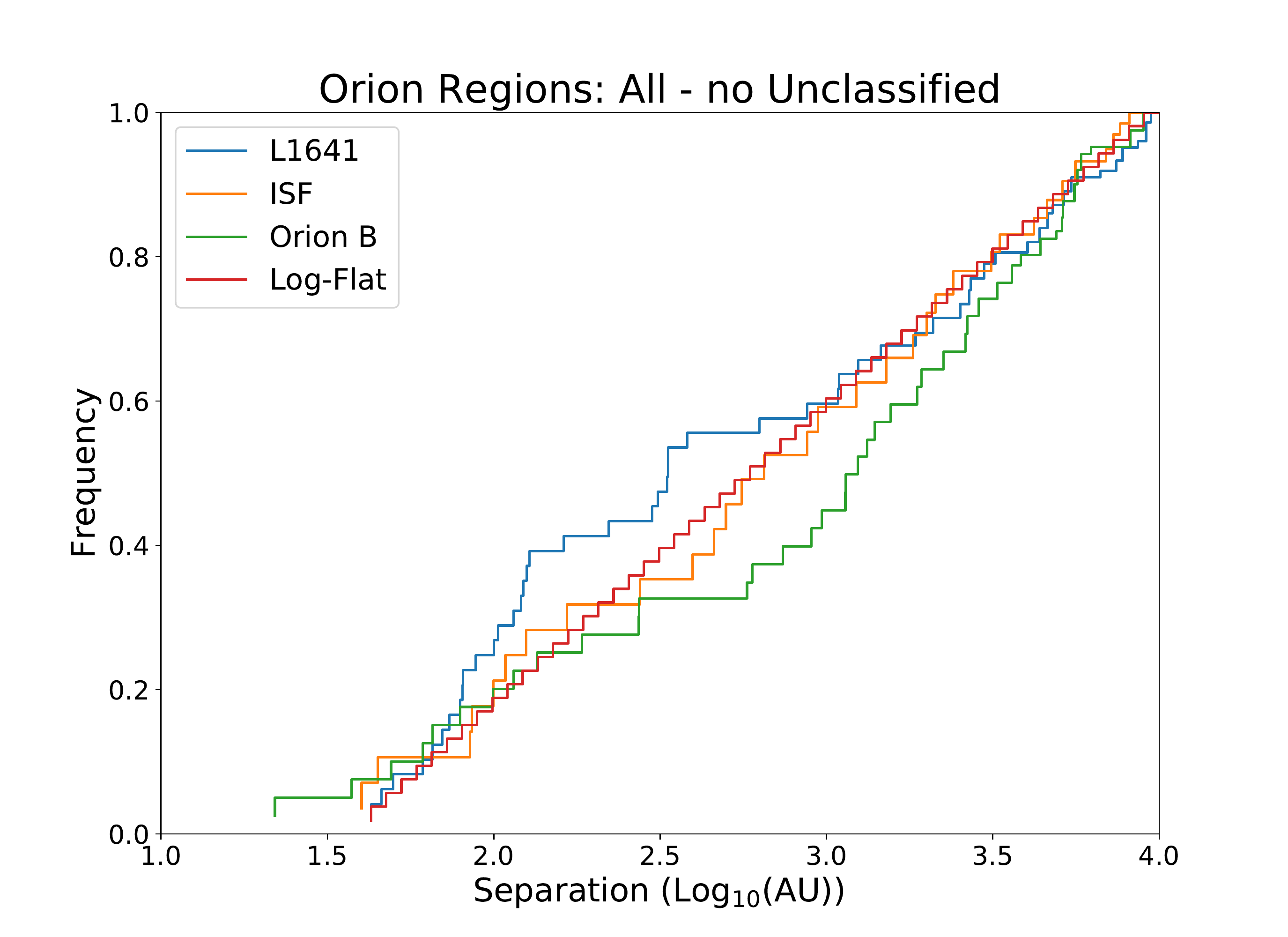}
\includegraphics[scale=0.3]{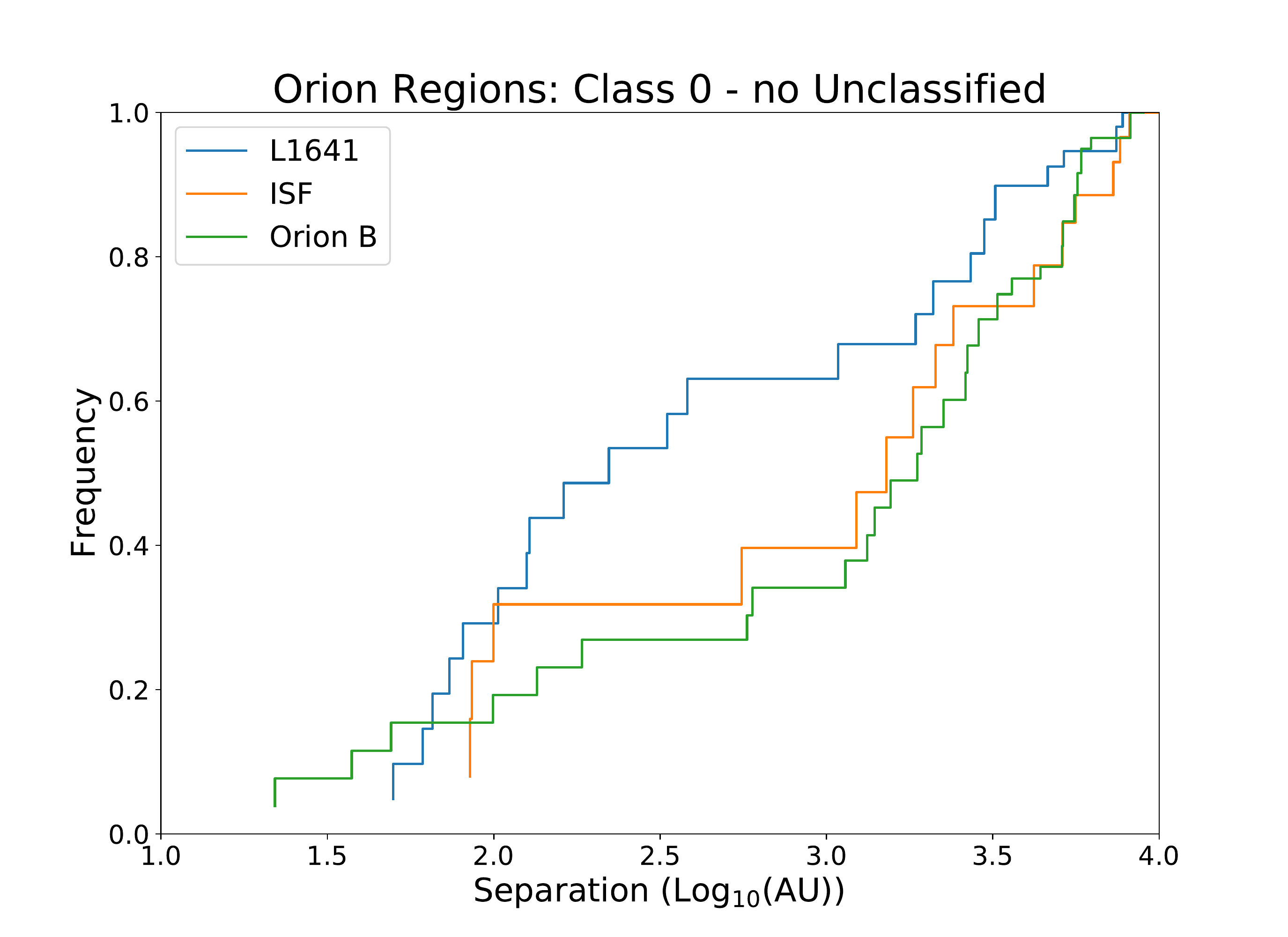}
\includegraphics[scale=0.3]{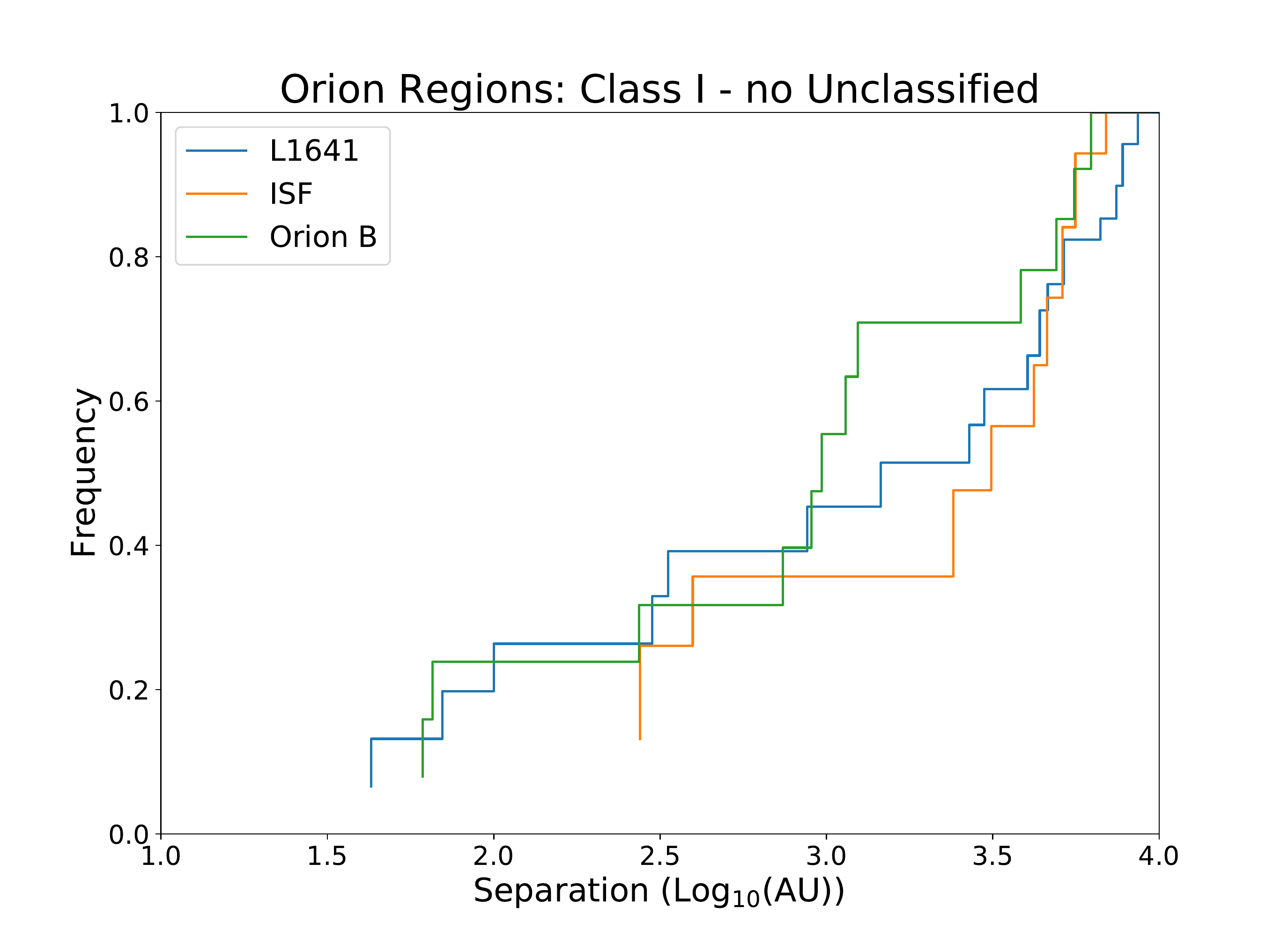}
\includegraphics[scale=0.3]{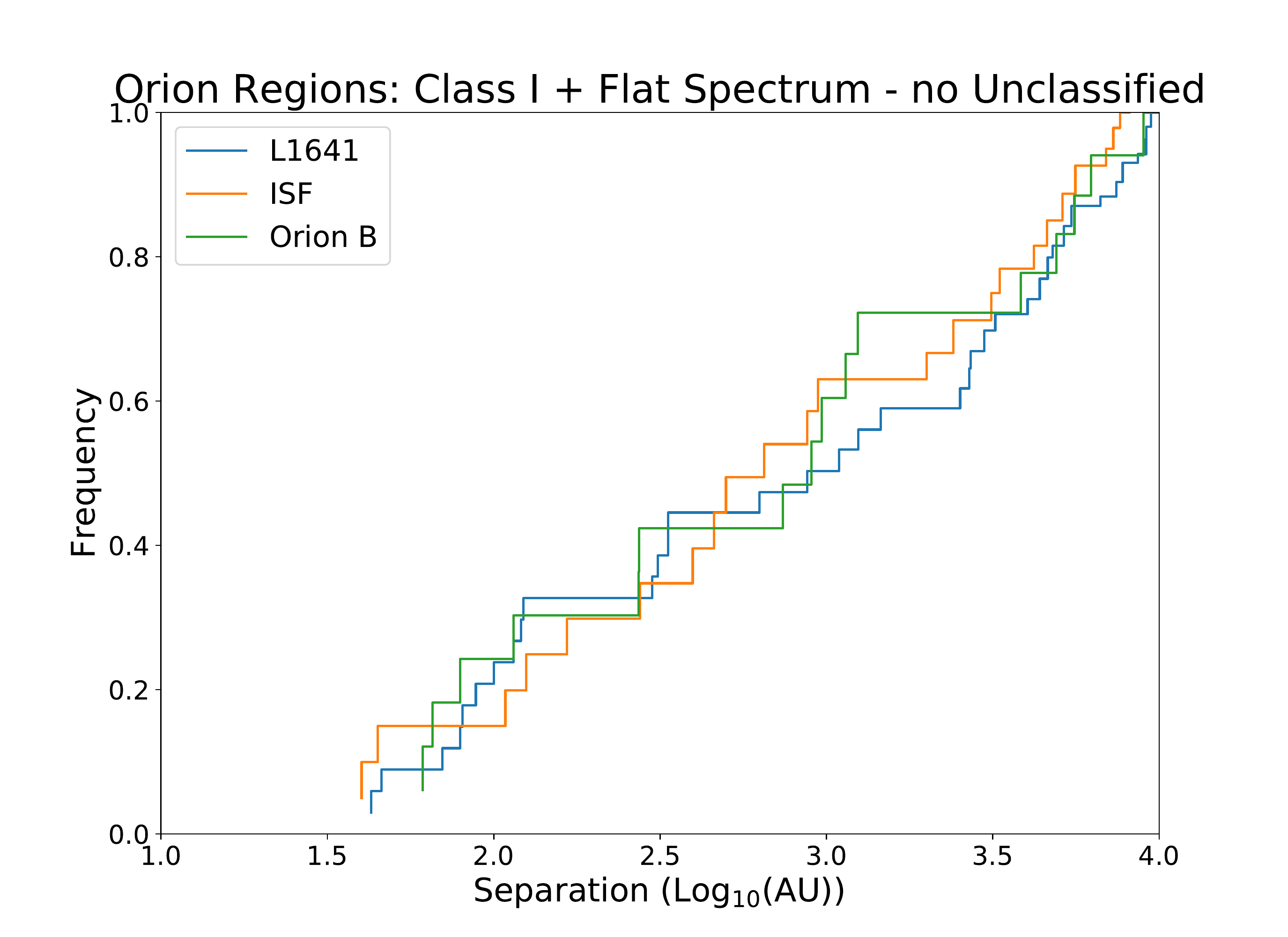}
\includegraphics[scale=0.3]{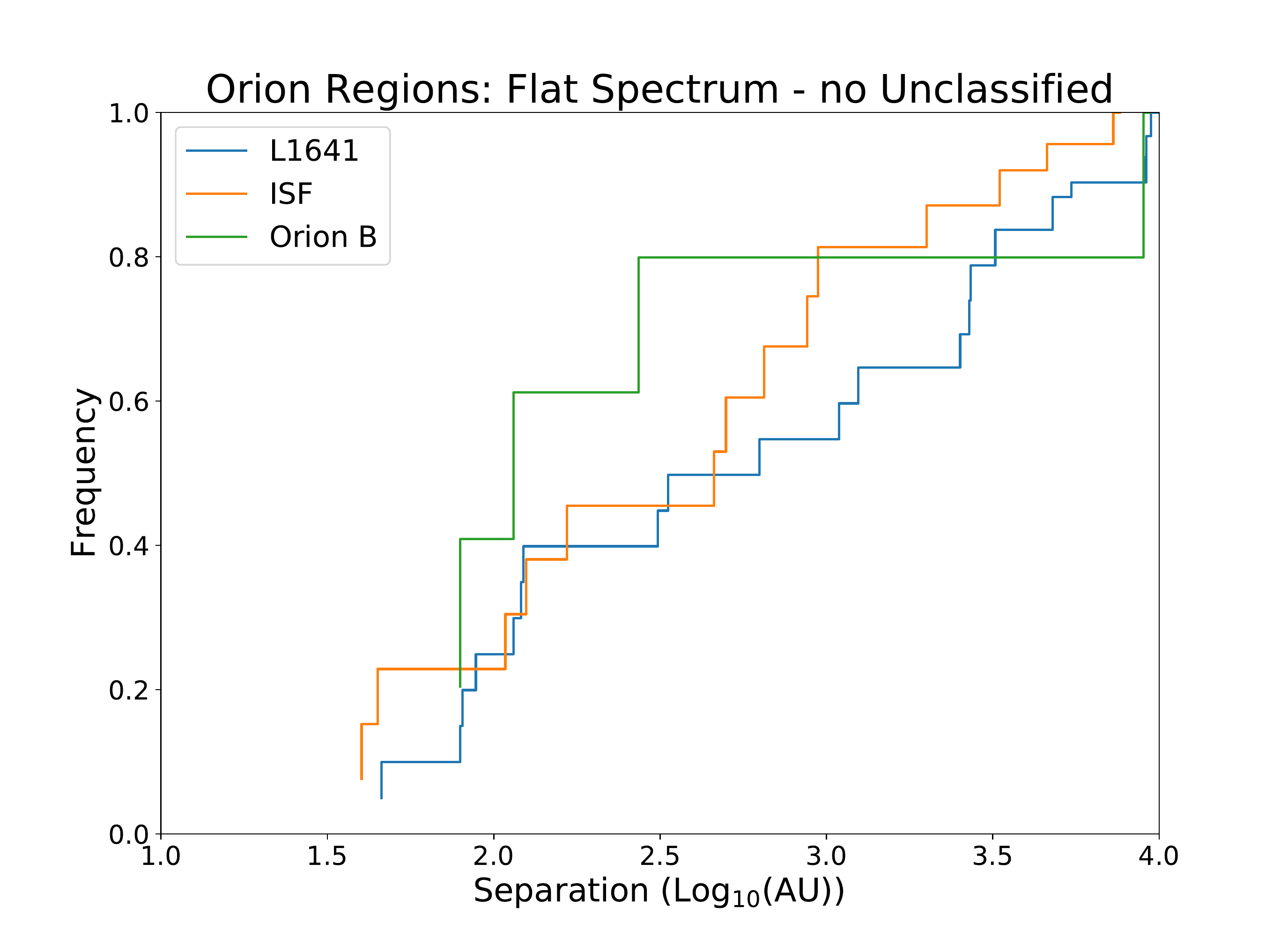}
\end{center}
\caption{
Cumulative separation distributions for the full samples of each region in Orion, and a log-flat
distribution is also drawn for comparison
for comparison to the full samples of companions.
In each protostellar class, all the regions are consistent with being drawn from the same
parent distribution, 
except for Class 0s in L1641 and Orion B.
}
\label{cumulative_regions}
\end{figure}

We compared the probability-weighted cumulative separation distributions 
for the regions in Figure \ref{cumulative_regions}. For the full sample and
samples divided by Class, only the 
Class 0--Class 0 separation distribution
in L1641 vs. Orion B regions (separations $<$ 10$^4$~au) reject the null 
hypothesis (that they are drawn from the same parent distribution) 
with p $<$ 0.01,
other classes do not show statistically significant differences
between regions. Also, when compared to a log-flat separation distribution,
only the Orion B Class 0s 
are inconsistent with log-flat.

\begin{figure}
\begin{center}
\includegraphics[scale=0.3]{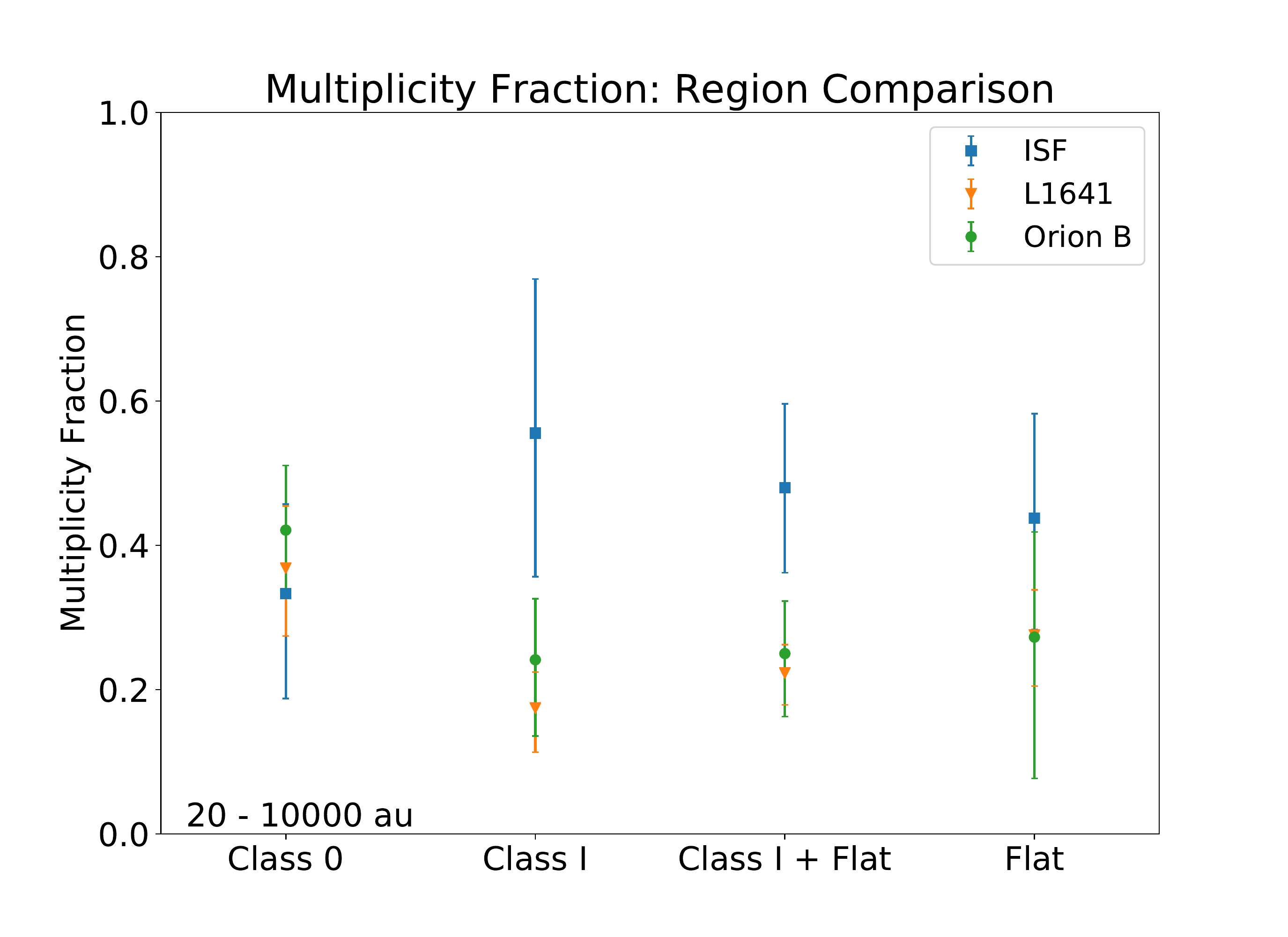}
\includegraphics[scale=0.3]{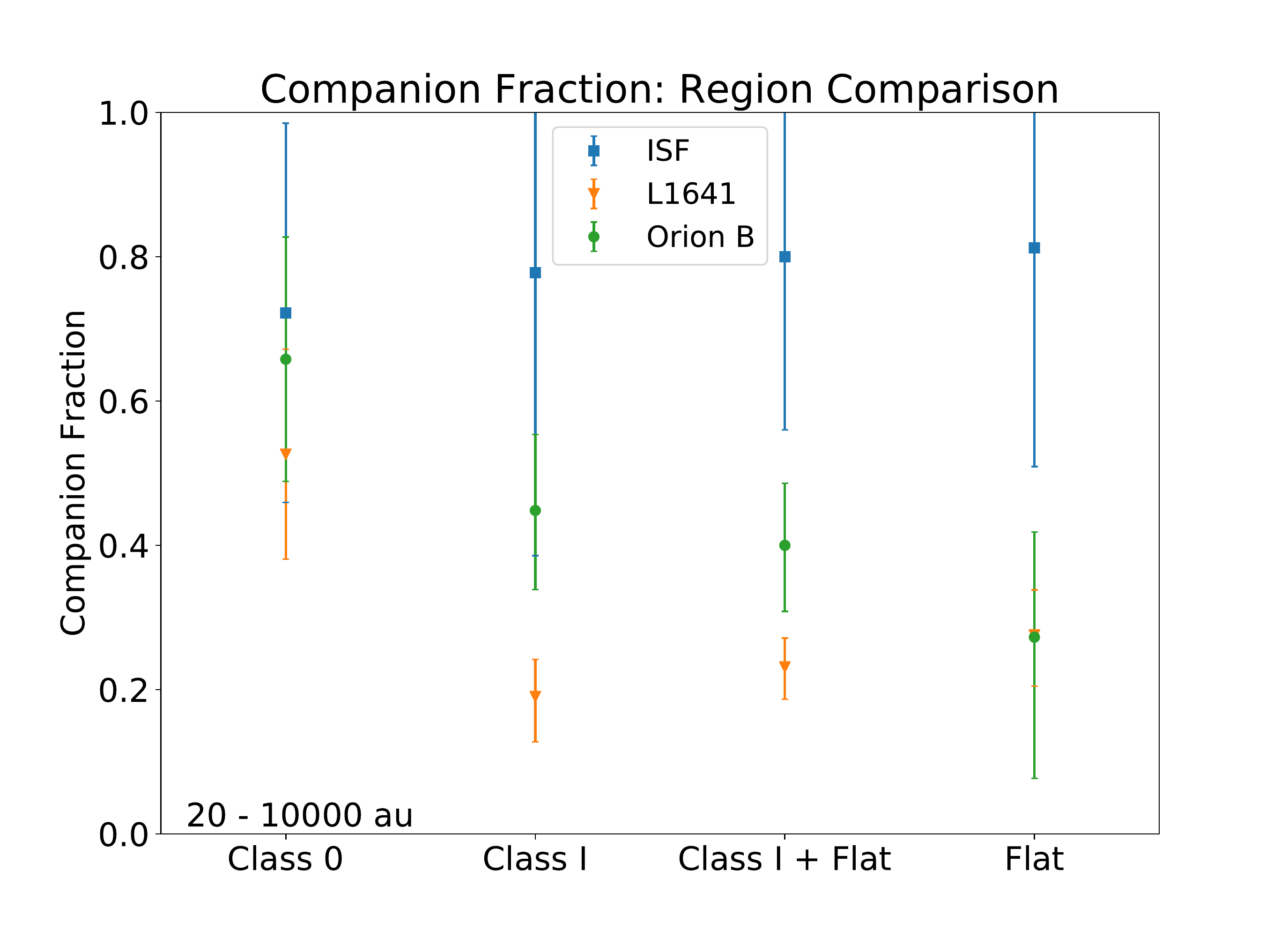}
\includegraphics[scale=0.3]{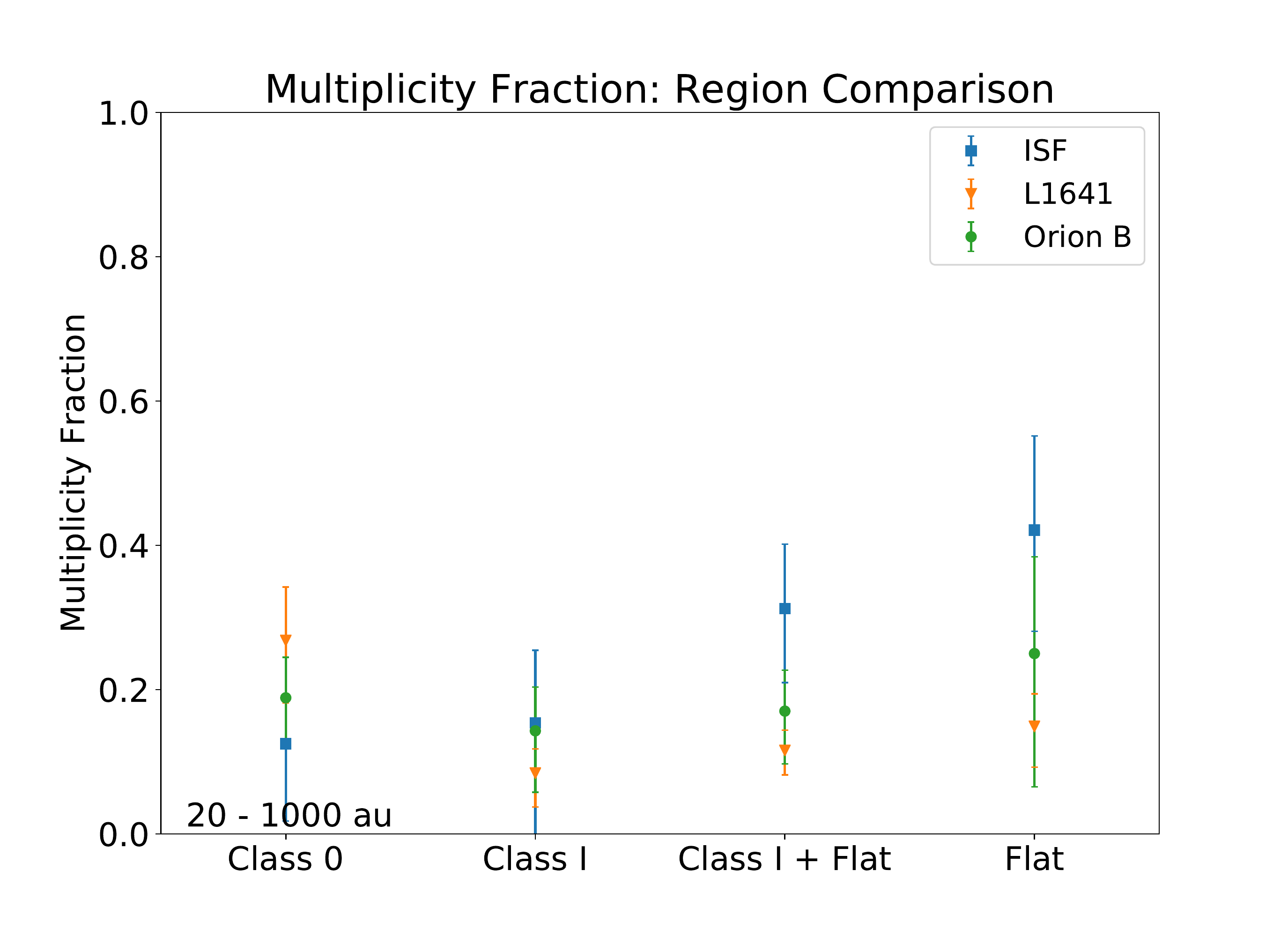}
\includegraphics[scale=0.3]{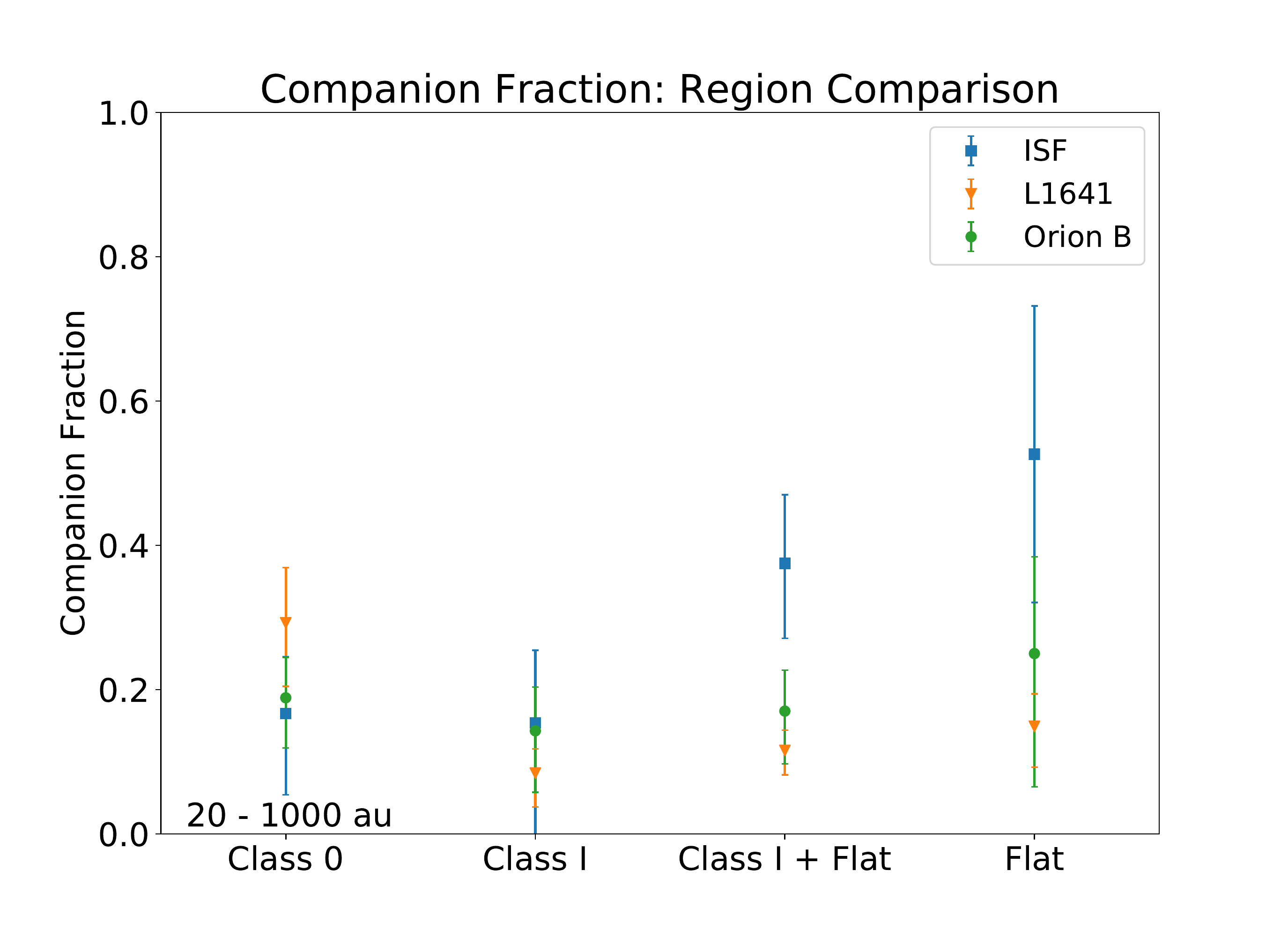}
\includegraphics[scale=0.3]{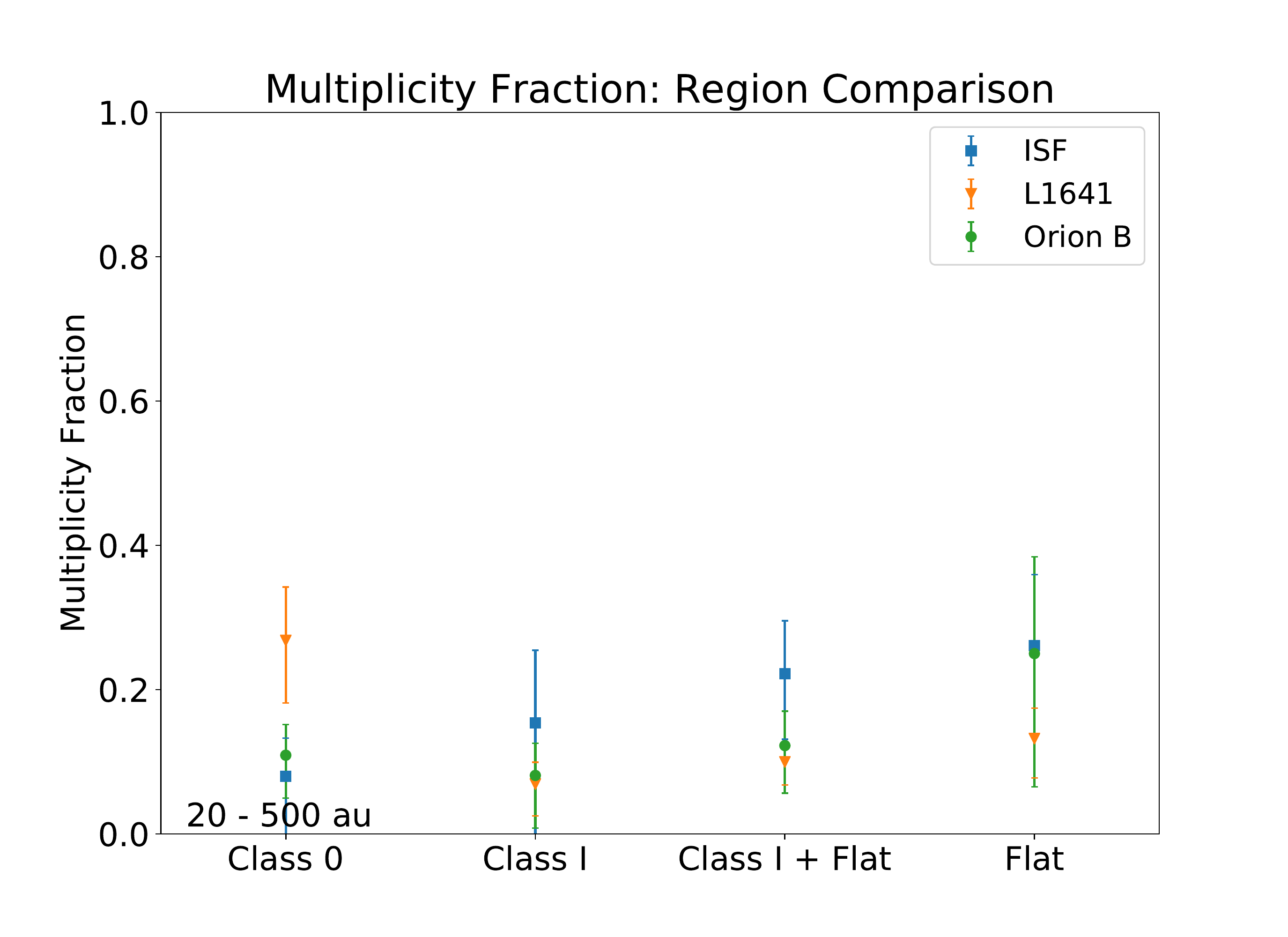}
\includegraphics[scale=0.3]{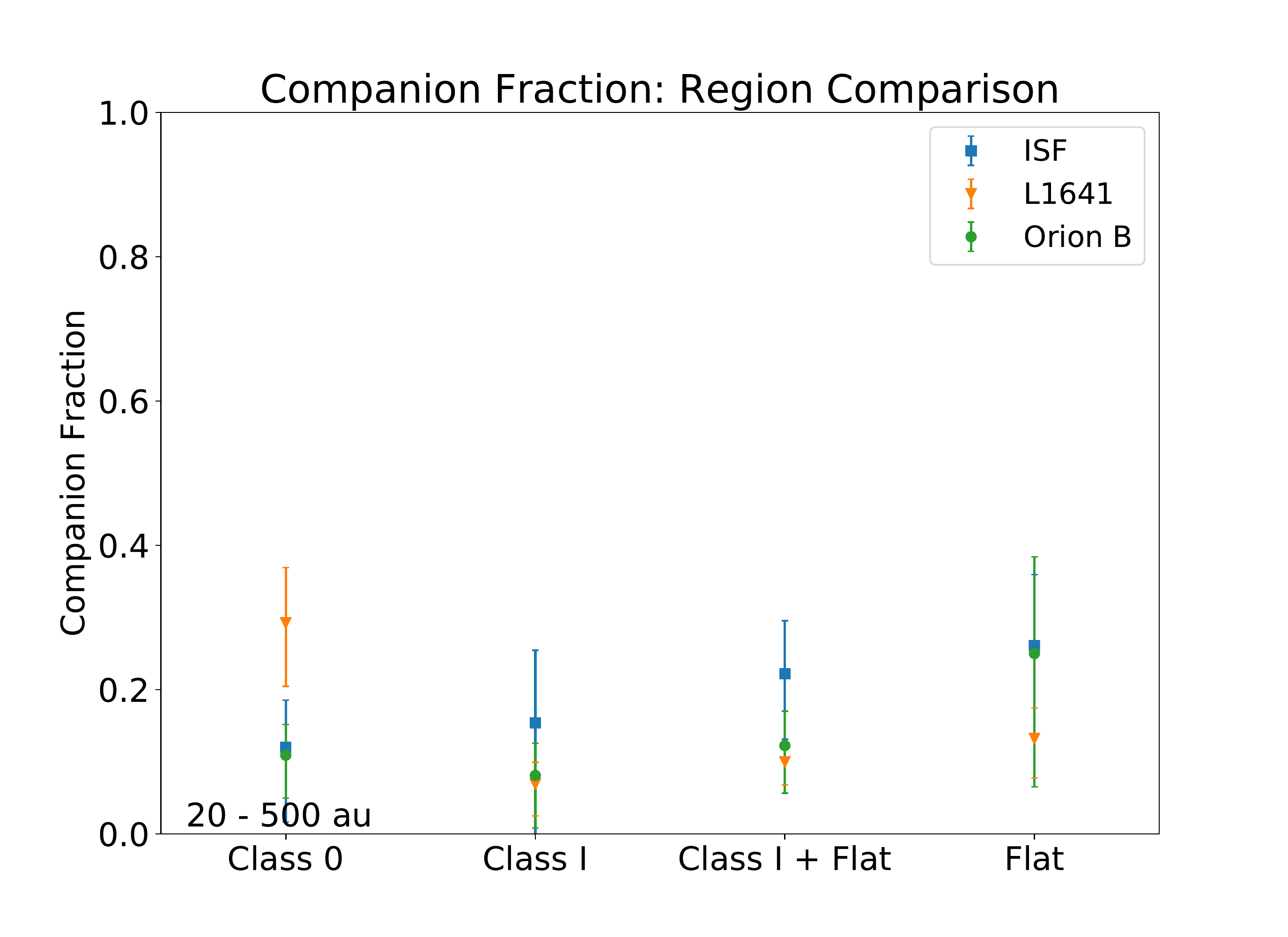}

\end{center}
\caption{
MFs and CFs on different scales as a function of protostar class for
different regions within Orion. 
The MFs and CFs of the non-Class 0 protostars in the ISF are systematically higher
than the other regions for separations 20 to 10$^4$~au, but the
differences are not statistically significant.
}
\label{mf_csf_regions}
\end{figure}

The MFs and CFs for the Orion regions are given in Table \ref{MF-CSFs-regions}, and 
we show them graphically as a function of protostellar class in Figure \ref{mf_csf_regions}.
For separations of 20 to 10$^4$ au, Class 0 protostars in all regions have MFs and 
CFs that are consistent within their uncertainties. 
Then, while the MFs of the ISF
are systematically higher for more-evolved protostars, the differences are all $<$2$\sigma$
and not definitive. 
Then the CFs for the more evolved protostars in the ISF,
the values are again systematically higher 
than the other regions, but are not statistically significant with
all $<$2$\sigma$. If the full sample of the regions are considered, the
difference between the CFs in the ISF and L1641 is $\sim$1.9$\sigma$.
Thus, the elevated CFs in the ISF suggest a potential dependence of multiplicity
difference between regions, but firm conclusions 
cannot be drawn.

In the 20 to 10$^3$~au separation range, the MFs and CFs for the Class 0 protostars in the three regions are consistent within the uncertainties. 
The ISF does have elevated CF values for Class I + Flat Spectrum and Flat Spectrum protostars, but
they are just beyond 1$\sigma$.
The 20 to 500~au separation range tells a similar story, without statistically significant
differences in the MFs and CFs for the different regions. However, there is a hint
that the L1641 Class 0 protostars could have a higher MF and CF in this separation range, but
the significance of this difference is only sightly greater than 1$\sigma$.

Overall, the only statistically significant difference between regions is found in the Class 0
separation distribution for L1641 vs. Orion B. There are no significant differences in the MFs or CFs
between regions, however, the ISF MFs and CFs are systematically higher than those of other regions, 
providing a hint that it might have a difference. The main difference between the ISF and the other
regions is YSO density, being much higher in the ISF, and we discussed in Sections 
4.5 and 5.3 that regions with high YSO densities could be more favorable to the formation of multiples.

\begin{small}
\bibliographystyle{apj}
\bibliography{ms}
\end{small}

\clearpage

\clearpage

\begin{deluxetable}{lllllllll}
\tablewidth{0pt}
\tabletypesize{\scriptsize}
\tablecaption{Orion Catalog}
\tablehead{
  \colhead{Source}  & \colhead{Main Source} & \colhead{RA} & \colhead{Decl.} & \colhead{L$_{\rm bol}$}  &   \colhead{T$_{\rm bol}$} & \colhead{Class} & \colhead{$\Sigma_{YSO}$} & \colhead{Distance}\\
  \colhead{}         & \colhead{}           & \colhead{(J2000)} & \colhead{(J2000)} & \colhead{(L$_{\sun}$)}  &   \colhead{(K)} & \colhead{} & \colhead{(pc$^{-2}$)} & \colhead{(pc)}
 }
\startdata
HH270VLA1-A & HH270VLA1 & 05:51:34.587 & +02:56:46.01 & 7.34 & 32.00 & 0 & 2.2 & 430.1\\
HH270VLA1-B & HH270VLA1 & 05:51:34.600 & +02:56:45.88 & 7.34 & 32.00 & 0 & 2.2 & 430.1\\
HH270mms1-B & HH270mms1 & 05:51:22.632 & +02:56:06.80 & 7.74 & 72.00 & I & 2.2 & 405.7\\
HH270mms1-A & HH270mms1 & 05:51:22.717 & +02:56:04.98 & 7.74 & 72.00 & I & 2.2 & 405.7\\
HH270mms2 & HH270mms2 & 05:51:22.572 & +02:55:43.16 & 4.55 & 249.00 & Flat & 2.2 & 413.3\\

\enddata
\tablecomments{Table 1 is published in its entirety in the machine-readable format.
 A portion is shown here for guidance regarding its form and content.}
\label{input-table-orion}
\end{deluxetable}

\begin{deluxetable}{llllllll}

\tablewidth{0pt}
\tabletypesize{\scriptsize}
\tablecaption{Perseus Catalog}
\tablehead{
  \colhead{Source}  & \colhead{Main Source} & \colhead{RA} & \colhead{Decl.} & \colhead{L$_{\rm bol}$}  &   \colhead{T$_{\rm bol}$} & \colhead{Class} & \colhead{$\Sigma_{YSO}$}\\
  \colhead{}         & \colhead{}           & \colhead{(J2000)} & \colhead{(J2000)} & \colhead{(L$_{\sun}$)}  &   \colhead{(K)} & \colhead{} & \colhead{(pc$^{-2}$)}
 }
\startdata
Per-emb-53 & Per-emb-53 & 03:47:41.591 & +32:51:43.67 & 8.00 & 287.00 & I & 0.6\\
Per-emb-56 & Per-emb-56 & 03:47:05.450 & +32:43:08.24 & 0.92 & 312.00 & I & 1.4\\
IRAS 03363+3207 & IRAS 03363+3207 & 03:39:25.546 & +32:17:07.08 & -99.00 & -99.00 & Unclassified & 0.6\\
EDJ2009-385 & EDJ2009-385 & 03:44:18.168 & +32:04:56.90 & 0.65 & 1200.00 & II & 93.7\\
Per-emb-16 & Per-emb-16 & 03:43:50.978 & +32:03:24.10 & 0.68 & 39.00 & 0 & 162.5\\

\enddata
\tablecomments{Table 2 is published in its entirety in the machine-readable format.
 A portion is shown here for guidance regarding its form and content.}
\label{input-table-perseus}
\end{deluxetable}

\begin{longrotatetable}
\begin{deluxetable}{lllllll}
\tablewidth{0pt}
\tabletypesize{\tiny}
\tablecaption{Orion Protostar Pairings}
\tablehead{
  \colhead{Source Pair}  & \colhead{Separation} & \colhead{Separation}  &   \colhead{0.87~mm Flux Ratio}  & \colhead{9~mm Flux Ratio} & \colhead{Prob.} & \colhead{Classes}\\
  \colhead{}  & \colhead{(\arcsec)} & \colhead{(au)}  &   \colhead{(Log $\frac{F1}{F2}$)} & \colhead{(Log $\frac{F1}{F2}$)} & \colhead{} & \colhead{}\\
 }
\startdata
H361-E-A+H361-E-B & 0.055$\pm$0.015 & 21.8$\pm$6.1 & \nodata & -0.49$\pm$0.15 & 1.00 & C0+C0\\ 
H361-C-A+H361-C-B & 0.087$\pm$0.002 & 37.6$\pm$1.0 & \nodata & -0.77$\pm$0.01 & 1.00 & C0+C0\\ 
H85-B+H85-A & 0.102$\pm$0.001 & 40.2$\pm$0.6 & -0.15$\pm$0.02 & -0.92$\pm$0.05 & 1.00 & FS+FS\\ 
H20-A+H20-B & 0.110$\pm$0.002 & 42.9$\pm$1.0 & -0.98$\pm$0.44 & -0.50$\pm$0.14 & 1.00 & CI+CI\\ 
H77-A-B+H77-A-A & 0.114$\pm$0.004 & 44.6$\pm$1.4 & -0.08$\pm$0.03 & \nodata & 1.00 & FS+FS\\ 
H158-B+H158-A & 0.119$\pm$0.007 & 46.0$\pm$2.5 & -0.02$\pm$0.07 & \nodata & 1.00 & FS+FS\\ 
H384-A+H384-A-B & 0.120$\pm$0.003 & 49.0$\pm$1.2 & \nodata & -0.83$\pm$0.03 & 1.00 & C0+C0\\ 
H203-B+H203-A & 0.130$\pm$0.003 & 49.7$\pm$1.1 & -0.89$\pm$0.02 & -0.86$\pm$0.02 & 1.00 & C0+C0\\
\enddata
\tablecomments{Classes are provided in abbreviated form: C0 - Class 0, CI - Class I, FS - Flat Spectrum,
CII - Class II. Components classified as `N' do not have a classification. 
Flux ratios are only provided for pairings of single components. The column labeled Prob. corresponds to the probability
of the pairing being a true, bound pair and not a line of sight association, based on the YSO surface density.
The overall probability of each member may be lower given the multiplicative nature
of companion probabilities when pairing two binary (or higher order systems) together.
}
\label{orion-seps}
\end{deluxetable}
\end{longrotatetable}

\begin{longrotatetable}
\begin{deluxetable}{llllll}
\tablewidth{0pt}
\tabletypesize{\tiny}
\tablecaption{Perseus Protostar Pairings}
\tablehead{
  \colhead{Source Pair}  & \colhead{Separation} & \colhead{Separation}  &  \colhead{9~mm Flux Ratio} & \colhead{Prob.} & \colhead{Classes}\\
  \colhead{}  & \colhead{(\arcsec)} & \colhead{(au)}  &  \colhead{} & \colhead{} & \colhead{}\\
 }
\startdata
P2-A-P2-B & 0.080$\pm$0.005 & 24.1$\pm$1.5 & -0.17$\pm$0.09 & 1.00 & C0+C0\\ 
P18-A-P18-B & 0.085$\pm$0.004 & 25.5$\pm$1.2 & -0.03$\pm$0.11 & 1.00 & C0+C0\\ 
P5-A-P5-B & 0.097$\pm$0.003 & 29.1$\pm$1.0 & -0.26$\pm$0.03 & 1.00 & C0+C0\\ 
L1448NW-A-L1448NW-B & 0.251$\pm$0.003 & 75.3$\pm$1.0 & -0.17$\pm$0.02 & 1.00 & C0+C0\\ 
P33-B-P33-C & 0.265$\pm$0.008 & 79.6$\pm$2.3 & -0.14$\pm$0.04 & 1.00 & C0+C0\\ 
P17-A-P17-B & 0.278$\pm$0.008 & 83.4$\pm$2.5 & -0.75$\pm$0.04 & 1.00 & C0+C0\\ 
P44-A-P44-B & 0.300$\pm$0.003 & 90.1$\pm$1.0 & -0.48$\pm$0.02 & 1.00 & C0+C0\\ 
P36-A-P36-B & 0.311$\pm$0.006 & 93.3$\pm$1.8 & -0.80$\pm$0.02 & 1.00 & CI+CI\\
\enddata
\tablecomments{Classes are provided in abbreviated form: C0 - Class 0, CI - Class I, FS - Flat Spectrum,
CII - Class II. Components classified as \textit{None} do not have a classification. 
Flux ratios are only provided for pairings of single components. The column labeled Prob. corresponds to the probability
of the pairing being a true, bound pair and not a line of sight association, based on the YSO surface density.}
\label{perseus-seps}
\end{deluxetable}
\end{longrotatetable}

\begin{deluxetable}{lllll}
\tablewidth{0pt}
\tabletypesize{\tiny}
\tablecaption{Orion and Perseus Multiplicity and Companion Fractions with Probabilities}
\tablehead{
  \colhead{Sample/Sub-sample}  & \colhead{Separation Range} & \colhead{S:B:T:Q:5:6:7:8:9:10:11}  &   \colhead{MF} & \colhead{CF}\\
 }
\startdata
Orion All                            & 20 - 10000 & 218:70:10:2:4:1:0:0:0:1:0:0 & 0.29$_{-0.03}^{+0.03}$ &  0.41$_{-0.03}^{+0.03}$  \\
Orion All - no Unclassified          & 20 - 10000 & 195:68:9:2:4:1:0:0:0:1:0:0 & 0.30$_{-0.03}^{+0.03}$ &  0.44$_{-0.03}^{+0.03}$  \\
Orion Class 0 - (Class 0, Class I, Flat) & 20 - 10000 &  58:27:5:1:1:1:0:0:0:1:0:0 & 0.38$_{-0.05}^{+0.06}$ &  0.62$_{-0.10}^{+0.10}$  \\
Orion Class 0 - Class 0              & 20 - 10000 &  58:16:4:0:0:0:0:0:0:1:0:0 & 0.27$_{-0.05}^{+0.06}$ &  0.42$_{-0.06}^{+0.06}$  \\
Orion Class I - (Class 0, Class I, Flat) & 20 - 10000 &  78:18:3:0:2:0:0:0:0:0:0:0 & 0.23$_{-0.04}^{+0.05}$ &  0.32$_{-0.05}^{+0.05}$  \\
Orion Class I - Class I              & 20 - 10000 &  78:16:1:0:1:0:0:0:0:0:0:0 & 0.19$_{-0.04}^{+0.05}$ &  0.23$_{-0.04}^{+0.05}$  \\
Orion Flat - (Class 0, Class I, Flat) & 20 - 10000 &  59:23:1:1:1:0:0:0:0:0:0:0 & 0.31$_{-0.05}^{+0.06}$ &  0.38$_{-0.06}^{+0.06}$  \\
Orion Flat - Flat                    & 20 - 10000 &  59:23:1:0:1:0:0:0:0:0:0:0 & 0.30$_{-0.05}^{+0.06}$ &  0.35$_{-0.06}^{+0.06}$  \\
\enddata
\tablecomments{Note that the uncertainties are calculated assuming binomial statistics (Wilson score
interval) for the MF and CF for CF $<$ 0.5. When the CF $>$ 0.5, we use Poisson uncertainties, see Section
2.4.2. The same uncertainties are applied for all tables in the text.
}
\label{MF-CSFs-all}
\end{deluxetable}

\begin{deluxetable}{lllll}
\tablewidth{0pt}
\tabletypesize{\scriptsize}
\tablecaption{Combined Multiplicity and Companion Fractions with Probabilities}
\tablehead{
  \colhead{Sample/Sub-sample}  & \colhead{Separation Range} & \colhead{S:B:T:Q:5:6:7:8:9:10:11}  &   \colhead{MF} & \colhead{CF}\\
 }
\startdata
Combined All                            & 20 - 10000 & 263:88:14:4:5:1:0:0:0:1:0:0 & 0.30$_{-0.02}^{+0.03}$ & 0.43$_{-0.03}^{+0.03}$ \\
Combined All - no Unclassified          & 20 - 10000 & 232:84:13:4:5:1:0:0:0:1:0:0 & 0.32$_{-0.03}^{+0.03}$ & 0.46$_{-0.03}^{+0.03}$ \\
Combined Class 0 - (Class 0, Class I, Flat) & 20 - 10000 &  76:38:7:3:2:1:0:0:0:1:0:0 & 0.41$_{-0.05}^{+0.05}$ & 0.65$_{-0.09}^{+0.09}$ \\
Combined Class 0 - Class 0              & 20 - 10000 &  76:25:5:1:0:0:0:0:0:1:0:0 & 0.30$_{-0.05}^{+0.05}$ & 0.44$_{-0.05}^{+0.05}$ \\
Combined (Class I, Flat) - (Class 0, Class I, Flat) & 20 - 10000 & 156:46:6:1:3:0:0:0:0:0:0:0 & 0.26$_{-0.03}^{+0.03}$ & 0.34$_{-0.03}^{+0.04}$ \\
Combined (Class I, Flat) - (Class I, Flat) & 20 - 10000 & 156:44:2:0:2:0:0:0:0:0:0:0 & 0.24$_{-0.03}^{+0.03}$ & 0.27$_{-0.03}^{+0.03}$ \\
\hline
Combined All                            & 20 - 1000 & 369:77:5:0:0:0:0:0:0:0:0:0 & 0.18$_{-0.02}^{+0.02}$ & 0.19$_{-0.02}^{+0.02}$ \\
Combined All - no Unclassified          & 20 - 1000 & 330:72:5:0:0:0:0:0:0:0:0:0 & 0.19$_{-0.02}^{+0.02}$ & 0.20$_{-0.02}^{+0.02}$ \\
\enddata
\tablecomments{}
\label{MF-CSFs-comb}
\end{deluxetable}

\clearpage

\begin{deluxetable}{lllll}

\tablewidth{0pt}
\tabletypesize{\scriptsize}
\tablecaption{Multiplicity and Companion Fractions at High and Low YSO Surface Densities}
\tablehead{
  \colhead{Sample/Sub-sample}  & \colhead{Separation Range} & \colhead{S:B:T:Q:5:6:7:8:9:10:11}  &   \colhead{MF} & \colhead{CF}\\
 }
\startdata
$>=$ 30 pc$^{-2}$ All                            & 20 - 10000 & 106:39:10:1:4:1:0:0:0:1:0 & 0.35$_{-0.04}^{+0.04}$ & 0.57$_{-0.07}^{+0.07}$ \\
$>=$ 30 pc$^{-2}$ All - no Unclassified          & 20 - 10000 &  83:38:9:1:4:1:0:0:0:1:0 & 0.39$_{-0.04}^{+0.05}$ & 0.65$_{-0.09}^{+0.09}$ \\
$>=$ 30 pc$^{-2}$ Class 0 - (Class 0, Class I, Flat) & 20 - 10000 &  35:15:5:0:1:1:0:0:0:1:0 & 0.40$_{-0.07}^{+0.07}$ & 0.74$_{-0.15}^{+0.15}$ \\
$>=$ 30 pc$^{-2}$ Class I - (Class 0, Class I, Flat) & 20 - 10000 &  22:9:3:0:2:0:0:0:0:0:0 & 0.39$_{-0.09}^{+0.10}$ & 0.64$_{-0.17}^{+0.17}$ \\
$>=$ 30 pc$^{-2}$ Flat - (Class 0, Class I, Flat) & 20 - 10000 &  26:14:1:1:1:0:0:0:0:0:0 & 0.40$_{-0.08}^{+0.09}$ & 0.53$_{-0.14}^{+0.14}$ \\
$>=$ 30 pc$^{-2}$ (Class I, Flat) - (Class 0, Class I, Flat) & 20 - 10000 &  48:23:4:1:3:0:0:0:0:0:0 & 0.39$_{-0.06}^{+0.06}$ & 0.58$_{-0.11}^{+0.11}$ \\
\hline
$>=$ 30 pc$^{-2}$ All                            & 20 - 1000 & 170:36:4:0:0:0:0:0:0:0:0 & 0.19$_{-0.03}^{+0.03}$ & 0.21$_{-0.03}^{+0.03}$ \\
$>=$ 30 pc$^{-2}$ All - no Unclassified          & 20 - 1000 & 141:34:4:0:0:0:0:0:0:0:0 & 0.21$_{-0.03}^{+0.04}$ & 0.23$_{-0.03}^{+0.04}$ \\
\enddata
\tablecomments{}
\label{MF-CSFs-density}
\end{deluxetable}

\begin{deluxetable}{lllll}

\tablewidth{0pt}
\tabletypesize{\tiny}
\tablecaption{Orion and Perseus Multiplicity and Companion Fractions \textbf{without Probabilities}}
\tablehead{
  \colhead{Sample/Sub-sample}  & \colhead{Separation Range} & \colhead{S:B:T:Q:5:6:7:8:9:10:11}  &   \colhead{MF} & \colhead{CF}\\
 }
\startdata
Orion All                            & 20 - 10000 & 191:62:9:5:4:4:1:1:0:0:1:0 & 0.31$_{-0.03}^{+0.03}$ &  0.55$_{-0.06}^{+0.06}$  \\
Orion All - no Unclassified          & 20 - 10000 & 185:57:8:4:4:3:1:0:0:0:1:0 & 0.30$_{-0.03}^{+0.03}$ &  0.50$_{-0.05}^{+0.05}$  \\
Orion Class 0 - (Class 0, Class I, Flat) & 20 - 10000 &  51:21:3:2:1:3:1:0:0:0:1:0 & 0.39$_{-0.06}^{+0.06}$ &  0.82$_{-0.13}^{+0.13}$  \\
Orion Class 0 - Class 0              & 20 - 10000 &  51:16:2:0:0:0:0:0:0:0:1:0 & 0.27$_{-0.06}^{+0.06}$ &  0.43$_{-0.06}^{+0.07}$  \\
Orion Class I - (Class 0, Class I, Flat) & 20 - 10000 &  77:16:3:0:2:0:0:0:0:0:0:0 & 0.21$_{-0.04}^{+0.05}$ &  0.31$_{-0.05}^{+0.05}$  \\
Orion Class I - Class I              & 20 - 10000 &  77:15:1:0:1:0:0:0:0:0:0:0 & 0.18$_{-0.04}^{+0.05}$ &  0.22$_{-0.04}^{+0.05}$  \\
Orion Flat - (Class 0, Class I, Flat) & 20 - 10000 &  57:20:2:2:1:0:0:0:0:0:0:0 & 0.30$_{-0.05}^{+0.06}$ &  0.41$_{-0.06}^{+0.06}$  \\
Orion Flat - Flat                    & 20 - 10000 &  57:20:1:0:1:0:0:0:0:0:0:0 & 0.28$_{-0.05}^{+0.06}$ &  0.33$_{-0.06}^{+0.06}$  \\
\enddata
\tablecomments{We only provide the 20 to 10$^4$~au ranges here because it is the only range
affected by the companion probabilities.}
\label{MF-CSFs-all-noweight}

\end{deluxetable}

\begin{deluxetable}{lllll}

\tablewidth{0pt}
\tabletypesize{\scriptsize}
\tablecaption{Combined Multiplicity and Companion Fractions \textbf{without Probabilities}}
\tablehead{
  \colhead{Sample/Sub-sample}  & \colhead{Separation Range} & \colhead{S:B:T:Q:5:6:7:8:9:10:11}  &   \colhead{MF} & \colhead{CF}\\
 }
\startdata
Combined All                            & 20 - 10000 & 231:80:12:5:7:5:1:1:0:0:1:0 & 0.33$_{-0.03}^{+0.03}$ & 0.57$_{-0.05}^{+0.05}$ \\
Combined All - no Unclassified          & 20 - 10000 & 217:73:11:4:7:4:1:0:0:0:1:0 & 0.32$_{-0.03}^{+0.03}$ & 0.54$_{-0.05}^{+0.05}$ \\
Combined Class 0 - (Class 0, Class I, Flat) & 20 - 10000 &  67:31:5:2:4:4:1:0:0:0:1:0 & 0.42$_{-0.05}^{+0.05}$ & 0.86$_{-0.12}^{+0.12}$ \\
Combined Class 0 - Class 0              & 20 - 10000 &  67:24:3:0:1:0:0:0:0:0:1:0 & 0.30$_{-0.05}^{+0.05}$ & 0.46$_{-0.06}^{+0.06}$ \\
Combined (Class I, Flat) - (Class 0, Class I, Flat) & 20 - 10000 & 150:42:6:2:3:0:0:0:0:0:0:0 & 0.26$_{-0.03}^{+0.03}$ & 0.35$_{-0.04}^{+0.04}$ \\
Combined (Class I, Flat) - (Class I, Flat) & 20 - 10000 & 150:41:2:0:2:0:0:0:0:0:0:0 & 0.23$_{-0.03}^{+0.03}$ & 0.27$_{-0.03}^{+0.04}$ \\
\hline

\enddata
\tablecomments{We only provide the 20 to 10$^4$~au ranges here because it is the only range
affected by the companion probabilities.}
\label{MF-CSFs-comb-noweight}

\end{deluxetable}

\clearpage
\startlongtable
\begin{deluxetable}{lllll}
\tablewidth{0pt}
\tabletypesize{\tiny}
\tablecaption{Multiplicity and Companion Fractions for Orion Regions}
\tablehead{
  \colhead{Sample/Sub-sample}  & \colhead{Separation Range} & \colhead{S:B:T:Q:5:6:7:8:9:10:11}  &   \colhead{MF} & \colhead{CF}\\
 }
\startdata
ISF All                            & 20 - 10000 &  43:12:3:1:2:1:0:0:0:0:0:0 & 0.31$_{-0.06}^{+0.07}$ & 0.55$_{-0.12}^{+0.12}$ \\
ISF All - no Unclassified          & 20 - 10000 &  25:11:3:1:2:1:0:0:0:0:0:0 & 0.42$_{-0.08}^{+0.09}$ & 0.77$_{-0.18}^{+0.18}$ \\
ISF Class 0 - (Class 0, Class I, Flat) & 20 - 10000 &  12:4:0:0:1:1:0:0:0:0:0:0 & 0.33$_{-0.12}^{+0.15}$ & 0.72$_{-0.26}^{+0.26}$ \\
ISF Class I - (Class 0, Class I, Flat) & 20 - 10000 &   4:3:2:0:0:0:0:0:0:0:0:0 & 0.56$_{-0.21}^{+0.20}$ & 0.78$_{-0.39}^{+0.39}$ \\
ISF Flat - (Class 0, Class I, Flat) & 20 - 10000 &   9:4:1:1:1:0:0:0:0:0:0:0 & 0.44$_{-0.14}^{+0.15}$ & 0.81$_{-0.30}^{+0.30}$ \\
ISF (Class I, Flat) - (Class 0, Class I, Flat) & 20 - 10000 &  13:7:3:1:1:0:0:0:0:0:0:0 & 0.48$_{-0.12}^{+0.12}$ & 0.80$_{-0.24}^{+0.24}$ \\
\hline
ISF All                            & 20 - 1000 &  65:11:3:0:0:0:0:0:0:0:0:0 & 0.18$_{-0.04}^{+0.05}$ & 0.22$_{-0.05}^{+0.06}$ \\
ISF All - no Unclassified          & 20 - 1000 &  43:10:3:0:0:0:0:0:0:0:0:0 & 0.23$_{-0.06}^{+0.07}$ & 0.29$_{-0.06}^{+0.07}$ \\
\enddata
\tablecomments{}
\label{MF-CSFs-regions}
\end{deluxetable}

\end{document}